# Cause-of-death estimates for the early and late neonatal periods for 194 countries from 2000-2013

Shefali Oza[a]; Joy E Lawn[a,b]; Daniel R Hogan[c], Colin Mathers[c], Simon N Cousens[a]

a. MARCH, London School of Hygiene and Tropical Medicine, Keppel Street, London, WC1N 7HT, United Kingdom
b. Saving Newborn Lives, Save the Children
c. Department of Health Statistics and Information Systems, World Health Organization, Avenue Appia 20, 1211 Geneva 27, Switzerland

Corresponding author:
Professor Simon Cousens
London School of Hygiene and Tropical Medicine
Keppel Street
London, UK WC1N 7HT
Email: simon.cousens@lshtm.ac.uk

**Word count:** approx 3900

**Abstract**

**Objective**
Cause-of-death distributions are important for prioritising interventions. We estimated proportions, risks, and numbers of deaths (with uncertainty) for programme-relevant causes of neonatal death for 194 countries for 2000-2013, differentiating between the early (days 0-6) and late (days 7-27) neonatal periods.

**Methods**
For 65 high-quality VR countries, we used the observed early and late neonatal proportional cause distributions. For the remaining 129 countries, we used multinomial logistic models to estimate the early and late proportional cause distributions. We used separate models, with different inputs, for low and high neonatal mortality countries. We applied these cause-specific proportions to United Nations' neonatal death estimates by country/year to estimate cause-specific risks and numbers of deaths.

**Findings**
Of the 2.76 million neonatal deaths in 2013, 0.99 (uncertainty: 0.70-1.31) million (35.7%) were estimated to be from preterm complications, 0.64 (uncertainty: 0.46-0.84) million (23.4%) from intrapartum-related complications, and 0.43 (0.22-0.66) million (15.6%) from sepsis. Preterm (40.8%) and intrapartum-related (27.0%) complications accounted for the majority of early neonatal deaths while infections caused nearly half of late neonatal deaths. In every region, preterm was the leading cause of neonatal death, with the highest risks in Southern Asia (11.9 per 1000 livebirths) and Sub-Saharan Africa (9.5).

**Conclusion**
The neonatal cause-of-death distribution differs between the early and late periods, and varies with NMR level and over time. To reduce neonatal deaths, this knowledge must be incorporated into policy decisions. The Every Newborn Action Plan provides stimulus for countries to update national strategies and include high-impact interventions to address these causes.

Words: 250

## Introduction
Most of the 2.76 million neonatal (first month of life) deaths in 2013 occurred from preventable causes (1). While the global neonatal mortality rate (NMR) is decreasing, its rate of reduction has been substantially slower than the decreases in under-5 and maternal mortality (2, 3). Neonatal deaths now constitute 44% of all deaths in children under 5 years old (1). Many countries are unlikely to meet the child mortality target of the 4th Millennium Development Goal (MDG) by 2015, at least partly because of inadequate neonatal mortality reductions (4). The Every Newborn Action Plan, launched in June 2014 (5), provides a stimulus to accelerate progress by implementing effective cause-specific interventions that can rapidly reduce neonatal mortality.

Understanding the neonatal cause-of-death (COD) distribution is important for identifying appropriate interventions and program priorities. Such a distribution should be as local as possible (national or even subnational in large countries), current, and distinguish programmatically relevant causes. Moreover, separate COD estimates are required for the early (days 0-6) and late (days 7-27) neonatal periods since both our understanding of pathology and empirical data suggest that the COD distribution differs substantially between these periods (6, 7). Around three-quarters of neonatal deaths occur during the early period (8), and most interventions to prevent these deaths need to be delivered within a very short window of time.

For countries with high-quality vital registration (VR) data by cause and age at death, such data provide the information needed to determine policies and priorities. VR data quality is dependent on the completeness of reporting and the quality of cause-of-death coding (9, 10). Unfortunately, high-quality VR data are available for only about one-third of countries (11), which account for only about 4% of neonatal deaths. Thus, statistical modeling remains necessary to estimate cause-of-death distributions in the majority of countries.

Systematic estimates of neonatal deaths classified into programmatically relevant cause categories were first published in 2005, for the year 2000 (12), by the Child Health Epidemiology Research Group (CHERG). These estimates were developed using data from high-quality VR systems and from research studies in high mortality/low resource settings in which high-quality VR data were lacking. Updated neonatal cause-of-death estimates using this approach were subsequently published for 2008 (13) and 2010 (14).

The current work goes beyond these previous exercises by estimating neonatal causes of death separately for the early and late neonatal periods, and by adding injuries as a distinct cause for low mortality countries. The separation of early and late neonatal deaths is an important advance which will aid policy makers and programme managers. The input data have also been updated and modifications made to the modelling strategy, particularly for the split of neonatal infections between pneumonia and sepsis.

Here, we present global, regional, and national estimates (with uncertainty ranges) of proportions, risks, and numbers of deaths for key programmatically relevant neonatal causes of death by the early and late neonatal periods for 194 countries for 2000-2013.

## Methods
### *Overview of cause-of-death estimation*
We divided 194 countries into three groups based on the quality of their VR data and their child mortality rates (appendix A). Different methods were used to estimate the proportional cause-of-death (COD) distributions for countries in each group (figure 1). For the 65 countries with high-quality VR data, the proportional cause distribution from 2000-2013 was obtained directly from the country's VR data. For the 49 countries without high-quality VR but with low child mortality, this distribution was estimated using a multi-cause model ("low mortality model") with input data from high-quality VR countries. For the high-quality VR countries and low mortality model, we used seven cause categories: complications of preterm birth ("preterm"), intrapartum-related complications ("intrapartum"), congenital disorders, pneumonia, sepsis and other severe infections ("sepsis"), injuries, and other causes (panel 1). For the 80 countries with inadequate VR and high child mortality, we used studies that identified neonatal COD in high mortality settings as input data in a separate multi-cause model ("high mortality model"). The eight cause categories for this model were preterm, intrapartum, congenital disorders, pneumonia, diarrhea, neonatal tetanus, sepsis, and other causes (panel 1). In appendix B, we detail the methodological differences between this work and previous estimates.

*[appendix A: countries in the 3 models]*
*[appendix B: key methodological differences between this work and previous estimates]*
*[figure 1: flowchart of methods]*

***Data inputs***
Cause-of-death data from vital registration
For the 65 high-quality VR countries, we obtained publicly available VR cause-of-death data by the early and late neonatal periods from the World Health Organization (WHO) for years 2000 and later. We mapped the reported causes of death to our cause categories (appendix C). We then generated a proportional cause distribution by dividing the number of deaths attributed to each cause by the total deaths across the seven causes. To create a full time series, we imputed the cause-specific proportions for years with missing VR data (appendix D). The imputed proportions were only used as estimates for the high-quality VR countries; the low mortality model input dataset only included the non-imputed data.

*[appendix C: ICD 9/10 to CHERG conversion codes]*
*[appendix D: table with information about the VR input data]*

Cause-of-death data from high mortality setting studies
The high mortality model input data consisted of neonatal COD distribution data from studies in high mortality settings. We updated a previously developed database of neonatal COD studies (15) by conducting an extensive literature review for relevant research published from January 2011 to May 2013 (appendix E). For each study, we extracted COD data and, when necessary, re-categorized the causes into our cause categories. While we included injuries as a separate category in the low mortality model, the study data lacked enough information on injuries for separate estimation in the high mortality model. We recorded deaths separately by early or late neonatal period whenever possible.

*[appendix E: study data inclusion criteria]*
*[appendix F: full study list with relevant info]*

Covariate data
We chose for investigation covariates that we believed might partly predict variation in the COD distribution across countries. We could only use covariates for which national time series are publically available. The covariates were used in two ways: 1) as inputs into the multinomial models, and 2) as predictor variables to which the final model coefficients were applied to estimate the national proportional cause distributions. We used national-level covariates as inputs to the low mortality model since the input COD data are national, and, whenever possible, local-level data extracted from the studies for the high mortality model. When local-level data were unavailable, we used subnational- or national-level covariate data instead. For the predictions, we used national-level covariate data, with the exception of India for which we used state-level data to produce state-level estimates. We applied the same rules for imputing missing covariate data as for the VR data (appendix D). For prediction purposes, we restricted covariate values to the input data ranges, and performed a sensitivity analysis without this restriction.

*Statistical modelling*
All statistical analyses were done using Stata (version 12). For each model, a baseline cause was chosen; this cause was preterm for the low mortality model (the most common cause) and intrapartum for the high mortality model (reported in all studies). Our overall estimation process had two stages. First, we selected covariates. Then, we estimated the log of the ratio of each of the other causes to the baseline cause (the "log-cause ratio") as a function of the selected covariates using a multinomial logistic regression model. For both the low and high mortality models, we ran separate models for the early and late neonatal periods. Since not all studies in the high mortality model reported deaths by period, we included the studies reporting only overall neonatal deaths in both the early and late high mortality models, and included a binary covariate for period in these models.

Covariate selection for models
For each of the four models, we used a previously developed jackknife procedure (14) to select the set of covariates that minimized the out-of-sample prediction error for each log-cause ratio separately. First, we determined if the relationship between each covariate and log-cause ratio in the input data was best represented by a linear, quadratic, or restricted cubic spline relationship. We did this by choosing the covariate relationship which yielded the smallest chi-squared statistic (sum of the squared differences between observed and expected deaths divided by expected deaths) for the given log-cause ratio. We then selected the covariate with the smallest chi-squared statistic as the first covariate in the model. Finally, we added one covariate at a time, retaining it in the model only if the chi-squared statistic decreased and cycling through all the remaining covariates again. In this way, we selected a set of covariates for each non-baseline cause in each model.

Multi-cause models
The multi-cause multinomial logistic regression models fit the data for all causes simultaneously. Each input observation received a weight inversely proportional to the square root of the total deaths contributed by that observation. This weighting is intermediate between giving equal weight to each death and equal weight to each study or country year in the input data. We made assumptions about the cause category into which deaths from an unreported cause would have been assigned (appendix G). The coefficient values from the multinomial logistic regression models were applied to the country- and year-specific national level predictor covariates to estimate the proportional cause distribution for each modelled country from 2000-2013.

*[appendix G: assumptions about missing causes in the high mortality model input data]*

*Estimation of cause-specific deaths and risks*
We applied our estimated cause fractions to the neonatal deaths and live births estimated for each country from 2000-2013 by the Inter-agency Group for Child Mortality Estimation (IGME) (1). We split the overall neonatal deaths from IGME into early and late neonatal deaths. For high-quality VR countries, we took this split directly from the data. For the other countries, we assumed that 74% of neonatal deaths occurred during the early period and 26% in the late period based on previous work (8). We also performed a sensitivity analysis in which we assumed the early proportion was 65% or 85% instead of 74% (appendix H). To determine the number of deaths by cause, neonatal period, and year for each country, we applied the cause-specific proportions derived from the VR data (for high-quality VR countries) or the relevant models to the period-specific neonatal deaths for each country and year. We then divided these by the relevant country-specific live births to obtain the risk (per 1,000 live births) for each cause by neonatal period and year.

*[appendix H: sensitivity analysis assuming 65% or 85% for proportion of early neonatal deathsl]*

*Uncertainty estimation*
For the modelled estimates, we generated uncertainty estimates by drawing 1,000 bootstrap samples with replacement from the input data and re-running the multi-cause models to produce new proportional cause distributions. We took the 2.5$^{th}$ and 97.5$^{th}$ percentile values for each cause as the uncertainty bounds. For the high-quality VR data, we developed uncertainty estimates by assuming a Poisson distribution for the number of deaths (i.e. the standard error equaled the square root of the reported number of deaths).

**Results**
*Data inputs*
The high-quality VR input dataset, which was also used in the low mortality model, included 65 countries with 1,267,404 neonatal deaths and 665 country-years for each neonatal period from 2000-2013. Of these deaths, 75.8% occurred in the early period. Preterm and congenital were the most common causes during both neonatal periods (figure 2a).

The high mortality model input dataset included 112 data points consisting of 98,222 deaths from 36 countries (figure 1). This includes the addition of nearly 4100 neonatal deaths from 15 new studies, representing 10 countries across five MDG regions. The overall dataset had 31 observations for the early period, 18 for the late period, and 63 for the overall neonatal period. Seventy-eight observations had missing information on one or more cause, with pneumonia and diarrhoea being the causes most commonly unreported (appendix G). Preterm and intrapartum were the most common causes of death during the early period, while infections (sepsis and pneumonia) dominated during the late period (figure 2b).

*[figures 2a and 2b: box and whisker plots of the proportional cause distributions for the VR and high mortality model input data]*

While most covariate values were within the ranges of the input data, a few prediction covariates had values substantially outside the input data range, especially in the low mortality model (appendix I). Notable examples were Monaco with an average GNI about 3.5 times larger than the maximum input data and Egypt with female literacy and ANC coverage 15-20 percentage points below the input ranges. While we capped the prediction covariate data to the input ranges in the analysis, a sensitivity analysis without these caps suggested that the decision to cap or not had minimal influence on the results (appendix J).

*[appendix I: comparison of the input and prediction datasets in the multinomial models]*
*[appendix J: sensitivity analysis when prediction data is not restricted to ranges of the input data]*

*Statistical modelling*
<u>Covariate selection for models</u>
The model equations varied substantially in their performance, from 0% reduction in the chi-squared statistic (injuries, low mortality model, late period) to 87% reduction (diarrhea, high mortality model, early period) (appendix K). Overall, equations in the high mortality model appeared to have better performance, with an average of 50% reduction in residuals across the equations compared to an average of 30% in the low mortality model equations.

*[appendix K: selected covariates and their performance for each cause equation]*

<u>Multi-cause models</u>

Model regression coefficients are in appendix L.

*[appendix L: Regression coefficients for the models]*

### *Cause-specific deaths and risks*

Globally, the leading causes of neonatal death in 2013 were preterm (35.7%), intrapartum (23.4%), and sepsis (15.6%), accounting for 2.1 (uncertainty range: 1.4-2.8) of the 2.8 million neonatal deaths (table 1). There is little difference in these numbers when the proportion of early deaths is assumed to be 65% or 85% instead of 74% for modelled countries (appendix H). The proportional cause distribution varied by both neonatal period and NMR level. In the early period, preterm (40.8%) and intrapartum (27.0%) accounted for the majority of deaths while in the late neonatal period nearly half of all deaths occurred from infectious causes (sepsis, pneumonia, tetanus, and diarrhoea) (table 1). The proportion of deaths from congenital disorders was relatively stable across the periods. Higher NMR and lower income were associated with a higher proportion of deaths attributable to intrapartum and infectious causes (appendix M). The variation between regions appears to largely reflect NMR differences between regions. In low mortality settings, injuries accounted for less than 1% of neonatal deaths, and this fraction increased slightly from the early to late period (appendix M). See appendix N for model-specific results and appendix O for country-specific results.

*[table 1: global cause-specific proportions and number of neonatal deaths].*
*[appendix M: results by NMR level and income category].*
*[appendix N: proportions/risk by high-quality VR, low mortality model, and high mortality model].*
*[appendix O: country-specific results of proportions, risk, and number of deaths]*

Globally, risks for all causes have been decreasing as the global NMR decreases over time. The risk of death from each cause is substantially higher in higher mortality settings, even for causes that dominate proportionally in low mortality settings (e.g. preterm and congenital disorders). The risks of death due to preterm, intrapartum, and sepsis are 10, 36, and 34 times greater in settings with NMR≥30 compared to NMR≤5. In every MDG region, preterm is the leading cause of neonatal death, with the highest risks in Southern Asia (11.9 per 1000 live births) and Sub-Saharan Africa (9.5) (figure 3). While the absolute risks of death due to intrapartum and preterm have been falling in the early period, they have decreased less in the late period. Risk of tetanus has declined in both periods (figure 4).

*[figure 3: stacked bars of risk by region for 2013]*
*[figure 4: global area chart of time trends from 2000-2013 by neonatal period]*

## Discussion

We developed systematic, globally comparable estimates of programmatically relevant neonatal causes of death. Our current estimates go beyond our previous estimates in providing separate decompositions for the early and late neonatal periods. The proportional neonatal cause distribution varies with a number of factors, including the early versus late neonatal periods, NMR level, and over time. To reduce neonatal deaths, such variations must be understood, and this knowledge must be incorporated into decisions regarding the selection of appropriate interventions. With the launch of the Every Newborn Action Plan in June, this is a particularly opportune time for affecting such change within countries.

The three leading categories of causes of neonatal death (preterm, intrapartum, and infections) are the same for the early and late neonatal periods, but their distribution is very different between the two periods. Globally, in the early period, preterm and intrapartum account for nearly 68% of deaths while infections (pneumonia, sepsis, tetanus, and diarrhea) account for around 14%. In the late period, around 34% of deaths are due to intrapartum or preterm while roughly 48% are from infections. Intrapartum-related complications are expected to occur in the minutes or hours after birth, and hence cause more deaths during the early period. Infections usually take some time to develop, and thus become more common during the late period.

Even within each neonatal period, considerable differences exist in the proportional cause distribution by NMR level. Generally, NMR is closely linked to the level of care available to neonates. Settings with very low NMR tend to have readily available intensive care for newborns, while areas with high NMRs often lack even simple interventions like clean delivery kits and resuscitation equipment. In this analysis, low mortality countries had higher proportions of deaths from congenital disorders and lower proportions from intrapartum and infections, while the opposite was true in high mortality settings. While infection deaths accounted for 51% of deaths in the late period in high mortality countries, they caused less than 30% of late neonatal deaths in low mortality countries. This may be because of better access to and availability of treatment and infection control in low mortality countries.

We used our model to predict trends in individual causes of death. Our model predicts that deaths due to intrapartum-related complications had the largest absolute risk reduction between 2000 and 2013, from 7.2 (4.8-9.5) to 4.7 (3.4-6.1) per 1000 live births, possibly because of increased coverage of skilled obstetric care. The largest relative decrease in risk was predicted for neonatal tetanus, which dropped by 73% (from 1.1 [0.3-2.8] to 0.3 [0.1-0.9]) between 2000 and 2013. This may be due to increases in clean deliveries, facility birth, cord care, and tetanus toxoid vaccination (a predictor in the model), as well as contextual changes in maternal education and social norms. Additionally, a few countries in the low mortality model eliminated neonatal tetanus after 2000. Since tetanus is not estimated in the low mortality model, we may thus be underestimating the relative decline in risk.

The smallest predicted relative decrease in risk was for congenital (13% drop; from 2.4 [1.4-4.1] to 2.1 [1.3-3.3]). While the predicted risk of death due to preterm complications fell by 2.4 per 1,000 live births between 2000 and 2013 (from 9.6 [6.7-12.8] to 7.2 [5.1-9.5]), this represented the second smallest relative decrease (25%), despite the existence of simple and cost-effective interventions such as antenatal corticosteroids and Kangaroo Mother Care (8, 16). In addition to prematurity, there is also evidence that babies that are small-for-gestational age (SGA) are at higher risk of death (17).

Broadly, our results are similar to those from 2010 (14), with the exception that we estimate substantially fewer pneumonia deaths than before. This is likely due to improvements to the estimation approach, namely the inclusion of additional studies that split pneumonia from sepsis and the inclusion of pneumonia directly in the multinomial model (appendix B).

We used the low mortality model to estimate the proportional cause distribution in China due to its low NMR (18.8 in 2000, 7.7 in 2013), obtaining around 14.4% of neonatal deaths attributable to intrapartum-related complications. Others have also estimated the neonatal COD distribution in China and their results differ somewhat from ours, most notably with higher proportions of intrapartum-related deaths (18, 19). Our approach differs in three important ways: 1) we used multi-cause instead of single-cause models, 2) we estimated results for the early and late neonatal periods separately, and 3) we included input data from outside of China. We believe the first two are advantages in our work, while using China-only data is an advantage of the other models. The global COD distribution changes little when we apply the WHO COD proportions (20) for China instead of ours, with the biggest differences in 2013 being intrapartum-related and other deaths increasing by about 2 percentage points. Alternative COD distributions for China are shown in Appendix P for comparison.

*[appendix P: Comparison of different China estimates]*

Although the quantity and quality of data has improved in recent years, enormous data gaps still exist. We now have nearly 100,000 deaths and over 90 studies in the high mortality model compared to <14,000 deaths and <60 studies when we produced the estimates in 2005 (12). However, while we used the high mortality model for 80 countries, the inputs only included data from 36 countries. Many of these studies were relatively small, and few were nationally representative. We could include data from only 13 Sub-Saharan African countries, the region with the highest risk of neonatal death. Excluding a large South African dataset, the studies from these countries contribute only 4,000 deaths to our database. It is an unfortunate reality that we know the least about the areas with the highest burdens.

As with all such modeling exercises, our estimates should be viewed as an interim measure to help policymakers, particularly in settings with little or no data currently. It is important to distinguish *estimates* from *data* and to recognize that not all estimates are "equal". We used UN-IGME estimates of the NMR and total number of neonatal deaths in each country. The UN-IGME estimates of all-cause mortality in each country are derived from data for that country and therefore can be said to "track" mortality in each country. For most countries, our cause-specific estimates are not based on data from that country, but from a model bringing together data from many countries. The model then predicts the cause-of-death distribution, and changes in the cause-of-death distribution, in individual countries based on covariate values for the individual country. Some countries contribute little or no input data to the modeling process. For example, only 24 deaths in our input data come from Nigeria, one of the most populous countries. Our estimates should not therefore be interpreted as "tracking" changes in causes – of-death for the majority of countries, but rather as predictions of what might be occurring in countries. To track changes in burden due to specific causes of death requires each country to collect representative and consistent data on cause of death on a continuing basis. We emphasise once again that our estimates are not a panacea for actual data collection and should not be an excuse for a failure to collect data.

Fortunately, rapid but sensible improvements in data collection are possible. Recent examples of countries like South Africa improving their data systems to the point where their VR data is considered high quality are encouraging. In table 2, we provide a set of proposals for improving the counting of births and neonatal deaths. As data systems improve, regression-based models should be replaced by reliable local COD data.

*[table 2: recommendations for data improvements based on neonatal mortality level]*

The validity of our estimates relies on the quality of the input/prediction data, and our modelling techniques. Quality is of particular concern for verbal autopsy studies, in which the reported cause distributions depend heavily on the case definitions and causal hierarchies used to attribute deaths (10). Accurate cause attribution using VA will always be problematic for causes that are difficult to distinguish, such as sepsis and pneumonia, or difficult to identify, such as internal congenital abnormalities. The potential lack of comparability between different VA studies can affect the ability of our model to predict variation between settings. By following standardized methods when conducting VA studies, some of these problems can be partially alleviated (21). Additionally, regression-based models inherently depend on the relationship between outcome variables and covariates, which should ideally come from the same population and time period. While we sought to include as much local covariate information as possible for the input studies, 52% of the total covariate information came from national data instead of from local/regional data. Finally, when re-categorizing reported VA causes of death (panel 1), we had to make choices, for example placing deaths reported as being due to "very low birth weight" into the preterm complication category. This may introduce a degree of misclassification as some "very low birth weight" deaths may be attributable to congenital abnormalities. We made those choices that we believed would introduce the least misclassification, but until VA methods improve, this will continue to be a challenge. Similar issues exist in ICD coding, but are more common in VA studies because of the limited and lower quality information collected.

Even high-quality VR data can have problems. Unfortunately, ICD-10 codes are not ideal for neonatal causes, particularly because several programmatically relevant causes are relegated to the often-unused fourth digit in the codes. Codes for the upcoming ICD-11 revision are currently being drafted, providing the opportunity to develop more appropriate, clinically relevant coding for neonatal causes. Additionally, ICD coding practices can vary between and even within countries and over time (22, 23). Such variations reduce our model's ability to predict true variation in causes of death. Other issues in VR coding include changes during the transition between ICD revisions, differences in relegating certain causes to non-specific/ill-defined cause categories, and the assumption inherent in our exclusion of such codes that the deaths attributed to them are a random sample of all deaths. Finally, the availability and quality of VR data in a country may change over time, especially in countries with newly emerging surveillance systems. Developing consistent time trend estimates given such changes remains a challenge.

We used multinomial models because they naturally allow proportional cause distributions to sum to one, so that the sum of cause-specific deaths equals total deaths. Single-cause models require post-hoc adjustments to retain this property, but there may be limited information on which to base such adjustments. An important concern for both types of model, however, is the attribution of death to a single cause. This does not allow for co-morbidities, which are a frequent occurrence in neonatal deaths, and may thus underestimate the impact of a given cause.

Given the considerable variations in health systems and contextual factors within individual countries, sub-national neonatal COD estimates are needed and should be a target for future estimation exercises and data collection. National-level estimates like ours aim to ascertain the average causal distribution for a country, which can help guide national priorities, but may mask subnational variation. Some countries are beginning to collect the necessary information for sub-national estimates. For example, our national India estimates were produced by aggregating state-level estimates. In the future, we also hope to further differentiate causes within the current broad categories like congenital disorders. However, such differentiation may only be possible for VR-based models, as VA-based data generally lack the detail needed to do this. Finally, we strongly believe that the production of such estimates should be transparent. In accordance with this, the datasets and Stata code we used for this analysis are available on the WHO Global Health Observatory website.

Neonates constitute an important component of the unfinished agenda of MDG-4, and reducing preventable neonatal deaths will be essential to achieve the "grand convergence" to which the global community now aspires (24). The *Every Newborn Action Plan*, which calls for all countries to reduce their NMRs to 10 or fewer deaths per 1000 live births by 2035, provides fresh impetus for ending preventable newborn deaths (5), and a future in which every baby has an equal chance of survival.

## Conflicts of Interest
We declare that we have no conflicts of interest.

## Acknowledgements
We thank the Bill & Melinda Gates Foundation for funding through the Child Health Epidemiology Reference Group (CHERG) and Save the Children's Saving Newborn Lives programme. We also thank those who have contributed to earlier rounds of this work, including Katarzyna Wilczynska-Ketende, Alma Adler, and Susana Scott. The named authors alone are responsible for the views expressed in this manuscript, which do not necessarily reflect the opinion of the World Health Organization, its Member States, or of sponsoring agencies.

*Role of the funding source*
The sponsors of this work had no role in study design, data collection, data analysis, data interpretation, or writing of the report. SO, SC, JEL, DH, and CM had full access to all of the data and final responsibility for the decision to submit for publication.

**Panel 1: Definitions for the neonatal period and causes of death**

*Neonatal period definitions*
Neonatal period: days 0-27, or first month after birth
Early neonatal period: days 0-6, or first week after birth
Late neonatal period: days 7-27, or weeks 2-4 after birth

*Neonatal causes of death*

|  | Case definitions for neonatal causes of death | |
|---|---|---|
|  | **Used in VR and sought in study data** | **Accepted in study data** |
| **Preterm complications** | Neonatal deaths from one or more of the following:<br>- specific complications of preterm birth like surfactant deficiency (Respiratory Distress Syndrome), intraventricular haemorrhage, necrotizing enterocolitis<br>- immaturity (<34 weeks) at which level preterm specific complications occur for the majority of babies<br>- neonatal death with birth weight < 2,000 g where gestational age is unknown | - "Prematurity"<br>- "Very low birth weight" |
| **Intrapartum-related complications** | Neonatal death due to:<br>- neonatal encephalopathy with criteria suggestive of intrapartum events<br>- early neonatal death in a term baby with no congenital malformations and a specific history of acute intrapartum insult or obstructed labor | - "Birth asphyxia" with Apgar-based definition but excluding preterm infants<br>- Fits and/or coma in the first two days of life in a term baby<br>- Acute intrapartum complications |
| **Congenital disorders** | - Neonatal death due to major or lethal congenital abnormalities<br>- Specific abnormality listed (e.g. neural tube defect, cardiac defect) | Congenital abnormality or malformation |
| **Sepsis** | Neonatal deaths due to sepsis/septicaemia, meningitis, or neonatal infection | Neonatal infection |
| **Pneumonia** | Neonatal deaths due to pneumonia or acute respiratory tract infection | Pneumonia |
| **Neonatal tetanus** | Neonatal deaths due to tetanus | Spasms and poor feeding after age of 3 days |
| **Diarrhoea** | Neonatal deaths due to diarrhoea | --- |
| **Injuries** | Neonatal deaths due to injuries | N/A |
| **Other** | Specific causes not included in the above-listed causes, including:<br>- neonatal jaundice<br>- haemorrhagic disease of the newborn<br>- term baby dying due to in-utero growth restriction<br>- injuries (in study data only) | Author grouping of "other" (not unknown) |
| Notes: 1) This table is slightly modified from Lawn (25), which adapted it from Wigglesworth (26, 27) and NICE (27); 2) specific ICD codes for causes of death in the VR data are included in appendix C. | | |

# Figure 1: Flowchart of modelling methods

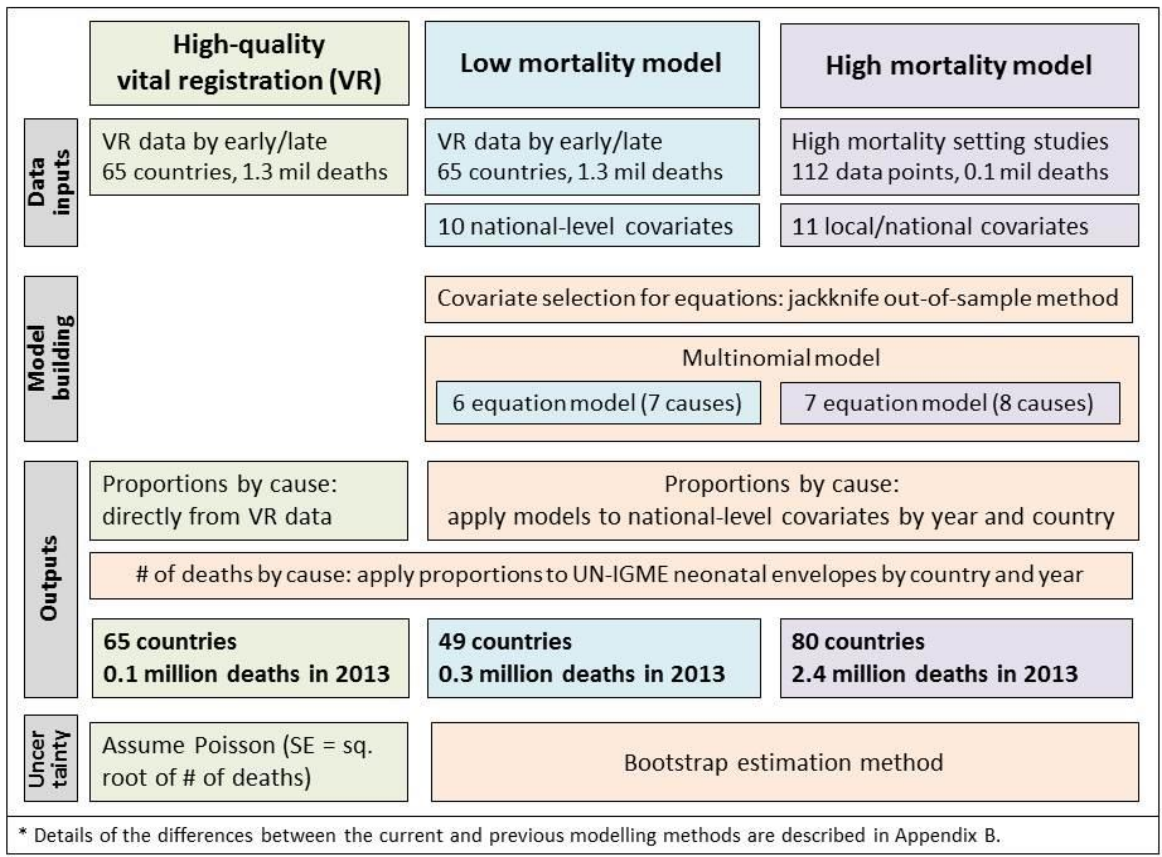

**Figure 2: The proportional cause distribution of neonatal deaths by neonatal period for the input data in the a) low mortality model (one observation per country by averaging across years) and b) high mortality model.**

a)

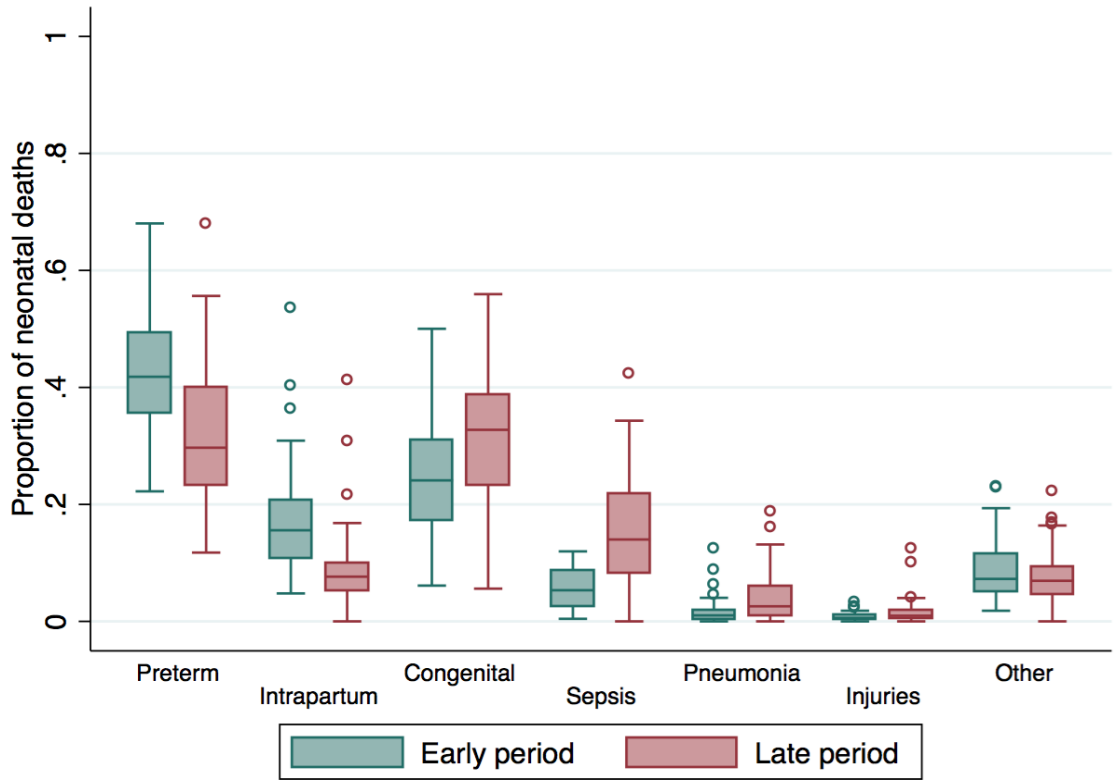

b)

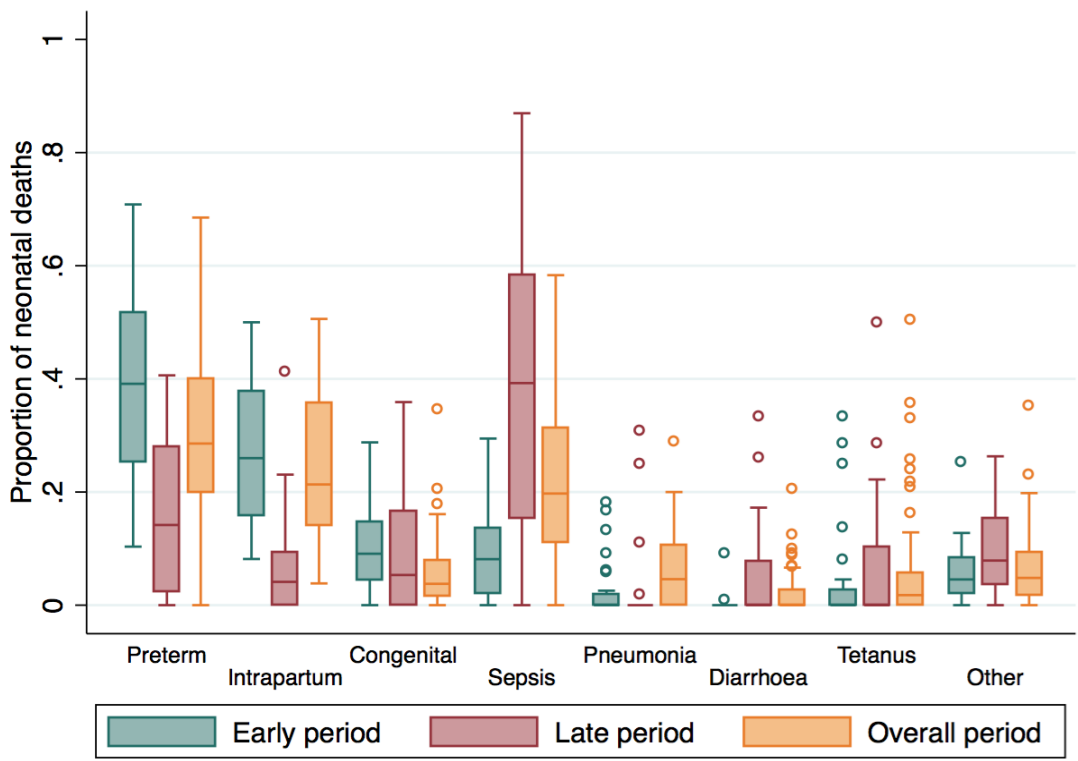

**Figure 3: Cause-specific risk of neonatal death by MDG region in 2013**

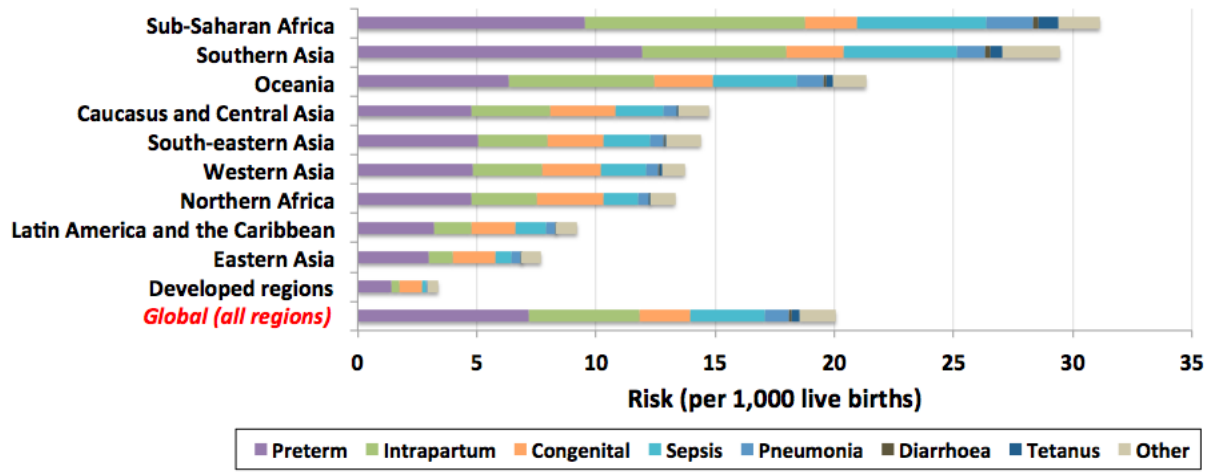

**Figure 4: Global cause-specific risks of death from 2000-2013 for the early and late neonatal periods**

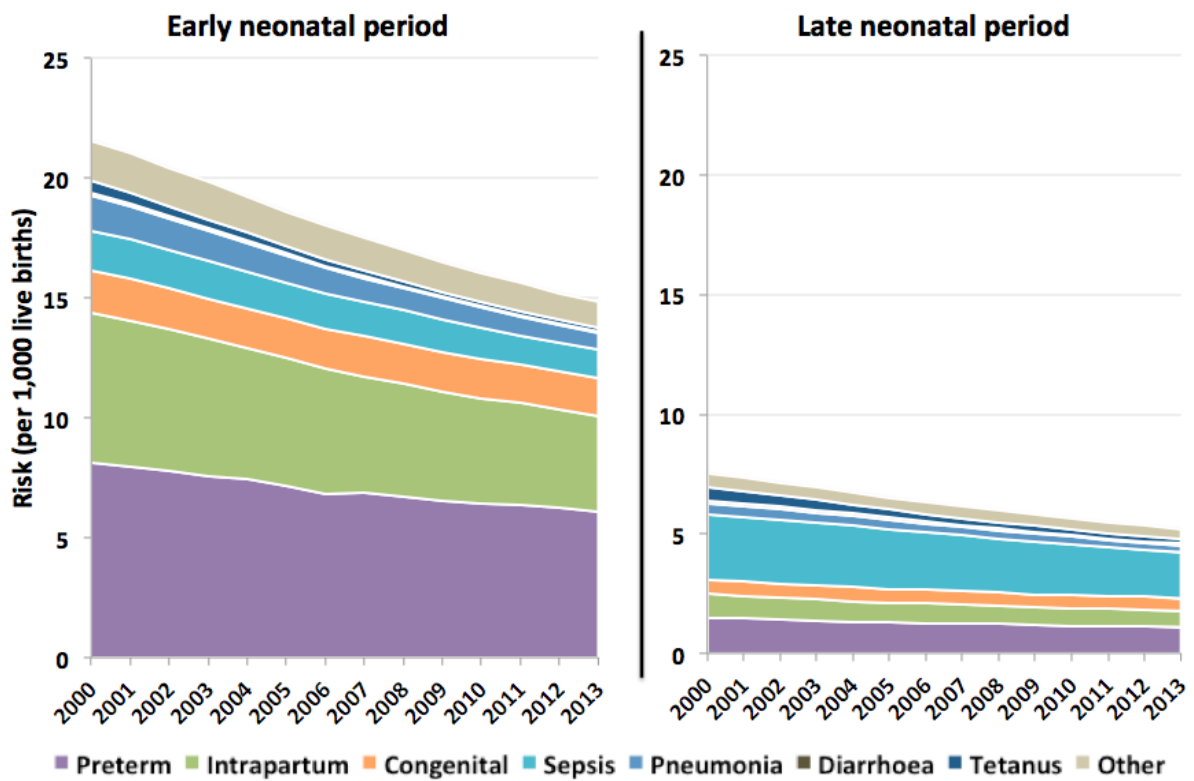

| | Early neonatal period | | Late neonatal period | | Overall neonatal period | | |
|---|---|---|---|---|---|---|---|
| | % | # of deaths in thousands (uncertainty) | % | # of deaths in thousands (uncertainty) | % | # of deaths in thousands (uncertainty) | Risk |
| Preterm | 40.8 | 834.8 (608.1-1083.5) | 21.2 | 152.1 (91.0-229.0) | 35.7 | 986.9 (699.1-1312.5) | 7.2 |
| Intrapartum | 27.0 | 552.7 (407.6-711.4) | 12.9 | 92.1 (54.8-133.4) | 23.4 | 644.8 (462.4-844.7) | 4.7 |
| Congenital | 10.6 | 217.0 (140.9-325.9) | 10.2 | 72.8 (42.5-124.5) | 10.5 | 289.8 (183.3-450.4) | 2.1 |
| Sepsis | 8.0 | 163.7 (62.4-271.6) | 37.2 | 266.7 (156.5-393.2) | 15.6 | 430.4 (218.9-664.8) | 3.1 |
| Pneumonia | 4.8 | 98.9 (48.8-200.3) | 5.2 | 37.6 (21.5-58.7) | 4.9 | 136.4 (70.3-259.0) | 1.0 |
| Diarrhoea | 0.3 | 6.7 (0-57.4) | 1.4 | 10.0 (3.2-25.6) | 0.6 | 16.6 (3.2-83.0) | 0.1 |
| Tetanus | 1.0 | 21.1 (7.4-53.2) | 3.8 | 27.1 (8.1-67.2) | 1.7 | 48.2 (15.5-120.4) | 0.3 |
| Other | 7.3 | 149.9 (72.7-250.3) | 8.1 | 57.9 (26.3-117.2) | 7.5 | 207.8 (99.0-367.4) | 1.5 |

Table 1: Global cause-specific proportions, risk, and numbers of neonatal deaths (with uncertainty ranges) in 2013

Notes: 1) diarrhoea and tetanus were estimated only for the 80 high mortality model countries; 2) injuries are included within the "other" category; 3) risk is per 1,000 live births.

| Table 2: Recommendations for data improvements based on neonatal mortality level | | | | |
|---|---|---|---|---|
| **Neonatal mortality rate (per 1,000 live births)** | **< 5** | **5 – <15** | **15 – <30** | **≥30** |
| **Main data collection platforms** | 1. Full coverage vital registration (VR) 2. National perinatal and maternal mortality audit 3. Strong Health Management Information Systems (HMIS) | 1. VR usually full coverage, quality may be lacking 2. Audit may not be full coverage 3. HMIS may be high coverage but inconsistent quality | 1. Limited to no VR 2. Usually dependent on 5 yearly national surveys and HMIS | |
| **Counting all births and neonatal deaths** | | | | |
| *Counting all births* | Consistent counting of all live birth and any birth over 500 g/22 weeks, noting if singleton or multiple birth | | | |
| *Comparable definitions* to count neonatal deaths | Reporting all deaths in live born babies using perinatal death certificates where possible (no gestational age or weight cut off) Specify early neonatal (day 0, days 1–6) and late neonatal (days 7–27) deaths separately | | | |
| *Collecting more detailed data* | 1. VR using specific neonatal death certificates with birth weight and gestational age 2. Improved cause coding in ICD-11 with clinically-relevant codes 3. Health facility surveillance with detailed dataset 4. Crosslink VR and health facility databases to maximise capture 5. Analysis to track and target disparities | | 1. Develop or enhance sentinel surveillance sites for pregnancy, child and other health outcomes (prospective), with a focus on enhancing national representativeness and coverage of the poorest 2. Improve VR systems. Use specific neonatal death certificates 3. Track urban/rural and other key disparities | |
| *Comparable cause-of-death categories* | Consensus on a minimum dataset to be collected on all neonatal deaths, with a limited number of programmatically relevant, comparable cause categories distinguishable through verbal autopsy, but which can be further specified by clinical data in mid-mortality settings and linked to complex classification systems/ICD codes. Include both direct fetal/neonatal causes and maternal condition so can cross tabulate | | | |
| *Categorising small babies* (weight and gestational age) | All babies to be weighed at birth and weight to be recorded on birth and death certificates | | | |
| | 1. Gestational age to be assessed using routine high-quality early pregnancy ultrasound and recorded on birth and death certificates 2. Track the % of births that are reported <28 weeks (noting that if <3% of preterm births are <28 weeks, the system may be under-recording preterm births) | | 1. Gestational age to be assessed in all babies using simplified clinical examination or last menstrual period (LMP) where early pregnancy ultrasound is not available 2. Improved technology and low-cost assessment tools are required to increase reliability | |
| Notes: 1) This table is adapted from the "Every Newborn: beyond survival" article published in the Lancet (3) | | | | |

SUPPLEMENTARY WEB MATERIAL

# Cause-of-death estimates for the early and late neonatal periods for 194 countries from 2000-2013

Shefali Oza, Joy Lawn, Daniel R Hogan, Colin Mathers, Simon Cousens

## Appendix A: Country groupings by estimatation method

The 194 countries in this analysis were separated into 3 groups based on the quality if their vital registration (VR) data and under-5 mortality rates (U5MR). U5MR was used instead of the neonatal mortality rate (NMR) in order to be consistent with researchers working in parallel to develop estimates for children 1-59 months old. We kept the same country groupings as the previous work (1), with the exception of a few countries (see "changes to country groupings" in Appendix B).

As defined in the previous work (1), countries were considered to have high-quality VR data if they 1) had 80% or higher coverage, 2) did not have excessive use of non-specific/ill-defined VR codes, and 3) provided sufficient details in the coding such that the deaths could be grouped in the programmatically-relevant categories used in this work. Countries that lacked high-quality VR data were included in the low mortality model if their U5MR from 2000-2010 was ≤35 (per 1,000 live births) and in the high mortality model if it their U5MR was >35.

### High-quality VR – 65 countries
Antigua and Barbuda, Argentina, Australia, Austria, Bahamas, Bahrain, Barbados, Belgium, Belize, Brazil, Bulgaria, Chile, Colombia, Costa Rica, Croatia, Cuba, Czech Republic, Denmark, Dominica, Estonia, Finland, France, Germany, Greece, Grenada, Guyana, Hungary, Iceland, Ireland, Israel, Italy, Japan, Kuwait, Latvia, Lithuania, Luxembourg, Macedonia (TFYR of), Malta, Mauritius, Mexico, Montenegro, Netherlands, New Zealand, Norway, Panama, Poland, Republic of Korea, Republic of Moldova, Romania, Saint Kitts and Nevis, Saint Lucia, Saint Vincent and the Grenadines, Serbia, Singapore, Slovakia, Slovenia, South Africa, Spain, Suriname, Sweden, Trinadad and Tobago, United Kingdom, United States, Uruguay, Venezuela

### Low mortality model – 49 countries
Albania, Andorra, Armenia, Belarus, Bosnia and Herzegovina, Brunei Darussalam, Canada, Cabo Verde, China, Cook Islands, Cyprus, Ecuador, Egypt, El Salvador, Fiji, Georgia, Hondorus, Jamaica, Jordan, Lebanon, Libya, Malaysia, Maldives, Monaco, Nicaragua, Niue, Oman, Palau, Paraguay, Peru, Portugal, Qatar, Russian Federation, Samoa, San Marino, Saudi Arabia, Seyechelles, Sri Lanka, Switzerland, Syrian Arab Republic, Thailand, Tongo, Tunisia, Turkey, Tuvalu, Ukraine, United Arab Emirates, Vanuatu, Vietnam

### High mortality model – 80 countries
Afghanistan, Algeria, Angola, Azerbaijan, Bangladesh, Benin, Bhutan, Bolivia, Botswana, Burkina Faso, Burundi, Cambodia, Cameroon, Central African Republic, Chad, Comoros, Congo, Cote d'Ivoir, Democratic People's Republic of Korea, Democratic Republic of the Congo, Djibouti, Dominican Republic, Equatorial Guinea, Eritrea, Ethiopia, Gabon, Gambia, Ghana, Guatemala, Guinea, Guinea-Bissau, Haiti, India, Indonesia, Iran, Iraq, Kazakhstan, Kenya, Kiribati, Kyrgyzstan, Lao People's Democratic Republic, Lesotho, Liberia, Madagascar, Malawi, Mali, Marshall Islands, Mauritania, Micronesia, Mongolia, Morocco, Mozambique, Myanmar, Namibia, Nauru, Nepal, Niger, Nigeria, Pakistan, Papau New Guinea, Philippines, Rwanda, Sao Tome and Principe, Senegal, Sierra Leone, Solomon Islands, Somalia, South Sudan, Sudan, Swaziland, Tajikistan, Timor-Leste, Togo, Turkmenistan, Uganda, United Republic of Tanzania, Uzbekistan, Yemen, Zambia, Zimbabwe

**Appendix B: Key methodological differences between current and previous estimates**
Some key methological changes since the last round of neonatal cause-of-death estimates are described below.

Early/late neonatal period estimates
We are now reporting the results by the early and late neonatal periods. As described in the paper, we run 4 separate models (early and late separately for both the low and high mortality models).

Changes to country groupings
Since the last iteration of this work, we have included Kuwait, Macedonia, Montenegro, Republic of Korea, and Saint Lucia in the high-quality VR countries instead of the low mortality model. We made this change because VR time series data became avaible from the WHO for these countries recently. Additionally, due to improved vital registration data collection in South Africa, we have now included South Africa in the group of high-quality VR countries instead of the high mortality model.

Covariate selection
Previously (1), we allowed the relationship between each covariate and the log of the cause/baseline cause ratio to be described either linearly or quadratically. In this work, we also included the possibility of a restricted cubic spline relationship to include potentially more accurate non-linear relationships. We also included more covariates to the low mortality model than were previously used.

Additional cause categories within the multinomial
In the low mortality model, we added injuries as a separate cause, while in the previous work injuries was included in the "other" category. In the high mortality model, we included sepsis, pneumonia, and tetanus as separate cause categories. In the previous work, the high mortality model included a broader "infections" category that included sepsis and pneumonia, and these were then split separately using the results from the multinomial. Additionally, in the previous work tetanus was estimated using a single-cause, but now is a part of the multinomial itself.

Indian estimates
Similar to the 2010 estimates (1), we produced national Indian estimates by aggregating state-level estimates. For the 2010 estimates, we had developed a separate multinomial using only Indian studies/surveys for the input data. This time, however, we estimated the state-level proportional cause distribution for each Indian state/year within the overall high mortality model. We did this because there were not enough Indian studies that reported sepsis and pneumonia separately in order to estimate these causes in an India-only multinomial, and because the cause-of-death distribution between Indian studies and non-Indian studies appeared to be quite similar.

Table S1 contains an overview of how the methods have changed since the first Child Health Epidemiology Research Group (CHERG) neonatal estimates were published in 2005 for the year 2000.

| Table S1: Methodological differences between the current and previous estimates | | | | |
|---|---|---|---|---|
| **Publication year** | **2005** | **2010** | **2012** | **Current** |
| **Estimation year(s)** | 2000 | 2008 | 2000-2010 | 2000-2013 |
| **Goal of estimation** | *Nonatal cause-of-death distribution for 193 countries in 2000* | Neonatal cause-of-death distribution for 193 countries in *2008* | Neonatal cause-of-death distribution for 193 countries from *2000-2010* | Neonatal cause-of-death distribution for 194 countries from *2000-2013 by early and late periods* |
| **Countries in each model** | VR: 45<br>Low mortality: 37<br>High mortality: 111 | VR: 73<br>Low mortality: 37<br>Low/high averaged: 22<br>High mortality: 61 | VR: 61<br>Low mortality: 51<br>High mortality: 81 | VR: 65<br>Low mortality: 49<br>High mortality: 80 |
| **Thresholds for classifying countries into high-quality VR, low mortality model, or high mortality model** | - High quality VR = >90% coverage<br>- Low mort model = countries without high-quality VR data and NMR<10 (or NMR<15 in WHO EURO/AMRO regions)<br>- High mort model = remaining countries | - High quality VR = >80% coverage<br>- Low mort model = countries without high-quality VR and NMR<15<br>- averaged model = countries without high-quality VR and NMR between 15-20: low/high mortality models averaged<br>- High mort model = countries without high-quality VR data and NMR>20 | - High quality VR = >80% coverage + quality criteria (e.g. limited non-specific/implausible codes)<br>- Low mort model = countries without adequate VR and with U5MR≤35 from 2000-2010.<br>- High mort model = countries without adequate VR and with U5MR>35 from 2000-2010. | - High quality VR = >80% coverage + quality criteria (e.g. limited non-specific/implausible codes)<br>- Low mort model = countries without adequate VR and with U5MR≤35 from 2000-2013.<br>- High mort model = countries without adequate VR and with U5MR>35 from 2000-2013. |
| **Input data** | High mortality model: 13685 deaths<br>Low mortality model: 96,797 deaths | High mortality model: 23,220 deaths<br>Low mortality model: 1,005,478 deaths | High mortality model: 56,890 deaths<br>Low mortality model: 1,013,599 deaths | High mortality model: 98,222 deaths<br>Low mortality model:1,267,404 deaths |
| **Covariate selection** | -High mortality model: expert opinion on which covariates may be associated with outcomes<br>-Low mortality model: forward stepwise with 5% sig. level | Same covariates used as previous round | -High and low mortality models: jackknife procedure to minimize out-of-sample prediction error.<br>- Allowed relationship between covariate and outcome to be linear or quadratic | -High and low mortality models: jackknife procedure to minimize out-of-sample prediction error.<br>- Allowed relationship between covariate and outcome to be linear, quadratic, or spline |
| **Multinomial model** | All causes in multinomial | Infections in multinomial with sepsis/pneumonia split done after; tetanus as single-cause | Infections in multinomial with sepsis/pneumonia split done after; tetanus as single-cause | All causes included in multinomial |
| **Causes** | Preterm, Intrapartum, Congenital, Infection, Diarrhoea, Tetanus, Other | Preterm, Intrapartum, Congenital, Sepsis, Pneumonia, Diarrhoea, Tetanus, Other | Preterm, Intrapartum, Congenital, Sepsis, Pneumonia, Diarrhoea, Tetanus, Other | Preterm, Intrapartum, Congenital, Sepsis, Pneumonia, Diarrhoea, Tetanus, Other (+ injuries for VR and low mortality model countries) |
| **Uncertainty** | Jackknife | Jackknife | Bootstrap | Bootstrap |

Differences in global estimates between previous Child Health Epidemiology Reference Group (CHERG) estimates and this round are shown in table S2. Since time trends are a recent addition, we have compared the data reported in the earlier studies with the relevant year from our current work. For example, the paper published in 2005 had estimates for 2000, so we have compared the 2000 predictions from that paper and our work.

| Table S2: Comparison of estimated proportions between current and previous estimates | | |
|---|---|---|
| | **Percentages from previous work (%)** | **Percentages from current work (%)** |
| **2000 estimates (2)** | | |
| **Preterm** | 27.9 (0.19-0.35) | 33.1 (23.1-44.0) |
| **Intrapartum** | 22.8 (0.15-0.27) | 24.8 (16.4-32.8) |
| **Infections** | 26.0 (0.17-0.31) | 21.5 (10.5-35.9) |
| **Congenital** | 7.4 (0.06-0.12) | 8.3 (5.0-14.2) |
| **Tetanus** | 6.5 (0.05-0.20) | 3.8 (1.0-9.7) |
| **Diarrhoea** | 2.8 (0.02-0.10) | 0.9 (0.1-4.7) |
| **Other** | 6.6 (0.05-0.16) | 7.6 (3.8-13.2) |
| **2008 estimates (3)** | | |
| **Preterm** | 28.9 (20.1-34.0) | 34.6 (24.4-46.2) |
| **Intrapartum** | 22.8 (15.7-27.9) | 23.8 (16.8-31.5) |
| **Sepsis** | 14.6 (10.0-20.6) | 16.0 (8.2-24.7) |
| **Pneumonia** | 10.8 (7.4-15.2) | 5.5 (2.8-10.4) |
| **Congenital** | 7.6 (5.7-10.7) | 9.6 (6.2-15.0) |
| **Tetanus** | 1.7 (0.9-2.3) | 2.1 (0.6-5.5) |
| **Diarrhoea** | 2.2 (1.6-5.9) | 0.7 (0.1-3.6) |
| **Other** | 11.4 (8.9-24.7) | 7.7 (3.7-13.2) |
| **2010 estimates (1)** | | |
| **Preterm** | 30.2 (25.6-37.1) | 35.0 (24.6-46.9) |
| **Intrapartum** | 20.1 (17.1-24.5) | 23.6 (16.8-31.0) |
| **Sepsis** | 11.0 (7.0-15.4) | 16.0 (8.2-24.7) |
| **Pneumonia** | 9.1 (5.8-13.1) | 5.2 (2.7-10.0) |
| **Congenital** | 7.6 (5.8-10.2) | 10.1 (6.4-15.6) |
| **Tetanus** | 1.6 (0.6-7.7) | 1.8 (0.5-4.8) |
| **Diarrhoea** | 1.4 (0.5-4.2) | 0.6 (0.2-3.3) |
| **Other** | 5.1 (3.2-7.9) | 7.6 (3.7-13.2) |
| Notes: 1) Intrapartum-related conditions were previously referred to as "birth asphyxia"; 2) infections include sepsis and other severe infections as well as pneumonia, which in recent years have been estimated separately. To compare with previous estimates, however, these are included in the aggregate infection category for estimates. | | |

Note that these estimates are not necessarily directly comparable. One reason for this is that UN-IGME updates their NMR, U5MR, and IMR time series each year, and so the values estimated for 2008 in one year may be different from those from another year. Since our model includes these as covariates, the covariate values for the same year may be different for these values, which could affect the proportions. But as is seen in Table S2, the previous and current estimates generally fall within each other's uncertainty bounds.

**Appendix C: Mapping from ICD codes to cause categories used in this work for VR data**

Here we have included the mapping from the 9[th] and 10[th] International Classification of Diseases (ICD-9 and ICD-10) to the 7 programatically relevant cause categories we used in our work for the VR data.

| Table S3: Mapping between ICD codes and cause categories used in this work | | |
|---|---|---|
| | **ICD-10 codes** | **ICD-9 codes** |
| **Complications of preterm birth** | P01.0-P01.1, P07, P22, P25-P28, P52, P61.2, P77 | 434.9, 518.1-518.9, 761.0-761.1, 765, 769-770.0, 770.2-770.9, 772.1, 774.2, 776.6, 777.5-777.6, 786.3 |
| **Intrapartum-related complications** | P01.7-P02.1, P02.4-P02.6, P03, P07, P10-P15, P20-P21, P24, P50, P90-P91 | 348.1-348.9, 437.1-437.9, 723.4, 761.7-762.1, 762.4-762.6, 763, 767-768, 770.1, 772.2, 779.0-779.2 |
| **Congenital disorders** | D55-D68.9, E01-E07, E70-E84, G10-G99, H, I, K, L, M, N, P35, P76, Q | 056, 240-243, 245-259, 272-277, 279.3-286, 288.2, 303, 330-348.0, 349-426, 429-434.0, 435-437.0, 438-451, 520-723.0, 724-728, 731-759, 775.2, 777.0, 795.2 |
| **Sepsis and other severe infections** | A00-A35, A38-A99, B, G00-G09, P36-P39 | 000-031, 034-055, 057-134, 136-139, 320-326, 491, 730, 771, 780.6, 785.4 |
| **Pneumonia** | A36-A37, J, P23 | 032-033, 460-490, 492-518.0 |
| **Injuries** | S, V, W, X, Y | 800-999 |
| **Other** | C, D00-D54.9, D69-D99, E00, E08-E69, E85-E99, P00, P01.2-P01.6, P02.2-P02.3, P02.7-P02.9, P04-P06, P08, P29, P51, P53-P61.1, P61.3-P74, P78, P80-P83, P93-P94 | 135, 140-239, 244, 260-271, 278-279.2, 287-288.1, 288.3-289, 427, 452-459, 760, 761.2-761.6, 762.2-762.3, 762.7-762.9, 764, 766, 772.0, 772.3-774.1, 774.3-775.1, 775.3-776.5, 776.7-776.9, 778.0, 779.5-779.6 |

We excluded the following ICD codes as non-specific/ill-defined:
1) ICD-10 – F, O, P92, P95-96, R
2) ICD-9 – 295.4, 305.6, 205.9, 308.9, 311.0, 317.0, 319.0, 779.3, 779.8-799 (except for 780.6, 785.4, 786.3, and 795.2 as mentioned above)

**Appendix D: Details of the high-quality VR data used in the low mortality model**

Below is a table describing the missing data years for the high-quality VR countries. This data was used as is for the low mortality model input dataset (i.e. no imputation of missing data). For the high-quality VR country estimates, we imputed data for the years with missing data. For years with missing data that were between years with existing data, we used linear interpolation to impute the missing proportions. For years that were before the earliest or after the latest available data year, we applied the proportions from the nearest year with available data to the missing data. None of the 7 causes we modelled for low mortality countries were missing in these data. The full dataset can be found at the WHO Global Health Observatory (www.who.int/gho).

| Table S4: List of years with missing data for high-quality VR countries | | | | | |
|---|---|---|---|---|---|
| **Country** | **Years with missing data** | **Country** | **Years with missing data** | **Country** | **Years with missing data** |
| **Antigua and Barbuda** | 2010-2012 | **Germany** | 2012 | **Panama** | 2010-2012 |
| **Argentina** | 2011-2012 | **Greece** | 2012 | **Poland** | 2012 |
| **Australia** | 2005, 2012 | **Grenada** | 2000, 2011-2012 | **Republic of Korea** | 2012 |
| **Austria** | 2012 | **Guyana** | 2000, 2009-2012 | **Republic of Moldova** | none |
| **Bahamas** | 2009-2012 | **Hungary** | 2012 | **Romania** | 2012 |
| **Bahrain** | 2010-2012 | **Iceland** | 2010-2012 | **Saint Kitts and Nevis** | 2009-2012 |
| **Barbados** | 2009-2012 | **Ireland** | 2000-2006, 2011-2012 | **Saint Lucia** | 2000-2001, 2007, 2009-2012 |
| **Belgium** | 2000-2002; 2010-2012 | **Israel** | 2011-2012 | **Saint Vincent and the Grenadines** | 2011-2012 |
| **Belize** | 2008-2012 | **Italy** | 2004-2005, 2011-2012 | **Serbia** | 2012 |
| **Brazil** | 2011-2012 | **Japan** | 2012 | **Singapore** | 2012 |
| **Bulgaria** | 2012 | **Kuwait** | 2000, 2012 | **Slovakia** | 2011-2012 |
| **Chile** | 2010-2012 | **Latvia** | 2011-2012 | **Slovenia** | 2011-2012 |
| **Colombia** | 2010-2012 | **Lithuania** | 2011-2012 | **South Africa** | 2000-2005, 2008, 2010-2012 |
| **Costa Rica** | 2010-2012 | **Luxembourg** | 2012 | **Spain** | 2012 |
| **Croatia** | 2012 | **Macedonia (TYFR of)** | 2001-2004, 2011-2012 | **Suriname** | 2010-2012 |
| **Cuba** | 2011-2012 | **Malta** | 2012 | **Sweden** | 2011-2012 |
| **Czech Republic** | 2012 | **Mauritius** | 2012 | **Trinadad and Tobago** | 2003, 2009-2012 |
| **Denmark** | 2000, 2010-2012 | **Mexico** | 2011-2012 | **United Kingdom** | 2011-2012 |
| **Dominica** | 2011-2012 | **Montenegro** | 2000-2004, 2010-2012 | **United States** | 2011-2012 |
| **Estonia** | 2009, 2012 | **Netherlands** | 2012 | **Uruguay** | 2002-2003, 2005-2006, 2010-2012 |
| **Finland** | 2012 | **New Zealand** | 2010-2012 | **Venezuela** | 2008-2012 |
| **France** | 2010-2012 | **Norway** | 2012 | | |

**Appendix E: Inclusion criteria for studies in the high mortality model input dataset**
For this work, we used the same criteria for selecting new studies to include in the input dataset as used in previous iterations of this work (4). The criteria are as follows:

- Publication in 1980 or later
- Study set in one of nine (of a total of 14) subregions with no or few countries with >90% VR coverage
- Community-based study or hospital based in populations with over 90% hospital delivery and defined catchment population
- Case ascertainment: follow up of newly born infants from birth to at least 7 or 28 days
- Number of deaths with known cause ≥ 20
- Study duration ≥ 12 months
- Included four or more of the eight selected programme relevant causes of neonatal death
- Deaths of unknown cause ≤ 25% of total deaths
- Cause attribution based on skilled clinical investigation, post mortem, or verbal autopsy
- Case definitions specified and comparable with other studies

Our search strategy involved doing a literature review in ten databases for articles published between January 2011 and May 2013. Previous searches covered periods from 1980 to December 2010. The databases we searched were Pubmed, Embase, Web of Knowledge, Medline, Global Health, Popline, and region-specific indices (LILACS, Africa-Wide Information, Western Pacific Region, Eastern Mediterranean, IndMed).

Our search covered the following areas (including variations, alternatives, and MESH for these terms):
*Cause-specific (one or more of these)*: haemorrhage, jaundice, abnormality, malformation, neural tube defect, sudden infant daeath syndrome, congenital malformations, congenital abnormalities, necrotic, respiratory distress, prematurity, preterm birth, asphyxia, tetanus, sepsis, birth injury, intrapartum, birth, cause, fetal alcohol syndrome, rubella, diarrhea, dysentery, cholera, gastroenteritis, digestive tract infection, pneumonia, respiratory infections, bronchitis, croup, meningitis, encephalitis, meningococcal
AND
*Age group (one or more of these):* infant/newborn, neonatal, perinatal
AND
*Mortality terms (one or more of these):* death, mortality, fatality
AND
*High mortality countries/regions:* Argentina, Bolivia, Brazil, Brasil, Chile, Colombia, Ecuador, French Guiana, Guyana, Paraguay, Peru, Suriname, Uruguay, Venezuela, Mexico, Belize, Costa Rica, El Salvador, Guatemala, Honduras, Nicaragua, Puerto Rico, Panama, West Indies, Antigua, Bahamas, Barbados, Cuba, Dominica, Dominican Republic, Grenada, Guadeloupe, Haiti, Jamaica, Martinique, Antilles, Anguilla, Saint Kitts, St Kitts, Saint Lucia, St Lucia, Saint Vincent, St Vincent, Trinidad, Tobago, Virgin Islands, Kazakhstan, Kyrgyzstan, Tajikistan, Turkmenistan, Uzbekistan, Borneo, Brunei, Cambodia, East Timor, Indonesia, Laos, Malaysia, Mekong Valley, Myanmar, Burma, Philippines, Singapore, Thailand, Vietnam, Bangladesh, Bhutan, India, Nepal, Pakistan, Sri Lanka, China, Korea, Macao, Mongolia, Taiwan, Afghanistan, Bahrain, Iran, Iraq, Israel, Jordan, Kuwait, Lebanon, Oman, Qatar, Saudi Arabia, Syria, Turkey, United Arab Emirates, Yemen, Fiji, New Caledonia, Papua New Guinea, Vanuatu, Micronesia, Melanesia, Guam, Palau, Polynesia, Samoa, Tonga, Armenia, Azerbaijan, Georgia, Albania, Estonia, Latvia, Lithuania, Bosnia, Herzegovina, Serbia, Bulgaria, Belarus, Croatia, Czech Republic, Hungary, Macedonia, Moldova, Montenegro, Poland, Romania, Russia, Bashkiria, Dagestan, Slovakia, Slovenia, Ukraine, Cameroon, Central African Republic, Chad, Congo, "Democratic Republic of the Congo", Equatorial Guinea, Gabon, Burundi, Djibouti, Eritrea, Ethiopia, Kenya, Rwanda, Somalia, Sudan, Tanzania, Uganda, Angola, Botswana, Lesotho, Malawi, Mozambique, Namibia, South Africa, Swaziland, Zambia, Zimbabwe, Benin, Burkina Faso, Cote d'Ivoire, Gambia, Ghana, Guinea, Guinea-Bissau, Liberia, Mali, Mauritania, Niger, Nigeria, Senegal, Sierra Leone, Togo, Algeria, Egypt, Libya, Morocco, Tunisia, Comoros, Madagascar, Mauritius, Reunion, Seychelles, developing country, third world country, less developed, sub-Saharan, Caribbean, Pacific Islands, Mexico, Latin America, South America, Indian Ocean Islands, Central America, Asia, Africa, Far East
AND
*NOT* "case reports", editorial, comment, practice guideline

**Appendix F: Details of the study data for the high mortality model input dataset**

Below is a table describing the studies and surveys included in the high mortality model input dataset. New studies have an asterisk (*) in front of the author name. While most studies were included in the dataset as a single observation, those stratified by neonatal period or in other ways (e.g. setting location) were included as multiple observations. The full dataset can be found at the WHO Global Health Observatory (www.who.int/gho).

| Table S5: Details of the high mortality model input dataset | | | | | | |
|---|---|---|---|---|---|---|
| First author | Year published | Country | Additional strata | Median data year | # of causes reported | # of deaths used in analysis |
| Adazu (5) | 2005 | Kenya | | 2002 | 5 | 75 |
| Aguilar (6) | 1998 | Bolivia | | 1995 | 8 | 79 |
| *Akgun (7) | 2012 | Turkey | | 2003 | 4 | 34 |
| Aleman (8) | 1998 | Nicaragua | | 1993 | 6 | 72 |
| Anand (9) | 2000 | India | | 1993 | 8 | 50 |
| Asling-Monemi (10) | 2003 | Nicaragua | | 1995 | 6 | 56 |
| Awasthi (11) | 1998 | India | | 1994 | 7 | 286 |
| Baiden (12) | 2006 | Ghana | | 1999 | 4 | 1001 |
| Balci (13) | 2008 | Turkey | early/late | 2006 | 5 | 68 |
| Bang (14) | 1999 | India | | 1994 | 6 | 36 |
| *Bapat (15) | 2012 | India | early/late | 2006 | 5 | 102 |
| Baqui (16) | 2006 | India | early/late | not given | 7 | 477 |
| Barros (17) | 1987 | Brazil | | 1982 | 7 | 113 |
| Barros (18) | 2008 | Brazil | | 2004 | 8 | 54 |
| Bassani (19) | 2010 | India | | 2002 | 7 | 10892 |
| Bezzaoucha (20) | 2010 | Algeria | early/late | 2003 | 5 | 2167 |
| Bhatia (21) | 1989 | Bangladesh | | 1982 | 7 | 513 |
| Bhutta (22) | - | Pakistan | | 2001 | 7 | 152 |
| Campbell (23) | 2004 | Egypt | | 2000 | 7 | 103 |
| Chowdhury (24) | 1996 | Bangladesh | | 1983 | 5 | 474 |
| Chowdhury (25) | 2005 | Bangladesh | early/late | 1992 | 5 | 43 |
| Chowdhury (26) | 2010 | Bangladesh | | 2004 | 8 | 332 |
| Darmstadt (27) | 2010 | Bangladesh | time periods | 2002 | 7 | 56 |
| Darmstadt (27) | 2010 | Bangladesh | time periods | 2005 | 7 | 125 |
| Datta (28) | 1988 | India | | not given | 8 | 163 |
| Deribew (29) | 2007 | Ethiopia | | 2005 | 8 | 45 |
| *DHS (30) | 2005 | Bangladesh | | 2002 | 8 | 302 |
| *DHS (31) | 2006 | Honduras | | 2004 | 7 | 142 |
| *DHS (32) | 2007 | Nepal | | 2004 | 8 | 220 |
| *DHS (33) | 2008 | Pakistan | | 2006 | 8 | 1484 |
| *DHS (34) | 2009 | Mozambique | | 2007 | 5 | 718 |
| Djaja (35) | 2003 | Indonesia | | 2001 | 8 | 178 |
| Dommisse (36) | 1991 | South Africa | early/late | 1988 | 7 | 276 |
| Edmond (37) | 2008 | Ghana | early/late | 2004 | 5 | 582 |
| Ekanem (38) | 1994 | Nigeria | | 1991 | 7 | 24 |

| Table S5: Details of the high mortality model input dataset | | | | | | |
|---|---|---|---|---|---|---|
| **First author** | **Year published** | **Country** | **Additional strata** | **Median data year** | **# of causes reported** | **# of deaths used in analysis** |
| **El-Zibdeh (39)** | 1988 | Saudi Arabia | | 1983 | 7 | 78 |
| **Fantahun (40)** | 1998 | Ethiopia | | 1992 | 7 | 47 |
| **Fauveau (41)** | 1990 | Bangladesh | | 1982 | 8 | 201 |
| **Fikree (42)** | 2002 | Pakistan | | 1992 | 8 | 497 |
| **Fonseka (43)** | 1994 | Sri Lanka | | 1987 | 7 | 253 |
| **Garenne (44)** | 2007 | Morocco | time periods | 1987 | 8 | 109 |
| **Garenne (44)** | 2007 | Morocco | time periods | 1987 | 8 | 329 |
| ***Gill (45)** | 2011 | Zambia | | 2007 | 7 | 58 |
| **Gomes (46)** | 1997 | Brazil | early/late | 1991 | 6 | 138 |
| **Greenwood (47)** | 1987 | Gambia | | 1982 | 8 | 32 |
| **Halder (48)** | 2009 | Bangladesh | | 2007 | 8 | 49 |
| **Hinderaker (49)** | 2003 | Tanzania | early/late | 1995 | 5 | 71 |
| **Jehan (50)** | 2009 | Pakistan | early/late | 2004 | 5 | 49 |
| **Karabulut (51)** | 2009 | Turkey | early/late | 2006 | 6 | 178 |
| **Khalique (52)** | 1993 | India | early/late | 1990 | 8 | 21 |
| **Khan (53)** | 1993 | Pakistan | | 1986 | 7 | 80 |
| ***Khanal (32)** | 2011 | Nepal | early/late | 2006 | 5 | 183 |
| **Khanjanasthiti (54)** | 1984 | Thailand | | not given | 8 | 22 |
| **Kishan (55)** | 1988 | Libya | | 1984 | 6 | 245 |
| ***Krishnan (56)** | 2012 | India | time periods | 1983 | 5 | 154 |
| ***Krishnan (56)** | 2012 | India | time periods | 2003 | 5 | 106 |
| **Leach (57)** | 1999 | Gambia | | 1992 | 8 | 130 |
| **Liu (58)** | 1985 | China | | 1980 | 7 | 956 |
| **Lucas (59)** | 1996 | Sri Lanka | | 1993 | 6 | 120 |
| **Manandhar (60)** | 2010 | Nepal | | 2007 | 5 | 640 |
| ***Matendo (61)** | 2011 | Congo | | 2005 | 6 | 56 |
| **Matijasevich (62)** | 2008 | Brazil | time periods | 1982 | 4 | 115 |
| **Matijasevich (62)** | 2008 | Brazil | time periods | 1993 | 4 | 88 |
| **Matijasevich (62)** | 2008 | Brazil | time periods | 2004 | 4 | 47 |
| **Mendieta (63)** | 2001 | Paraguay | | 1996 | 6 | 3638 |
| **Nandan (64)** | 2005 | India | | 2001 | 8 | 299 |
| ***Nga (65)** | 2012 | Viet Nam | | 2009 | 7 | 225 |
| **Pattison (66)** | 2013 | South Africa | | 2006 | 7 | 45848 |
| **Perry (67)** | 2003 | Bangladesh | | 1997 | 8 | 102 |
| **Perry (67)** | 2003 | Haiti | | 1997 | 8 | 28 |
| **Phukan (68)** | 1998 | India | | 1994 | 7 | 101 |
| **Pison (69)** | 1993 | Senegal | | 1987 | 6 | 26 |
| **Pratinidhi (70)** | 1986 | India | | 1982 | 7 | 135 |
| **Rahman (71)** | 1989 | Bangladesh | early/late | 1985 | 7 | 69 |
| **Rajindrajith (72)** | 2009 | Sri Lanka | | 1999 | 6 | 17946 |
| ***RHS (31)** | 1997 | Honduras | | 1994 | 7 | 121 |

| Table S5: Details of the high mortality model input dataset | | | | | | |
|---|---|---|---|---|---|---|
| First author | Year published | Country | Additional strata | Median data year | # of causes reported | # of deaths used in analysis |
| *RHS (31) | 2002 | Honduras | | 1999 | 7 | 143 |
| Samms-Vaughan (73) | 1990 | Jamaica | | 1986 | 7 | 885 |
| Schumacher (74) | 2002 | Guinea | | 1998 | 8 | 94 |
| Settel (75) | 2004 | Tanzania | location | 2000 | 8 | 75 |
| Settel (75) | 2004 | Tanzania | location | 2000 | 8 | 119 |
| Settel (75) | 2004 | Tanzania | location | 2000 | 8 | 124 |
| Shah (76) | 2010 | India | | 2006 | 8 | 22 |
| Sharifzadeh (77) | 2008 | Iran | | 2005 | 7 | 87 |
| Shrivastava (78) | 2001 | India | early/late | 1995 | 8 | 895 |
| Singhal (79) | 1990 | India | early/late | 1984 | 7 | 50 |
| Sivagnanasundram (80) | 1985 | Sri Lanka | early/late | 1982 | 8 | 46 |
| Tikmani (81) | 2010 | Pakistan | | 2005 | 7 | 41 |
| *Turnbull (82) | 2011 | Zambia | | 2008 | 6 | 35 |
| Vaid (83) | 2007 | India | early/late | 2000 | 8 | 102 |
| Waiswa (84) | 2010 | Uganda | | 2007 | 7 | 58 |
| Walraven (85) | 2003 | Gambia | | 2000 | 6 | 70 |
| Woods (86) | 2001 | South Africa | | 2001 | 7 | 253 |
| Yassin (87) | 2000 | Egypt | | 1994 | 8 | 39 |

## Appendix G: Assumptions about missing causes in the high mortality model input data

Of the 112 observations in the high mortality model input data, 34 had no causes missing, 37 had one cause missing, 13 had two causes missing, 23 had three causes missing, and 5 had four causes missing. No observations were missing data for the "intrapartum" or "other" cause categories. For the other causes, the assumptions we made about the cause category into which any deaths from an unreported cause would have been assigned are shown in table S6.

| Table S6: Assumptions about missing causes in the high mortality model input data | | |
|---|---|---|
| | # of observations with this cause missing | If missing, assumed to be in… |
| Preterm | 1 | Other |
| Congenital | 13 | Other |
| Sepsis | 3 | Other |
| Pneumonia | 62 | Sepsis |
| Diarrhoea | 42 | Sepsis |
| Tetanus | 31 | Sepsis |
| Notes: 1) observations with sepsis missing were also missing all other infection categories (diarrhea, pneumonia, tetatanus) | | |

**Appendix H: Sensitivity analysis assuming 65% and 85% for early period proportion**

Based on previous work (88), we assumed that 74% and 26% of neonatal deaths occurred in the early and late periods, respectively, for countries without adequate VR data. The three-quarters/one-quarter split is quite consistent across countries, including ones with widely varying contexts (88). However, to test how this affected our results, we estimated the global cause-of-death distribution for the overall neonatal period if the proportion of early neonatal deaths was 65% or 85% instead of 74%. The results are shown in table S7.

| | Assuming 74% deaths in early neonatal period (current method)* | | Assuming 65% deaths in early neonatal period* | | Assuming 85% deaths in early neonatal period* | |
|---|---|---|---|---|---|---|
| | % | # of deaths in thousands (uncertainty) | % | # of deaths in thousands | % | # of deaths in thousands |
| Preterm | 35.7 | 986.9 (699.1-1312.5) | 34.0 | 939.8 | 37.8 | 1044.4 |
| Intrapartum | 23.4 | 644.8 (462.4-844.7) | 22.1 | 610.4 | 24.9 | 686.9 |
| Congenital | 10.5 | 289.8 (183.3-450.4) | 10.4 | 288.1 | 10.6 | 291.8 |
| Sepsis | 15.6 | 430.4 (218.9-664.8) | 18.2 | 501.5 | 12.4 | 343.6 |
| Pneumonia | 4.9 | 136.4 (70.3-259.0) | 5.0 | 137.0 | 4.9 | 135.8 |
| Diarrhoea | 0.6 | 16.6 (3.2-83.0) | 0.7 | 19.3 | 0.5 | 13.4 |
| Tetanus | 1.7 | 48.2 (15.5-120.4) | 2.0 | 55.0 | 1.4 | 39.8 |
| Other | 7.5 | 207.8 (99.0-367.4) | 7.6 | 209.8 | 7.4 | 205.4 |

Table S7: Global cause-specific proportions and numbers of neonatal deaths in 2013 assuming different proportions of early neonatal deaths

* this split is only applied to countries for which modelled estimates are needed – the split for countries with high-quality VR data is taken from the data as is. Additional notes: 1) diarrhoea and tetanus were estimated only for the 80 high mortality model countries; and 2) injuries are included within the "other" category.

**Appendix I: Comparison of input and prediction covariates in the multinomial models**

| Table S8a: Comparison of input and prediction covariates in the low and high mortality models | | | | | | |
|---|---|---|---|---|---|---|
| | Input data | | | Prediction data | | |
| | Mean (SD) | Median (IQR) | Range (min-max) | Mean (SD) | Median (IQR) | Range (min-max) |
| Low mortality model | | | | | | |
| NMR | 6.2 (4.3) | 4.7 (2.8-8.8) | 1.1-24.4 | 9.2 (4.5) | 9.1 (5.4-12.4) | 1.1-22.5 |
| IMR | 9.6 (7.2) | 7.0 (4.2-13.6) | 1.8-49.9 | 15.0 (7.7) | 15.1 (8.4-20.6) | 2.2-35.9 |
| U5MR | 11.4 (8.8) | 8.2 (5.0-15.6) | 2.3-76.9 | 17.8 (9.3) | 17.6 (9.9-24.3) | 3.0-44.8 |
| LBW | 7.3 (2.5) | 7.0 (5.7-8.2) | 3.8-23.2 | 8.7 (3.6) | 8.4 (6.3-10.2) | **0.0**-22.3 |
| GNI | 21026 (13292) | 18130 (11021-30090) | 1490-67970 | 20450 (38356) | 7302 (4450-19600) | **1400-327346** |
| GFR | 0.055 (0.018) | 0.050 (0.040-0.065) | 0.033-0.119 | 0.076 (0.029) | 0.075 (0.050-0.096) | **0.032-0.171** |
| GINI | 37.6 (9.7) | 34.3 (30.9-45.3) | 24.2-67.4 | 39.3 (7.7) | 39.4 (32.6-42.1) | 25.6-**69.2** |
| ANC | 96.7 (4.3) | 97.9 (96.3-99.4) | 71.3-100 | 93.6 (7.2) | 96.3 (90.3-98.9) | **52.9**-100.0 |
| DPT | 94.2 (5.6) | 96.0 (93.0-98.0) | 62.0-99.0 | 93.1 (8.6) | 96.0 (92.0-98.0) | **41.0**-99.0 |
| femlit | 93.0 (4.7) | 92.4 (90.7-97.0) | 70.3-99.8 | 87.8 (9.0) | 90.1 (81.9-93.5) | **50.6**-99.7 |
| High mortality model | | | | | | |
| NMR | 33.0 (15.9) | 30.2 (18.8-47.2) | 10.5-70.1 | 30.2 (10.2) | 29.7 (21.8-37.3) | **8.8**-55.1 |
| IMR | 62.1 (30.0) | 58.7 (35.7-81.6) | 14.7-142.0 | 60.3 (24.9) | 57.2 (40.0-76.5) | **14.4**-141.3 |
| U5MR | 88.6 (45.8) | 89.1 (52.9-125.4) | 17.1-227.0 | 89.2 (45.1) | 83.0 (52.0-116.7) | **16.3-231.5** |
| LBW | 18.9 (11.3) | 15.9 (10.6-27.6) | 2.5-50.0 | 14.2 (6.7) | 12.8 (10.2-16.7) | 4.2-35.6 |
| GFR | 0.127 (0.043) | 0.118 (0.092-0.158) | 0.057-0.235 | 0.140 (0.046) | 0.145 (0.101-0.175) | **0.054-0.246** |
| ANC | 67.5 (26.5) | 73.9 (50.0-92.0) | 5.0-98.3 | 78.6 (18.5) | 84.1 (71.0-92.7) | 16.1-**100.0** |
| DPT | 67.9 (24.6) | 73.5 (60.5-83.5) | 0.0-99.0 | 76.3 (18.5) | 80.0 (66.0-91.0) | 3.0-99.0 |
| BCG | 78.8 (23.9) | 87.0 (73.0-93.0) | 0.0-100.0 | 85.1 (14.2) | 90.0 (78.0-96.0) | 24.0-99.0 |
| PAB | 63.3 | 68.0 | 0.0-98.5 | 76.0 | 79.7 | 24.0-97.0 |
| | (25.0) | (51.4-83.5) | | (13.1) | (68.2-85.0) | |
| femlit | 51.9 (24.7) | 48.7 (34.6-77.3) | 4.0-94.0 | 62.2 (23.6) | 61.8 (43.9-83.6) | 9.4-**100.0** |
| SBA | 48.7 (33.1) | 45.3 (18.9-83.7) | 0.0-100.0 | 60.1 (25.3) | 60.2 (42.0-82.0) | 5.6-100.0 |
| Region | East Asia and Pacific: n = 4<br>Europe and Central Asia: n = 5<br>Latin America/Caribbean: n = 16<br>Middle East and North Africa: n = 8<br>South Asia: n = 52<br>Sub-Saharan Africa: n = 26<br>High income: n = 1 | | | East Asia and Pacific: n = 14<br>Europe and Central Asia: n = 6<br>Latin America/Caribbean: n = 4<br>Middle East and North Africa: n = 6<br>South Asia: n = 5<br>Sub-Saharan Africa: n = 44<br>High income: n = 0 | | |
| Notes: 1) bolded values are those outside the input data range; 2) these do not include Indian states (see table S8b for comparison of Indian state data); 3) SD = standard deviation, IQR = interquartile range; 4) region was only included as a covariate in the high mortality model; 5) covariate acronyms are as follows: NMR = neonatal mortality rate (per 1,000 live births); IMR = infant mortality rate (per 1,000 live births); U5MR = under-5 mortality rate (per 1,000 live births); LBW = low birth weight rate (%); GNI = gross national income; GFR = general fertility rate; GINI = gini coefficient; ANC = antenatal care (%); DPT = diphtheria/pertussis/tetanus vaccine (%); femlit = female literacy (%); BCG = Bacille Calmette-Guerin vaccine (%); PAB = Protected at birth (against neonatal tetanus) (%); SBA = skilled birth attendance (%). | | | | | | |

Table S8b table is similar the table above, but the prediction data is now for Indian states. In the analysis, we predicted the proportional cause-of-death distribution at the state-level for India and aggregated the estimates to develop national estimates.

| Table S8b: Comparison of high mortality model input covariates and covariates for Indian states | | | | | | |
|---|---|---|---|---|---|---|
| | High mortality model input data | | | Indian state-level prediction data | | |
| | Mean (SD) | Median (IQR) | Range (min-max) | Mean (SD) | Median (IQR) | Range (min-max) |
| NMR | 33.0 (15.9) | 30.2 (18.8-47.2) | 10.5-70.1 | 26.8 (11.6) | 26.8 (18.1-33.9) | **6.3**-59.9 |
| IMR | 62.1 (30.0) | 58.7 (35.7-81.6) | 14.7-142.0 | 40.2 (17.3) | 38.8 (27.1-52.2) | **9.2**-95.5 |
| U5MR | 88.6 (45.8) | 89.1 (52.9-125.4) | 17.1-227.0 | 54.9 (25.2) | 51.0 (36.1-72.5) | **11.8**-122.0 |
| LBW | 18.9 (11.3) | 15.9 (10.6-27.6) | 2.5-50.0 | 20.4 (8.4) | 21.0 (15.0-26.0) | **0.0**-43.0 |
| GFR | 0.127 (0.043) | 0.118 (0.092-0.158) | 0.057-0.235 | 0.089 (0.007) | 0.088 (0.083-0.096) | 0.080-0.103 |
| ANC | 67.5 (26.5) | 73.9 (50.0-92.0) | 5.0-98.3 | 77.8 (13.2) | 75.2 (70.1-88.8) | 33.1-**99.5** |
| DPT | 67.9 (24.6) | 73.5 (60.5-83.5) | 0.0-99.0 | 68.9 (14.4) | 70.6 (61.2-78.1) | 30.2-96.9 |
| BCG | 78.8 (23.9) | 87.0 (73.0-93.0) | 0.0-100.0 | 87.2 (10.9) | 89.7 (85.0-96.1) | 45.4-99.5 |
| PAB | 63.3 (25.0) | 68.0 (51.4-83.5) | 0.0-98.5 | 72.6 (15.5) | 76.9 (60.6-85.4) | 31.3-97.1 |
| **femlit** | 51.9 (24.7) | 48.7 (34.6-77.3) | 4.0-94.0 | 62.9 (14.7) | 63.8 (52.7-70.3) | 35.4-94.0 |
| **SBA** | 48.7 (33.1) | 45.3 (18.9-83.7) | 0.0-100.0 | 59.8 (16.8) | 59.4 (48.1-70.0) | 25.5-99.6 |

Notes: 1) bolded values are those outside the input data range; 2) the region covariate was "Southern Asia" for all states; 3) SD = standard deviation, IQR = interquartile range; 4) covariate acronyms are in the footnotes of table S5a.

## Appendix J: Sensitivity analysis – uncapped prediction covariate values

As noted in the paper, we capped the predication covariate values to the min/max of the input covariate values in both the low and high mortality models to avoid predicting on covariates outside fo the input data range. Capping the prediction data will tend to "shrink" the predicted cause distribution towards that seen in the input data. Here, we present the results of a sensitivity analysis in which we ran the low and high mortality models without the caps on the covariate values.

*Low mortality model*

**Table S9a: Low mortality model sensitivity analysis of uncapped prediction covariate values for 2013**

|  | Early | | | Late | | |
|---|---|---|---|---|---|---|
|  | % of deaths (capped) | % of deaths (uncapped) | Difference | % of deaths (capped) | % of deaths (uncapped) | Difference |
| Preterm | 43.0 | 44.2 | +1.2 | 27.9 | 27.7 | -0.2 |
| Intrapartum | 15.8 | 16.5 | +0.7 | 9.0 | 9.2 | +0.2 |
| Congenital | 21.8 | 19.1 | -2.7 | 27.6 | 27.4 | -0.2 |
| Sepsis | 5.8 | 6.1 | +0.3 | 17.6 | 17.9 | +0.3 |
| Pneumonia | 2.9 | 3.0 | +0.1 | 9.9 | 9.8 | -0.1 |
| Injuries | 0.7 | 0.7 | 0 | 1.5 | 1.5 | 0 |
| Other | 10.1 | 10.4 | +0.3 | 6.7 | 6.6 | -0.1 |

Notes: 1) the "capped" analysis is the main analysis presented in the paper; the "uncapped" is the sensitivity analysis.

Excluding countries with few deaths (<50), the key differences (greater than 5 percentage point difference) in countries were:
1) Egypt (both periods, 2000-2006) – intrapartum was 5-9 percentage points higher in the sensitivity analysis
2) Syria (xxx periods, 2012-2013) – sepsis was 9-11 percentage points higher and congenital was 6 percentage points lower in the sensitivity analysis
3) Honduras (early period, 2000) – injuries was 8 percentage points higher in the sensitivity analysis
4) Jordan (early period, 2000) – injuries was 5 percentage points higher in the sensitivity analysis

*High mortality model*

**Table S9b: High mortality model sensitivity analysis of uncapped prediction covariate values**

|  | Early | | | Late | | |
|---|---|---|---|---|---|---|
|  | % of deaths (capped) | % of deaths (uncapped) | Difference | % of deaths (capped) | % of deaths (uncapped) | Difference |
| Preterm | 40.6 | 40.6 | 0 | 20.3 | 20.3 | 0 |
| Intrapartum | 28.8 | 28.7 | -0.1 | 13.5 | 13.5 | 0 |
| Congenital | 9.0 | 9.0 | 0 | 7.6 | 7.6 | 0 |
| Sepsis | 8.3 | 8.3 | 0 | 40.0 | 40.1 | +0.1 |
| Pneumonia | 5.2 | 5.2 | 0 | 4.7 | 4.7 | 0 |
| Tetanus | 1.2 | 1.2 | 0 | 4.4 | 4.3 | -0.1 |
| Diarrhoea | 0.4 | 0.4 | 0 | 1.6 | 1.6 | 0 |
| Other | 6.7 | 6.7 | 0 | 7.9 | 7.9 | 0 |

Notes: 1) the "capped" analysis is the main analysis presented in the paper; the "uncapped" is the sensitivity analysis.

In the early period, both Niger (2000, "other") and Kazhakstan (2013, "congenital") were 2 percentage points higher in the sensitivity analysis. No other countries had more than a one percentage point difference between the sensitivity versus main analysis.

**Appendix K: Selected covariates in model equations and performance of the equations**

The following table lists the covariates that were selected for each of the four models, as well as the performance of each equation in reducing the out-of-sample residuals. The % reduction in residuals were calculated based on comparing the chi-squared statistic (sum of the squared differences between the observed and expected deaths divided by expected deaths) of the final equation compared to the null model with no covariates (see main paper for further methods). While for some causes our models seemed to explain the majority of variation in the input data, for others the models performed less well. The poorer performance for some causes may be due to a number of factors, including the limited range of covariates available for inclusion, inaccurate measurement of included covariates, or the possibility that there is no pattern that can be predicted based on the input data.

**Table S10: Selected covariates in equations of the low and high mortality models and % reduction in residuals**

| | Early neonatal period | | Late neonatal period | |
|---|---|---|---|---|
| | Selected covariates[1] | % reduction in residuals | Selected covariates[1] | % reduction in residuals |
| **Low mortality model** | | | | |
| Intrapartum: Preterm | L: femlit | 16% | S: femlit, DPT | 21% |
| Congenital: Preterm | L: GINI, DPT, femlit<br>S: IMR, U5MR, LBW | 62% | L: NMR, DPT | 10% |
| Sepsis: Preterm | L: GNI, GINI, ANC<br>S: IMR, DPT | 66% | L: femlit, GINI<br>Q: IMR<br>S: DPT | 67% |
| Pneumonia: Preterm | L: GNI | 25% | L: GNI<br>S: ANC | 40% |
| Injuries: Preterm | Q: GFR | 17% | none | 0% |
| Other: Preterm | Q: GFR | 19% | L: LBW<br>Q: DPT<br>S: NMR | 16% |
| **High mortality model** | | | | |
| Preterm: Intrapartum | L: BCG, PAB, SBA, DPT<br>S: LBW, GFR | 47% | S: LBW, PAB, GFR<br>B: reg_SSA | 61% |
| Congenital: Intrapartum | L: LBW<br>Q: NMR, U5MR<br>S: BCG<br>B: period | 77% | Q: SBA, U5MR<br>B: period, reg_SSA | 71% |
| Sepsis: Intrapartum | L: LBW<br>Q: BCG<br>B: period, reg_SA | 81% | Q: PAB<br>S: LBW<br>B: period | 13% |
| Pneumonia: Intrapartum | L: U5MR, LBW<br>B: period | 16% | Q: PAB | 23% |
| Diarrhoea: Intrapartum | L: DPT, GFR<br>Q: NMR<br>B: period, reg_SA, reg_SSA | 87% | L: DPT, BCG, GFR, femlit<br>S: LBW<br>B: reg_LAC | 45% |
| Tetanus: Intrapartum | L: PAB, ANC, NMR<br>B: period | 86% | L: NMR, IMR, U5MR, PAB<br>B: period | 44% |

| Other: Intrapartum | S: GFR<br>B: reg_SSA | 6% | S: GFR<br>B: period, reg_SSA | 49% |

Notes: 1) L = linear; Q = quadratic; S = restricted cubic spline; B = binary; 2) reg_SSA = sub-Saharan Africa, reg_SA = South Asia, reg_LAC = Latin America and the Caribbean; 3) see the "notes" at the end of table S7a for the other covariate acronym definitions.

**Appendix L: Regression coefficients for the low and high mortality models**

The following table includes the regression coefficients from the multinomial regressions for the two models used in this work.

| Table S11: Regression coefficients for the low and high mortality models | | |
|---|---|---|
| | **Early neonatal period** | **Late neonatal period** |
| | **Coveraites with regression coefficients** | **Covariate with regression coefficients** |
| **Low mortality model** | | |
| Intrapartum: Preterm | femlit (-0.018); const (0.572) | femlit_S1 (-0.030); femlit_S2 (-0.025); femlit_S3 (0.134); DPT_S1 (0.002); DPT_S2 (0.006); DPT_S3 (0.064); const (1.173) |
| Congenital: Preterm | GINI (-0.008); DPT (0.015); femlit (0.004); IMR_S1 (-1.669); IMR_S2 (11.454); IMR_S3 (-13.670); U5MR_S1 (1.417); U5MR_S2 (-9.443); U5MR_S3 (13.523); LBW_S1 (0.185); LBW_S2 (-1.133); LBW_S3 (2.726); const | NMR (-0.039); DPT (0.007); const (-0.307) |
| Sepsis: Preterm | GNI ($-2.6 \times 10^{-5}$); GINI (0.025); ANC (0.029); IMR_S1 (-0.274); IMR_S2 (2.647); IMR_S3 (-4.009); DPT_S1 (-0.013); DPT_S2 (0.034); DPT_S3 (-0.555); const (-3.267) | femlit (-0.009); GINI (0.031); DPT_S1 (-0.026); DPT_S2 (0.062); DPT_S3 (-1.125); IMR (0.049); IMR_Q (-0.001); const (0.643) |
| Pneumonia: Preterm | GNI ($-6.7 \times 10^{-5}$); const (-2.078) | GNI ($-4.8 \times 10^{-5}$); ANC_S1 (0.072); ANC_S2 (-0.205); ANC_S3 (2.250); const (-6.979) |
| Injuries: Preterm | GFR (-104.414); GFR_Q (763.714); const (-0.903) | const (-2.942) |
| Other: Preterm | GFR (103.863); GFR_Q (-908.192); const (-4.231) | LBW (0.029); NMR_S1 (-0.212); NMR_S2 (1.117); NMR_S3 (-1.514); DPT (-0.090); DPT_Q (0.001); const (2.433) |
| **High mortality model** | | |
| Preterm: Intrapartum | BCG (0.007); PAB (-0.007); SBA (0.010); DPT (-0.004); LBW_S1 (0.028); LBW_S2 (-0.006); GFR_S1 (-11.980); GFR_S2 (14.787); const (0.602) | LBW_S1 (0.038); LBW_S2 (-0.054); PAB_S1 (-0.011); PAB_S2 (0.000); GFR_S1 (-21.739); GFR_S2 (19.502); reg_SSA (0.305); const (2.764) |
| Congenital: Intrapartum | LBW (0.009); NMR (-0.073); NMR_Q (0.001); U5MR (-0.022); U5MR_Q (0.000); BCG_S1 (0.004); BCG_S2 (0.003); early (0.223); const (0.364) | SBA (-0.021); SBA_Q (0.000); U5MR (-0.021); U5MR_Q (0.000); late (0.739); reg_SSA (-0.186); const (-0.070) |
| Sepsis: Intrapartum | LBW (0.013); BCG (0.022); BCG_Q (0.000); early (-0.765); reg_SA (0.253); const (-2.303) | PAB (0.019); PAB_Q (0.000); LBW_S1 (0.023); LBW_S2 (-0.018); late (1.510); const (-1.708) |
| Pneumonia: Intrapartum | U5MR (0.006); LBW (0.009); early (-0.674); const (-1.713) | PAB (-0.029); PAB_Q (0.000); const (-0.092) |
| Diarrhoea: Intrapartum | DPT (-0.005); GFR (12.725); NMR (0.282); NMR_Q (-0.003); early (-1.725); reg_SA (-0.226); reg_SSA (-1.491); const (-8.746) | DPT (-0.003); BCG (-0.011); GFR (-3.800); femlit (-0.022); LBW_S1 (0.139); LBW_S2 (-0.192); const (-1.332) |
| Tetanus: Intrapartum | PAB (-0.011); ANC (-0.018); NMR (0.043); early (-0.985); const (-1.642) | NMR (0.038); IMR (-0.010); U5MR (0.011); PAB (-0.020); late (0.839); const (-2.186) |
| Other: Intrapartum | GFR_S1 (-22.942); GFR_S2 (33.572); reg_SSA (-0.746); const (0.824) | GFR_S1 (-25.195); GFR_S2 (25.625); late (0.474); reg_SSA (0.317); const (1.317) |

Notes: 1) reg_SSA = sub-Saharan Africa, reg_SA = South Asia, reg_LAC = Latin America and the Caribbean; 2) see the "notes" at the end of table 1 for covariate acronym definitions; 3) while most regression coefficients have been rounded to the third decimal place in this table, GNI has been rounded to the 6[th] decimal place because the GNI values are on the order of thousands instead of the much smaller values of the other coefficients.

**Appendix M: Results by NMR level and World Bank income category**

Table S12a: Cause-specific proportions (median with IQR) and risk (with uncertainty) for 2013 by NMR level

| | Median proportion (with IQR) by NMR level | | | | Total risk (with uncertainty) by NMR level | | | |
|---|---|---|---|---|---|---|---|---|
| | <5 | 5 to <15 | 15 to <30 | 30+ | <5 | 5 to <15 | 15 to <30 | 30+ |
| **Early neonatal period** | | | | | | | | |
| Preterm | 0.35 (0.29-0.45) | 0.31 (0.26-0.42) | 0.27 (0.15-0.35) | 0.25 (0.16-0.37) | 1.1 (1.0-1.2) | 2.9 (2.5-3.4) | 8.4 (5.4-11.8) | 11.0 (8.6-13.1) |
| Intrapartum | 0.10 (0.07-0.14) | 0.13 (0.09-0.18) | 0.18 (0.15-0.32) | 0.22 (0.15-0.34) | 0.3 (0.2-0.3) | 1.3 (1.1-1.6) | 5.4 (3.7-7.5) | 9.3 (7.3-10.9) |
| Congenital | 0.31 (0.24-0.38) | 0.24 (0.20-0.29) | 0.11 (0.08-0.14) | 0.05 (0.05-0.07) | 0.6 (0.5-0.7) | 1.5 (1.1-1.9) | 1.9 (1.1-3.1) | 1.7 (1.1-2.7) |
| Sepsis | 0.05 (0.02-0.11) | 0.10 (0.06-0.21) | 0.11 (0.09-0.43) | 0.23 (0.09-0.45) | 0.1 (0.0-0.1) | 0.5 (0.3-0.6) | 1.7 (0.5-3.0) | 2.4 (0.9-3.8) |
| Pneumonia | 0.01 (0.00-0.02) | 0.04 (0.02-0.07) | 0.05 (0.04-0.06) | 0.05 (0.05-0.07) | 0.0 (0.0-0.0) | 0.2 (0.1-0.3) | 0.9 (0.4-1.9) | 2.0 (1.1-3.7) |
| Tetanus | --- | 0.00 (0.00-0.00) | 0.01 (0.00-0.02) | 0.03 (0.01-0.06) | --- | 0.0 (0.0-0.0) | 0.2 (0.0-0.5) | 0.6 (0.2-1.3) |
| Diarrhoea | --- | 0.00 (0.00-0.00) | 0.00 (0.00-0.01) | 0.01 (0.00-0.01) | --- | 0.0 (0.0-0.0) | 0.0 (0.0-0.4) | 0.2 (0.0-1.7) |
| Other | 0.10 (0.06-0.14) | 0.08 (0.06-0.10) | 0.07 (0.05-0.09) | 0.06 (0.04-0.07) | 0.3 (0.3-0.4) | 0.7 (0.5-0.9) | 1.4 (0.5-2.6) | 1.6 (0.9-2.8) |
| *Total* | --- | --- | --- | --- | 2.4 (2.0-2.7) | 7.1 (5.6-8.7) | 19.9 (11.6-30.8) | 28.8 (20.1-40.0) |
| **Late neonatal period** | | | | | | | | |
| Preterm | 0.42 (0.32-0.50) | 0.42 (0.38-0.46) | 0.35 (0.33-0.41) | 0.37 (0.33-0.40) | 0.2 (0.2-0.3) | 0.6 (0.5-0.8) | 1.6 (0.8-2.6) | 1.7 (1.1-2.4) |
| Intrapartum | 0.14 (0.11-0.17) | 0.18 (0.15-0.22) | 0.32 (0.26-0.35) | 0.34 (0.33-0.36) | 0.1 (0.0-0.1) | 0.2 (0.2-0.3) | 0.9 (0.5-1.4) | 1.5 (1.0-2.1) |
| Congenital | 0.26 (0.21-0.31) | 0.22 (0.18-0.25) | 0.12 (0.09-0.15) | 0.06 (0.05-0.07) | 0.3 (0.2-0.3) | 0.6 (0.5-0.7) | 0.6 (0.2-1.2) | 0.5 (0.2-1.0) |
| Sepsis | 0.03 (0.02- | 0.07 (0.05- | 0.09 (0.08- | 0.09 (0.07- | 0.1 (0.1- | 0.6 (0.4- | 2.7 (1.5-4.3) | 4.3 (2.7-5.8) |
| | 0.05) | 0.08) | 0.10) | 0.10) | 0.1) | 0.8) | | |
| Pneumonia | 0.00 (0.00-0.02) | 0.03 (0.02-0.04) | 0.05 (0.04-0.06) | 0.07 (0.06-0.08) | 0.0 (0.0-0.0) | 0.2 (0.1-0.3) | 0.3 (0.2-0.5) | 0.5 (0.3-0.8) |
| Tetanus | --- | 0.00 (0.00-0.00) | 0.00 (0.00-0.01) | 0.01 (0.01-0.02) | --- | 0.0 (0.0-0.0) | 0.2 (0.1-0.6) | 0.7 (0.2-1.6) |
| Diarrhoea | --- | 0.00 (0.00-0.00) | 0.00 (0.00-0.00) | 0.00 (0.00-0.01) | --- | 0.0 (0.0-0.0) | 0.1 (0.0-0.4) | 0.2 (0.1-0.3) |
| Other | 0.11 (0.06-0.14) | 0.08 (0.05-0.11) | 0.06 (0.04-0.09) | 0.04 (0.03-0.07) | 0.1 (0.0-0.1) | 0.2 (0.1-0.3) | 0.6 (0.2-1.3) | 0.7 (0.3-1.5) |
| *Total* | --- | --- | --- | --- | 0.8 (0.5-0.9) | 2.4 (1.8-3.2) | 7.0 (3.5-12.3) | 10.1 (5.9-15.5) |
| **Overall neonatal period** | | | | | | | | |
| Preterm | 0.32 (0.27-0.40) | 0.27 (0.19-0.30) | 0.15 (0.14-0.18) | 0.16 (0.15-0.18) | 1.3 (1.2-1.4) | 3.5 (3.0-4.1) | 9.9 (6.2-14.4) | 12.7 (9.6-15.6) |
| Intrapartum | 0.07 (0.04-0.09) | 0.09 (0.07-0.11) | 0.15 (0.12-0.16) | 0.15 (0.14-0.16) | 0.3 (0.2-0.4) | 1.6 (1.2-2.0) | 6.3 (4.2-8.8) | 10.8 (8.3-13.0) |
| Congenital | 0.36 (0.30-0.39) | 0.27 (0.23-0.32) | 0.09 (0.07-0.13) | 0.05 (0.04-0.06) | 0.8 (0.7-1.0) | 2.1 (1.5-2.7) | 2.5 (1.3-4.2) | 2.1 (1.3-3.7) |
| Sepsis | 0.10 (0.07-0.13) | 0.21 (0.16-0.28) | 0.43 (0.38-0.47) | 0.45 (0.40-0.46) | 0.2 (0.1-0.2) | 1.1 (0.7-1.4) | 4.4 (2.0-7.3) | 6.7 (3.6-9.7) |
| Pneumonia | 0.01 (0.00-0.03) | 0.06 (0.04-0.09) | 0.05 (0.05-0.06) | 0.05 (0.05-0.05) | 0.0 (0.0-0.1) | 0.4 (0.3-0.6) | 1.2 (0.5-2.5) | 2.5 (1.4-4.5) |
| Tetanus | --- | 0.00 (0.00-0.00) | 0.02 (0.01-0.03) | 0.05 (0.04-0.09) | --- | 0.0 (0.0-0.0) | 0.4 (0.1-1.1) | 1.3 (0.4-2.9) |
| Diarrhoea | --- | 0.00 (0.00-0.00) | 0.01 (0.00-0.01) | 0.01 (0.01-0.02) | --- | 0.0 (0.0-0.0) | 0.2 (0.0-0.8) | 0.3 (0.1-2.0) |
| Other | 0.10 (0.06-0.13) | 0.09 (0.07-0.10) | 0.07 (0.06-0.09) | 0.06 (0.06-0.07) | 0.4 (0.3-0.4) | 1.0 (0.6-1.3) | 2.0 (0.7-3.9) | 2.3 (1.2-4.3) |
| *Total* | --- | --- | --- | --- | 3.0 (2.5-3.5) | 9.7 (7.3-12.1) | 26.9 (15.0-43.0) | 38.7 (25.9-55.7) |

Notes: 1) no high mortality model countries had an NMR<5 so there were no tetanus/diarrhoea estimates for this NMR category; 2) injuries are included in the "other" category; 3) risk uncertainty estimates assume no uncertainty in live birth estimates.

| Table S12b: Cause-specific proportions (median with IQR) and risk (with uncertainty) for 2012 by World Bank income level | | | | | | | | |
|---|---|---|---|---|---|---|---|---|
| | Median proportion (with IQR) by income level | | | | Total risk (with uncertainty) by income level | | | |
| | **High** | **Upper-middle** | **Lower-middle** | **Low** | **High** | **Upper-middle** | **Lower-middle** | **Low** |
| Early neonatal period | | | | | | | | |
| Preterm | 0.42 (0.31-0.46) | 0.42 (0.38-0.46) | 0.39 (0.34-0.42) | 0.34 (0.31-0.38) | 1.2 (1.1-1.3) | 2.9 (2.5-3.3) | 8.8 (6.0-11.8) | 7.3 (5.5-9.3) |
| Intrapartum | 0.14 (0.11-0.19) | 0.17 (0.14-0.23) | 0.26 (0.19-0.34) | 0.34 (0.31-0.37) | 0.3 (0.3-0.4) | 1.4 (1.1-1.7) | 5.2 (3.7-7.1) | 7.0 (5.4-8.4) |
| Congenital | 0.27 (0.22-0.31) | 0.22 (0.14-0.25) | 0.15 (0.09-0.19) | 0.09 (0.07-0.13) | 0.7 (0.6-0.8) | 1.4 (1.0-1.8) | 1.8 (1.0-2.9) | 1.9 (1.3-2.9) |
| Sepsis | 0.03 (0.02-0.05) | 0.06 (0.05-0.08) | 0.08 (0.07-0.09) | 0.09 (0.08-0.10) | 0.1 (0.1-0.1) | 0.5 (0.3-0.7) | 1.6 (0.5-2.9) | 1.8 (0.6-2.8) |
| Pneumonia | 0.00 (0.00-0.01) | 0.03 (0.02-0.04) | 0.04 (0.04-0.05) | 0.06 (0.05-0.07) | 0.0 (0.0-0.0) | 0.3 (0.2-0.4) | 0.9 (0.4-2.0) | 1.3 (0.6-2.4) |
| Tetanus | 0.00 (0.00-0.00) | 0.00 (0.00-0.00) | 0.00 (0.00-0.01) | 0.01 (0.01-0.01) | 0.0 (0.0-0.0) | 0.0 (0.0-0.1) | 0.2 (0.1-0.6) | 0.3 (0.1-0.7) |
| Diarrhoea | 0.00 (0.00-0.00) | 0.00 (0.00-0.00) | 0.00 (0.00-0.00) | 0.00 (0.00-0.00) | 0.0 (0.0-0.0) | 0.0 (0.0-0.1) | 0.1 (0.0-0.6) | 0.1 (0.0-0.8) |
| Other | 0.11 (0.06-0.17) | 0.08 (0.05-0.11) | 0.07 (0.04-0.10) | 0.05 (0.04-0.08) | 0.3 (0.3-0.4) | 0.7 (0.5-0.9) | 1.4 (0.5-2.4) | 1.3 (0.6-2.5) |
| *Total* | --- | --- | --- | --- | 2.6 (2.4-3.0) | 7.2 (5.6-9.0) | 20.0 (12.2-30.3) | 21.0 (14.1-29.8) |
| Late neonatal period | | | | | | | | |
| Preterm | 0.30 (0.23-0.37) | 0.27 (0.21-0.30) | 0.18 (0.15-0.27) | 0.15 (0.14-0.18) | 0.3 (0.2-0.3) | 0.6 (0.5-0.7) | 1.6 (0.9-2.5) | 1.3 (0.7-1.9) |
| Intrapartum | 0.07 (0.04-0.10) | 0.09 (0.07-0.11) | 0.13 (0.10-0.15) | 0.15 (0.14-0.16) | 0.1 (0.0-0.1) | 0.2 (0.2-0.3) | 0.9 (0.5-1.4) | 1.1 (0.7-1.5) |
| Congenital | 0.37 (0.32-0.39) | 0.27 (0.22-0.31) | 0.14 (0.08-0.23) | 0.06 (0.05-0.08) | 0.3 (0.3-0.3) | 0.6 (0.5-0.7) | 0.6 (0.3-1.1) | 0.5 (0.2-1.1) |
| Sepsis | 0.11 (0.08-0.15) | 0.20 (0.15-0.32) | 0.38 (0.24-0.46) | 0.45 (0.40-0.47) | 0.1 (0.1-0.2) | 0.6 (0.4-0.8) | 2.7 (1.5-4.1) | 3.2 (1.9-4.5) |
| Pneumonia | 0.01 (0.00-0.02) | 0.05 (0.04-0.08) | 0.05 (0.05-0.07) | 0.05 (0.05-0.05) | 0.0 (0.0-0.0) | 0.2 (0.1-0.3) | 0.3 (0.2-0.5) | 0.4 (0.2-0.6) |
| Tetanus | 0.00 (0.00-0.00) | 0.00 (0.00-0.00) | 0.01 (0.00-0.03) | 0.03 (0.02-0.05) | 0.0 (0.0-0.0) | 0.0 (0.0-0.1) | 0.3 (0.1-0.7) | 0.3 (0.1-0.7) |
| Diarrhoea | 0.00 (0.00-0.00) | 0.00 (0.00-0.00) | 0.00 (0.00-0.01) | 0.01 (0.01-0.02) | 0.0 (0.0-0.0) | 0.0 (0.0-0.0) | 0.1 (0.0-0.3) | 0.1 (0.0-0.3) |
| Other | 0.10 (0.06-0.14) | 0.09 (0.07-0.10) | 0.08 (0.07-0.09) | 0.07 (0.06-0.07) | 0.1 (0.1-0.1) | 0.2 (0.1-0.3) | 0.6 (0.2-1.2) | 0.5 (0.2-1.3) |
| *Total* | --- | --- | --- | --- | 0.9 (0.7-1.0) | 2.4 (1.8-3.2) | 7.1 (3.7-11.8) | 7.4 (4.0-11.9) |
| **Overall neonatal period** | | | | | | | | |
| Preterm | 0.33 (0.28-0.45) | 0.34 (0.27-0.43) | 0.30 (0.18-0.39) | 0.25 (0.15-0.34) | 1.4 (1.3-1.6) | 3.5 (3.0-4.1) | 10.4 (6.9-14.3) | 8.5 (6.2-11.2) |
| Intrapartum | 0.10 (0.07-0.15) | 0.13 (0.09-0.17) | 0.16 (0.13-0.26) | 0.19 (0.15-0.34) | 0.4 (0.3-0.5) | 1.7 (1.3-2.0) | 6.1 (4.2-8.5) | 8.1 (6.1-9.9) |
| Congenital | 0.31 (0.24-0.38) | 0.24 (0.16-0.28) | 0.15 (0.09-0.22) | 0.07 (0.06-0.11) | 1.0 (0.8-1.2) | 2.0 (1.5-2.5) | 2.3 (1.3-4.0) | 2.4 (1.5-3.9) |
| Sepsis | 0.06 (0.02-0.12) | 0.09 (0.06-0.20) | 0.11 (0.08-0.38) | 0.19 (0.09-0.45) | 0.2 (0.1-0.3) | 1.1 (0.8-1.5) | 4.3 (2.0-7.0) | 5.0 (2.6-7.3) |
| Pneumonia | 0.00 (0.00-0.02) | 0.04 (0.03-0.05) | 0.05 (0.04-0.06) | 0.05 (0.05-0.06) | 0.0 (0.0-0.1) | 0.5 (0.3-0.7) | 1.3 (0.6-2.6) | 1.6 (0.9-3.0) |
| Tetanus | 0.00 (0.00-0.00) | 0.00 (0.00-0.00) | 0.00 (0.00-0.01) | 0.02 (0.01-0.03) | 0.0 (0.0-0.0) | 0.1 (0.0-0.2) | 0.5 (0.2-1.3) | 0.6 (0.2-1.4) |
| Diarrhoea | 0.00 (0.00-0.00) | 0.00 (0.00-0.00) | 0.00 (0.00-0.01) | 0.00 (0.00-0.01) | 0.0 (0.0-0.0) | 0.0 (0.0-0.1) | 0.2 (0.0-0.9) | 0.2 (0.0-1.1) |
| Other | 0.10 (0.06-0.14) | 0.08 (0.06-0.11) | 0.07 (0.05-0.09) | 0.06 (0.05-0.07) | 0.4 (0.3-0.5) | 1.0 (0.6-1.3) | 2.0 (0.8-3.6) | 1.9 (0.8-3.8) |
| *Total* | --- | --- | --- | --- | 3.4 (2.8-4.2) | 9.9 (7.5-12.4) | 27.1 (16.0-42.2) | 28.3 (18.3-41.6) |

Notes: 1) injuries are included in the "other" category; 2) risk uncertainty estimates assume no uncertainty in live birth estimates.

**Appendix N: Cause-specific results by estimation method**

Here, we present the results by estimation method (i.e. high-quality VR data, low mortality model, and high mortality model). Table S14 includes the cause-specific proprtions, risks, and numbers of deaths by estimation method, and Figures S1a-c show the 2000-2013 time trends in cause-specific risks by neonatal period for each estimation method.

| Table S14: Cause-specific proportions, risks, and numbers of deaths (with uncertainty) in 2013 by estimation method | | | | | | |
|---|---|---|---|---|---|---|
| | **Early period** | | | **Late period** | | |
| | % | # of deaths in 1000s (uncertainty) | Risk | % | # of deaths in 1000s (uncertainty) | Risk |
| **High-quality VR countries** | | | | | | |
| Preterm | 41.5 | 35.9 (34.0-37.7) | 1.7 | 25.9 | 7.3 (6.4-8.2) | 0.3 |
| Intrapartum | 16.1 | 13.9 (12.8-15.0) | 0.7 | 7.5 | 2.1 (1.7-2.6) | 0.1 |
| Congenital | 20.2 | 17.5 (16.1-18.8) | 0.8 | 26.4 | 7.5 (6.6-8.4) | 0.4 |
| Sepsis | 7.3 | 6.3 (5.6-7.0) | 0.3 | 21.3 | 6.0 (5.4-6.7) | 0.3 |
| Pneumonia | 2.2 | 1.9 (1.5-2.2) | 0.1 | 6.8 | 1.9 (1.6-2.3) | 0.1 |
| Injuries | 0.5 | 0.4 (0.3-0.6) | <0.05 | 1.6 | 0.4 (0.3-0.6) | <0.05 |
| Other | 12.3 | 10.6 (9.7-11.5) | 0.5 | 10.5 | 3.0 (2.5-3.5) | 0.1 |
| *Total* | *100* | *86.4* | *4.1* | *100* | *28.2* | *1.3* |
| **Low mortality model countries** | | | | | | |
| Preterm | 43.0 | 80.8 (68.8-91.8) | 2.6 | 27.9 | 18.4 (16.0-22.2) | 0.6 |
| Intrapartum | 15.8 | 29.7 (22.3-36.5) | 0.9 | 9.0 | 5.9 (3.4-8.1) | 0.2 |
| Congenital | 21.8 | 41.0 (30.3-57.1) | 1.3 | 27.6 | 18.2 (15.9-21.6) | 0.6 |
| Sepsis | 5.8 | 11.0 (7.0-14.5) | 0.4 | 17.6 | 11.6 (7.6-14.8) | 0.4 |
| Pneumonia | 2.9 | 5.4 (3.5-8.3) | 0.2 | 9.9 | 6.5 (4.1-9.7) | 0.2 |
| Injuries | 0.7 | 1.2 (0.9-1.8) | <0.05 | 1.5 | 1.0 (0.8-1.3) | <0.05 |
| Other | 10.1 | 19.0 (13.8-23.3) | 0.6 | 6.7 | 4.4 (2.6-7.4) | 0.1 |
| *Total* | *100* | *188.1* | *6.0* | *100* | *66.1* | *2.1* |
| **High mortality model countries** | | | | | | |
| Preterm | 40.6 | 718.1 (505.3-954.0) | 8.4 | 20.3 | 126.4 (68.6-198.6) | 1.5 |
| Intrapartum | 28.8 | 509.2 (372.6-659.9) | 6.0 | 13.5 | 84.0 (49.7-122.7) | 1.0 |
| Congenital | 9.0 | 158.6 (94.5-250.0) | 1.9 | 7.6 | 47.1 (20.0-94.6) | 0.6 |
| Sepsis | 8.3 | 146.5 (49.7-250.2) | 1.7 | 40.0 | 249.0 (143.6-371.7) | 2.9 |
| Pneumonia | 5.2 | 91.5 (43.8-189.7) | 1.1 | 4.7 | 29.2 (15.8-46.7) | 0.3 |
| Tetanus | 1.2 | 21.1 (7.4-53.2) | 0.2 | 4.4 | 27.1 (8.1-67.2) | 0.3 |
| Diarrhoea | 0.4 | 6.7 (0-57.4) | 0.1 | 1.6 | 10.0 (3.2-25.6) | 0.1 |
| Other | 6.7 | 118.6 (48.1-213.1) | 1.4 | 7.9 | 49.1 (20.2-104.5) | 0.6 |
| *Total* | *100* | *1770.2* | *20.8* | *100* | *622.0* | *7.3* |
| Notes: 1) risk is per 1,000 live births | | | | | | |

**Figure S1: Cause-specific risk from 2000-2013 by neonatal period for:** a) high-quality VR countries, b) low mortality model countries, and c) high mortality model countries. *Important note*: the y-axes (risk) are different on the 3 graphs due to the very different mortality risks between these estimation categories.

**a)**

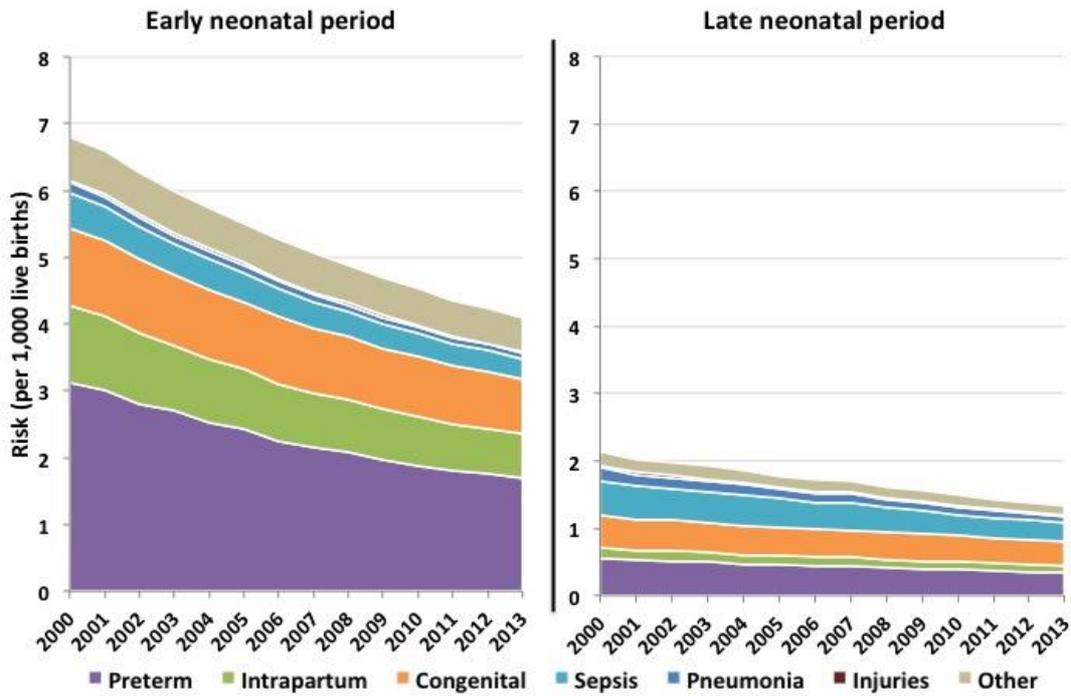

b)

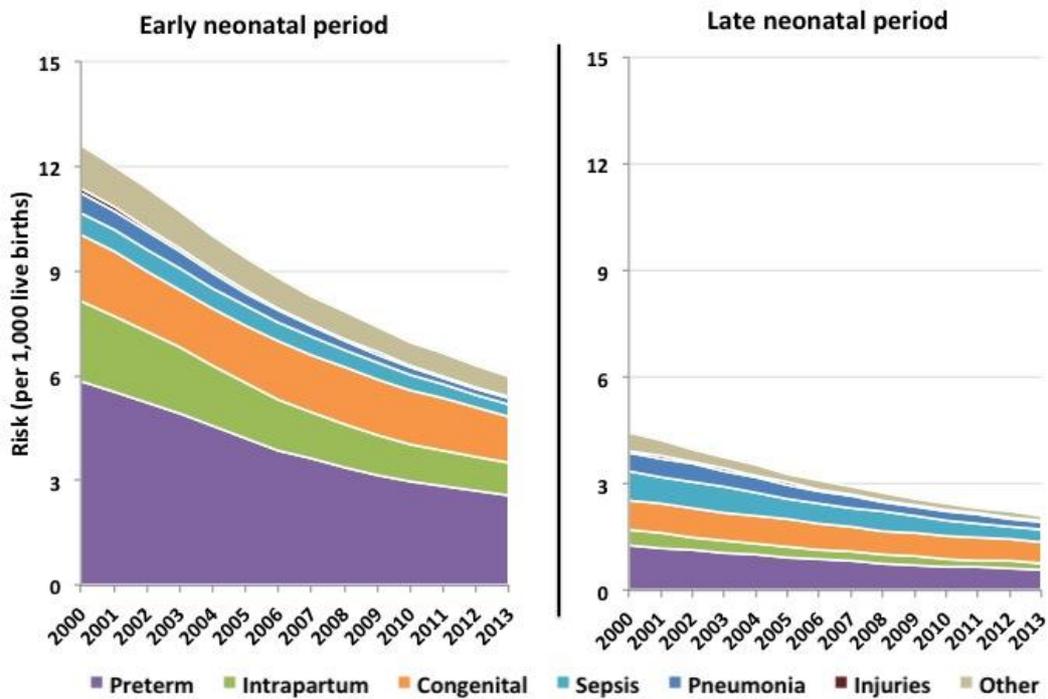

c)

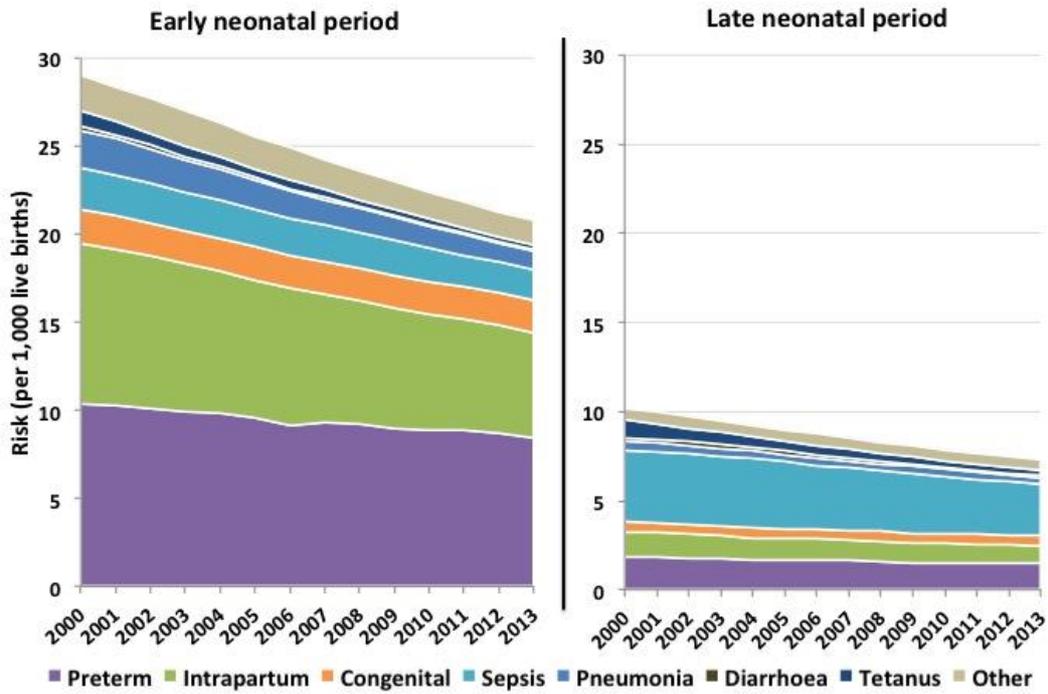

**Appendix O: Country-specific proportions, risks, and number of deaths for 2013**

The following pages (pages 19-83) contain estimates of the proportions, risks, and numbers of deaths in 2013 for 194 countries. The estimation technique (high-quality VR, low mortality model, and high mortality model) is listed below the country name in the left-most column. In the table, the low mortality model is called "low mort model" and the high mortality model is called "high mort model". The estimates are given by the early (days 0-6), late (days 7-27) and overall (days 0-27) neonatal periods. As noted in the paper, injuries were only estimated for high-quality VR and low mortality model countries, while diarrhoea and tetanus were only estimated for high mortality model countries.

| Table S14: Cause-specific proportions, risks, and numbers of deaths (with uncertainty) in 2013 for 194 countries by neonatal period ||||||||||||
| Country | Period | Stat* | Preterm | Intrapartum | Congenital | Sepsis | Pneumonia | Tetanus | Diarrhoea | Injuries | Other | Total |
|---|---|---|---|---|---|---|---|---|---|---|---|---|
| **Afghanistan** (high mort model) | Early | prop | 0.33 | 0.33 | 0.06 | 0.1 | 0.07 | 0.02 | 0.01 | --- | 0.08 | 1 |
| | | risk | 8.8 | 8.9 | 1.6 | 2.7 | 1.8 | 0.6 | 0.4 | --- | 2 | 26.9 |
| | | num | 89.3 | 90.5 | 16.3 | 26.9 | 18 | 6.5 | 4 | --- | 20.6 | 272.1 |
| | | | (67.7-117.4) | (69.1-114.5) | (11.7-27.2) | (8.7-45.3) | (9.7-34.3) | (2.8-15.2) | (0.0-39.2) | --- | (10.1-31.0) | |
| | Late | prop | 0.15 | 0.16 | 0.05 | 0.46 | 0.05 | 0.06 | 0.02 | --- | 0.05 | 1 |
| | | risk | 1.4 | 1.5 | 0.5 | 4.4 | 0.5 | 0.6 | 0.2 | --- | 0.4 | 9.4 |
| | | num | 14.2 | 15 | 5 | 44.4 | 5.1 | 5.9 | 1.6 | --- | 4.5 | 95.6 |
| | | | (8.1-24.2) | (9.4-22.2) | (2.7-10.6) | (27.5-65.8) | (2.9-8.1) | (2.1-13.5) | (0.5-3.4) | --- | (1.8-13.7) | |
| | Overall | prop | 0.29 | 0.28 | 0.06 | 0.19 | 0.06 | 0.04 | 0.02 | 0 | 0.07 | 1 |
| | | risk | 10.7 | 10.4 | 2.1 | 6.9 | 2.3 | 1.4 | 0.6 | 0 | 2.5 | 36.9 |
| | | num | 110.6 | 108.1 | 21.4 | 71.9 | 24 | 14.3 | 5.9 | 0 | 25.7 | 381.9 |
| | | | (82.8-149.3) | (81.3-140.0) | (14.4-38.1) | (35.9-108.5) | (13.2-43.9) | (5.4-32.7) | (0.5-42.6) | (0.0-0.0) | (12.2-45.9) | |
| **Albania** (low mort model) | Early | prop | 0.39 | 0.13 | 0.28 | 0.06 | 0.03 | --- | --- | 0.01 | 0.11 | 1 |
| | | risk | 2.2 | 0.7 | 1.5 | 0.3 | 0.1 | --- | --- | 0 | 0.6 | 5.5 |
| | | num | 0.9 | 0.3 | 0.7 | 0.1 | 0.1 | --- | --- | 0 | 0.3 | 2.4 |
| | | | (0.8-1.2) | (0.3-0.4) | (0.5-0.9) | (0.1-0.2) | (0.1-0.1) | --- | --- | (0.0-0.0) | (0.2-0.3) | |
| | Late | prop | 0.3 | 0.08 | 0.32 | 0.14 | 0.08 | --- | --- | 0.02 | 0.07 | 1 |
| | | risk | 0.6 | 0.2 | 0.6 | 0.3 | 0.1 | --- | --- | 0 | 0.1 | 1.9 |
| | | num | 0.2 | 0.1 | 0.3 | 0.1 | 0.1 | --- | --- | 0 | 0.1 | 0.8 |
| | | | (0.2-0.3) | (0.1-0.1) | (0.2-0.3) | (0.1-0.2) | (0.0-0.1) | --- | --- | (0.0-0.0) | (0.0-0.1) | |
| | Overall | prop | 0.37 | 0.12 | 0.29 | 0.08 | 0.04 | 0 | 0 | 0.01 | 0.1 | 1 |
| | | risk | 2.9 | 0.9 | 2.2 | 0.6 | 0.3 | 0 | 0 | 0.1 | 0.7 | 7.7 |
| | | num | 1.2 | 0.4 | 0.9 | 0.3 | 0.1 | 0 | 0 | 0 | 0.3 | 3.3 |
| | | | (1.1-1.6) | (0.3-0.5) | (0.8-1.3) | (0.2-0.4) | (0.1-0.2) | (0.0-0.0) | (0.0-0.0) | (0.0-0.0) | (0.2-0.5) | |
| **Algeria** (high mort model) | Early | prop | 0.36 | 0.25 | 0.22 | 0.07 | 0.03 | 0 | 0 | --- | 0.08 | 1 |
| | | risk | 3.7 | 2.6 | 2.2 | 0.7 | 0.3 | 0 | 0 | --- | 0.8 | 10.3 |
| | | num | 36 | 25.1 | 22.1 | 6.7 | 2.8 | 0.2 | 0 | --- | 8 | 101 |
| | | | (24.2-44.7) | (16.9-30.9) | (11.7-28.6) | (2.1-10.3) | (1.1-5.9) | (0.1-0.6) | (0.0-0.2) | --- | (3.3-10.4) | |
| | Late | prop | 0.14 | 0.12 | 0.27 | 0.32 | 0.04 | 0.01 | 0 | --- | 0.08 | 1 |
| | | risk | 0.5 | 0.4 | 1 | 1.2 | 0.2 | 0 | 0 | --- | 0.3 | 3.6 |
| | | num | 5 | 4.4 | 9.6 | 11.5 | 1.6 | 0.3 | 0.1 | --- | 3 | 35.5 |
| | | | (2.7-7.5) | (2.5-6.0) | (4.4-13.3) | (5.4-16.6) | (0.9-2.4) | (0.1-0.9) | (0.0-0.2) | --- | (1.5-4.6) | |
| | Overall | prop | 0.3 | 0.22 | 0.23 | 0.13 | 0.03 | 0 | 0 | 0 | 0.08 | 1 |
| | | risk | 4.3 | 3.1 | 3.2 | 1.9 | 0.5 | 0.1 | 0 | 0 | 1.1 | 14.1 |
| | | num | 42.5 | 30.6 | 32.3 | 18.8 | 4.6 | 0.6 | 0.2 | 0 | 11.4 | 141 |
| | | | (27.7-54.1) | (20.0-37.9) | (16.9-42.2) | (7.9-27.4) | (2.0-8.9) | (0.2-1.6) | (0.0-0.5) | (0.0-0.0) | (4.9-15.8) | |

* prop = proportion; num = number of deaths (in 100s).

| Table S14: Cause-specific proportions, risks, and numbers of deaths (with uncertainty) for 194 countries by neonatal period | | | | | | | | | | | |
|---|---|---|---|---|---|---|---|---|---|---|---|
| Country | Period | Stat* | Preterm | Intrapartum | Congenital | Sepsis | Pneumonia | Tetanus | Diarrhoea | Injuries | Other | Total |
| **Andorra** (low mort model) | Early | prop | 0.38 | 0.13 | 0.37 | 0.02 | 0 | --- | --- | 0.01 | 0.1 | 1 |
| | | risk | 0.4 | 0.1 | 0.4 | 0 | 0 | --- | --- | 0 | 0.1 | 1 |
| | | num | 0 | 0 | 0 | 0 | 0 | --- | --- | 0 | 0 | 0 |
| | | | (0.0-0.0) | (0.0-0.0) | (0.0-0.0) | (0.0-0.0) | (0.0-0.0) | --- | --- | (0.0-0.0) | (0.0-0.0) | |
| | Late | prop | 0.3 | 0.08 | 0.39 | 0.1 | 0 | --- | --- | 0.02 | 0.11 | 1 |
| | | risk | 0.1 | 0 | 0.1 | 0 | 0 | --- | --- | 0 | 0 | 0.4 |
| | | num | 0 | 0 | 0 | 0 | 0 | --- | --- | 0 | 0 | 0 |
| | | | (0.0-0.0) | (0.0-0.0) | (0.0-0.0) | (0.0-0.0) | (0.0-0.0) | --- | --- | (0.0-0.0) | (0.0-0.0) | |
| | Overall | prop | 0.36 | 0.11 | 0.38 | 0.04 | 0 | 0 | 0 | 0.01 | 0.1 | 1 |
| | | risk | 0.5 | 0.2 | 0.6 | 0.1 | 0 | 0 | 0 | 0 | 0.2 | 1.5 |
| | | num | 0 | 0 | 0 | 0 | 0 | 0 | 0 | 0 | 0 | 0 |
| | | | (0.0-0.0) | (0.0-0.0) | (0.0-0.0) | (0.0-0.0) | (0.0-0.0) | (0.0-0.0) | (0.0-0.0) | (0.0-0.0) | (0.0-0.0) | |
| **Angola** (high mort model) | Early | prop | 0.33 | 0.34 | 0.06 | 0.08 | 0.1 | 0.02 | 0.01 | --- | 0.07 | 1 |
| | | risk | 11.5 | 11.6 | 1.9 | 2.8 | 3.3 | 0.7 | 0.2 | --- | 2.3 | 34.5 |
| | | num | 105.3 | 106.5 | 17.7 | 25.7 | 30.6 | 6 | 2.2 | --- | 21.4 | 315.4 |
| | | | (73.9-140.0) | (77.2-131.3) | (9.4-30.2) | (9.9-41.9) | (15.9-60.0) | (2.0-15.8) | (0.0-21.1) | --- | (10.8-38.4) | |
| | Late | prop | 0.15 | 0.15 | 0.04 | 0.42 | 0.05 | 0.11 | 0.01 | --- | 0.07 | 1 |
| | | risk | 1.9 | 1.8 | 0.5 | 5.1 | 0.6 | 1.3 | 0.1 | --- | 0.9 | 12.1 |
| | | num | 17 | 16.5 | 4.2 | 46.4 | 5.6 | 12.3 | 0.9 | --- | 7.9 | 110.8 |
| | | | (9.4-24.6) | (9.9-22.9) | (1.6-9.1) | (27.0-67.3) | (3.2-8.6) | (2.2-32.4) | (0.2-2.6) | --- | (3.4-19.8) | |
| | Overall | prop | 0.29 | 0.28 | 0.05 | 0.17 | 0.09 | 0.05 | 0.01 | 0 | 0.07 | 1 |
| | | risk | 13.9 | 13.4 | 2.5 | 7.9 | 4 | 2.2 | 0.3 | 0 | 3.2 | 47.4 |
| | | num | 126.2 | 121.7 | 22.7 | 71.7 | 36.7 | 20.3 | 3 | 0 | 29 | 431.4 |
| | | | (88.0-168.3) | (88.8-155.4) | (10.9-42.4) | (36.4-108.7) | (18.7-71.4) | (4.5-53.1) | (0.2-22.6) | (0.0-0.0) | (14.4-57.6) | |
| **Antigua and Barbuda** (high-quality VR) | Early | prop | 0 | 0 | 0 | 0 | 0 | --- | --- | 0 | 1 | 1 |
| | | risk | 0 | 0 | 0 | 0 | 0 | --- | --- | 0 | 1.3 | 1.3 |
| | | num | 0 | 0 | 0 | 0 | 0 | --- | --- | 0 | 0 | 0 |
| | | | (0.0-0.0) | (0.0-0.0) | (0.0-0.0) | (0.0-0.0) | (0.0-0.0) | --- | --- | (0.0-0.0) | (0.0-0.0) | |
| | Late | prop | 0.33 | 0.67 | 0 | 0 | 0 | --- | --- | 0 | 0 | 1 |
| | | risk | 1.3 | 2.5 | 0 | 0 | 0 | --- | --- | 0 | 0 | 3.8 |
| | | num | 0 | 0 | 0 | 0 | 0 | --- | --- | 0 | 0 | 0.1 |
| | | | (0.0-0.0) | (0.0-0.1) | (0.0-0.0) | (0.0-0.0) | (0.0-0.0) | --- | --- | (0.0-0.0) | (0.0-0.0) | |
| | Overall | prop | 0.25 | 0.5 | 0 | 0 | 0 | 0 | 0 | 0 | 0.25 | 1 |
| | | risk | 1.3 | 2.6 | 0 | 0 | 0 | 0 | 0 | 0 | 1.3 | 5.2 |
| | | num | 0 | 0 | 0 | 0 | 0 | 0 | 0 | 0 | 0 | 0.1 |

| Table S14: Cause-specific proportions, risks, and numbers of deaths (with uncertainty) for 194 countries by neonatal period |||||||||||||
|---|---|---|---|---|---|---|---|---|---|---|---|---|
| Country | Period | Stat* | Preterm | Intrapartum | Congenital | Sepsis | Pneumonia | Tetanus | Diarrhoea | Injuries | Other | Total |
|  |  |  | (0.0-0.0) | (0.0-0.1) | (0.0-0.0) | (0.0-0.0) | (0.0-0.0) | (0.0-0.0) | (0.0-0.0) | (0.0-0.0) | (0.0-0.0) |  |

* prop = proportion; num = number of deaths (in 100s).

| Table S14: Cause-specific proportions, risks, and numbers of deaths (with uncertainty) for 194 countries by neonatal period |||||||||||||
|---|---|---|---|---|---|---|---|---|---|---|---|---|
| Country | Period | Stat* | Preterm | Intrapartum | Congenital | Sepsis | Pneumonia | Tetanus | Diarrhoea | Injuries | Other | Total |
| Argentina (high-quality VR) | Early | prop | 0.49 | 0.08 | 0.27 | 0.08 | 0.02 | --- | --- | 0 | 0.06 | 1 |
|  |  | risk | 2.5 | 0.4 | 1.4 | 0.4 | 0.1 | --- | --- | 0 | 0.3 | 5.1 |
|  |  | num | 17.4 | 3 | 9.5 | 3 | 0.5 | --- | --- | 0.1 | 2.2 | 35.9 |
|  |  |  | (16.6-18.3) | (2.7-3.4) | (8.9-10.1) | (2.6-3.3) | (0.4-0.7) | --- | --- | (0.1-0.2) | (2.0-2.5) |  |
|  | Late | prop | 0.34 | 0.05 | 0.3 | 0.2 | 0.03 | --- | --- | 0.01 | 0.06 | 1 |
|  |  | risk | 0.7 | 0.1 | 0.6 | 0.4 | 0.1 | --- | --- | 0 | 0.1 | 2 |
|  |  | num | 4.7 | 0.7 | 4.2 | 2.7 | 0.5 | --- | --- | 0.2 | 0.8 | 13.7 |
|  |  |  | (4.2-5.1) | (0.6-0.9) | (3.8-4.6) | (2.4-3.0) | (0.3-0.6) | --- | --- | (0.1-0.3) | (0.6-1.0) |  |
|  | Overall | prop | 0.45 | 0.08 | 0.28 | 0.11 | 0.02 | 0 | 0 | 0.01 | 0.06 | 1 |
|  |  | risk | 3.3 | 0.6 | 2 | 0.8 | 0.1 | 0 | 0 | 0 | 0.5 | 7.3 |
|  |  | num | 23 | 3.9 | 14.2 | 5.9 | 1.1 | 0 | 0 | 0.3 | 3.2 | 51.5 |
|  |  |  | (21.7-24.2) | (3.4-4.4) | (13.2-15.2) | (5.2-6.5) | (0.8-1.3) | (0.0-0.0) | (0.0-0.0) | (0.2-0.5) | (2.7-3.7) |  |
| Armenia (low mort model) | Early | prop | 0.4 | 0.12 | 0.26 | 0.06 | 0.03 | --- | --- | 0.01 | 0.11 | 1 |
|  |  | risk | 2.8 | 0.9 | 1.8 | 0.4 | 0.2 | --- | --- | 0 | 0.8 | 7 |
|  |  | num | 1.1 | 0.3 | 0.7 | 0.2 | 0.1 | --- | --- | 0 | 0.3 | 2.7 |
|  |  |  | (1.0-1.3) | (0.3-0.4) | (0.5-0.9) | (0.1-0.2) | (0.1-0.1) | --- | --- | (0.0-0.0) | (0.2-0.4) |  |
|  | Late | prop | 0.33 | 0.07 | 0.31 | 0.14 | 0.06 | --- | --- | 0.02 | 0.08 | 1 |
|  |  | risk | 0.8 | 0.2 | 0.8 | 0.3 | 0.1 | --- | --- | 0 | 0.2 | 2.5 |
|  |  | num | 0.3 | 0.1 | 0.3 | 0.1 | 0.1 | --- | --- | 0 | 0.1 | 1 |
|  |  |  | (0.3-0.3) | (0.1-0.1) | (0.3-0.3) | (0.1-0.2) | (0.0-0.1) | --- | --- | (0.0-0.0) | (0.1-0.1) |  |
|  | Overall | prop | 0.38 | 0.11 | 0.27 | 0.08 | 0.04 | 0 | 0 | 0.01 | 0.1 | 1 |
|  |  | risk | 3.8 | 1.1 | 2.7 | 0.8 | 0.4 | 0 | 0 | 0.1 | 1 | 10 |
|  |  | num | 1.6 | 0.5 | 1.1 | 0.3 | 0.2 | 0 | 0 | 0 | 0.4 | 4.2 |
|  |  |  | (1.4-1.8) | (0.4-0.6) | (0.9-1.3) | (0.2-0.5) | (0.1-0.3) | (0.0-0.0) | (0.0-0.0) | (0.0-0.0) | (0.3-0.5) |  |
| Australia (high-quality VR) | Early | prop | 0.3 | 0.2 | 0.27 | 0.02 | 0 | --- | --- | 0 | 0.21 | 1 |
|  |  | risk | 0.6 | 0.4 | 0.5 | 0 | 0 | --- | --- | 0 | 0.4 | 2 |
|  |  | num | 2 | 1.3 | 1.8 | 0.1 | 0 | --- | --- | 0 | 1.3 | 6.5 |
|  |  |  | (1.7-2.3) | (1.1-1.5) | (1.5-2.0) | (0.0-0.2) | (0.0-0.0) | --- | --- | (0.0-0.0) | (1.1-1.6) |  |
|  | Late | prop | 0.32 | 0.08 | 0.37 | 0.08 | 0.03 | --- | --- | 0 | 0.12 | 1 |
|  |  | risk | 0.1 | 0 | 0.1 | 0 | 0 | --- | --- | 0 | 0 | 0.4 |
|  |  | num | 0.4 | 0.1 | 0.4 | 0.1 | 0 | --- | --- | 0 | 0.1 | 1.2 |
|  |  |  | (0.3-0.5) | (0.0-0.2) | (0.3-0.6) | (0.0-0.2) | (0.0-0.1) | --- | --- | (0.0-0.0) | (0.1-0.2) |  |

| Table S14: Cause-specific proportions, risks, and numbers of deaths (with uncertainty) for 194 countries by neonatal period |||||||||||||
|---|---|---|---|---|---|---|---|---|---|---|---|---|
| Country | Period | Stat* | Preterm | Intrapartum | Congenital | Sepsis | Pneumonia | Tetanus | Diarrhoea | Injuries | Other | Total |
| | | prop | 0.31 | 0.18 | 0.29 | 0.03 | 0.01 | 0 | 0 | 0 | 0.19 | 1 |
| | Overall | risk | 0.8 | 0.5 | 0.7 | 0.1 | 0 | 0 | 0 | 0 | 0.5 | 2.5 |
| | | num | 2.4 | 1.4 | 2.3 | 0.2 | 0 | 0 | 0 | 0 | 1.5 | 8 |
| | | | (2.0-2.9) | (1.2-1.7) | (1.9-2.7) | (0.1-0.3) | (0.0-0.1) | (0.0-0.0) | (0.0-0.0) | (0.0-0.0) | (1.2-1.8) | |

* prop = proportion; num = number of deaths (in 100s).

| Country | Period | Stat* | Preterm | Intrapartum | Congenital | Sepsis | Pneumonia | Tetanus | Diarrhoea | Injuries | Other | Total |
|---|---|---|---|---|---|---|---|---|---|---|---|---|
| **Austria** (high-quality VR) | Early | prop | 0.33 | 0.17 | 0.34 | 0.03 | 0 | --- | --- | 0.01 | 0.13 | 1 |
| | | risk | 0.6 | 0.3 | 0.6 | 0 | 0 | --- | --- | 0 | 0.2 | 1.8 |
| | | num | 0.5 | 0.3 | 0.5 | 0 | 0 | --- | --- | 0 | 0.2 | 1.5 |
| | | | (0.4-0.6) | (0.2-0.4) | (0.4-0.7) | (0.0-0.1) | (0.0-0.0) | --- | --- | (0.0-0.0) | (0.1-0.3) | |
| | Late | prop | 0.3 | 0.03 | 0.54 | 0.05 | 0 | --- | --- | 0 | 0.08 | 1 |
| | | risk | 0.1 | 0 | 0.2 | 0 | 0 | --- | --- | 0 | 0 | 0.5 |
| | | num | 0.1 | 0 | 0.2 | 0 | 0 | --- | --- | 0 | 0 | 0.4 |
| | | | (0.0-0.2) | (0.0-0.0) | (0.1-0.3) | (0.0-0.0) | (0.0-0.0) | --- | --- | (0.0-0.0) | (0.0-0.1) | |
| | Overall | prop | 0.32 | 0.14 | 0.38 | 0.03 | 0 | 0 | 0 | 0.01 | 0.12 | 1 |
| | | risk | 0.8 | 0.3 | 0.9 | 0.1 | 0 | 0 | 0 | 0 | 0.3 | 2.4 |
| | | num | 0.6 | 0.3 | 0.7 | 0.1 | 0 | 0 | 0 | 0 | 0.2 | 2 |
| | | | (0.4-0.8) | (0.2-0.4) | (0.5-1.0) | (0.0-0.1) | (0.0-0.0) | (0.0-0.0) | (0.0-0.0) | (0.0-0.0) | (0.1-0.4) | |
| **Azerbaijan** (high mort model) | Early | prop | 0.46 | 0.2 | 0.14 | 0.06 | 0.03 | 0 | 0 | --- | 0.11 | 1 |
| | | risk | 5.4 | 2.4 | 1.7 | 0.7 | 0.3 | 0 | 0 | --- | 1.3 | 11.8 |
| | | num | 9.3 | 4.1 | 2.9 | 1.1 | 0.5 | 0.1 | 0 | --- | 2.3 | 20.3 |
| | | | (7.3-12.6) | (3.1-5.5) | (2.0-3.9) | (0.5-1.8) | (0.2-1.1) | (0.0-0.2) | (0.0-0.1) | --- | (0.5-4.1) | |
| | Late | prop | 0.22 | 0.1 | 0.22 | 0.28 | 0.04 | 0.01 | 0 | --- | 0.13 | 1 |
| | | risk | 0.9 | 0.4 | 0.9 | 1.2 | 0.1 | 0 | 0 | --- | 0.5 | 4.1 |
| | | num | 1.6 | 0.7 | 1.6 | 2 | 0.3 | 0.1 | 0 | --- | 0.9 | 7.1 |
| | | | (1.0-2.5) | (0.5-1.0) | (0.8-2.4) | (1.2-3.2) | (0.1-0.4) | (0.0-0.2) | (0.0-0.0) | --- | (0.3-1.7) | |
| | Overall | prop | 0.4 | 0.18 | 0.16 | 0.12 | 0.03 | 0 | 0 | 0 | 0.12 | 1 |
| | | risk | 6.6 | 2.9 | 2.6 | 1.9 | 0.5 | 0.1 | 0 | 0 | 1.9 | 16.4 |
| | | num | 11.3 | 4.9 | 4.4 | 3.2 | 0.8 | 0.1 | 0 | 0 | 3.3 | 28.1 |
| | | | (8.7-15.5) | (3.7-6.8) | (2.7-6.3) | (1.6-5.1) | (0.4-1.6) | (0.0-0.3) | (0.0-0.1) | (0.0-0.0) | (0.9-5.8) | |
| **Bahamas** (high-quality VR) | Early | prop | 0.31 | 0.21 | 0.07 | 0.03 | 0.21 | --- | --- | 0 | 0.17 | 1 |
| | | risk | 1.6 | 1 | 0.3 | 0.2 | 1 | --- | --- | 0 | 0.9 | 5 |
| | | num | 0.1 | 0.1 | 0 | 0 | 0.1 | --- | --- | 0 | 0.1 | 0.3 |
| | | | (0.0-0.2) | (0.0-0.1) | (0.0-0.0) | (0.0-0.0) | (0.0-0.1) | --- | --- | (0.0-0.0) | (0.0-0.1) | |
| | Late | prop | 0.18 | 0 | 0.18 | 0.36 | 0.27 | --- | --- | 0 | 0 | 1 |

| Country | Period | Stat* | Preterm | Intrapartum | Congenital | Sepsis | Pneumonia | Tetanus | Diarrhoea | Injuries | Other | Total |
|---|---|---|---|---|---|---|---|---|---|---|---|---|
| | | risk | 0.3 | 0 | 0.3 | 0.7 | 0.5 | --- | --- | 0 | 0 | 1.9 |
| | | num | 0 | 0 | 0 | 0 | 0 | --- | --- | 0 | 0 | 0.1 |
| | | | (0.0-0.0) | (0.0-0.0) | (0.0-0.0) | (0.0-0.1) | (0.0-0.1) | --- | --- | (0.0-0.0) | (0.0-0.0) | |
| | Overall | prop | 0.28 | 0.15 | 0.1 | 0.13 | 0.23 | 0 | 0 | 0 | 0.13 | 1 |
| | | risk | 2 | 1.1 | 0.7 | 0.9 | 1.6 | 0 | 0 | 0 | 0.9 | 7.1 |
| | | num | 0.1 | 0.1 | 0 | 0.1 | 0.1 | 0 | 0 | 0 | 0.1 | 0.4 |
| | | | (0.0-0.2) | (0.0-0.1) | (0.0-0.1) | (0.0-0.1) | (0.0-0.2) | (0.0-0.0) | (0.0-0.0) | (0.0-0.0) | (0.0-0.1) | |

\* prop = proportion; num = number of deaths (in 100s).

**Table S14: Cause-specific proportions, risks, and numbers of deaths (with uncertainty) for 194 countries by neonatal period**

| Country | Period | Stat* | Preterm | Intrapartum | Congenital | Sepsis | Pneumonia | Tetanus | Diarrhoea | Injuries | Other | Total |
|---|---|---|---|---|---|---|---|---|---|---|---|---|
| **Bahrain** (high-quality VR) | Early | prop | 0.38 | 0.06 | 0.5 | 0 | 0 | --- | --- | 0 | 0.06 | 1 |
| | | risk | 0.5 | 0.1 | 0.7 | 0 | 0 | --- | --- | 0 | 0.1 | 1.4 |
| | | num | 0.1 | 0 | 0.2 | 0 | 0 | --- | --- | 0 | 0 | 0.3 |
| | | | (0.0-0.2) | (0.0-0.0) | (0.1-0.2) | (0.0-0.0) | (0.0-0.0) | --- | --- | (0.0-0.0) | (0.0-0.0) | |
| | Late | prop | 0.25 | 0.1 | 0.3 | 0.3 | 0 | --- | --- | 0 | 0.05 | 1 |
| | | risk | 0.2 | 0.1 | 0.3 | 0.3 | 0 | --- | --- | 0 | 0 | 0.9 |
| | | num | 0 | 0 | 0.1 | 0.1 | 0 | --- | --- | 0 | 0 | 0.2 |
| | | | (0.0-0.1) | (0.0-0.0) | (0.0-0.1) | (0.0-0.1) | (0.0-0.0) | --- | --- | (0.0-0.0) | (0.0-0.0) | |
| | Overall | prop | 0.33 | 0.08 | 0.42 | 0.12 | 0 | 0 | 0 | 0 | 0.06 | 1 |
| | | risk | 0.8 | 0.2 | 1.1 | 0.3 | 0 | 0 | 0 | 0 | 0.1 | 2.5 |
| | | num | 0.2 | 0 | 0.2 | 0.1 | 0 | 0 | 0 | 0 | 0 | 0.5 |
| | | | (0.1-0.3) | (0.0-0.1) | (0.1-0.3) | (0.0-0.1) | (0.0-0.0) | (0.0-0.0) | (0.0-0.0) | (0.0-0.0) | (0.0-0.1) | |
| **Bangladesh** (high mort model) | Early | prop | 0.34 | 0.26 | 0.14 | 0.11 | 0.04 | 0.01 | 0 | --- | 0.11 | 1 |
| | | risk | 6.1 | 4.7 | 2.4 | 2 | 0.7 | 0.1 | 0 | --- | 2 | 17.9 |
| | | num | 192.2 | 147.5 | 77.4 | 61.9 | 21.1 | 4 | 0.7 | --- | 63 | 567.7 |
| | | | (155.1-253.6) | (120.3-192.1) | (47.3-102.9) | (19.4-96.8) | (10.2-44.2) | (1.4-12.9) | (0.0-5.2) | --- | (21.5-93.8) | |
| | Late | prop | 0.24 | 0.11 | 0.07 | 0.4 | 0.04 | 0.01 | 0.02 | --- | 0.11 | 1 |
| | | risk | 1.5 | 0.7 | 0.4 | 2.5 | 0.3 | 0.1 | 0.1 | --- | 0.7 | 6.3 |
| | | num | 47 | 22.5 | 14.1 | 80.2 | 8.5 | 2.4 | 3.3 | --- | 21.4 | 199.5 |
| | | | (30.6-71.7) | (14.1-29.2) | (5.7-25.4) | (48.3-117.7) | (4.2-13.3) | (0.7-6.2) | (1.1-8.4) | --- | (7.1-36.8) | |
| | Overall | prop | 0.31 | 0.22 | 0.11 | 0.19 | 0.04 | 0.01 | 0.01 | 0 | 0.11 | 1 |
| | | risk | 7.9 | 5.6 | 2.9 | 4.7 | 1 | 0.2 | 0.1 | 0 | 2.8 | 25.2 |
| | | num | 246.3 | 175.1 | 88.9 | 145.9 | 30.7 | 7 | 4.2 | 0 | 86.9 | 785 |
| | | | (191.6-331.7) | (137.9-227.6) | (51.2-125.4) | (71.0-221.0) | (15.3-60.2) | (2.2-20.1) | (1.1-14.7) | (0.0-0.0) | (29.7-133.5) | |

| Table S14: Cause-specific proportions, risks, and numbers of deaths (with uncertainty) for 194 countries by neonatal period | | | | | | | | | | | | |
|---|---|---|---|---|---|---|---|---|---|---|---|---|
| Country | Period | Stat* | Preterm | Intrapartum | Congenital | Sepsis | Pneumonia | Tetanus | Diarrhoea | Injuries | Other | Total |
| **Barbados** (high-quality VR) | Early | prop | 0.29 | 0.29 | 0.29 | 0.07 | 0 | --- | --- | 0 | 0.07 | 1 |
| | | risk | 1.2 | 1.2 | 1.2 | 0.3 | 0 | --- | --- | 0 | 0.3 | 4.4 |
| | | num | 0 (0.0-0.1) | 0 (0.0-0.1) | 0 (0.0-0.1) | 0 (0.0-0.0) | 0 (0.0-0.0) | --- | --- | 0 (0.0-0.0) | 0 (0.0-0.0) | 0.2 |
| | Late | prop | 0.17 | 0.25 | 0.33 | 0 | 0 | --- | --- | 0 | 0.25 | 1 |
| | | risk | 0.6 | 0.9 | 1.2 | 0 | 0 | --- | --- | 0 | 0.9 | 3.7 |
| | | num | 0 (0.0-0.1) | 0 (0.0-0.1) | 0 (0.0-0.1) | 0 (0.0-0.0) | 0 (0.0-0.0) | --- | --- | 0 (0.0-0.0) | 0 (0.0-0.1) | 0.1 |
| | Overall | prop | 0.23 | 0.27 | 0.31 | 0.04 | 0 | 0 | 0 | 0 | 0.15 | 1 |
| | | risk | 1.9 | 2.2 | 2.5 | 0.3 | 0 | 0 | 0 | 0 | 1.3 | 8.2 |
| | | num | 0.1 (0.0-0.1) | 0.1 (0.0-0.2) | 0.1 (0.0-0.2) | 0 (0.0-0.0) | 0 (0.0-0.0) | 0 (0.0-0.0) | 0 (0.0-0.0) | 0 (0.0-0.0) | 0 (0.0-0.1) | 0.3 |

* prop = proportion; num = number of deaths (in 100s).

| Table S14: Cause-specific proportions, risks, and numbers of deaths (with uncertainty) for 194 countries by neonatal period | | | | | | | | | | | | |
|---|---|---|---|---|---|---|---|---|---|---|---|---|
| Country | Period | Stat* | Preterm | Intrapartum | Congenital | Sepsis | Pneumonia | Tetanus | Diarrhoea | Injuries | Other | Total |
| **Belarus** (low mort model) | Early | prop | 0.36 | 0.11 | 0.37 | 0.05 | 0.02 | --- | --- | 0.01 | 0.09 | 1 |
| | | risk | 0.6 | 0.2 | 0.6 | 0.1 | 0 | --- | --- | 0 | 0.1 | 1.7 |
| | | num | 0.6 (0.5-0.9) | 0.2 (0.1-0.3) | 0.6 (0.4-1.0) | 0.1 (0.1-0.1) | 0 (0.0-0.1) | --- | --- | 0 (0.0-0.0) | 0.1 (0.1-0.2) | 1.7 |
| | Late | prop | 0.31 | 0.07 | 0.39 | 0.09 | 0.03 | --- | --- | 0.02 | 0.09 | 1 |
| | | risk | 0.2 | 0 | 0.2 | 0.1 | 0 | --- | --- | 0 | 0.1 | 0.6 |
| | | num | 0.2 (0.2-0.3) | 0 (0.0-0.1) | 0.2 (0.2-0.3) | 0.1 (0.0-0.1) | 0 (0.0-0.0) | --- | --- | 0 (0.0-0.0) | 0.1 (0.1-0.1) | 0.6 |
| | Overall | prop | 0.36 | 0.1 | 0.37 | 0.06 | 0.02 | 0 | 0 | 0.01 | 0.09 | 1 |
| | | risk | 0.9 | 0.2 | 0.9 | 0.1 | 0.1 | 0 | 0 | 0 | 0.2 | 2.4 |
| | | num | 0.9 (0.8-1.3) | 0.3 (0.2-0.4) | 0.9 (0.7-1.4) | 0.1 (0.1-0.2) | 0.1 (0.0-0.1) | 0 (0.0-0.0) | 0 (0.0-0.0) | 0 (0.0-0.0) | 0.2 (0.2-0.3) | 2.5 |
| **Belgium** (high-quality VR) | Early | prop | 0.35 | 0.17 | 0.26 | 0.03 | 0 | --- | --- | 0 | 0.19 | 1 |
| | | risk | 0.6 | 0.3 | 0.5 | 0.1 | 0 | --- | --- | 0 | 0.3 | 1.7 |
| | | num | 0.8 (0.6-0.9) | 0.4 (0.2-0.5) | 0.6 (0.4-0.7) | 0.1 (0.0-0.1) | 0 (0.0-0.0) | --- | --- | 0 (0.0-0.0) | 0.4 (0.3-0.5) | 2.2 |
| | Late | prop | 0.22 | 0.11 | 0.41 | 0.08 | 0 | --- | --- | 0.01 | 0.16 | 1 |
| | | risk | 0.1 | 0.1 | 0.2 | 0 | 0 | --- | --- | 0 | 0.1 | 0.6 |
| | | num | 0.2 (0.1-0.2) | 0.1 (0.0-0.1) | 0.3 (0.2-0.4) | 0.1 (0.0-0.1) | 0 (0.0-0.0) | --- | --- | 0 (0.0-0.0) | 0.1 (0.1-0.2) | 0.8 |
| | Overall | prop | 0.31 | 0.15 | 0.3 | 0.05 | 0 | 0 | 0 | 0 | 0.18 | 1 |

| Table S14: Cause-specific proportions, risks, and numbers of deaths (with uncertainty) for 194 countries by neonatal period ||||||||||||
| Country | Period | Stat* | Preterm | Intrapartum | Congenital | Sepsis | Pneumonia | Tetanus | Diarrhoea | Injuries | Other | Total |
|---|---|---|---|---|---|---|---|---|---|---|---|---|
| | | risk | 0.7 | 0.4 | 0.7 | 0.1 | 0 | 0 | 0 | 0 | 0.4 | 2.3 |
| | | num | 0.9 | 0.5 | 0.9 | 0.1 | 0 | 0 | 0 | 0 | 0.5 | 3 |
| | | | (0.7-1.2) | (0.3-0.6) | (0.6-1.2) | (0.0-0.2) | (0.0-0.0) | (0.0-0.0) | (0.0-0.0) | (0.0-0.0) | (0.3-0.7) | |
| **Belize** (high-quality VR) | Early | prop | 0.28 | 0.24 | 0.22 | 0.07 | 0.02 | --- | --- | 0 | 0.17 | 1 |
| | | risk | 1.5 | 1.3 | 1.2 | 0.3 | 0.1 | --- | --- | 0 | 0.9 | 5.3 |
| | | num | 0.1 | 0.1 | 0.1 | 0 | 0 | --- | --- | 0 | 0.1 | 0.4 |
| | | | (0.1-0.2) | (0.0-0.2) | (0.0-0.2) | (0.0-0.1) | (0.0-0.0) | --- | --- | (0.0-0.0) | (0.0-0.1) | |
| | Late | prop | 0.27 | 0.14 | 0.09 | 0.23 | 0.23 | --- | --- | 0 | 0.05 | 1 |
| | | risk | 0.7 | 0.3 | 0.2 | 0.6 | 0.6 | --- | --- | 0 | 0.1 | 2.6 |
| | | num | 0.1 | 0 | 0 | 0 | 0 | --- | --- | 0 | 0 | 0.2 |
| | | | (0.0-0.1) | (0.0-0.1) | (0.0-0.0) | (0.0-0.1) | (0.0-0.1) | --- | --- | (0.0-0.0) | (0.0-0.0) | |
| | Overall | prop | 0.28 | 0.21 | 0.18 | 0.12 | 0.09 | 0 | 0 | 0 | 0.13 | 1 |
| | | risk | 2.3 | 1.7 | 1.4 | 1 | 0.7 | 0 | 0 | 0 | 1.1 | 8.1 |
| | | num | 0.2 | 0.1 | 0.1 | 0.1 | 0.1 | 0 | 0 | 0 | 0.1 | 0.6 |
| | | | (0.1-0.3) | (0.0-0.2) | (0.0-0.2) | (0.0-0.1) | (0.0-0.1) | (0.0-0.0) | (0.0-0.0) | (0.0-0.0) | (0.0-0.2) | |

* prop = proportion; num = number of deaths (in 100s).

| Table S14: Cause-specific proportions, risks, and numbers of deaths (with uncertainty) for 194 countries by neonatal period ||||||||||||
| Country | Period | Stat* | Preterm | Intrapartum | Congenital | Sepsis | Pneumonia | Tetanus | Diarrhoea | Injuries | Other | Total |
|---|---|---|---|---|---|---|---|---|---|---|---|---|
| **Benin** (high mort model) | Early | prop | 0.4 | 0.33 | 0.08 | 0.09 | 0.06 | 0.01 | 0 | --- | 0.03 | 1 |
| | | risk | 8.1 | 6.6 | 1.6 | 1.7 | 1.2 | 0.1 | 0 | --- | 0.7 | 19.9 |
| | | num | 29.2 | 23.8 | 5.7 | 6.2 | 4.2 | 0.4 | 0.1 | --- | 2.4 | 72.1 |
| | | | (24.3-34.7) | (19.3-27.6) | (4.3-8.2) | (2.6-9.7) | (2.2-7.9) | (0.2-1.0) | (0.0-1.2) | --- | (1.3-5.9) | |
| | Late | prop | 0.13 | 0.15 | 0.09 | 0.47 | 0.05 | 0.02 | 0.02 | --- | 0.06 | 1 |
| | | risk | 0.9 | 1 | 0.6 | 3.3 | 0.4 | 0.2 | 0.2 | --- | 0.4 | 7 |
| | | num | 3.3 | 3.7 | 2.2 | 12 | 1.4 | 0.6 | 0.6 | --- | 1.5 | 25.3 |
| | | | (1.6-5.9) | (2.3-5.0) | (0.8-5.1) | (7.6-17.0) | (0.7-2.3) | (0.1-1.6) | (0.3-1.0) | --- | (0.6-3.8) | |
| | Overall | prop | 0.33 | 0.29 | 0.08 | 0.19 | 0.06 | 0.01 | 0.01 | 0 | 0.04 | 1 |
| | | risk | 9.1 | 7.9 | 2.2 | 5.2 | 1.6 | 0.3 | 0.2 | 0 | 1.1 | 27.5 |
| | | num | 32.3 | 28.1 | 7.8 | 18.5 | 5.8 | 1.1 | 0.7 | 0 | 4 | 98.2 |
| | | | (25.6-41.0) | (22.0-33.3) | (4.9-13.2) | (10.5-27.4) | (3.1-10.7) | (0.3-2.7) | (0.3-2.4) | (0.0-0.0) | (2.0-9.7) | |
| **Bhutan** (high mort model) | Early | prop | 0.35 | 0.26 | 0.16 | 0.09 | 0.03 | 0 | 0 | --- | 0.1 | 1 |
| | | risk | 4.7 | 3.5 | 2.1 | 1.2 | 0.4 | 0 | 0 | --- | 1.4 | 13.4 |
| | | num | 0.7 | 0.5 | 0.3 | 0.2 | 0.1 | 0 | 0 | --- | 0.2 | 2 |
| | | | (0.6-0.9) | (0.4-0.7) | (0.2-0.4) | (0.1-0.3) | (0.0-0.1) | (0.0-0.0) | (0.0-0.0) | --- | (0.1-0.3) | |
| | Late | prop | 0.23 | 0.13 | 0.11 | 0.36 | 0.04 | 0.01 | 0.01 | --- | 0.11 | 1 |
| | | risk | 1.1 | 0.6 | 0.5 | 1.7 | 0.2 | 0.1 | 0.1 | --- | 0.5 | 4.7 |

| Table S14: Cause-specific proportions, risks, and numbers of deaths (with uncertainty) for 194 countries by neonatal period | | | | | | | | | | | |
|---|---|---|---|---|---|---|---|---|---|---|---|
| Country | Period | Stat* | Preterm | Intrapartum | Congenital | Sepsis | Pneumonia | Tetanus | Diarrhoea | Injuries | Other | Total |
| | | num | 0.2 | 0.1 | 0.1 | 0.2 | 0 | 0 | 0 | --- | 0.1 | 0.7 |
| | | | (0.1-0.2) | (0.1-0.1) | (0.0-0.1) | (0.2-0.4) | (0.0-0.1) | (0.0-0.0) | (0.0-0.0) | --- | (0.0-0.1) | |
| | | prop | 0.31 | 0.23 | 0.14 | 0.16 | 0.04 | 0.01 | 0 | 0 | 0.11 | 1 |
| | Overall | risk | 5.9 | 4.3 | 2.6 | 3.1 | 0.7 | 0.1 | 0.1 | 0 | 2 | 18.8 |
| | | num | 0.9 | 0.6 | 0.4 | 0.5 | 0.1 | 0 | 0 | 0 | 0.3 | 2.8 |
| | | | (0.7-1.2) | (0.5-0.9) | (0.3-0.6) | (0.2-0.7) | (0.1-0.2) | (0.0-0.1) | (0.0-0.0) | (0.0-0.0) | (0.1-0.5) | |
| | | prop | 0.32 | 0.3 | 0.18 | 0.08 | 0.04 | 0 | 0 | --- | 0.07 | 1 |
| | Early | risk | 4.3 | 4 | 2.4 | 1.1 | 0.5 | 0.1 | 0 | --- | 0.9 | 13.2 |
| | | num | 11.5 | 10.8 | 6.4 | 2.9 | 1.3 | 0.2 | 0 | --- | 2.5 | 35.7 |
| | | | (9.5-14.5) | (9.0-13.3) | (4.5-8.9) | (1.1-4.6) | (0.6-3.1) | (0.1-0.4) | (0.0-0.3) | --- | (1.5-3.5) | |
| **Bolivia** (high mort model) | Late | prop | 0.15 | 0.16 | 0.14 | 0.39 | 0.05 | 0.02 | 0 | --- | 0.08 | 1 |
| | | risk | 0.7 | 0.7 | 0.7 | 1.8 | 0.2 | 0.1 | 0 | --- | 0.4 | 4.7 |
| | | num | 1.9 | 2 | 1.8 | 4.9 | 0.7 | 0.2 | 0 | --- | 0.9 | 12.5 |
| | | | (1.4-2.8) | (1.4-2.7) | (1.0-2.8) | (2.9-6.7) | (0.4-1.0) | (0.1-0.6) | (0.0-0.1) | --- | (0.6-1.7) | |
| | | prop | 0.28 | 0.27 | 0.16 | 0.16 | 0.04 | 0.01 | 0 | 0 | 0.07 | 1 |
| | Overall | risk | 5.2 | 4.9 | 3 | 3 | 0.8 | 0.2 | 0 | 0 | 1.3 | 18.4 |
| | | num | 13.7 | 13 | 7.9 | 7.9 | 2 | 0.4 | 0.1 | 0 | 3.5 | 48.5 |
| | | | (11.0-17.4) | (10.5-16.1) | (5.3-11.5) | (4.0-11.3) | (1.1-4.2) | (0.1-1.0) | (0.0-0.3) | (0.0-0.0) | (2.0-5.2) | |

* prop = proportion; num = number of deaths (in 100s).

| Table S14: Cause-specific proportions, risks, and numbers of deaths (with uncertainty) for 194 countries by neonatal period | | | | | | | | | | | |
|---|---|---|---|---|---|---|---|---|---|---|---|
| Country | Period | Stat* | Preterm | Intrapartum | Congenital | Sepsis | Pneumonia | Tetanus | Diarrhoea | Injuries | Other | Total |
| | | prop | 0.46 | 0.14 | 0.22 | 0.05 | 0.03 | --- | --- | 0.01 | 0.09 | 1 |
| | Early | risk | 1.5 | 0.5 | 0.7 | 0.2 | 0.1 | --- | --- | 0 | 0.3 | 3.3 |
| | | num | 0.5 | 0.2 | 0.3 | 0.1 | 0 | --- | --- | 0 | 0.1 | 1.2 |
| | | | (0.5-0.6) | (0.1-0.2) | (0.2-0.3) | (0.0-0.1) | (0.0-0.1) | --- | --- | (0.0-0.0) | (0.1-0.1) | |
| **Bosnia and Herzegovina** (low mort model) | Late | prop | 0.32 | 0.07 | 0.36 | 0.11 | 0.05 | --- | --- | 0.02 | 0.06 | 1 |
| | | risk | 0.4 | 0.1 | 0.4 | 0.1 | 0.1 | --- | --- | 0 | 0.1 | 1.1 |
| | | num | 0.1 | 0 | 0.2 | 0 | 0 | --- | --- | 0 | 0 | 0.4 |
| | | | (0.1-0.2) | (0.0-0.0) | (0.1-0.2) | (0.0-0.1) | (0.0-0.0) | --- | --- | (0.0-0.0) | (0.0-0.0) | |
| | | prop | 0.42 | 0.13 | 0.25 | 0.07 | 0.04 | 0 | 0 | 0.01 | 0.08 | 1 |
| | Overall | risk | 2 | 0.6 | 1.2 | 0.3 | 0.2 | 0 | 0 | 0.1 | 0.4 | 4.6 |
| | | num | 0.7 | 0.2 | 0.4 | 0.1 | 0.1 | 0 | 0 | 0 | 0.1 | 1.5 |
| | | | (0.6-0.8) | (0.2-0.2) | (0.3-0.5) | (0.1-0.1) | (0.0-0.1) | (0.0-0.0) | (0.0-0.0) | (0.0-0.0) | (0.1-0.2) | |
| **Botswana** (high mort model) | Early | prop | 0.45 | 0.27 | 0.12 | 0.08 | 0.04 | 0 | 0 | --- | 0.04 | 1 |
| | | risk | 8.3 | 5 | 2.2 | 1.5 | 0.7 | 0.1 | 0 | --- | 0.7 | 18.4 |
| | | num | 3.9 | 2.4 | 1.1 | 0.7 | 0.3 | 0 | 0 | --- | 0.3 | 8.8 |

| Table S14: Cause-specific proportions, risks, and numbers of deaths (with uncertainty) for 194 countries by neonatal period | | | | | | | | | | | |
|---|---|---|---|---|---|---|---|---|---|---|---|
| Country | Period | Stat* | Preterm | Intrapartum | Congenital | Sepsis | Pneumonia | Tetanus | Diarrhoea | Injuries | Other | Total |
| | | | (3.1-5.1) | (1.8-3.1) | (0.6-1.6) | (0.2-1.2) | (0.2-0.7) | (0.0-0.1) | (0.0-0.1) | --- | (0.2-1.2) | |
| | Late | prop | 0.21 | 0.12 | 0.15 | 0.36 | 0.04 | 0.01 | 0 | --- | 0.1 | 1 |
| | | risk | 1.4 | 0.8 | 1 | 2.3 | 0.3 | 0.1 | 0 | --- | 0.6 | 6.5 |
| | | num | 0.6 | 0.4 | 0.5 | 1.1 | 0.1 | 0 | 0 | --- | 0.3 | 3.1 |
| | | | (0.3-1.3) | (0.2-0.5) | (0.1-1.0) | (0.6-1.9) | (0.1-0.2) | (0.0-0.1) | (0.0-0.0) | --- | (0.1-0.8) | |
| | Overall | prop | 0.39 | 0.23 | 0.13 | 0.15 | 0.04 | 0.01 | 0 | 0 | 0.05 | 1 |
| | | risk | 9.7 | 5.8 | 3.1 | 3.9 | 1 | 0.2 | 0 | 0 | 1.4 | 25.1 |
| | | num | 4.7 | 2.8 | 1.5 | 1.9 | 0.5 | 0.1 | 0 | 0 | 0.7 | 12.1 |
| | | | (3.4-6.5) | (2.0-3.7) | (0.7-2.6) | (0.9-3.0) | (0.2-1.0) | (0.0-0.2) | (0.0-0.1) | (0.0-0.0) | (0.3-2.0) | |
| **Brazil** (high-quality VR) | Early | prop | 0.36 | 0.18 | 0.18 | 0.1 | 0.01 | --- | --- | 0 | 0.16 | 1 |
| | | risk | 2.4 | 1.2 | 1.2 | 0.7 | 0.1 | --- | --- | 0 | 1 | 6.5 |
| | | num | 71.2 | 35.2 | 34.8 | 19.9 | 2.8 | --- | --- | 0.6 | 30.9 | 195.4 |
| | | | (69.6-72.9) | (34.1-36.4) | (33.7-36.0) | (19.0-20.7) | (2.5-3.1) | --- | --- | (0.4-0.7) | (29.8-32.0) | |
| | Late | prop | 0.2 | 0.09 | 0.24 | 0.28 | 0.04 | --- | --- | 0.01 | 0.14 | 1 |
| | | risk | 0.4 | 0.2 | 0.5 | 0.5 | 0.1 | --- | --- | 0 | 0.3 | 1.9 |
| | | num | 11.7 | 5.3 | 13.6 | 16.2 | 2.2 | --- | --- | 0.6 | 7.9 | 57.5 |
| | | | (11.1-12.4) | (4.8-5.7) | (12.9-14.4) | (15.5-17.0) | (1.9-2.4) | --- | --- | (0.5-0.8) | (7.3-8.4) | |
| | Overall | prop | 0.33 | 0.16 | 0.19 | 0.14 | 0.02 | 0 | 0 | 0 | 0.15 | 1 |
| | | risk | 2.9 | 1.4 | 1.7 | 1.3 | 0.2 | 0 | 0 | 0 | 1.4 | 8.8 |
| | | num | 83.6 | 40.8 | 48.8 | 36.4 | 5 | 0 | 0 | 1.2 | 39.1 | 254.8 |
| | | | (81.3-85.9) | (39.2-42.4) | (46.9-50.7) | (34.7-38.1) | (4.4-5.6) | (0.0-0.0) | (0.0-0.0) | (0.9-1.5) | (37.5-40.7) | |

* prop = proportion; num = number of deaths (in 100s).

| Table S14: Cause-specific proportions, risks, and numbers of deaths (with uncertainty) for 194 countries by neonatal period | | | | | | | | | | | |
|---|---|---|---|---|---|---|---|---|---|---|---|
| Country | Period | Stat* | Preterm | Intrapartum | Congenital | Sepsis | Pneumonia | Tetanus | Diarrhoea | Injuries | Other | Total |
| **Brunei Darussalam** (low mort model) | Early | prop | 0.46 | 0.15 | 0.23 | 0.01 | 0 | --- | --- | 0.01 | 0.13 | 1 |
| | | risk | 1.8 | 0.6 | 0.9 | 0.1 | 0 | --- | --- | 0 | 0.5 | 3.9 |
| | | num | 0.1 | 0 | 0.1 | 0 | 0 | --- | --- | 0 | 0 | 0.2 |
| | | | (0.1-0.1) | (0.0-0.0) | (0.1-0.1) | (0.0-0.0) | (0.0-0.0) | --- | --- | (0.0-0.0) | (0.0-0.0) | |
| | Late | prop | 0.34 | 0.08 | 0.37 | 0.11 | 0.01 | --- | --- | 0.02 | 0.07 | 1 |
| | | risk | 0.5 | 0.1 | 0.5 | 0.2 | 0 | --- | --- | 0 | 0.1 | 1.4 |
| | | num | 0 | 0 | 0 | 0 | 0 | --- | --- | 0 | 0 | 0.1 |
| | | | (0.0-0.0) | (0.0-0.0) | (0.0-0.0) | (0.0-0.0) | (0.0-0.0) | --- | --- | (0.0-0.0) | (0.0-0.0) | |
| | Overall | prop | 0.43 | 0.13 | 0.27 | 0.04 | 0 | 0 | 0 | 0.01 | 0.12 | 1 |
| | | risk | 2.2 | 0.7 | 1.4 | 0.2 | 0 | 0 | 0 | 0 | 0.6 | 5.2 |
| | | num | 0.1 | 0 | 0.1 | 0 | 0 | 0 | 0 | 0 | 0 | 0.3 |
| | | | (0.1-0.2) | (0.0-0.1) | (0.1-0.1) | (0.0-0.0) | (0.0-0.0) | (0.0-0.0) | (0.0-0.0) | (0.0-0.0) | (0.0-0.1) | |

| Table S14: Cause-specific proportions, risks, and numbers of deaths (with uncertainty) for 194 countries by neonatal period ||||||||||||
|---|---|---|---|---|---|---|---|---|---|---|---|
| Country | Period | Stat* | Preterm | Intrapartum | Congenital | Sepsis | Pneumonia | Tetanus | Diarrhoea | Injuries | Other | Total |
| **Bulgaria** (high-quality VR) | Early | prop | 0.41 | 0.23 | 0.25 | 0.03 | 0.04 | --- | --- | 0 | 0.05 | 1 |
| | | risk | 1.9 | 1.1 | 1.2 | 0.1 | 0.2 | --- | --- | 0 | 0.2 | 4.7 |
| | | num | 1.3 | 0.7 | 0.8 | 0.1 | 0.1 | --- | --- | 0 | 0.2 | 3.2 |
| | | | (1.1-1.5) | (0.6-0.9) | (0.6-1.0) | (0.0-0.1) | (0.1-0.2) | --- | --- | (0.0-0.0) | (0.1-0.2) | |
| | Late | prop | 0.39 | 0.09 | 0.31 | 0.07 | 0.04 | --- | --- | 0.01 | 0.08 | 1 |
| | | risk | 0.7 | 0.2 | 0.5 | 0.1 | 0.1 | --- | --- | 0 | 0.1 | 1.7 |
| | | num | 0.5 | 0.1 | 0.4 | 0.1 | 0 | --- | --- | 0 | 0.1 | 1.2 |
| | | | (0.3-0.6) | (0.0-0.2) | (0.2-0.5) | (0.0-0.1) | (0.0-0.1) | --- | --- | (0.0-0.0) | (0.0-0.2) | |
| | Overall | prop | 0.4 | 0.19 | 0.27 | 0.04 | 0.04 | 0 | 0 | 0.01 | 0.06 | 1 |
| | | risk | 2.7 | 1.3 | 1.7 | 0.3 | 0.3 | 0 | 0 | 0 | 0.4 | 6.6 |
| | | num | 1.8 | 0.9 | 1.2 | 0.2 | 0.2 | 0 | 0 | 0 | 0.3 | 4.6 |
| | | | (1.5-2.2) | (0.6-1.1) | (0.9-1.5) | (0.1-0.3) | (0.1-0.3) | (0.0-0.0) | (0.0-0.0) | (0.0-0.1) | (0.1-0.4) | |
| **Burkina Faso** (high mort model) | Early | prop | 0.39 | 0.33 | 0.08 | 0.1 | 0.06 | 0.01 | 0 | --- | 0.05 | 1 |
| | | risk | 7.7 | 6.5 | 1.5 | 1.9 | 1.3 | 0.1 | 0 | --- | 0.9 | 19.9 |
| | | num | 51 | 43.1 | 10.3 | 12.6 | 8.3 | 0.7 | 0.3 | --- | 6 | 132.4 |
| | | | (41.4-63.8) | (35.4-52.8) | (6.3-16.0) | (4.7-20.8) | (4.6-16.3) | (0.3-1.7) | (0.0-2.8) | --- | (3.7-11.4) | |
| | Late | prop | 0.14 | 0.15 | 0.05 | 0.48 | 0.05 | 0.03 | 0.03 | --- | 0.06 | 1 |
| | | risk | 1 | 1 | 0.4 | 3.4 | 0.4 | 0.2 | 0.2 | --- | 0.5 | 7 |
| | | num | 6.6 | 7 | 2.5 | 22.5 | 2.5 | 1.3 | 1.2 | --- | 3 | 46.5 |
| | | | (3.8-10.2) | (4.4-9.5) | (0.9-6.2) | (14.5-31.7) | (1.4-4.0) | (0.3-3.6) | (0.4-3.0) | --- | (1.5-6.5) | |
| | Overall | prop | 0.32 | 0.28 | 0.07 | 0.2 | 0.06 | 0.01 | 0.01 | 0 | 0.05 | 1 |
| | | risk | 8.9 | 7.8 | 1.9 | 5.4 | 1.7 | 0.3 | 0.2 | 0 | 1.4 | 27.7 |
| | | num | 58.3 | 51.1 | 12.4 | 35.7 | 11.3 | 2.1 | 1.5 | 0 | 9.2 | 181.7 |
| | | | (45.5-74.6) | (40.3-62.6) | (6.8-21.5) | (19.5-53.5) | (6.3-20.7) | (0.6-5.9) | (0.4-6.5) | (0.0-0.0) | (5.4-18.3) | |

* prop = proportion; num = number of deaths (in 100s).

| Table S14: Cause-specific proportions, risks, and numbers of deaths (with uncertainty) for 194 countries by neonatal period ||||||||||||
|---|---|---|---|---|---|---|---|---|---|---|---|
| Country | Period | Stat* | Preterm | Intrapartum | Congenital | Sepsis | Pneumonia | Tetanus | Diarrhoea | Injuries | Other | Total |
| **Burundi** (high mort model) | Early | prop | 0.35 | 0.35 | 0.08 | 0.09 | 0.06 | 0.01 | 0 | --- | 0.06 | 1 |
| | | risk | 7.7 | 7.7 | 1.9 | 2.1 | 1.3 | 0.1 | 0.1 | --- | 1.2 | 22.1 |
| | | num | 33.4 | 33.3 | 8 | 9 | 5.6 | 0.6 | 0.3 | --- | 5.3 | 95.6 |
| | | | (24.8-45.1) | (24.7-43.2) | (4.8-13.8) | (3.3-14.5) | (2.9-11.7) | (0.2-1.5) | (0.0-3.4) | --- | (3.2-10.1) | |
| | Late | prop | 0.14 | 0.16 | 0.06 | 0.47 | 0.06 | 0.03 | 0.01 | --- | 0.07 | 1 |
| | | risk | 1.1 | 1.3 | 0.5 | 3.6 | 0.4 | 0.3 | 0.1 | --- | 0.6 | 7.7 |
| | | num | 4.7 | 5.4 | 2 | 15.8 | 1.9 | 1.1 | 0.2 | --- | 2.4 | 33.6 |
| | | | (2.7-7.5) | (3.4-7.7) | (0.7-5.3) | (9.6-23.3) | (1.1-3.0) | (0.3-2.8) | (0.1-0.6) | --- | (1.3-5.6) | |
| | Overall | prop | 0.3 | 0.3 | 0.08 | 0.19 | 0.06 | 0.01 | 0 | 0 | 0.06 | 1 |

| Table S14: Cause-specific proportions, risks, and numbers of deaths (with uncertainty) for 194 countries by neonatal period ||||||||||||
| Country | Period | Stat* | Preterm | Intrapartum | Congenital | Sepsis | Pneumonia | Tetanus | Diarrhoea | Injuries | Other | Total |
|---|---|---|---|---|---|---|---|---|---|---|---|---|
| | | risk | 9.1 | 9.1 | 2.3 | 5.9 | 1.8 | 0.4 | 0.1 | 0 | 1.8 | 30.5 |
| | | num | 38.5 | 38.4 | 9.8 | 25.1 | 7.6 | 1.8 | 0.6 | 0 | 7.7 | 129.5 |
| | | | (27.3-54.0) | (28.2-50.4) | (5.2-19.0) | (13.1-38.7) | (4.0-14.8) | (0.6-4.6) | (0.1-4.1) | (0.0-0.0) | (4.3-15.5) | |
| **Cambodia** (high mort model) | Early | prop | 0.34 | 0.29 | 0.17 | 0.08 | 0.04 | 0 | 0 | --- | 0.08 | 1 |
| | | risk | 4.4 | 3.8 | 2.3 | 1 | 0.5 | 0 | 0 | --- | 1 | 13 |
| | | num | 16.4 | 14.2 | 8.4 | 3.8 | 1.8 | 0.2 | 0 | --- | 3.8 | 48.7 |
| | | | (13.9-19.4) | (11.7-16.2) | (5.9-10.6) | (1.5-5.7) | (0.9-3.8) | (0.1-0.4) | (0.0-0.3) | --- | (2.0-4.9) | |
| | Late | prop | 0.16 | 0.14 | 0.13 | 0.42 | 0.05 | 0.01 | 0.01 | --- | 0.08 | 1 |
| | | risk | 0.7 | 0.6 | 0.6 | 1.9 | 0.2 | 0.1 | 0 | --- | 0.4 | 4.6 |
| | | num | 2.8 | 2.4 | 2.2 | 7.2 | 0.9 | 0.2 | 0.1 | --- | 1.4 | 17.1 |
| | | | (1.7-4.1) | (1.5-3.0) | (1.1-3.5) | (4.4-9.9) | (0.5-1.3) | (0.1-0.5) | (0.1-0.2) | --- | (0.7-2.2) | |
| | Overall | prop | 0.29 | 0.25 | 0.16 | 0.17 | 0.04 | 0.01 | 0 | 0 | 0.08 | 1 |
| | | risk | 5.4 | 4.5 | 2.9 | 3.1 | 0.7 | 0.1 | 0 | 0 | 1.4 | 18.2 |
| | | num | 19.4 | 16.4 | 10.6 | 11.2 | 2.7 | 0.4 | 0.2 | 0 | 5.1 | 65.9 |
| | | | (15.6-23.7) | (13.1-19.4) | (6.7-14.3) | (5.8-15.9) | (1.4-5.1) | (0.1-1.0) | (0.1-0.5) | (0.0-0.0) | (2.7-7.1) | |
| **Cameroon** (high mort model) | Early | prop | 0.33 | 0.39 | 0.08 | 0.09 | 0.07 | 0.01 | 0 | --- | 0.04 | 1 |
| | | risk | 6.8 | 8.1 | 1.6 | 1.8 | 1.5 | 0.2 | 0.1 | --- | 0.9 | 20.9 |
| | | num | 54.5 | 64.6 | 12.8 | 14.7 | 11.7 | 1.4 | 0.4 | --- | 6.8 | 166.8 |
| | | | (43.4-75.0) | (51.9-81.1) | (9.9-23.2) | (6.2-24.9) | (6.2-23.9) | (0.5-3.2) | (0.0-3.7) | --- | (3.7-17.4) | |
| | Late | prop | 0.14 | 0.16 | 0.06 | 0.47 | 0.06 | 0.03 | 0.01 | --- | 0.07 | 1 |
| | | risk | 1 | 1.2 | 0.4 | 3.5 | 0.4 | 0.2 | 0.1 | --- | 0.5 | 7.3 |
| | | num | 8.2 | 9.6 | 3.4 | 27.7 | 3.4 | 2 | 0.5 | --- | 3.9 | 58.6 |
| | | | (4.6-14.3) | (6.7-13.8) | (1.3-9.1) | (18.0-41.3) | (2.1-5.3) | (0.6-5.7) | (0.1-1.1) | --- | (1.9-10.1) | |
| | Overall | prop | 0.28 | 0.33 | 0.07 | 0.19 | 0.07 | 0.02 | 0 | 0 | 0.05 | 1 |
| | | risk | 8 | 9.4 | 2 | 5.3 | 1.9 | 0.4 | 0.1 | 0 | 1.4 | 28.5 |
| | | num | 63 | 73.8 | 15.6 | 42 | 15.2 | 3.4 | 0.9 | 0 | 10.7 | 224.6 |
| | | | (48.9-88.4) | (57.7-96.1) | (10.7-31.4) | (24.0-65.0) | (8.4-29.0) | (1.1-8.9) | (0.2-4.9) | (0.0-0.0) | (5.6-26.7) | |

* prop = proportion; num = number of deaths (in 100s).

| Table S14: Cause-specific proportions, risks, and numbers of deaths (with uncertainty) for 194 countries by neonatal period ||||||||||||
| Country | Period | Stat* | Preterm | Intrapartum | Congenital | Sepsis | Pneumonia | Tetanus | Diarrhoea | Injuries | Other | Total |
|---|---|---|---|---|---|---|---|---|---|---|---|---|
| **Canada** (low mort model) | Early | prop | 0.46 | 0.15 | 0.23 | 0.03 | 0 | --- | --- | 0.01 | 0.12 | 1 |
| | | risk | 1.2 | 0.4 | 0.6 | 0.1 | 0 | --- | --- | 0 | 0.3 | 2.5 |
| | | num | 4.7 | 1.6 | 2.3 | 0.3 | 0 | --- | --- | 0.1 | 1.2 | 10.2 |
| | | | (4.2-5.2) | (1.3-1.8) | (1.9-3.1) | (0.2-0.4) | (0.0-0.1) | --- | --- | (0.1-0.1) | (0.9-1.4) | |
| | Late | prop | 0.32 | 0.08 | 0.38 | 0.12 | 0.01 | --- | --- | 0.02 | 0.08 | 1 |
| | | risk | 0.3 | 0.1 | 0.3 | 0.1 | 0 | --- | --- | 0 | 0.1 | 0.9 |

| Table S14: Cause-specific proportions, risks, and numbers of deaths (with uncertainty) for 194 countries by neonatal period ||||||||||||
| Country | Period | Stat* | Preterm | Intrapartum | Congenital | Sepsis | Pneumonia | Tetanus | Diarrhoea | Injuries | Other | Total |
| --- | --- | --- | --- | --- | --- | --- | --- | --- | --- | --- | --- | --- |
| | | num | 1.1 | 0.3 | 1.4 | 0.4 | 0 | --- | --- | 0.1 | 0.3 | 3.6 |
| | | | (1.0-1.3) | (0.2-0.4) | (1.2-1.5) | (0.3-0.5) | (0.0-0.1) | --- | --- | (0.1-0.1) | (0.2-0.3) | |
| | Overall | prop | 0.43 | 0.14 | 0.26 | 0.05 | 0 | 0 | 0 | 0.01 | 0.11 | 1 |
| | | risk | 1.5 | 0.5 | 0.9 | 0.2 | 0 | 0 | 0 | 0 | 0.4 | 3.4 |
| | | num | 5.9 | 1.9 | 3.6 | 0.7 | 0.1 | 0 | 0 | 0.1 | 1.5 | 13.7 |
| | | | (5.2-6.5) | (1.5-2.2) | (3.1-4.5) | (0.5-0.9) | (0.0-0.1) | (0.0-0.0) | (0.0-0.0) | (0.1-0.2) | (1.1-1.8) | |
| | Early | prop | 0.41 | 0.18 | 0.19 | 0.09 | 0.04 | --- | --- | 0 | 0.09 | 1 |
| | | risk | 3.5 | 1.5 | 1.6 | 0.8 | 0.3 | --- | --- | 0 | 0.7 | 8.4 |
| | | num | 0.4 | 0.2 | 0.2 | 0.1 | 0 | --- | --- | 0 | 0.1 | 0.9 |
| | | | (0.3-0.4) | (0.1-0.2) | (0.1-0.2) | (0.1-0.1) | (0.0-0.1) | --- | --- | (0.0-0.0) | (0.1-0.1) | |
| **Cabo Verde** (low mort model) | Late | prop | 0.27 | 0.1 | 0.23 | 0.25 | 0.08 | --- | --- | 0.01 | 0.06 | 1 |
| | | risk | 0.8 | 0.3 | 0.7 | 0.7 | 0.2 | --- | --- | 0 | 0.2 | 3 |
| | | num | 0.1 | 0 | 0.1 | 0.1 | 0 | --- | --- | 0 | 0 | 0.3 |
| | | | (0.1-0.1) | (0.0-0.0) | (0.1-0.1) | (0.1-0.1) | (0.0-0.0) | --- | --- | (0.0-0.0) | (0.0-0.0) | |
| | Overall | prop | 0.37 | 0.16 | 0.2 | 0.13 | 0.05 | 0 | 0 | 0.01 | 0.08 | 1 |
| | | risk | 4.3 | 1.8 | 2.3 | 1.6 | 0.6 | 0 | 0 | 0.1 | 0.9 | 11.6 |
| | | num | 0.4 | 0.2 | 0.2 | 0.2 | 0.1 | 0 | 0 | 0 | 0.1 | 1.2 |
| | | | (0.3-0.6) | (0.1-0.3) | (0.2-0.3) | (0.1-0.2) | (0.0-0.1) | (0.0-0.0) | (0.0-0.0) | (0.0-0.0) | (0.1-0.1) | |
| | Early | prop | 0.35 | 0.4 | 0.05 | 0.05 | 0.1 | 0.03 | 0.01 | --- | 0.04 | 1 |
| | | risk | 11.1 | 12.6 | 1.5 | 1.5 | 3.1 | 0.8 | 0.2 | --- | 1.1 | 31.8 |
| | | num | 17.2 | 19.5 | 2.4 | 2.3 | 4.8 | 1.3 | 0.3 | --- | 1.7 | 49.3 |
| | | | (11.9-23.9) | (12.8-23.5) | (1.3-4.0) | (0.6-3.7) | (2.4-8.0) | (0.5-2.5) | (0.0-2.6) | --- | (0.8-4.9) | |
| **Central African Republic** (high mort model) | Late | prop | 0.18 | 0.15 | 0.04 | 0.4 | 0.05 | 0.09 | 0.03 | --- | 0.06 | 1 |
| | | risk | 2 | 1.6 | 0.4 | 4.5 | 0.6 | 1 | 0.4 | --- | 0.7 | 11.2 |
| | | num | 3.2 | 2.5 | 0.7 | 6.9 | 0.9 | 1.6 | 0.6 | --- | 1.1 | 17.3 |
| | | | (1.7-5.1) | (1.5-3.5) | (0.3-1.3) | (3.8-9.6) | (0.5-1.3) | (0.4-3.8) | (0.2-1.5) | --- | (0.4-2.7) | |
| | Overall | prop | 0.32 | 0.3 | 0.05 | 0.16 | 0.08 | 0.04 | 0.01 | 0 | 0.04 | 1 |
| | | risk | 13.9 | 13.1 | 2.1 | 7 | 3.4 | 1.8 | 0.4 | 0 | 1.7 | 43.5 |
| | | num | 21 | 19.9 | 3.2 | 10.6 | 5.2 | 2.8 | 0.6 | 0 | 2.6 | 65.8 |
| | | | (14.4-27.2) | (13.7-24.3) | (1.9-5.4) | (5.3-15.4) | (2.7-8.8) | (0.8-6.2) | (0.1-2.7) | (0.0-0.0) | (1.1-7.2) | |

* prop = proportion; num = number of deaths (in 100s).

| Table S14: Cause-specific proportions, risks, and numbers of deaths (with uncertainty) for 194 countries by neonatal period ||||||||||||
| Country | Period | Stat* | Preterm | Intrapartum | Congenital | Sepsis | Pneumonia | Tetanus | Diarrhoea | Injuries | Other | Total |
| --- | --- | --- | --- | --- | --- | --- | --- | --- | --- | --- | --- | --- |
| **Chad** (high mort model) | Early | prop | 0.38 | 0.31 | 0.04 | 0.06 | 0.08 | 0.03 | 0.01 | --- | 0.08 | 1 |
| | | risk | 11.3 | 9.2 | 1.3 | 1.8 | 2.5 | 0.8 | 0.3 | --- | 2.3 | 29.5 |
| | | num | 63.9 | 52.2 | 7.3 | 9.9 | 14.1 | 4.7 | 1.6 | --- | 12.9 | 166.6 |

| Table S14: Cause-specific proportions, risks, and numbers of deaths (with uncertainty) for 194 countries by neonatal period | | | | | | | | | | | |
|---|---|---|---|---|---|---|---|---|---|---|---|
| Country | Period | Stat* | Preterm | Intrapartum | Congenital | Sepsis | Pneumonia | Tetanus | Diarrhoea | Injuries | Other | Total |
| | | | (48.3-87.3) | (40.9-67.2) | (4.8-12.1) | (3.8-18.1) | (8.1-26.7) | (2.1-9.6) | (0.0-16.5) | --- | (6.3-25.5) | |
| | Late | prop | 0.21 | 0.13 | 0.03 | 0.35 | 0.05 | 0.12 | 0.04 | --- | 0.07 | 1 |
| | | risk | 2.2 | 1.4 | 0.3 | 3.6 | 0.5 | 1.2 | 0.4 | --- | 0.7 | 10.3 |
| | | num | 12.4 | 7.9 | 2 | 20.3 | 2.9 | 6.9 | 2.1 | --- | 4 | 58.5 |
| | | | (7.1-19.5) | (5.1-11.7) | (0.9-3.8) | (11.4-33.2) | (1.6-4.7) | (1.5-17.7) | (0.9-4.6) | --- | (1.6-12.3) | |
| | Overall | prop | 0.35 | 0.26 | 0.04 | 0.13 | 0.08 | 0.06 | 0.02 | 0 | 0.07 | 1 |
| | | risk | 14.2 | 10.4 | 1.6 | 5.1 | 3.1 | 2.4 | 0.7 | 0 | 2.9 | 40.4 |
| | | num | 79 | 58 | 9 | 28.1 | 17 | 13.3 | 3.7 | 0 | 16.3 | 224.3 |
| | | | (56.5-109.6) | (44.9-77.2) | (5.4-15.8) | (13.6-49.7) | (9.8-30.8) | (3.9-31.6) | (0.8-20.1) | (0.0-0.0) | (7.5-35.9) | |
| **Chile** (high-quality VR) | Early | prop | 0.44 | 0.06 | 0.37 | 0.04 | 0.01 | --- | --- | 0.01 | 0.07 | 1 |
| | | risk | 1.7 | 0.2 | 1.4 | 0.2 | 0 | --- | --- | 0 | 0.3 | 3.9 |
| | | num | 4.2 | 0.6 | 3.5 | 0.4 | 0.1 | --- | --- | 0.1 | 0.7 | 9.5 |
| | | | (3.8-4.6) | (0.4-0.7) | (3.2-3.9) | (0.3-0.5) | (0.0-0.1) | --- | --- | (0.0-0.1) | (0.5-0.9) | |
| | Late | prop | 0.31 | 0.03 | 0.45 | 0.1 | 0.05 | --- | --- | 0.01 | 0.05 | 1 |
| | | risk | 0.3 | 0 | 0.5 | 0.1 | 0 | --- | --- | 0 | 0.1 | 1 |
| | | num | 0.8 | 0.1 | 1.1 | 0.2 | 0.1 | --- | --- | 0 | 0.1 | 2.5 |
| | | | (0.6-0.9) | (0.0-0.1) | (0.9-1.3) | (0.1-0.3) | (0.1-0.2) | --- | --- | (0.0-0.1) | (0.1-0.2) | |
| | Overall | prop | 0.41 | 0.05 | 0.39 | 0.05 | 0.02 | 0 | 0 | 0.01 | 0.07 | 1 |
| | | risk | 2.1 | 0.3 | 1.9 | 0.3 | 0.1 | 0 | 0 | 0 | 0.4 | 5 |
| | | num | 5.1 | 0.7 | 4.8 | 0.7 | 0.2 | 0 | 0 | 0.1 | 0.9 | 12.3 |
| | | | (4.5-5.6) | (0.4-0.9) | (4.2-5.3) | (0.4-0.9) | (0.1-0.3) | (0.0-0.0) | (0.0-0.0) | (0.0-0.2) | (0.6-1.1) | |
| **China** (low mort model) | Early | prop | 0.42 | 0.15 | 0.22 | 0.06 | 0.03 | --- | --- | 0.01 | 0.11 | 1 |
| | | risk | 2.4 | 0.8 | 1.3 | 0.3 | 0.2 | --- | --- | 0 | 0.6 | 5.7 |
| | | num | 449.2 | 155.5 | 235.1 | 63.4 | 30.3 | --- | --- | 6.8 | 119.8 | 1060.2 |
| | | | (389.5-497.5) | (128.2-177.8) | (180.6-313.7) | (43.9-79.8) | (20.3-44.3) | --- | --- | (5.5-8.3) | (89.4-140.3) | |
| | Late | prop | 0.28 | 0.08 | 0.29 | 0.16 | 0.12 | --- | --- | 0.01 | 0.06 | 1 |
| | | risk | 0.6 | 0.2 | 0.6 | 0.3 | 0.2 | --- | --- | 0 | 0.1 | 2 |
| | | num | 102.7 | 30.7 | 107.5 | 61 | 43.2 | --- | --- | 5.4 | 22 | 372.5 |
| | | | (91.1-119.8) | (22.6-42.0) | (97.3-117.8) | (44.6-69.9) | (28.4-61.5) | --- | --- | (4.3-6.8) | (13.2-31.8) | |
| | Overall | prop | 0.38 | 0.13 | 0.24 | 0.09 | 0.05 | 0 | 0 | 0.01 | 0.1 | 1 |
| | | risk | 3.2 | 1.1 | 1.9 | 0.7 | 0.4 | 0 | 0 | 0.1 | 0.8 | 8.2 |
| | | num | 583.6 | 196.9 | 360.4 | 136.2 | 77.9 | 0 | 0 | 13 | 150.3 | 1518.2 |
| | | | (510.9-650.7) | (159.4-233.1) | (290.4-455.1) | (97.7-164.0) | (51.5-112.5) | (0.0-0.0) | (0.0-0.0) | (10.3-16.0) | (108.1-183.8) | |

* prop = proportion; num = number of deaths (in 100s).

| Table S14: Cause-specific proportions, risks, and numbers of deaths (with uncertainty) for 194 countries by neonatal period | | | | | | | | | | | |
|---|---|---|---|---|---|---|---|---|---|---|---|
| Country | Period | Stat* | Preterm | Intrapartum | Congenital | Sepsis | Pneumonia | Tetanus | Diarrhoea | Injuries | Other | Total |
| **Colombia** (high-quality VR) | Early | prop | 0.38 | 0.14 | 0.22 | 0.09 | 0.03 | --- | --- | 0.01 | 0.13 | 1 |
| | | risk | 2.8 | 1.1 | 1.6 | 0.7 | 0.2 | --- | --- | 0.1 | 0.9 | 7.4 |
| | | num | 25.7 | 9.7 | 14.8 | 6 | 1.8 | --- | --- | 0.6 | 8.4 | 67 |
| | | | (24.7-26.7) | (9.1-10.3) | (14.1-15.6) | (5.5-6.5) | (1.5-2.0) | --- | --- | (0.4-0.7) | (7.8-9.0) | |
| | Late | prop | 0.27 | 0.07 | 0.27 | 0.21 | 0.07 | --- | --- | 0.01 | 0.09 | 1 |
| | | risk | 0.8 | 0.2 | 0.8 | 0.6 | 0.2 | --- | --- | 0 | 0.3 | 3 |
| | | num | 7.4 | 1.9 | 7.4 | 5.7 | 2 | --- | --- | 0.2 | 2.4 | 27.1 |
| | | | (6.9-8.0) | (1.6-2.1) | (6.9-8.0) | (5.2-6.1) | (1.7-2.3) | --- | --- | (0.1-0.3) | (2.1-2.8) | |
| | Overall | prop | 0.35 | 0.12 | 0.24 | 0.12 | 0.04 | 0 | 0 | 0.01 | 0.12 | 1 |
| | | risk | 3.7 | 1.3 | 2.5 | 1.3 | 0.4 | 0 | 0 | 0.1 | 1.2 | 10.6 |
| | | num | 34.1 | 11.9 | 23 | 12.1 | 3.9 | 0 | 0 | 0.8 | 11.2 | 96.9 |
| | | | (32.6-35.7) | (11.0-12.8) | (21.7-24.3) | (11.1-13.0) | (3.4-4.5) | (0.0-0.0) | (0.0-0.0) | (0.6-1.0) | (10.3-12.1) | |
| **Comoros** (high mort model) | Early | prop | 0.39 | 0.34 | 0.08 | 0.09 | 0.06 | 0.01 | 0 | --- | 0.03 | 1 |
| | | risk | 8.9 | 7.7 | 1.8 | 2 | 1.4 | 0.2 | 0.1 | --- | 0.7 | 22.8 |
| | | num | 2.2 | 1.9 | 0.5 | 0.5 | 0.4 | 0.1 | 0 | --- | 0.2 | 5.7 |
| | | | (1.5-2.6) | (1.3-2.1) | (0.3-0.6) | (0.2-0.8) | (0.2-0.6) | (0.0-0.1) | (0.0-0.1) | --- | (0.1-0.5) | |
| | Late | prop | 0.15 | 0.14 | 0.06 | 0.5 | 0.05 | 0.03 | 0.01 | --- | 0.06 | 1 |
| | | risk | 1.2 | 1.1 | 0.4 | 4 | 0.4 | 0.2 | 0.1 | --- | 0.5 | 8 |
| | | num | 0.3 | 0.3 | 0.1 | 1 | 0.1 | 0.1 | 0 | --- | 0.1 | 2 |
| | | | (0.1-0.4) | (0.2-0.4) | (0.0-0.3) | (0.6-1.3) | (0.1-0.1) | (0.0-0.1) | (0.0-0.1) | --- | (0.0-0.3) | |
| | Overall | prop | 0.33 | 0.29 | 0.07 | 0.2 | 0.06 | 0.01 | 0.01 | 0 | 0.04 | 1 |
| | | risk | 10.2 | 9 | 2.3 | 6.2 | 1.9 | 0.5 | 0.2 | 0 | 1.2 | 31.4 |
| | | num | 2.6 | 2.2 | 0.6 | 1.5 | 0.5 | 0.1 | 0 | 0 | 0.3 | 7.8 |
| | | | (1.6-3.1) | (1.5-2.6) | (0.3-0.9) | (0.8-2.1) | (0.2-0.8) | (0.0-0.3) | (0.0-0.2) | (0.0-0.0) | (0.1-0.7) | |
| **Congo** (high mort model) | Early | prop | 0.42 | 0.29 | 0.14 | 0.08 | 0.04 | 0 | 0 | --- | 0.03 | 1 |
| | | risk | 6 | 4.2 | 2 | 1.1 | 0.6 | 0.1 | 0 | --- | 0.4 | 14.4 |
| | | num | 9.7 | 6.8 | 3.2 | 1.8 | 0.9 | 0.1 | 0 | --- | 0.7 | 23.3 |
| | | | (7.8-12.8) | (5.4-8.8) | (2.3-4.6) | (0.7-2.9) | (0.5-2.0) | (0.0-0.2) | (0.0-0.1) | --- | (0.4-1.8) | |
| | Late | prop | 0.13 | 0.14 | 0.17 | 0.42 | 0.05 | 0.02 | 0.01 | --- | 0.06 | 1 |
| | | risk | 0.7 | 0.7 | 0.8 | 2.1 | 0.2 | 0.1 | 0 | --- | 0.3 | 5 |
| | | num | 1.1 | 1.2 | 1.4 | 3.5 | 0.4 | 0.1 | 0.1 | --- | 0.5 | 8.2 |
| | | | (0.6-1.8) | (0.8-1.6) | (0.5-2.9) | (2.1-5.3) | (0.2-0.6) | (0.0-0.4) | (0.0-0.1) | --- | (0.2-1.2) | |
| | Overall | prop | 0.35 | 0.25 | 0.14 | 0.17 | 0.04 | 0.01 | 0 | 0 | 0.04 | 1 |
| | | risk | 7 | 5.1 | 2.8 | 3.4 | 0.9 | 0.2 | 0.1 | 0 | 0.8 | 20.2 |
| | | num | 11.2 | 8.2 | 4.4 | 5.4 | 1.4 | 0.3 | 0.1 | 0 | 1.2 | 32.2 |
| | | | (8.7-15.2) | (6.3-10.7) | (2.7-7.2) | (3.0-8.3) | (0.7-2.8) | (0.1-0.7) | (0.0-0.3) | (0.0-0.0) | (0.6-3.1) | |

* prop = proportion; num = number of deaths (in 100s).

| Table S14: Cause-specific proportions, risks, and numbers of deaths (with uncertainty) for 194 countries by neonatal period | | | | | | | | | | | |
|---|---|---|---|---|---|---|---|---|---|---|---|
| Country | Period | Stat* | Preterm | Intrapartum | Congenital | Sepsis | Pneumonia | Tetanus | Diarrhoea | Injuries | Other | Total |
| **Cook Islands** (low mort model) | Early | prop | 0.47 | 0.18 | 0.25 | 0.03 | 0.01 | --- | --- | 0.01 | 0.06 | 1 |
| | | risk | 1.6 | 0.6 | 0.9 | 0.1 | 0 | --- | --- | 0 | 0.2 | 3.4 |
| | | num | 0 | 0 | 0 | 0 | 0 | --- | --- | 0 | 0 | 0 |
| | | | (0.0-0.0) | (0.0-0.0) | (0.0-0.0) | (0.0-0.0) | (0.0-0.0) | --- | --- | (0.0-0.0) | (0.0-0.0) | |
| | Late | prop | 0.28 | 0.1 | 0.33 | 0.16 | 0.05 | --- | --- | 0.01 | 0.07 | 1 |
| | | risk | 0.3 | 0.1 | 0.4 | 0.2 | 0.1 | --- | --- | 0 | 0.1 | 1.2 |
| | | num | 0 | 0 | 0 | 0 | 0 | --- | --- | 0 | 0 | 0 |
| | | | (0.0-0.0) | (0.0-0.0) | (0.0-0.0) | (0.0-0.0) | (0.0-0.0) | --- | --- | (0.0-0.0) | (0.0-0.0) | |
| | Overall | prop | 0.42 | 0.16 | 0.27 | 0.07 | 0.02 | 0 | 0 | 0.01 | 0.06 | 1 |
| | | risk | 2 | 0.8 | 1.3 | 0.3 | 0.1 | 0 | 0 | 0 | 0.3 | 4.8 |
| | | num | 0 | 0 | 0 | 0 | 0 | 0 | 0 | 0 | 0 | 0 |
| | | | (0.0-0.0) | (0.0-0.0) | (0.0-0.0) | (0.0-0.0) | (0.0-0.0) | (0.0-0.0) | (0.0-0.0) | (0.0-0.0) | (0.0-0.0) | |
| **Costa Rica** (high-quality VR) | Early | prop | 0.38 | 0.13 | 0.41 | 0.02 | 0.01 | --- | --- | 0 | 0.05 | 1 |
| | | risk | 2.2 | 0.8 | 2.3 | 0.1 | 0 | --- | --- | 0 | 0.3 | 5.8 |
| | | num | 1.6 | 0.6 | 1.7 | 0.1 | 0 | --- | --- | 0 | 0.2 | 4.3 |
| | | | (1.4-1.9) | (0.4-0.7) | (1.5-2.0) | (0.0-0.2) | (0.0-0.1) | --- | --- | (0.0-0.0) | (0.1-0.3) | |
| | Late | prop | 0.22 | 0.08 | 0.47 | 0.08 | 0.05 | --- | --- | 0.03 | 0.07 | 1 |
| | | risk | 0.1 | 0 | 0.3 | 0 | 0 | --- | --- | 0 | 0 | 0.6 |
| | | num | 0.1 | 0 | 0.2 | 0 | 0 | --- | --- | 0 | 0 | 0.5 |
| | | | (0.0-0.2) | (0.0-0.1) | (0.1-0.3) | (0.0-0.1) | (0.0-0.1) | --- | --- | (0.0-0.0) | (0.0-0.1) | |
| | Overall | prop | 0.37 | 0.13 | 0.41 | 0.03 | 0.01 | 0 | 0 | 0 | 0.05 | 1 |
| | | risk | 2.4 | 0.8 | 2.7 | 0.2 | 0.1 | 0 | 0 | 0 | 0.3 | 6.5 |
| | | num | 1.7 | 0.6 | 1.9 | 0.1 | 0.1 | 0 | 0 | 0 | 0.2 | 4.7 |
| | | | (1.4-2.0) | (0.4-0.8) | (1.6-2.3) | (0.0-0.2) | (0.0-0.1) | (0.0-0.0) | (0.0-0.0) | (0.0-0.0) | (0.1-0.4) | |
| **Cote d'Ivoire** (high mort model) | Early | prop | 0.35 | 0.36 | 0.07 | 0.09 | 0.07 | 0.01 | 0 | --- | 0.04 | 1 |
| | | risk | 9.8 | 10.1 | 1.8 | 2.5 | 2 | 0.3 | 0.1 | --- | 1 | 27.7 |
| | | num | 73.2 | 75.1 | 13.8 | 18.9 | 14.8 | 2.4 | 0.8 | --- | 7.6 | 206.6 |
| | | | (58.1-87.3) | (59.1-84.3) | (9.7-20.7) | (7.9-30.1) | (7.8-27.5) | (0.8-5.6) | (0.0-8.0) | --- | (3.8-18.4) | |
| | Late | prop | 0.16 | 0.15 | 0.05 | 0.47 | 0.05 | 0.04 | 0.02 | --- | 0.06 | 1 |
| | | risk | 1.5 | 1.5 | 0.5 | 4.6 | 0.5 | 0.4 | 0.2 | --- | 0.6 | 9.8 |
| | | num | 11.3 | 10.8 | 3.4 | 34.4 | 3.7 | 3.2 | 1.4 | --- | 4.4 | 72.6 |
| | | | (6.2-17.7) | (6.8-14.5) | (1.3-8.3) | (21.8-45.0) | (2.1-5.3) | (1.0-7.3) | (0.5-3.1) | --- | (2.0-10.6) | |
| | Overall | prop | 0.31 | 0.29 | 0.06 | 0.2 | 0.06 | 0.02 | 0.01 | 0 | 0.04 | 1 |
| | | risk | 12 | 11.3 | 2.4 | 7.5 | 2.5 | 0.8 | 0.3 | 0 | 1.6 | 38.4 |
| | | num | 85.6 | 80.7 | 17 | 53.6 | 17.7 | 5.7 | 2.1 | 0 | 11.4 | 273.7 |

| Table S14: Cause-specific proportions, risks, and numbers of deaths (with uncertainty) for 194 countries by neonatal period | | | | | | | | | | | | |
|---|---|---|---|---|---|---|---|---|---|---|---|---|
| Country | Period | Stat* | Preterm | Intrapartum | Congenital | Sepsis | Pneumonia | Tetanus | Diarrhoea | Injuries | Other | Total |
| | | | (66.7-101.6) | (62.9-93.0) | (9.8-28.3) | (28.8-75.8) | (9.6-30.9) | (1.8-13.2) | (0.5-10.4) | (0.0-0.0) | (5.6-27.7) | |

* prop = proportion; num = number of deaths (in 100s).

| Table S14: Cause-specific proportions, risks, and numbers of deaths (with uncertainty) for 194 countries by neonatal period | | | | | | | | | | | | |
|---|---|---|---|---|---|---|---|---|---|---|---|---|
| Country | Period | Stat* | Preterm | Intrapartum | Congenital | Sepsis | Pneumonia | Tetanus | Diarrhoea | Injuries | Other | Total |
| Croatia (high-quality VR) | Early | prop | 0.21 | 0.12 | 0.21 | 0.03 | 0.01 | --- | --- | 0 | 0.41 | 1 |
| | | risk | 0.4 | 0.2 | 0.4 | 0.1 | 0 | --- | --- | 0 | 0.8 | 1.9 |
| | | num | 0.2 | 0.1 | 0.2 | 0 | 0 | --- | --- | 0 | 0.3 | 0.8 |
| | | | (0.1-0.3) | (0.0-0.2) | (0.1-0.3) | (0.0-0.1) | (0.0-0.0) | --- | --- | (0.0-0.0) | (0.2-0.4) | |
| | Late | prop | 0.29 | 0.04 | 0.36 | 0.13 | 0.02 | --- | --- | 0 | 0.16 | 1 |
| | | risk | 0.3 | 0 | 0.3 | 0.1 | 0 | --- | --- | 0 | 0.1 | 0.9 |
| | | num | 0.1 | 0 | 0.1 | 0 | 0 | --- | --- | 0 | 0.1 | 0.4 |
| | | | (0.0-0.2) | (0.0-0.0) | (0.1-0.2) | (0.0-0.1) | (0.0-0.0) | --- | --- | (0.0-0.0) | (0.0-0.1) | |
| | Overall | prop | 0.24 | 0.1 | 0.26 | 0.06 | 0.01 | 0 | 0 | 0 | 0.33 | 1 |
| | | risk | 0.7 | 0.3 | 0.8 | 0.2 | 0 | 0 | 0 | 0 | 1 | 3 |
| | | num | 0.3 | 0.1 | 0.4 | 0.1 | 0 | 0 | 0 | 0 | 0.4 | 1.4 |
| | | | (0.2-0.5) | (0.0-0.2) | (0.2-0.5) | (0.0-0.2) | (0.0-0.1) | (0.0-0.0) | (0.0-0.0) | (0.0-0.0) | (0.3-0.6) | |
| Cuba (high-quality VR) | Early | prop | 0.31 | 0.21 | 0.13 | 0.14 | 0.1 | --- | --- | 0.01 | 0.1 | 1 |
| | | risk | 0.6 | 0.4 | 0.2 | 0.3 | 0.2 | --- | --- | 0 | 0.2 | 1.9 |
| | | num | 0.6 | 0.4 | 0.3 | 0.3 | 0.2 | --- | --- | 0 | 0.2 | 2 |
| | | | (0.5-0.8) | (0.3-0.6) | (0.2-0.4) | (0.2-0.4) | (0.1-0.3) | --- | --- | (0.0-0.0) | (0.1-0.3) | |
| | Late | prop | 0.25 | 0.06 | 0.31 | 0.2 | 0.05 | --- | --- | 0.04 | 0.09 | 1 |
| | | risk | 0.2 | 0.1 | 0.3 | 0.2 | 0.1 | --- | --- | 0 | 0.1 | 1 |
| | | num | 0.3 | 0.1 | 0.3 | 0.2 | 0.1 | --- | --- | 0 | 0.1 | 1 |
| | | | (0.2-0.4) | (0.0-0.1) | (0.2-0.4) | (0.1-0.3) | (0.0-0.1) | --- | --- | (0.0-0.1) | (0.0-0.1) | |
| | Overall | prop | 0.29 | 0.16 | 0.19 | 0.16 | 0.09 | 0 | 0 | 0.02 | 0.09 | 1 |
| | | risk | 0.9 | 0.5 | 0.6 | 0.5 | 0.3 | 0 | 0 | 0.1 | 0.3 | 3.1 |
| | | num | 0.9 | 0.5 | 0.6 | 0.5 | 0.3 | 0 | 0 | 0.1 | 0.3 | 3.2 |
| | | | (0.7-1.2) | (0.3-0.7) | (0.4-0.8) | (0.3-0.7) | (0.1-0.4) | (0.0-0.0) | (0.0-0.0) | (0.0-0.1) | (0.2-0.4) | |
| Cyprus (low mort model) | Early | prop | 0.38 | 0.12 | 0.34 | 0.05 | 0.01 | --- | --- | 0.01 | 0.09 | 1 |
| | | risk | 0.5 | 0.2 | 0.5 | 0.1 | 0 | --- | --- | 0 | 0.1 | 1.3 |
| | | num | 0.1 | 0 | 0.1 | 0 | 0 | --- | --- | 0 | 0 | 0.2 |
| | | | (0.1-0.1) | (0.0-0.0) | (0.0-0.1) | (0.0-0.0) | (0.0-0.0) | --- | --- | (0.0-0.0) | (0.0-0.0) | |
| | Late | prop | 0.3 | 0.07 | 0.39 | 0.09 | 0.02 | --- | --- | 0.02 | 0.11 | 1 |
| | | risk | 0.1 | 0 | 0.2 | 0 | 0 | --- | --- | 0 | 0.1 | 0.5 |
| | | num | 0 | 0 | 0 | 0 | 0 | --- | --- | 0 | 0 | 0.1 |
| | | | (0.0-0.0) | (0.0-0.0) | (0.0-0.0) | (0.0-0.0) | (0.0-0.0) | --- | --- | (0.0-0.0) | (0.0-0.0) | |

| Table S14: Cause-specific proportions, risks, and numbers of deaths (with uncertainty) for 194 countries by neonatal period | | | | | | | | | | | | |
|---|---|---|---|---|---|---|---|---|---|---|---|---|
| Country | Period | Stat* | Preterm | Intrapartum | Congenital | Sepsis | Pneumonia | Tetanus | Diarrhoea | Injuries | Other | Total |
| | Overall | prop | 0.36 | 0.11 | 0.35 | 0.06 | 0.01 | 0 | 0 | 0.01 | 0.1 | 1 |
| | | risk | 0.7 | 0.2 | 0.6 | 0.1 | 0 | 0 | 0 | 0 | 0.2 | 1.8 |
| | | num | 0.1 (0.1-0.1) | 0 (0.0-0.0) | 0.1 (0.1-0.1) | 0 (0.0-0.0) | 0 (0.0-0.0) | 0 (0.0-0.0) | 0 (0.0-0.0) | 0 (0.0-0.0) | 0 (0.0-0.0) | 0.2 |

* prop = proportion; num = number of deaths (in 100s).

| Table S14: Cause-specific proportions, risks, and numbers of deaths (with uncertainty) for 194 countries by neonatal period | | | | | | | | | | | | |
|---|---|---|---|---|---|---|---|---|---|---|---|---|
| Country | Period | Stat* | Preterm | Intrapartum | Congenital | Sepsis | Pneumonia | Tetanus | Diarrhoea | Injuries | Other | Total |
| **Czech Republic** (high-quality VR) | Early | prop | 0.36 | 0.14 | 0.29 | 0.08 | 0.02 | --- | --- | 0.02 | 0.09 | 1 |
| | | risk | 0.5 | 0.2 | 0.4 | 0.1 | 0 | --- | --- | 0 | 0.1 | 1.4 |
| | | num | 0.6 (0.4-0.7) | 0.2 (0.1-0.3) | 0.5 (0.3-0.6) | 0.1 (0.1-0.2) | 0 (0.0-0.1) | --- | --- | 0 (0.0-0.1) | 0.1 (0.1-0.2) | 1.6 |
| | Late | prop | 0.32 | 0.16 | 0.17 | 0.19 | 0.02 | --- | --- | 0 | 0.14 | 1 |
| | | risk | 0.2 | 0.1 | 0.1 | 0.1 | 0 | --- | --- | 0 | 0.1 | 0.7 |
| | | num | 0.3 (0.2-0.4) | 0.1 (0.1-0.2) | 0.1 (0.1-0.2) | 0.2 (0.1-0.2) | 0 (0.0-0.0) | --- | --- | 0 (0.0-0.0) | 0.1 (0.1-0.2) | 0.8 |
| | Overall | prop | 0.34 | 0.15 | 0.25 | 0.12 | 0.02 | 0 | 0 | 0.01 | 0.11 | 1 |
| | | risk | 0.8 | 0.3 | 0.5 | 0.3 | 0 | 0 | 0 | 0 | 0.2 | 2.2 |
| | | num | 0.9 (0.6-1.2) | 0.4 (0.2-0.6) | 0.6 (0.4-0.9) | 0.3 (0.2-0.5) | 0 (0.0-0.1) | 0 (0.0-0.0) | 0 (0.0-0.0) | 0 (0.0-0.1) | 0.3 (0.1-0.4) | 2.6 |
| **Democratic People's Republic of Korea** (high mort model) | Early | prop | 0.43 | 0.21 | 0.16 | 0.05 | 0.02 | 0 | 0 | --- | 0.13 | 1 |
| | | risk | 4.7 | 2.3 | 1.8 | 0.6 | 0.3 | 0 | 0 | --- | 1.4 | 11 |
| | | num | 16.9 (12.5-22.8) | 8.1 (5.9-11.5) | 6.5 (4.2-8.9) | 2.1 (0.8-3.4) | 0.9 (0.4-2.2) | 0.1 (0.0-0.2) | 0 (0.0-0.1) | --- | 5 (1.1-9.5) | 39.6 |
| | Late | prop | 0.2 | 0.1 | 0.25 | 0.26 | 0.04 | 0.01 | 0 | --- | 0.14 | 1 |
| | | risk | 0.8 | 0.4 | 1 | 1 | 0.1 | 0 | 0 | --- | 0.5 | 3.9 |
| | | num | 2.8 (1.6-4.8) | 1.4 (0.8-2.1) | 3.5 (1.7-5.3) | 3.7 (1.9-6.3) | 0.5 (0.3-0.9) | 0.1 (0.0-0.3) | 0 (0.0-0.1) | --- | 2 (0.5-3.7) | 13.9 |
| | Overall | prop | 0.37 | 0.18 | 0.18 | 0.11 | 0.03 | 0 | 0 | 0 | 0.13 | 1 |
| | | risk | 5.7 | 2.8 | 2.8 | 1.7 | 0.4 | 0.1 | 0 | 0 | 2 | 15.5 |
| | | num | 19.8 (14.0-28.0) | 9.7 (7.0-13.5) | 9.7 (5.8-13.8) | 5.9 (2.8-9.8) | 1.5 (0.7-3.1) | 0.2 (0.1-0.5) | 0 (0.0-0.1) | 0 (0.0-0.0) | 7 (1.7-13.3) | 53.7 |
| **Democratic Republic of the Congo** (high mort | Early | prop | 0.41 | 0.33 | 0.05 | 0.07 | 0.07 | 0.01 | 0.01 | --- | 0.05 | 1 |
| | | risk | 11.6 | 9.4 | 1.4 | 2 | 1.9 | 0.3 | 0.2 | --- | 1.5 | 28.3 |
| | | num | 316.6 (215.9-380.2) | 256.6 (169.8-288.1) | 38.1 (24.0-64.8) | 54.3 (18.8-79.6) | 53.1 (24.9-90.6) | 8.6 (2.6-19.2) | 4.9 (0.0-43.7) | --- | 41.8 (22.2-67.4) | 774.1 |

| Table S14: Cause-specific proportions, risks, and numbers of deaths (with uncertainty) for 194 countries by neonatal period |||||||||||||
|---|---|---|---|---|---|---|---|---|---|---|---|---|
| Country | Period | Stat* | Preterm | Intrapartum | Congenital | Sepsis | Pneumonia | Tetanus | Diarrhoea | Injuries | Other | Total |
| model) | Late | prop | 0.15 | 0.16 | 0.07 | 0.43 | 0.05 | 0.06 | 0.01 | --- | 0.07 | 1 |
| | | risk | 1.5 | 1.6 | 0.7 | 4.2 | 0.5 | 0.6 | 0.1 | --- | 0.7 | 9.9 |
| | | num | 40.4 | 43.8 | 19 | 115.8 | 14.8 | 16.4 | 2.1 | --- | 19.8 | 272 |
| | | | (22.1-55.1) | (26.6-57.4) | (7.2-38.5) | (63.0-153.2) | (8.1-21.0) | (4.3-40.2) | (0.5-4.0) | --- | (9.5-41.3) | |
| | Overall | prop | 0.35 | 0.28 | 0.05 | 0.16 | 0.07 | 0.03 | 0.01 | 0 | 0.06 | 1 |
| | | risk | 13.6 | 11 | 2.1 | 6.1 | 2.5 | 1 | 0.3 | 0 | 2.3 | 38.9 |
| | | num | 365.3 | 296.5 | 55.7 | 165.2 | 68.2 | 28.2 | 7 | 0 | 61 | 1047.1 |
| | | | (248.1-439.5) | (192.4-348.3) | (29.7-100.2) | (80.0-221.9) | (31.6-115.5) | (7.8-64.3) | (0.5-41.7) | (0.0-0.0) | (30.4-108.4) | |

* prop = proportion; num = number of deaths (in 100s).

| Table S14: Cause-specific proportions, risks, and numbers of deaths (with uncertainty) for 194 countries by neonatal period |||||||||||||
|---|---|---|---|---|---|---|---|---|---|---|---|---|
| Country | Period | Stat* | Preterm | Intrapartum | Congenital | Sepsis | Pneumonia | Tetanus | Diarrhoea | Injuries | Other | Total |
| **Denmark** (high-quality VR) | Early | prop | 0.61 | 0.1 | 0.08 | 0.02 | 0 | --- | --- | 0.02 | 0.17 | 1 |
| | | risk | 1 | 0.2 | 0.1 | 0 | 0 | --- | --- | 0 | 0.3 | 1.6 |
| | | num | 0.6 | 0.1 | 0.1 | 0 | 0 | --- | --- | 0 | 0.2 | 1 |
| | | | (0.5-0.8) | (0.0-0.2) | (0.0-0.1) | (0.0-0.0) | (0.0-0.0) | --- | --- | (0.0-0.0) | (0.1-0.3) | |
| | Late | prop | 0.56 | 0.04 | 0.3 | 0 | 0 | --- | --- | 0 | 0.11 | 1 |
| | | risk | 0.4 | 0 | 0.2 | 0 | 0 | --- | --- | 0 | 0.1 | 0.8 |
| | | num | 0.3 | 0 | 0.1 | 0 | 0 | --- | --- | 0 | 0.1 | 0.5 |
| | | | (0.2-0.4) | (0.0-0.0) | (0.1-0.2) | (0.0-0.0) | (0.0-0.0) | --- | --- | (0.0-0.0) | (0.0-0.1) | |
| | Overall | prop | 0.59 | 0.08 | 0.15 | 0.01 | 0 | 0 | 0 | 0.01 | 0.15 | 1 |
| | | risk | 1.5 | 0.2 | 0.4 | 0 | 0 | 0 | 0 | 0 | 0.4 | 2.5 |
| | | num | 1 | 0.1 | 0.2 | 0 | 0 | 0 | 0 | 0 | 0.2 | 1.6 |
| | | | (0.7-1.2) | (0.0-0.2) | (0.1-0.4) | (0.0-0.0) | (0.0-0.0) | (0.0-0.0) | (0.0-0.0) | (0.0-0.0) | (0.1-0.4) | |
| **Djibouti** (high mort model) | Early | prop | 0.42 | 0.3 | 0.07 | 0.07 | 0.05 | 0.01 | 0.01 | --- | 0.07 | 1 |
| | | risk | 9.7 | 7 | 1.7 | 1.7 | 1.1 | 0.2 | 0.1 | --- | 1.6 | 23.1 |
| | | num | 2.3 | 1.7 | 0.4 | 0.4 | 0.3 | 0 | 0 | --- | 0.4 | 5.4 |
| | | | (1.9-2.8) | (1.3-2.0) | (0.3-0.7) | (0.2-0.6) | (0.1-0.5) | (0.0-0.1) | (0.0-0.3) | --- | (0.2-0.5) | |
| | Late | prop | 0.15 | 0.14 | 0.15 | 0.4 | 0.05 | 0.03 | 0.01 | --- | 0.07 | 1 |
| | | risk | 1.2 | 1.2 | 1.3 | 3.2 | 0.4 | 0.2 | 0.1 | --- | 0.5 | 8.1 |
| | | num | 0.3 | 0.3 | 0.3 | 0.8 | 0.1 | 0.1 | 0 | --- | 0.1 | 1.9 |
| | | | (0.2-0.4) | (0.2-0.4) | (0.2-0.5) | (0.5-1.0) | (0.1-0.1) | (0.0-0.1) | (0.0-0.0) | --- | (0.1-0.2) | |
| | Overall | prop | 0.35 | 0.26 | 0.09 | 0.16 | 0.05 | 0.01 | 0.01 | 0 | 0.07 | 1 |
| | | risk | 11.2 | 8.3 | 3 | 5 | 1.5 | 0.4 | 0.2 | 0 | 2.2 | 31.8 |
| | | num | 2.6 | 2 | 0.7 | 1.2 | 0.4 | 0.1 | 0 | 0 | 0.5 | 7.5 |
| | | | (2.2-3.3) | (1.5-2.4) | (0.5-1.2) | (0.6-1.7) | (0.2-0.7) | (0.0-0.2) | (0.0-0.4) | (0.0-0.0) | (0.3-0.8) | |

| Table S14: Cause-specific proportions, risks, and numbers of deaths (with uncertainty) for 194 countries by neonatal period ||||||||||||
| Country | Period | Stat* | Preterm | Intrapartum | Congenital | Sepsis | Pneumonia | Tetanus | Diarrhoea | Injuries | Other | Total |
| --- | --- | --- | --- | --- | --- | --- | --- | --- | --- | --- | --- | --- |
| **Dominica** (high-quality VR) | Early | prop | 0.08 | 0.77 | 0 | 0.08 | 0 | --- | --- | 0 | 0.08 | 1 |
| | | risk | 0.5 | 5.1 | 0 | 0.5 | 0 | --- | --- | 0 | 0.5 | 6.6 |
| | | num | 0 | 0.1 | 0 | 0 | 0 | --- | --- | 0 | 0 | 0.1 |
| | | | (0.0-0.0) | (0.0-0.1) | (0.0-0.0) | (0.0-0.0) | (0.0-0.0) | --- | --- | (0.0-0.0) | (0.0-0.0) | |
| | Late | prop | 0.33 | 0 | 0 | 0.67 | 0 | --- | --- | 0 | 0 | 1 |
| | | risk | 0.5 | 0 | 0 | 1 | 0 | --- | --- | 0 | 0 | 1.5 |
| | | num | 0 | 0 | 0 | 0 | 0 | --- | --- | 0 | 0 | 0 |
| | | | (0.0-0.0) | (0.0-0.0) | (0.0-0.0) | (0.0-0.0) | (0.0-0.0) | --- | --- | (0.0-0.0) | (0.0-0.0) | |
| | Overall | prop | 0.13 | 0.63 | 0 | 0.19 | 0 | 0 | 0 | 0 | 0.06 | 1 |
| | | risk | 1 | 5.2 | 0 | 1.6 | 0 | 0 | 0 | 0 | 0.5 | 8.3 |
| | | num | 0 | 0.1 | 0 | 0 | 0 | 0 | 0 | 0 | 0 | 0.1 |
| | | | (0.0-0.0) | (0.0-0.1) | (0.0-0.0) | (0.0-0.1) | (0.0-0.0) | (0.0-0.0) | (0.0-0.0) | (0.0-0.0) | (0.0-0.0) | |

* prop = proportion; num = number of deaths (in 100s).

| Table S14: Cause-specific proportions, risks, and numbers of deaths (with uncertainty) for 194 countries by neonatal period ||||||||||||
| Country | Period | Stat* | Preterm | Intrapartum | Congenital | Sepsis | Pneumonia | Tetanus | Diarrhoea | Injuries | Other | Total |
| --- | --- | --- | --- | --- | --- | --- | --- | --- | --- | --- | --- | --- |
| **Dominican Republic** (high mort model) | Early | prop | 0.42 | 0.23 | 0.18 | 0.06 | 0.03 | 0 | 0 | --- | 0.08 | 1 |
| | | risk | 5.1 | 2.7 | 2.1 | 0.8 | 0.3 | 0 | 0 | --- | 1 | 11.9 |
| | | num | 10.8 | 5.8 | 4.5 | 1.6 | 0.7 | 0.1 | 0 | --- | 2 | 25.5 |
| | | | (9.3-13.2) | (4.7-7.1) | (2.8-6.0) | (0.6-2.5) | (0.3-1.5) | (0.0-0.1) | (0.0-0.1) | --- | (0.9-2.9) | |
| | Late | prop | 0.17 | 0.11 | 0.25 | 0.33 | 0.04 | 0.01 | 0 | --- | 0.09 | 1 |
| | | risk | 0.7 | 0.5 | 1.1 | 1.4 | 0.2 | 0 | 0 | --- | 0.4 | 4.2 |
| | | num | 1.5 | 1 | 2.3 | 2.9 | 0.4 | 0.1 | 0 | --- | 0.8 | 9 |
| | | | (1.0-2.4) | (0.6-1.3) | (1.1-3.4) | (1.7-4.5) | (0.2-0.6) | (0.0-0.2) | (0.0-0.1) | --- | (0.4-1.3) | |
| | Overall | prop | 0.36 | 0.2 | 0.19 | 0.13 | 0.03 | 0 | 0 | 0 | 0.08 | 1 |
| | | risk | 5.9 | 3.2 | 3.2 | 2.2 | 0.5 | 0.1 | 0 | 0 | 1.4 | 16.5 |
| | | num | 12.7 | 7 | 6.9 | 4.7 | 1.1 | 0.1 | 0 | 0 | 2.9 | 35.4 |
| | | | (10.5-15.9) | (5.5-8.8) | (4.0-9.6) | (2.4-7.3) | (0.6-2.1) | (0.1-0.4) | (0.0-0.1) | (0.0-0.0) | (1.3-4.3) | |
| **Ecuador** (low mort model) | Early | prop | 0.45 | 0.16 | 0.22 | 0.07 | 0.03 | --- | --- | 0.01 | 0.08 | 1 |
| | | risk | 3.5 | 1.2 | 1.7 | 0.5 | 0.2 | --- | --- | 0 | 0.6 | 7.8 |
| | | num | 11.4 | 4 | 5.5 | 1.7 | 0.7 | --- | --- | 0.1 | 1.9 | 25.4 |
| | | | (10.2-14.4) | (3.3-5.3) | (4.8-7.5) | (1.2-2.4) | (0.5-1.2) | --- | --- | (0.1-0.2) | (1.6-2.6) | |
| | Late | prop | 0.27 | 0.08 | 0.25 | 0.24 | 0.07 | --- | --- | 0.01 | 0.07 | 1 |
| | | risk | 0.7 | 0.2 | 0.7 | 0.6 | 0.2 | --- | --- | 0 | 0.2 | 2.7 |
| | | num | 2.4 | 0.7 | 2.3 | 2.1 | 0.6 | --- | --- | 0.1 | 0.7 | 8.9 |
| | | | (2.1-3.3) | (0.6-1.2) | (2.0-3.0) | (1.6-2.8) | (0.3-0.9) | --- | --- | (0.1-0.2) | (0.4-1.0) | |
| | Overall | prop | 0.4 | 0.14 | 0.22 | 0.11 | 0.04 | 0 | 0 | 0.01 | 0.08 | 1 |

| Table S14: Cause-specific proportions, risks, and numbers of deaths (with uncertainty) for 194 countries by neonatal period | | | | | | | | | | | | |
|---|---|---|---|---|---|---|---|---|---|---|---|---|
| Country | Period | Stat* | Preterm | Intrapartum | Congenital | Sepsis | Pneumonia | Tetanus | Diarrhoea | Injuries | Other | Total |
| | | risk | 4.3 | 1.5 | 2.4 | 1.2 | 0.4 | 0 | 0 | 0.1 | 0.8 | 10.7 |
| | | num | 14 | 4.8 | 7.7 | 3.8 | 1.4 | 0 | 0 | 0.3 | 2.7 | 34.7 |
| | | | (12.3-17.9) | (3.9-6.4) | (6.7-10.5) | (2.8-5.2) | (0.8-2.0) | (0.0-0.0) | (0.0-0.0) | (0.2-0.4) | (2.0-3.8) | |
| **Egypt** (low mort model) | Early | prop | 0.45 | 0.22 | 0.21 | 0.03 | 0.04 | --- | --- | 0.01 | 0.04 | 1 |
| | | risk | 3.9 | 2 | 1.8 | 0.3 | 0.3 | --- | --- | 0.1 | 0.4 | 8.7 |
| | | num | 74.2 | 37.4 | 34.7 | 5.6 | 6 | --- | --- | 1.3 | 7.2 | 166.3 |
| | | | (59.5-89.2) | (18.3-56.1) | (19.0-56.6) | (2.6-10.0) | (3.6-10.0) | --- | --- | (0.8-2.4) | (4.4-12.0) | |
| | Late | prop | 0.28 | 0.15 | 0.25 | 0.18 | 0.04 | --- | --- | 0.01 | 0.09 | 1 |
| | | risk | 0.9 | 0.5 | 0.8 | 0.6 | 0.1 | --- | --- | 0 | 0.3 | 3.1 |
| | | num | 16.4 | 8.8 | 14.4 | 10.5 | 2.1 | --- | --- | 0.9 | 5.4 | 58.4 |
| | | | (13.6-21.7) | (0.6-12.1) | (11.9-18.8) | (5.2-19.0) | (0.3-4.5) | --- | --- | (0.6-1.2) | (3.5-7.7) | |
| | Overall | prop | 0.41 | 0.21 | 0.21 | 0.06 | 0.04 | 0 | 0 | 0.01 | 0.06 | 1 |
| | | risk | 5 | 2.5 | 2.5 | 0.8 | 0.4 | 0 | 0 | 0.1 | 0.7 | 12 |
| | | num | 95.7 | 48.3 | 49 | 14.6 | 8.6 | 0 | 0 | 2.3 | 13 | 231.5 |
| | | | (77.2-115.8) | (19.8-71.6) | (30.7-74.7) | (7.5-27.0) | (4.0-15.3) | (0.0-0.0) | (0.0-0.0) | (1.5-3.8) | (8.2-20.3) | |

* prop = proportion; num = number of deaths (in 100s)

| Table S14: Cause-specific proportions, risks, and numbers of deaths (with uncertainty) for 194 countries by neonatal period | | | | | | | | | | | | |
|---|---|---|---|---|---|---|---|---|---|---|---|---|
| Country | Period | Stat* | Preterm | Intrapartum | Congenital | Sepsis | Pneumonia | Tetanus | Diarrhoea | Injuries | Other | Total |
| **El Salvador** (low mort model) | Early | prop | 0.42 | 0.17 | 0.19 | 0.07 | 0.03 | --- | --- | 0 | 0.1 | 1 |
| | | risk | 2 | 0.8 | 0.9 | 0.4 | 0.2 | --- | --- | 0 | 0.5 | 4.8 |
| | | num | 2.7 | 1.1 | 1.2 | 0.5 | 0.2 | --- | --- | 0 | 0.6 | 6.4 |
| | | | (2.3-3.0) | (0.7-1.4) | (1.0-1.6) | (0.4-0.6) | (0.1-0.3) | --- | --- | (0.0-0.0) | (0.5-0.8) | |
| | Late | prop | 0.26 | 0.09 | 0.27 | 0.19 | 0.12 | --- | --- | 0.01 | 0.05 | 1 |
| | | risk | 0.4 | 0.2 | 0.5 | 0.3 | 0.2 | --- | --- | 0 | 0.1 | 1.7 |
| | | num | 0.6 | 0.2 | 0.6 | 0.4 | 0.3 | --- | --- | 0 | 0.1 | 2.2 |
| | | | (0.5-0.7) | (0.1-0.3) | (0.5-0.7) | (0.3-0.6) | (0.2-0.4) | --- | --- | (0.0-0.0) | (0.1-0.2) | |
| | Overall | prop | 0.38 | 0.15 | 0.21 | 0.11 | 0.06 | 0 | 0 | 0.01 | 0.09 | 1 |
| | | risk | 2.6 | 1 | 1.4 | 0.7 | 0.4 | 0 | 0 | 0 | 0.6 | 6.7 |
| | | num | 3.5 | 1.4 | 1.9 | 1 | 0.5 | 0 | 0 | 0.1 | 0.8 | 9.1 |
| | | | (3.0-3.9) | (0.9-1.7) | (1.6-2.5) | (0.7-1.2) | (0.3-0.8) | (0.0-0.0) | (0.0-0.0) | (0.1-0.1) | (0.6-1.0) | |
| **Equatorial Guinea** (high mort model) | Early | prop | 0.42 | 0.34 | 0.06 | 0.07 | 0.06 | 0.01 | 0 | --- | 0.03 | 1 |
| | | risk | 10.4 | 8.4 | 1.4 | 1.7 | 1.5 | 0.2 | 0.1 | --- | 0.8 | 24.6 |
| | | num | 2.7 | 2.1 | 0.4 | 0.4 | 0.4 | 0.1 | 0 | --- | 0.2 | 6.3 |
| | | | (1.7-4.2) | (1.3-2.9) | (0.2-0.7) | (0.1-0.7) | (0.2-0.8) | (0.0-0.1) | (0.0-0.3) | --- | (0.1-0.6) | |
| | Late | prop | 0.16 | 0.16 | 0.06 | 0.45 | 0.05 | 0.05 | 0.01 | --- | 0.06 | 1 |
| | | risk | 1.4 | 1.4 | 0.5 | 3.9 | 0.5 | 0.4 | 0.1 | --- | 0.6 | 8.6 |

| Table S14: Cause-specific proportions, risks, and numbers of deaths (with uncertainty) for 194 countries by neonatal period | | | | | | | | | | | |
|---|---|---|---|---|---|---|---|---|---|---|---|
| Country | Period | Stat* | Preterm | Intrapartum | Congenital | Sepsis | Pneumonia | Tetanus | Diarrhoea | Injuries | Other | Total |
| **Eritrea** (high mort model) | Early | prop | 0.35 | 0.3 | 0.06 | 0.17 | 0.06 | 0.02 | 0.01 | 0 | 0.04 | 1 |
| | | risk | 11.7 | 10.1 | 1.9 | 5.7 | 2.1 | 0.7 | 0.2 | 0 | 1.4 | 33.8 |
| | | num | 0.4 | 0.4 | 0.1 | 1 | 0.1 | 0.1 | 0 | --- | 0.1 | 2.2 |
| | | | (0.2-0.6) | (0.2-0.5) | (0.1-0.3) | (0.6-1.4) | (0.1-0.2) | (0.0-0.3) | (0.0-0.1) | --- | (0.1-0.4) | |
| | | num | 3 | 2.6 | 0.5 | 1.5 | 0.5 | 0.2 | 0 | 0 | 0.4 | 8.6 |
| | | | (2.0-4.5) | (1.7-3.4) | (0.3-1.0) | (0.8-2.2) | (0.3-1.0) | (0.1-0.4) | (0.0-0.4) | (0.0-0.0) | (0.2-0.9) | |
| | Early | prop | 0.24 | 0.37 | 0.19 | 0.11 | 0.05 | 0.01 | 0 | --- | 0.04 | 1 |
| | | risk | 3.1 | 4.8 | 2.5 | 1.4 | 0.7 | 0.1 | 0 | --- | 0.5 | 13.1 |
| | | num | 7.1 | 10.9 | 5.8 | 3.2 | 1.5 | 0.2 | 0 | --- | 1.1 | 29.7 |
| | | | (5.7-10.8) | (9.4-15.8) | (4.2-8.8) | (1.3-5.6) | (0.9-3.5) | (0.1-0.5) | (0.0-0.1) | --- | (0.6-3.2) | |
| | Late | prop | 0.13 | 0.15 | 0.07 | 0.49 | 0.06 | 0.01 | 0.01 | --- | 0.06 | 1 |
| | | risk | 0.6 | 0.7 | 0.3 | 2.3 | 0.3 | 0.1 | 0 | --- | 0.3 | 4.6 |
| | | num | 1.4 | 1.6 | 0.7 | 5.2 | 0.6 | 0.1 | 0.1 | --- | 0.7 | 10.4 |
| | | | (0.7-2.9) | (1.0-2.5) | (0.2-2.2) | (3.4-8.6) | (0.3-1.1) | (0.0-0.5) | (0.0-0.3) | --- | (0.3-1.8) | |
| | Overall | prop | 0.21 | 0.31 | 0.16 | 0.21 | 0.05 | 0.01 | 0 | 0 | 0.04 | 1 |
| | | risk | 3.9 | 5.7 | 2.8 | 3.8 | 1 | 0.1 | 0.1 | 0 | 0.8 | 18.2 |
| | | num | 8.6 | 12.7 | 6.3 | 8.4 | 2.2 | 0.3 | 0.1 | 0 | 1.7 | 40.3 |
| | | | (6.3-13.8) | (10.5-18.1) | (4.2-10.5) | (4.6-14.1) | (1.1-4.7) | (0.1-1.0) | (0.0-0.4) | (0.0-0.0) | (0.9-5.1) | |

* prop = proportion; num = number of deaths (in 100s)

| Table S14: Cause-specific proportions, risks, and numbers of deaths (with uncertainty) for 194 countries by neonatal period | | | | | | | | | | | |
|---|---|---|---|---|---|---|---|---|---|---|---|
| Country | Period | Stat* | Preterm | Intrapartum | Congenital | Sepsis | Pneumonia | Tetanus | Diarrhoea | Injuries | Other | Total |
| **Estonia** (high-quality VR) | Early | prop | 0.14 | 0.29 | 0.29 | 0.21 | 0 | --- | --- | 0 | 0.07 | 1 |
| | | risk | 0.2 | 0.3 | 0.3 | 0.2 | 0 | --- | --- | 0 | 0.1 | 1.1 |
| | | num | 0 | 0 | 0 | 0 | 0 | --- | --- | 0 | 0 | 0.2 |
| | | | (0.0-0.1) | (0.0-0.1) | (0.0-0.1) | (0.0-0.1) | (0.0-0.0) | --- | --- | (0.0-0.0) | (0.0-0.0) | |
| | Late | prop | 0.14 | 0 | 0.43 | 0.29 | 0 | --- | --- | 0.14 | 0 | 1 |
| | | risk | 0.1 | 0 | 0.2 | 0.2 | 0 | --- | --- | 0.1 | 0 | 0.6 |
| | | num | 0 | 0 | 0 | 0 | 0 | --- | --- | 0 | 0 | 0.1 |
| | | | (0.0-0.0) | (0.0-0.0) | (0.0-0.1) | (0.0-0.1) | (0.0-0.0) | --- | --- | (0.0-0.0) | (0.0-0.0) | |
| | Overall | prop | 0.14 | 0.19 | 0.33 | 0.24 | 0 | 0 | 0 | 0.05 | 0.05 | 1 |
| | | risk | 0.3 | 0.3 | 0.6 | 0.4 | 0 | 0 | 0 | 0.1 | 0.1 | 1.8 |
| | | num | 0 | 0 | 0.1 | 0.1 | 0 | 0 | 0 | 0 | 0 | 0.3 |
| | | | (0.0-0.1) | (0.0-0.1) | (0.0-0.2) | (0.0-0.1) | (0.0-0.0) | (0.0-0.0) | (0.0-0.0) | (0.0-0.0) | (0.0-0.0) | |
| **Ethiopia** (high mort model) | Early | prop | 0.27 | 0.41 | 0.11 | 0.09 | 0.07 | 0.02 | 0 | --- | 0.04 | 1 |
| | | risk | 5.5 | 8.3 | 2.2 | 1.8 | 1.3 | 0.4 | 0 | --- | 0.8 | 20.4 |
| | | num | 168.4 | 253.5 | 68.9 | 55.6 | 41.3 | 12.7 | 1.3 | --- | 23.1 | 624.8 |

| Table S14: Cause-specific proportions, risks, and numbers of deaths (with uncertainty) for 194 countries by neonatal period |||||||||||
|---|---|---|---|---|---|---|---|---|---|---|---|
| Country | Period | Stat* | Preterm | Intrapartum | Congenital | Sepsis | Pneumonia | Tetanus | Diarrhoea | Injuries | Other | Total |
| | | | (117.0-199.9) | (186.4-278.9) | (45.5-95.5) | (21.6-85.6) | (19.9-71.5) | (4.5-33.7) | (0.0-11.3) | --- | (10.3-62.8) | |
| | Late | prop | 0.18 | 0.14 | 0.07 | 0.44 | 0.05 | 0.03 | 0.04 | --- | 0.06 | 1 |
| | | risk | 1.3 | 1 | 0.5 | 3.2 | 0.3 | 0.2 | 0.3 | --- | 0.4 | 7.2 |
| | | num | 38.9 | 30.9 | 14.3 | 97.2 | 10.4 | 6.8 | 8.3 | --- | 12.7 | 219.5 |
| | | | (21.7-59.5) | (19.4-39.7) | (3.6-33.8) | (56.2-127.0) | (5.8-14.3) | (2.3-14.8) | (3.2-17.3) | --- | (4.9-30.8) | |
| | Overall | prop | 0.26 | 0.33 | 0.09 | 0.18 | 0.06 | 0.03 | 0.01 | 0 | 0.04 | 1 |
| | | risk | 7.3 | 9.3 | 2.7 | 5.1 | 1.7 | 0.7 | 0.3 | 0 | 1.2 | 28.4 |
| | | num | 221.2 | 282.7 | 80.7 | 155.9 | 52.3 | 21.6 | 9.6 | 0 | 35.8 | 859.8 |
| | | | (146.7-271.1) | (209.1-315.1) | (48.2-126.2) | (79.2-218.8) | (26.1-88.1) | (7.6-50.7) | (3.5-28.6) | (0.0-0.0) | (15.6-92.2) | |
| **Fiji** (low mort model) | Early | prop | 0.43 | 0.17 | 0.19 | 0.09 | 0.04 | --- | --- | 0.01 | 0.07 | 1 |
| | | risk | 3.3 | 1.3 | 1.4 | 0.7 | 0.3 | --- | --- | 0 | 0.6 | 7.6 |
| | | num | 0.6 | 0.2 | 0.3 | 0.1 | 0.1 | --- | --- | 0 | 0.1 | 1.4 |
| | | | (0.5-0.8) | (0.1-0.3) | (0.2-0.4) | (0.1-0.2) | (0.0-0.1) | --- | --- | (0.0-0.0) | (0.1-0.1) | |
| | Late | prop | 0.28 | 0.1 | 0.27 | 0.21 | 0.04 | --- | --- | 0.01 | 0.08 | 1 |
| | | risk | 0.8 | 0.3 | 0.7 | 0.6 | 0.1 | --- | --- | 0 | 0.2 | 2.7 |
| | | num | 0.1 | 0 | 0.1 | 0.1 | 0 | --- | --- | 0 | 0 | 0.5 |
| | | | (0.1-0.2) | (0.0-0.1) | (0.1-0.2) | (0.1-0.1) | (0.0-0.0) | --- | --- | (0.0-0.0) | (0.0-0.1) | |
| | Overall | prop | 0.39 | 0.15 | 0.21 | 0.13 | 0.04 | 0 | 0 | 0.01 | 0.08 | 1 |
| | | risk | 4.1 | 1.6 | 2.1 | 1.3 | 0.4 | 0 | 0 | 0.1 | 0.8 | 10.4 |
| | | num | 0.8 | 0.3 | 0.4 | 0.2 | 0.1 | 0 | 0 | 0 | 0.1 | 1.9 |
| | | | (0.6-1.0) | (0.2-0.4) | (0.3-0.6) | (0.2-0.3) | (0.0-0.1) | (0.0-0.0) | (0.0-0.0) | (0.0-0.0) | (0.1-0.2) | |

* prop = proportion; num = number of deaths (in 100s)

| Table S14: Cause-specific proportions, risks, and numbers of deaths (with uncertainty) for 194 countries by neonatal period |||||||||||
|---|---|---|---|---|---|---|---|---|---|---|---|
| Country | Period | Stat* | Preterm | Intrapartum | Congenital | Sepsis | Pneumonia | Tetanus | Diarrhoea | Injuries | Other | Total |
| **Finland** (high-quality VR) | Early | prop | 0.31 | 0.14 | 0.37 | 0.04 | 0.03 | --- | --- | 0.01 | 0.1 | 1 |
| | | risk | 0.3 | 0.1 | 0.4 | 0 | 0 | --- | --- | 0 | 0.1 | 1 |
| | | num | 0.2 | 0.1 | 0.2 | 0 | 0 | --- | --- | 0 | 0.1 | 0.6 |
| | | | (0.1-0.3) | (0.0-0.1) | (0.1-0.3) | (0.0-0.1) | (0.0-0.0) | --- | --- | (0.0-0.0) | (0.0-0.1) | |
| | Late | prop | 0.37 | 0 | 0.32 | 0.11 | 0 | --- | --- | 0 | 0.21 | 1 |
| | | risk | 0.1 | 0 | 0.1 | 0 | 0 | --- | --- | 0 | 0.1 | 0.3 |
| | | num | 0.1 | 0 | 0 | 0 | 0 | --- | --- | 0 | 0 | 0.2 |
| | | | (0.0-0.1) | (0.0-0.0) | (0.0-0.1) | (0.0-0.0) | (0.0-0.0) | --- | --- | (0.0-0.0) | (0.0-0.1) | |
| | Overall | prop | 0.32 | 0.11 | 0.36 | 0.05 | 0.02 | 0 | 0 | 0.01 | 0.12 | 1 |
| | | risk | 0.4 | 0.2 | 0.5 | 0.1 | 0 | 0 | 0 | 0 | 0.2 | 1.4 |
| | | num | 0.3 | 0.1 | 0.3 | 0 | 0 | 0 | 0 | 0 | 0.1 | 0.9 |

| Table S14: Cause-specific proportions, risks, and numbers of deaths (with uncertainty) for 194 countries by neonatal period | | | | | | | | | | | |
|---|---|---|---|---|---|---|---|---|---|---|---|
| Country | Period | Stat* | Preterm | Intrapartum | Congenital | Sepsis | Pneumonia | Tetanus | Diarrhoea | Injuries | Other | Total |
| | | | (0.1-0.4) | (0.0-0.2) | (0.2-0.5) | (0.0-0.1) | (0.0-0.0) | (0.0-0.0) | (0.0-0.0) | (0.0-0.0) | (0.0-0.2) | |
| **France** (high-quality VR) | Early | prop | 0.27 | 0.18 | 0.26 | 0.05 | 0 | --- | --- | 0 | 0.23 | 1 |
| | | risk | 0.4 | 0.3 | 0.4 | 0.1 | 0 | --- | --- | 0 | 0.4 | 1.6 |
| | | num | 3.4 | 2.3 | 3.3 | 0.6 | 0 | --- | --- | 0.1 | 3 | 12.7 |
| | | | (3.0-3.8) | (2.0-2.6) | (3.0-3.7) | (0.4-0.7) | (0.0-0.1) | --- | --- | (0.0-0.1) | (2.6-3.3) | |
| | Late | prop | 0.2 | 0.2 | 0.36 | 0.08 | 0.01 | --- | --- | 0 | 0.15 | 1 |
| | | risk | 0.1 | 0.1 | 0.3 | 0.1 | 0 | --- | --- | 0 | 0.1 | 0.7 |
| | | num | 1.2 | 1.1 | 2 | 0.5 | 0 | --- | --- | 0 | 0.8 | 5.6 |
| | | | (0.9-1.4) | (0.9-1.3) | (1.7-2.3) | (0.3-0.6) | (0.0-0.1) | --- | --- | (0.0-0.0) | (0.6-1.0) | |
| | Overall | prop | 0.25 | 0.19 | 0.29 | 0.06 | 0 | 0 | 0 | 0 | 0.21 | 1 |
| | | risk | 0.6 | 0.4 | 0.7 | 0.1 | 0 | 0 | 0 | 0 | 0.5 | 2.3 |
| | | num | 4.5 | 3.4 | 5.3 | 1.1 | 0.1 | 0 | 0 | 0.1 | 3.8 | 18.2 |
| | | | (4.0-5.1) | (2.9-3.9) | (4.7-6.0) | (0.8-1.3) | (0.0-0.2) | (0.0-0.0) | (0.0-0.0) | (0.0-0.1) | (3.2-4.3) | |
| **Gabon** (high mort model) | Early | prop | 0.4 | 0.31 | 0.12 | 0.08 | 0.05 | 0 | 0 | --- | 0.03 | 1 |
| | | risk | 6.8 | 5.3 | 2 | 1.4 | 0.8 | 0.1 | 0 | --- | 0.5 | 16.9 |
| | | num | 3.6 | 2.8 | 1.1 | 0.8 | 0.4 | 0 | 0 | --- | 0.2 | 8.9 |
| | | | (2.9-4.5) | (2.1-3.4) | (0.7-1.5) | (0.3-1.2) | (0.2-0.8) | (0.0-0.1) | (0.0-0.1) | --- | (0.1-0.7) | |
| | Late | prop | 0.15 | 0.14 | 0.13 | 0.44 | 0.05 | 0.02 | 0.01 | --- | 0.06 | 1 |
| | | risk | 0.9 | 0.8 | 0.8 | 2.6 | 0.3 | 0.1 | 0 | --- | 0.4 | 5.9 |
| | | num | 0.5 | 0.4 | 0.4 | 1.4 | 0.2 | 0.1 | 0 | --- | 0.2 | 3.1 |
| | | | (0.2-0.8) | (0.3-0.6) | (0.1-0.9) | (0.8-2.0) | (0.1-0.2) | (0.0-0.2) | (0.0-0.0) | --- | (0.1-0.5) | |
| | Overall | prop | 0.36 | 0.26 | 0.12 | 0.17 | 0.05 | 0.01 | 0 | 0 | 0.04 | 1 |
| | | risk | 8.3 | 6 | 2.8 | 4 | 1.1 | 0.2 | 0.1 | 0 | 0.8 | 23.3 |
| | | num | 4.3 | 3.1 | 1.5 | 2.1 | 0.5 | 0.1 | 0 | 0 | 0.4 | 12.1 |
| | | | (3.2-5.5) | (2.4-3.9) | (0.8-2.4) | (1.0-3.2) | (0.3-1.0) | (0.0-0.3) | (0.0-0.1) | (0.0-0.0) | (0.2-1.2) | |

* prop = proportion; num = number of deaths (in 100s)

| Table S14: Cause-specific proportions, risks, and numbers of deaths (with uncertainty) for 194 countries by neonatal period | | | | | | | | | | | |
|---|---|---|---|---|---|---|---|---|---|---|---|
| Country | Period | Stat* | Preterm | Intrapartum | Congenital | Sepsis | Pneumonia | Tetanus | Diarrhoea | Injuries | Other | Total |
| **Gambia** (high mort model) | Early | prop | 0.33 | 0.35 | 0.1 | 0.1 | 0.06 | 0.01 | 0 | --- | 0.05 | 1 |
| | | risk | 6.9 | 7.4 | 2.1 | 2 | 1.2 | 0.1 | 0.1 | --- | 1.1 | 20.8 |
| | | num | 5.3 | 5.7 | 1.6 | 1.6 | 0.9 | 0.1 | 0 | --- | 0.8 | 16 |
| | | | (3.9-7.2) | (4.3-7.4) | (0.9-2.7) | (0.6-2.7) | (0.4-1.9) | (0.0-0.3) | (0.0-0.4) | --- | (0.5-1.6) | |
| | Late | prop | 0.14 | 0.16 | 0.06 | 0.46 | 0.06 | 0.03 | 0.01 | --- | 0.07 | 1 |
| | | risk | 1 | 1.2 | 0.5 | 3.4 | 0.4 | 0.2 | 0.1 | --- | 0.5 | 7.3 |
| | | num | 0.8 | 0.9 | 0.4 | 2.6 | 0.3 | 0.2 | 0.1 | --- | 0.4 | 5.6 |
| | | | (0.5-1.3) | (0.6-1.3) | (0.1-0.9) | (1.5-3.7) | (0.2-0.5) | (0.1-0.5) | (0.0-0.1) | --- | (0.2-0.9) | |

| Table S14: Cause-specific proportions, risks, and numbers of deaths (with uncertainty) for 194 countries by neonatal period | | | | | | | | | | | |
|---|---|---|---|---|---|---|---|---|---|---|---|
| Country | Period | Stat* | Preterm | Intrapartum | Congenital | Sepsis | Pneumonia | Tetanus | Diarrhoea | Injuries | Other | Total |
| **Georgia** (low mort model) | Overall | prop | 0.27 | 0.31 | 0.09 | 0.2 | 0.06 | 0.01 | 0 | 0 | 0.06 | 1 |
| | | risk | 7.7 | 9 | 2.5 | 5.7 | 1.7 | 0.3 | 0.1 | 0 | 1.6 | 28.7 |
| | | num | 5.7 | 6.7 | 1.9 | 4.3 | 1.3 | 0.2 | 0.1 | 0 | 1.2 | 21.4 |
| | | | (3.9-8.5) | (4.9-8.7) | (1.0-3.5) | (2.2-6.5) | (0.6-2.5) | (0.1-0.7) | (0.0-0.6) | (0.0-0.0) | (0.7-2.5) | |
| | Early | prop | 0.4 | 0.12 | 0.26 | 0.07 | 0.03 | --- | --- | 0.01 | 0.11 | 1 |
| | | risk | 2.9 | 0.9 | 1.8 | 0.5 | 0.2 | --- | --- | 0 | 0.8 | 7.2 |
| | | num | 1.6 | 0.5 | 1 | 0.3 | 0.1 | --- | --- | 0 | 0.4 | 4 |
| | | | (1.4-1.9) | (0.4-0.6) | (0.7-1.3) | (0.2-0.4) | (0.1-0.2) | --- | --- | (0.0-0.0) | (0.3-0.6) | |
| | Late | prop | 0.3 | 0.07 | 0.29 | 0.18 | 0.08 | --- | --- | 0.02 | 0.07 | 1 |
| | | risk | 0.8 | 0.2 | 0.7 | 0.4 | 0.2 | --- | --- | 0 | 0.2 | 2.5 |
| | | num | 0.4 | 0.1 | 0.4 | 0.3 | 0.1 | --- | --- | 0 | 0.1 | 1.4 |
| | | | (0.3-0.5) | (0.1-0.1) | (0.3-0.4) | (0.2-0.3) | (0.1-0.2) | --- | --- | (0.0-0.0) | (0.1-0.1) | |
| | Overall | prop | 0.39 | 0.11 | 0.25 | 0.09 | 0.05 | 0 | 0 | 0.01 | 0.1 | 1 |
| | | risk | 4 | 1.1 | 2.6 | 0.9 | 0.5 | 0 | 0 | 0.1 | 1.1 | 10.3 |
| | | num | 2.4 | 0.7 | 1.6 | 0.6 | 0.3 | 0 | 0 | 0.1 | 0.6 | 6.3 |
| | | | (2.1-2.8) | (0.5-0.9) | (1.2-1.9) | (0.5-0.7) | (0.2-0.5) | (0.0-0.0) | (0.0-0.0) | (0.0-0.1) | (0.5-0.8) | |
| **Germany** (high-quality VR) | Early | prop | 0.45 | 0.11 | 0.27 | 0.02 | 0 | --- | --- | 0.01 | 0.13 | 1 |
| | | risk | 0.8 | 0.2 | 0.5 | 0 | 0 | --- | --- | 0 | 0.2 | 1.7 |
| | | num | 5.6 | 1.3 | 3.3 | 0.3 | 0 | --- | --- | 0.2 | 1.6 | 12.3 |
| | | | (5.1-6.0) | (1.1-1.6) | (2.9-3.6) | (0.2-0.4) | (0.0-0.1) | --- | --- | (0.1-0.2) | (1.4-1.9) | |
| | Late | prop | 0.33 | 0.07 | 0.41 | 0.09 | 0.01 | --- | --- | 0.02 | 0.06 | 1 |
| | | risk | 0.2 | 0 | 0.2 | 0 | 0 | --- | --- | 0 | 0 | 0.5 |
| | | num | 1.1 | 0.2 | 1.4 | 0.3 | 0 | --- | --- | 0.1 | 0.2 | 3.4 |
| | | | (0.9-1.3) | (0.1-0.3) | (1.2-1.6) | (0.2-0.4) | (0.0-0.1) | --- | --- | (0.0-0.1) | (0.1-0.3) | |
| | Overall | prop | 0.43 | 0.1 | 0.3 | 0.04 | 0 | 0 | 0 | 0.01 | 0.12 | 1 |
| | | risk | 1 | 0.2 | 0.7 | 0.1 | 0 | 0 | 0 | 0 | 0.3 | 2.3 |
| | | num | 6.9 | 1.6 | 4.8 | 0.6 | 0.1 | 0 | 0 | 0.2 | 1.9 | 16.1 |
| | | | (6.2-7.6) | (1.3-2.0) | (4.2-5.4) | (0.4-0.8) | (0.0-0.1) | (0.0-0.0) | (0.0-0.0) | (0.1-0.3) | (1.5-2.2) | |

* prop = proportion; num = number of deaths (in 100s)

| Table S14: Cause-specific proportions, risks, and numbers of deaths (with uncertainty) for 194 countries by neonatal period | | | | | | | | | | | |
|---|---|---|---|---|---|---|---|---|---|---|---|
| Country | Period | Stat* | Preterm | Intrapartum | Congenital | Sepsis | Pneumonia | Tetanus | Diarrhoea | Injuries | Other | Total |
| **Ghana** (high mort model) | Early | prop | 0.36 | 0.35 | 0.09 | 0.1 | 0.06 | 0.01 | 0 | --- | 0.03 | 1 |
| | | risk | 7.8 | 7.6 | 2 | 2.2 | 1.3 | 0.1 | 0 | --- | 0.7 | 21.7 |
| | | num | 61.6 | 59.6 | 15.9 | 17.3 | 10 | 1 | 0.3 | --- | 5.3 | 171 |
| | | | (49.9-71.4) | (48.7-68.7) | (9.3-24.4) | (6.1-26.6) | (5.2-19.3) | (0.3-2.4) | (0.0-2.9) | --- | (2.5-17.4) | |
| | Late | prop | 0.17 | 0.15 | 0.06 | 0.46 | 0.05 | 0.03 | 0.01 | --- | 0.07 | 1 |

| Table S14: Cause-specific proportions, risks, and numbers of deaths (with uncertainty) for 194 countries by neonatal period | | | | | | | | | | | |
|---|---|---|---|---|---|---|---|---|---|---|---|
| Country | Period | Stat* | Preterm | Intrapartum | Congenital | Sepsis | Pneumonia | Tetanus | Diarrhoea | Injuries | Other | Total |
| | | risk | 1.3 | 1.1 | 0.5 | 3.5 | 0.4 | 0.2 | 0.1 | --- | 0.5 | 7.6 |
| | | num | 10 | 8.9 | 3.9 | 27.5 | 3.2 | 1.6 | 0.6 | --- | 4.3 | 60.1 |
| | | | (4.6-18.2) | (5.6-11.8) | (1.3-9.9) | (16.2-38.0) | (1.8-4.6) | (0.5-3.9) | (0.2-1.1) | --- | (1.4-10.9) | |
| | | prop | 0.31 | 0.3 | 0.08 | 0.19 | 0.06 | 0.01 | 0 | 0 | 0.04 | 1 |
| | Overall | risk | 9.2 | 8.8 | 2.5 | 5.8 | 1.7 | 0.4 | 0.1 | 0 | 1.2 | 29.7 |
| | | num | 72 | 69.5 | 19.6 | 45.3 | 13.5 | 2.8 | 0.9 | 0 | 9.7 | 233.2 |
| | | | (54.6-90.0) | (54.3-80.8) | (10.3-33.7) | (22.7-65.8) | (7.1-24.5) | (0.9-6.6) | (0.2-4.1) | (0.0-0.0) | (3.9-28.1) | |
| | | prop | 0.55 | 0.09 | 0.3 | 0.01 | 0 | --- | --- | 0 | 0.05 | 1 |
| | Early | risk | 0.9 | 0.2 | 0.5 | 0 | 0 | --- | --- | 0 | 0.1 | 1.7 |
| | | num | 1 | 0.2 | 0.5 | 0 | 0 | --- | --- | 0 | 0.1 | 1.8 |
| | | | (0.8-1.2) | (0.1-0.2) | (0.4-0.7) | (0.0-0.0) | (0.0-0.0) | --- | --- | (0.0-0.0) | (0.0-0.1) | |
| **Greece** (high-quality VR) | | prop | 0.45 | 0.1 | 0.43 | 0 | 0 | --- | --- | 0 | 0.02 | 1 |
| | Late | risk | 0.5 | 0.1 | 0.5 | 0 | 0 | --- | --- | 0 | 0 | 1.1 |
| | | num | 0.5 | 0.1 | 0.5 | 0 | 0 | --- | --- | 0 | 0 | 1.1 |
| | | | (0.4-0.7) | (0.0-0.2) | (0.4-0.6) | (0.0-0.0) | (0.0-0.0) | --- | --- | (0.0-0.0) | (0.0-0.1) | |
| | | prop | 0.51 | 0.09 | 0.35 | 0 | 0 | 0 | 0 | 0 | 0.04 | 1 |
| | Overall | risk | 1.4 | 0.3 | 1 | 0 | 0 | 0 | 0 | 0 | 0.1 | 2.8 |
| | | num | 1.6 | 0.3 | 1.1 | 0 | 0 | 0 | 0 | 0 | 0.1 | 3.2 |
| | | | (1.3-2.0) | (0.2-0.5) | (0.8-1.4) | (0.0-0.0) | (0.0-0.0) | (0.0-0.0) | (0.0-0.0) | (0.0-0.0) | (0.0-0.2) | |
| | | prop | 0.33 | 0.33 | 0.08 | 0.08 | 0 | --- | --- | 0 | 0.17 | 1 |
| | Early | risk | 1.7 | 1.7 | 0.4 | 0.4 | 0 | --- | --- | 0 | 0.8 | 5 |
| | | num | 0 | 0 | 0 | 0 | 0 | --- | --- | 0 | 0 | 0.1 |
| | | | (0.0-0.1) | (0.0-0.1) | (0.0-0.0) | (0.0-0.0) | (0.0-0.0) | --- | --- | (0.0-0.0) | (0.0-0.0) | |
| **Grenada** (high-quality VR) | | prop | 0 | 0 | 0.5 | 0.5 | 0 | --- | --- | 0 | 0 | 1 |
| | Late | risk | 0 | 0 | 0.4 | 0.4 | 0 | --- | --- | 0 | 0 | 0.8 |
| | | num | 0 | 0 | 0 | 0 | 0 | --- | --- | 0 | 0 | 0 |
| | | | (0.0-0.0) | (0.0-0.0) | (0.0-0.0) | (0.0-0.0) | (0.0-0.0) | --- | --- | (0.0-0.0) | (0.0-0.0) | |
| | | prop | 0.29 | 0.29 | 0.14 | 0.14 | 0 | 0 | 0 | 0 | 0.14 | 1 |
| | Overall | risk | 1.7 | 1.7 | 0.9 | 0.9 | 0 | 0 | 0 | 0 | 0.9 | 6 |
| | | num | 0 | 0 | 0 | 0 | 0 | 0 | 0 | 0 | 0 | 0.1 |
| | | | (0.0-0.1) | (0.0-0.1) | (0.0-0.1) | (0.0-0.1) | (0.0-0.0) | (0.0-0.0) | (0.0-0.0) | (0.0-0.0) | (0.0-0.0) | |

* prop = proportion; num = number of deaths (in 100s)

| Table S14: Cause-specific proportions, risks, and numbers of deaths (with uncertainty) for 194 countries by neonatal period | | | | | | | | | | | |
|---|---|---|---|---|---|---|---|---|---|---|---|
| Country | Period | Stat* | Preterm | Intrapartum | Congenital | Sepsis | Pneumonia | Tetanus | Diarrhoea | Injuries | Other | Total |
| **Guatemala** (high mort | Early | prop | 0.26 | 0.33 | 0.23 | 0.08 | 0.04 | 0 | 0 | --- | 0.06 | 1 |
| | | risk | 2.8 | 3.6 | 2.5 | 0.9 | 0.4 | 0 | 0 | --- | 0.7 | 11 |

| Table S14: Cause-specific proportions, risks, and numbers of deaths (with uncertainty) for 194 countries by neonatal period ||||||||||||
| Country | Period | Stat* | Preterm | Intrapartum | Congenital | Sepsis | Pneumonia | Tetanus | Diarrhoea | Injuries | Other | Total |
|---|---|---|---|---|---|---|---|---|---|---|---|---|
| model) | | num | 13.4 | 16.9 | 12.1 | 4.2 | 2.1 | 0.2 | 0 | --- | 3.1 | 52 |
| | | | (10.8-16.6) | (14.1-19.4) | (7.8-16.5) | (1.8-6.6) | (1.0-4.3) | (0.1-0.4) | (0.0-0.3) | --- | (1.5-4.8) | |
| | Late | prop | 0.12 | 0.16 | 0.12 | 0.46 | 0.06 | 0.01 | 0.01 | --- | 0.05 | 1 |
| | | risk | 0.5 | 0.6 | 0.5 | 1.8 | 0.2 | 0.1 | 0 | --- | 0.2 | 3.8 |
| | | num | 2.2 | 2.9 | 2.3 | 8.5 | 1 | 0.3 | 0.1 | --- | 1 | 18.3 |
| | | | (1.3-3.4) | (1.9-3.9) | (0.9-4.1) | (5.4-11.2) | (0.6-1.5) | (0.1-0.7) | (0.1-0.3) | --- | (0.5-2.2) | |
| | Overall | prop | 0.22 | 0.28 | 0.21 | 0.18 | 0.04 | 0.01 | 0 | 0 | 0.06 | 1 |
| | | risk | 3.3 | 4.2 | 3.1 | 2.8 | 0.7 | 0.1 | 0 | 0 | 0.9 | 15.2 |
| | | num | 15.5 | 19.8 | 14.6 | 13.1 | 3.1 | 0.4 | 0.2 | 0 | 4.1 | 70.8 |
| | | | (11.8-20.6) | (16.2-23.5) | (9.0-19.9) | (7.3-18.2) | (1.6-6.0) | (0.1-1.1) | (0.1-0.6) | (0.0-0.0) | (2.0-7.1) | |
| **Guinea** (high mort model) | Early | prop | 0.31 | 0.41 | 0.07 | 0.08 | 0.08 | 0.01 | 0 | --- | 0.04 | 1 |
| | | risk | 7.5 | 10 | 1.6 | 1.8 | 1.9 | 0.3 | 0.1 | --- | 1.1 | 24.3 |
| | | num | 31.6 | 41.9 | 6.7 | 7.7 | 8 | 1.1 | 0.4 | --- | 4.5 | 101.8 |
| | | | (24.7-39.0) | (31.1-47.2) | (5.0-11.6) | (3.0-11.9) | (4.0-14.0) | (0.4-2.2) | (0.0-4.1) | --- | (2.4-9.9) | |
| | Late | prop | 0.15 | 0.16 | 0.04 | 0.46 | 0.05 | 0.04 | 0.02 | --- | 0.06 | 1 |
| | | risk | 1.3 | 1.4 | 0.4 | 3.9 | 0.5 | 0.4 | 0.2 | --- | 0.6 | 8.5 |
| | | num | 5.4 | 5.7 | 1.6 | 16.4 | 1.9 | 1.6 | 0.9 | --- | 2.3 | 35.8 |
| | | | (3.1-8.0) | (3.7-7.4) | (0.6-3.9) | (10.2-21.5) | (1.2-2.8) | (0.5-3.8) | (0.4-1.5) | --- | (1.1-5.5) | |
| | Overall | prop | 0.27 | 0.35 | 0.06 | 0.17 | 0.07 | 0.02 | 0.01 | 0 | 0.05 | 1 |
| | | risk | 9 | 11.6 | 2 | 5.9 | 2.5 | 0.7 | 0.3 | 0 | 1.7 | 33.6 |
| | | num | 37.4 | 48 | 8.1 | 24.3 | 10.2 | 2.8 | 1.4 | 0 | 6.9 | 139.1 |
| | | | (28.1-47.5) | (36.1-55.3) | (5.4-15.8) | (13.3-34.0) | (5.3-17.1) | (0.9-6.6) | (0.4-6.0) | (0.0-0.0) | (3.5-15.4) | |
| **Guinea-Bissau** (high mort model) | Early | prop | 0.31 | 0.38 | 0.07 | 0.1 | 0.08 | 0.01 | 0.01 | --- | 0.04 | 1 |
| | | risk | 10 | 12.4 | 2.2 | 3.3 | 2.7 | 0.5 | 0.2 | --- | 1.3 | 32.6 |
| | | num | 6.1 | 7.6 | 1.3 | 2 | 1.6 | 0.3 | 0.1 | --- | 0.8 | 19.9 |
| | | | (4.5-7.9) | (5.6-9.7) | (0.8-2.3) | (0.7-3.3) | (0.8-3.1) | (0.1-0.8) | (0.0-0.9) | --- | (0.4-2.0) | |
| | Late | prop | 0.15 | 0.16 | 0.04 | 0.45 | 0.05 | 0.08 | 0.01 | --- | 0.06 | 1 |
| | | risk | 1.7 | 1.8 | 0.4 | 5.1 | 0.6 | 0.9 | 0.1 | --- | 0.7 | 11.4 |
| | | num | 1 | 1.1 | 0.3 | 3.1 | 0.4 | 0.5 | 0.1 | --- | 0.5 | 7 |
| | | | (0.6-1.7) | (0.7-1.6) | (0.1-0.7) | (1.9-4.6) | (0.2-0.6) | (0.2-1.3) | (0.0-0.2) | --- | (0.2-1.1) | |
| | Overall | prop | 0.26 | 0.32 | 0.06 | 0.19 | 0.08 | 0.03 | 0.01 | 0 | 0.05 | 1 |
| | | risk | 11.9 | 14.5 | 2.7 | 8.6 | 3.4 | 1.4 | 0.3 | 0 | 2.1 | 44.9 |
| | | num | 7.1 | 8.7 | 1.6 | 5.2 | 2.1 | 0.9 | 0.2 | 0 | 1.3 | 27 |
| | | | (5.1-9.6) | (6.3-11.3) | (0.9-2.8) | (2.6-8.1) | (1.1-3.9) | (0.3-2.3) | (0.0-1.0) | (0.0-0.0) | (0.6-3.1) | |

* prop = proportion; num = number of deaths (in 100s)

| Country | Period | Stat* | Preterm | Intrapartum | Congenital | Sepsis | Pneumonia | Tetanus | Diarrhoea | Injuries | Other | Total |
|---|---|---|---|---|---|---|---|---|---|---|---|---|
| **Guyana** (high-quality VR) | Early | prop | 0.4 | 0.24 | 0.09 | 0.1 | 0.02 | --- | --- | 0 | 0.15 | 1 |
| | | risk | 6.2 | 3.7 | 1.4 | 1.6 | 0.3 | --- | --- | 0 | 2.4 | 15.6 |
| | | num | 0.9 | 0.6 | 0.2 | 0.2 | 0 | --- | --- | 0 | 0.4 | 2.3 |
| | | | (0.7-1.1) | (0.4-0.7) | (0.1-0.3) | (0.1-0.3) | (0.0-0.1) | --- | --- | (0.0-0.0) | (0.2-0.5) | |
| | Late | prop | 0.19 | 0.16 | 0.13 | 0.19 | 0.06 | --- | --- | 0 | 0.26 | 1 |
| | | risk | 0.9 | 0.7 | 0.6 | 0.9 | 0.3 | --- | --- | 0 | 1.1 | 4.4 |
| | | num | 0.1 | 0.1 | 0.1 | 0.1 | 0 | --- | --- | 0 | 0.2 | 0.7 |
| | | | (0.1-0.2) | (0.0-0.2) | (0.0-0.1) | (0.1-0.2) | (0.0-0.1) | --- | --- | (0.0-0.0) | (0.1-0.3) | |
| | Overall | prop | 0.35 | 0.22 | 0.1 | 0.12 | 0.03 | 0 | 0 | 0 | 0.18 | 1 |
| | | risk | 7.2 | 4.5 | 2 | 2.5 | 0.6 | 0 | 0 | 0 | 3.6 | 20.4 |
| | | num | 1.2 | 0.7 | 0.3 | 0.4 | 0.1 | 0 | 0 | 0 | 0.6 | 3.3 |
| | | | (0.9-1.4) | (0.5-0.9) | (0.2-0.5) | (0.2-0.6) | (0.0-0.2) | (0.0-0.0) | (0.0-0.0) | (0.0-0.0) | (0.4-0.8) | |
| **Haiti** (high mort model) | Early | prop | 0.38 | 0.3 | 0.09 | 0.08 | 0.05 | 0.01 | 0 | --- | 0.07 | 1 |
| | | risk | 7.1 | 5.6 | 1.7 | 1.5 | 1 | 0.1 | 0.1 | --- | 1.4 | 18.4 |
| | | num | 18.7 | 14.8 | 4.4 | 4.1 | 2.6 | 0.3 | 0.2 | --- | 3.6 | 48.8 |
| | | | (16.1-21.3) | (12.6-17.3) | (3.3-6.3) | (1.7-7.1) | (1.4-5.0) | (0.1-0.7) | (0.0-1.1) | --- | (2.0-4.8) | |
| | Late | prop | 0.2 | 0.13 | 0.06 | 0.45 | 0.05 | 0.02 | 0.03 | --- | 0.07 | 1 |
| | | risk | 1.3 | 0.9 | 0.4 | 2.9 | 0.3 | 0.2 | 0.2 | --- | 0.4 | 6.5 |
| | | num | 3.4 | 2.3 | 0.9 | 7.7 | 0.8 | 0.4 | 0.5 | --- | 1.2 | 17.1 |
| | | | (2.4-4.5) | (1.5-3.1) | (0.5-1.7) | (5.4-10.1) | (0.5-1.1) | (0.1-1.0) | (0.2-0.9) | --- | (0.7-1.9) | |
| | Overall | prop | 0.34 | 0.26 | 0.08 | 0.18 | 0.05 | 0.01 | 0.01 | 0 | 0.07 | 1 |
| | | risk | 8.5 | 6.6 | 2 | 4.5 | 1.3 | 0.3 | 0.2 | 0 | 1.9 | 25.4 |
| | | num | 22.7 | 17.5 | 5.3 | 12.1 | 3.5 | 0.8 | 0.6 | 0 | 4.9 | 67.4 |
| | | | (19.1-26.5) | (14.4-20.8) | (3.7-8.0) | (7.2-17.6) | (2.0-6.3) | (0.3-1.8) | (0.2-2.2) | (0.0-0.0) | (2.8-6.9) | |
| **Honduras** (low mort model) | Early | prop | 0.47 | 0.18 | 0.16 | 0.1 | 0.05 | --- | --- | 0.01 | 0.03 | 1 |
| | | risk | 4.1 | 1.6 | 1.3 | 0.9 | 0.4 | --- | --- | 0.1 | 0.2 | 8.6 |
| | | num | 8.5 | 3.3 | 2.8 | 1.9 | 0.8 | --- | --- | 0.2 | 0.4 | 17.9 |
| | | | (7.5-9.3) | (2.4-3.9) | (2.4-3.7) | (1.5-2.4) | (0.5-1.4) | --- | --- | (0.1-0.6) | (0.2-1.0) | |
| | Late | prop | 0.25 | 0.08 | 0.21 | 0.25 | 0.13 | --- | --- | 0.01 | 0.06 | 1 |
| | | risk | 0.8 | 0.2 | 0.6 | 0.7 | 0.4 | --- | --- | 0 | 0.2 | 3 |
| | | num | 1.6 | 0.5 | 1.3 | 1.6 | 0.8 | --- | --- | 0.1 | 0.4 | 6.3 |
| | | | (1.3-1.9) | (0.3-0.6) | (1.1-1.6) | (1.2-2.0) | (0.6-1.2) | --- | --- | (0.1-0.1) | (0.3-0.5) | |
| | Overall | prop | 0.42 | 0.16 | 0.17 | 0.14 | 0.07 | 0 | 0 | 0.01 | 0.04 | 1 |
| | | risk | 5 | 1.9 | 2 | 1.7 | 0.8 | 0 | 0 | 0.1 | 0.4 | 11.9 |
| | | num | 10.2 | 3.9 | 4.2 | 3.4 | 1.7 | 0 | 0 | 0.3 | 0.9 | 24.6 |
| | | | (9.0-11.4) | (2.7-4.7) | (3.6-5.3) | (2.8-4.5) | (1.1-2.7) | (0.0-0.0) | (0.0-0.0) | (0.2-0.7) | (0.5-1.6) | |

* prop = proportion; num = number of deaths (in 100s)

| Table S14: Cause-specific proportions, risks, and numbers of deaths (with uncertainty) for 194 countries by neonatal period | | | | | | | | | | | | |
|---|---|---|---|---|---|---|---|---|---|---|---|---|
| Country | Period | Stat* | Preterm | Intrapartum | Congenital | Sepsis | Pneumonia | Tetanus | Diarrhoea | Injuries | Other | Total |
| **Hungary** (high-quality VR) | Early | prop | 0.56 | 0.11 | 0.24 | 0.03 | 0.01 | --- | --- | 0.02 | 0.03 | 1 |
| | | risk | 1.4 | 0.3 | 0.6 | 0.1 | 0 | --- | --- | 0 | 0.1 | 2.5 |
| | | num | 1.4 (1.2-1.6) | 0.3 (0.2-0.4) | 0.6 (0.4-0.8) | 0.1 (0.0-0.1) | 0 (0.0-0.1) | --- | --- | 0 (0.0-0.1) | 0.1 (0.0-0.1) | 2.5 |
| | Late | prop | 0.44 | 0.05 | 0.3 | 0.09 | 0.04 | --- | --- | 0.01 | 0.07 | 1 |
| | | risk | 0.5 | 0.1 | 0.3 | 0.1 | 0 | --- | --- | 0 | 0.1 | 1.1 |
| | | num | 0.5 (0.3-0.6) | 0.1 (0.0-0.1) | 0.3 (0.2-0.4) | 0.1 (0.0-0.2) | 0 (0.0-0.1) | --- | --- | 0 (0.0-0.0) | 0.1 (0.0-0.1) | 1.1 |
| | Overall | prop | 0.53 | 0.09 | 0.26 | 0.04 | 0.02 | 0 | 0 | 0.01 | 0.04 | 1 |
| | | risk | 2 | 0.3 | 1 | 0.2 | 0.1 | 0 | 0 | 0.1 | 0.2 | 3.7 |
| | | num | 1.9 (1.5-2.3) | 0.3 (0.2-0.5) | 0.9 (0.7-1.2) | 0.2 (0.1-0.3) | 0.1 (0.0-0.1) | 0 (0.0-0.0) | 0 (0.0-0.0) | 0.1 (0.0-0.1) | 0.1 (0.0-0.3) | 3.6 |
| **Iceland** (high-quality VR) | Early | prop | 0.67 | 0 | 0 | 0 | 0 | --- | --- | 0 | 0.33 | 1 |
| | | risk | 0.6 | 0 | 0 | 0 | 0 | --- | --- | 0 | 0.3 | 0.9 |
| | | num | 0 (0.0-0.1) | 0 (0.0-0.0) | 0 (0.0-0.0) | 0 (0.0-0.0) | 0 (0.0-0.0) | --- | --- | 0 (0.0-0.0) | 0 (0.0-0.0) | 0 |
| | Late | prop | --- | --- | --- | --- | --- | --- | --- | --- | --- | 1 |
| | | risk | 0 | 0 | 0 | 0 | 0 | --- | --- | 0 | 0 | 0 |
| | | num | 0 (0.0-0.0) | 0 (0.0-0.0) | 0 (0.0-0.0) | 0 (0.0-0.0) | 0 (0.0-0.0) | --- | --- | 0 (0.0-0.0) | 0 (0.0-0.0) | 0 |
| | Overall | prop | 0.67 | 0 | 0 | 0 | 0 | 0 | 0 | 0 | 0.33 | 1 |
| | | risk | 0.7 | 0 | 0 | 0 | 0 | 0 | 0 | 0 | 0.3 | 1 |
| | | num | 0 (0.0-0.1) | 0 (0.0-0.0) | 0 (0.0-0.0) | 0 (0.0-0.0) | 0 (0.0-0.0) | 0 (0.0-0.0) | 0 (0.0-0.0) | 0 (0.0-0.0) | 0 (0.0-0.0) | 0 |
| **India** (high mort model) | Early | prop | 0.5 | 0.22 | 0.08 | 0.08 | 0.04 | 0.01 | 0 | --- | 0.08 | 1 |
| | | risk | 10.7 | 4.7 | 1.7 | 1.7 | 0.8 | 0.2 | 0.1 | --- | 1.7 | 21.6 |
| | | num | 2742.2 (1586.8-4064.8) | 1210.3 (676.5-1959.1) | 447.5 (202.1-806.2) | 435.2 (107.6-849.0) | 198.1 (66.6-510.3) | 43.9 (9.0-137.1) | 15.8 (0.0-120.7) | --- | 438.5 (120.1-800.5) | 5531.5 |
| | Late | prop | 0.27 | 0.11 | 0.08 | 0.35 | 0.04 | 0.04 | 0.02 | --- | 0.09 | 1 |
| | | risk | 2 | 0.9 | 0.6 | 2.6 | 0.3 | 0.3 | 0.1 | --- | 0.7 | 7.6 |
| | | num | 522.6 (236.5-891.3) | 222.5 (100.4-384.6) | 159.1 (65.1-321.5) | 674.5 (305.5-1204.9) | 77.9 (33.3-145.4) | 71.9 (19.3-200.0) | 38.3 (9.4-122.4) | --- | 176.8 (57.8-366.3) | 1943.5 |
| | Overall | prop | 0.44 | 0.19 | 0.08 | 0.15 | 0.04 | 0.02 | 0.01 | 0 | 0.08 | 1 |

| Table S14: Cause-specific proportions, risks, and numbers of deaths (with uncertainty) for 194 countries by neonatal period | | | | | | | | | | | |
|---|---|---|---|---|---|---|---|---|---|---|---|
| Country | Period | Stat* | Preterm | Intrapartum | Congenital | Sepsis | Pneumonia | Tetanus | Diarrhoea | Injuries | Other | Total |
| | | risk | 13.2 | 5.8 | 2.4 | 4.5 | 1.1 | 0.5 | 0.2 | 0 | 2.5 | 30.1 |
| | | num | 3323.7 (1833.8-5081.7) | 1452.4 (777.7-2381.7) | 593.8 (263.6-1106.8) | 1127.2 (414.8-2094.7) | 283.1 (100.9-673.8) | 121.7 (28.7-354.6) | 55.2 (9.3-251.4) | 0 (0.0-0.0) | 623.8 (178.3-1194.0) | 7580.9 |

| Country | Period | Stat* | Preterm | Intrapartum | Congenital | Sepsis | Pneumonia | Tetanus | Diarrhoea | Injuries | Other | Total |
|---|---|---|---|---|---|---|---|---|---|---|---|
| **Indonesia** (high mort model) | Early | prop | 0.41 | 0.22 | 0.18 | 0.06 | 0.03 | 0 | 0 | --- | 0.1 | 1 |
| | | risk | 4.4 | 2.3 | 1.9 | 0.7 | 0.3 | 0 | 0 | --- | 1 | 10.7 |
| | | num | 200.4 (169.9-255.2) | 106.9 (87.0-143.2) | 88.4 (59.0-118.0) | 30 (12.5-47.9) | 13 (6.0-28.8) | 1.1 (0.4-2.8) | 0.1 (0.0-1.0) | --- | 47.2 (16.1-75.1) | 487.1 |
| | Late | prop | 0.23 | 0.11 | 0.16 | 0.33 | 0.04 | 0.01 | 0 | --- | 0.11 | 1 |
| | | risk | 0.9 | 0.4 | 0.6 | 1.2 | 0.1 | 0 | 0 | --- | 0.4 | 3.7 |
| | | num | 38.9 (28.6-56.6) | 19.5 (13.7-27.2) | 27.7 (14.4-43.0) | 56.6 (34.3-85.3) | 6.9 (4.4-10.4) | 1.7 (0.5-4.8) | 0.7 (0.3-1.4) | --- | 19.3 (7.8-34.4) | 171.2 |
| | Overall | prop | 0.37 | 0.19 | 0.17 | 0.13 | 0.03 | 0 | 0 | 0 | 0.1 | 1 |
| | | risk | 5.5 | 2.9 | 2.5 | 2 | 0.5 | 0.1 | 0 | 0 | 1.5 | 14.9 |
| | | num | 264.7 (220.1-349.7) | 138 (110.4-185.6) | 123 (77.7-171.3) | 95.2 (52.2-146.8) | 21.8 (11.5-43.3) | 3.1 (1.0-8.6) | 0.9 (0.3-2.7) | 0 (0.0-0.0) | 72.7 (26.3-118.8) | 719.4 |
| **Iran** (high mort model) | Early | prop | 0.38 | 0.2 | 0.24 | 0.05 | 0.02 | 0 | 0 | --- | 0.11 | 1 |
| | | risk | 2.9 | 1.5 | 1.9 | 0.4 | 0.2 | 0 | 0 | --- | 0.8 | 7.6 |
| | | num | 42.5 (30.9-57.3) | 22.3 (16.1-30.3) | 27.4 (14.6-36.1) | 6 (2.1-9.5) | 2.4 (1.0-5.5) | 0.2 (0.1-0.4) | 0 (0.0-0.1) | --- | 11.9 (3.1-20.1) | 112.7 |
| | Late | prop | 0.18 | 0.1 | 0.27 | 0.28 | 0.04 | 0.01 | 0 | --- | 0.12 | 1 |
| | | risk | 0.5 | 0.3 | 0.7 | 0.7 | 0.1 | 0 | 0 | --- | 0.3 | 2.7 |
| | | num | 7.3 (4.2-12.4) | 3.9 (2.3-5.8) | 10.9 (4.4-16.1) | 10.9 (5.5-18.4) | 1.5 (0.8-2.6) | 0.2 (0.1-0.7) | 0.1 (0.0-0.2) | --- | 4.8 (1.6-8.3) | 39.6 |
| | Overall | prop | 0.33 | 0.17 | 0.25 | 0.11 | 0.03 | 0 | 0 | 0 | 0.11 | 1 |
| | | risk | 3.5 | 1.8 | 2.6 | 1.2 | 0.3 | 0 | 0 | 0 | 1.2 | 10.6 |
| | | num | 51.8 (36.4-71.8) | 27.5 (19.2-38.0) | 39.5 (19.1-53.4) | 17.7 (8.2-28.8) | 4.1 (1.8-8.4) | 0.4 (0.1-1.2) | 0.1 (0.0-0.3) | 0 (0.0-0.0) | 17.4 (4.9-29.4) | 158.4 |
| **Iraq** (high mort model) | Early | prop | 0.41 | 0.27 | 0.16 | 0.07 | 0.03 | 0 | 0 | --- | 0.05 | 1 |
| | | risk | 5.6 | 3.7 | 2.2 | 1 | 0.5 | 0.1 | 0 | --- | 0.7 | 13.8 |
| | | num | 58.5 (46.1-67.8) | 38.5 (30.1-43.4) | 23.3 (14.2-30.8) | 10.3 (4.2-15.5) | 5 (2.3-9.8) | 0.6 (0.2-1.5) | 0.2 (0.0-1.5) | --- | 7.1 (3.4-10.4) | 143.5 |
| | Late | prop | 0.13 | 0.14 | 0.22 | 0.39 | 0.05 | 0.02 | 0.01 | --- | 0.04 | 1 |

| Table S14: Cause-specific proportions, risks, and numbers of deaths (with uncertainty) for 194 countries by neonatal period | | | | | | | | | | | | |
|---|---|---|---|---|---|---|---|---|---|---|---|---|
| Country | Period | Stat* | Preterm | Intrapartum | Congenital | Sepsis | Pneumonia | Tetanus | Diarrhoea | Injuries | Other | Total |
| | | risk | 0.6 | 0.7 | 1.1 | 1.9 | 0.2 | 0.1 | 0 | --- | 0.2 | 4.9 |
| | | num | 6.5 | 6.9 | 11.3 | 19.8 | 2.3 | 0.9 | 0.5 | --- | 2.2 | 50.4 |
| | | | (4.0-9.5) | (4.4-8.8) | (5.8-16.7) | (12.2-26.6) | (1.4-3.2) | (0.3-2.2) | (0.2-0.8) | --- | (1.0-5.1) | |
| | | prop | 0.32 | 0.24 | 0.18 | 0.16 | 0.04 | 0.01 | 0 | 0 | 0.05 | 1 |
| | Overall | risk | 6 | 4.6 | 3.4 | 3.1 | 0.7 | 0.1 | 0.1 | 0 | 0.9 | 19.1 |
| | | num | 61.4 | 47.2 | 34.7 | 32 | 7.6 | 1.3 | 0.8 | 0 | 9.6 | 194.6 |
| | | | (46.3-74.4) | (36.2-54.2) | (19.9-47.1) | (17.4-44.8) | (3.7-13.5) | (0.4-3.0) | (0.2-2.5) | (0.0-0.0) | (4.6-16.2) | |

* prop = proportion; num = number of deaths (in 100s)

| Table S14: Cause-specific proportions, risks, and numbers of deaths (with uncertainty) for 194 countries by neonatal period | | | | | | | | | | | | |
|---|---|---|---|---|---|---|---|---|---|---|---|---|
| Country | Period | Stat* | Preterm | Intrapartum | Congenital | Sepsis | Pneumonia | Tetanus | Diarrhoea | Injuries | Other | Total |
| | | prop | 0.33 | 0.13 | 0.46 | 0.01 | 0.02 | --- | --- | 0 | 0.06 | 1 |
| | Early | risk | 0.6 | 0.2 | 0.9 | 0 | 0 | --- | --- | 0 | 0.1 | 1.9 |
| | | num | 0.5 | 0.2 | 0.6 | 0 | 0 | --- | --- | 0 | 0.1 | 1.4 |
| | | | (0.3-0.6) | (0.1-0.3) | (0.5-0.8) | (0.0-0.0) | (0.0-0.1) | --- | --- | (0.0-0.0) | (0.0-0.1) | |
| **Ireland** (high-quality VR) | Late | prop | 0.38 | 0.09 | 0.38 | 0.06 | 0.03 | --- | --- | 0 | 0.06 | 1 |
| | | risk | 0.2 | 0 | 0.2 | 0 | 0 | --- | --- | 0 | 0 | 0.4 |
| | | num | 0.1 | 0 | 0.1 | 0 | 0 | --- | --- | 0 | 0 | 0.3 |
| | | | (0.0-0.2) | (0.0-0.1) | (0.0-0.2) | (0.0-0.0) | (0.0-0.0) | --- | --- | (0.0-0.0) | (0.0-0.0) | |
| | Overall | prop | 0.34 | 0.12 | 0.45 | 0.02 | 0.02 | 0 | 0 | 0 | 0.06 | 1 |
| | | risk | 0.8 | 0.3 | 1 | 0 | 0.1 | 0 | 0 | 0 | 0.1 | 2.3 |
| | | num | 0.6 | 0.2 | 0.8 | 0 | 0 | 0 | 0 | 0 | 0.1 | 1.7 |
| | | | (0.4-0.8) | (0.1-0.3) | (0.5-1.0) | (0.0-0.1) | (0.0-0.1) | (0.0-0.0) | (0.0-0.0) | (0.0-0.0) | (0.0-0.2) | |
| | Early | prop | 0.45 | 0.07 | 0.33 | 0.03 | 0 | --- | --- | 0.01 | 0.12 | 1 |
| | | risk | 0.7 | 0.1 | 0.5 | 0 | 0 | --- | --- | 0 | 0.2 | 1.6 |
| | | num | 1.1 | 0.2 | 0.8 | 0.1 | 0 | --- | --- | 0 | 0.3 | 2.5 |
| | | | (0.9-1.3) | (0.1-0.3) | (0.6-1.0) | (0.0-0.1) | (0.0-0.0) | --- | --- | (0.0-0.0) | (0.2-0.4) | |
| **Israel** (high-quality VR) | Late | prop | 0.49 | 0.05 | 0.34 | 0.1 | 0 | --- | --- | 0 | 0.01 | 1 |
| | | risk | 0.2 | 0 | 0.2 | 0 | 0 | --- | --- | 0 | 0 | 0.4 |
| | | num | 0.4 | 0 | 0.2 | 0.1 | 0 | --- | --- | 0 | 0 | 0.7 |
| | | | (0.2-0.5) | (0.0-0.1) | (0.1-0.3) | (0.0-0.1) | (0.0-0.0) | --- | --- | (0.0-0.0) | (0.0-0.0) | |
| | Overall | prop | 0.46 | 0.07 | 0.33 | 0.05 | 0 | 0 | 0 | 0.01 | 0.09 | 1 |
| | | risk | 1 | 0.1 | 0.7 | 0.1 | 0 | 0 | 0 | 0 | 0.2 | 2.1 |
| | | num | 1.5 | 0.2 | 1.1 | 0.2 | 0 | 0 | 0 | 0 | 0.3 | 3.3 |
| | | | (1.2-1.9) | (0.1-0.3) | (0.8-1.4) | (0.0-0.3) | (0.0-0.0) | (0.0-0.0) | (0.0-0.0) | (0.0-0.0) | (0.2-0.4) | |
| **Italy** (high- | Early | prop | 0.41 | 0.14 | 0.22 | 0.04 | 0 | --- | --- | 0 | 0.19 | 1 |
| | | risk | 0.6 | 0.2 | 0.3 | 0.1 | 0 | --- | --- | 0 | 0.3 | 1.6 |

| Country | Period | Stat* | Preterm | Intrapartum | Congenital | Sepsis | Pneumonia | Tetanus | Diarrhoea | Injuries | Other | Total |
|---|---|---|---|---|---|---|---|---|---|---|---|---|
| quality VR) | | num | 3.6 | 1.2 | 2 | 0.4 | 0 | --- | --- | 0 | 1.7 | 8.8 |
| | | | (3.2-4.0) | (1.0-1.4) | (1.7-2.2) | (0.2-0.5) | (0.0-0.0) | --- | --- | (0.0-0.0) | (1.4-2.0) | |
| | Late | prop | 0.4 | 0.07 | 0.24 | 0.11 | 0.01 | --- | --- | 0.01 | 0.15 | 1 |
| | | risk | 0.3 | 0 | 0.2 | 0.1 | 0 | --- | --- | 0 | 0.1 | 0.6 |
| | | num | 1.5 | 0.3 | 0.9 | 0.4 | 0 | --- | --- | 0 | 0.6 | 3.7 |
| | | | (1.2-1.7) | (0.2-0.4) | (0.7-1.1) | (0.3-0.5) | (0.0-0.1) | --- | --- | (0.0-0.1) | (0.4-0.7) | |
| | Overall | prop | 0.41 | 0.12 | 0.23 | 0.06 | 0 | 0 | 0 | 0 | 0.18 | 1 |
| | | risk | 0.9 | 0.3 | 0.5 | 0.1 | 0 | 0 | 0 | 0 | 0.4 | 2.2 |
| | | num | 5.1 | 1.5 | 2.8 | 0.7 | 0 | 0 | 0 | 0 | 2.3 | 12.5 |
| | | | (4.4-5.7) | (1.1-1.8) | (2.4-3.3) | (0.5-1.0) | (0.0-0.1) | (0.0-0.0) | (0.0-0.0) | (0.0-0.1) | (1.9-2.7) | |

\* prop = proportion; num = number of deaths (in 100s)

Table S14: Cause-specific proportions, risks, and numbers of deaths (with uncertainty) for 194 countries by neonatal period

| Country | Period | Stat* | Preterm | Intrapartum | Congenital | Sepsis | Pneumonia | Tetanus | Diarrhoea | Injuries | Other | Total |
|---|---|---|---|---|---|---|---|---|---|---|---|---|
| **Jamaica** (low mort model) | Early | prop | 0.42 | 0.15 | 0.21 | 0.08 | 0.03 | --- | --- | 0 | 0.11 | 1 |
| | | risk | 3.3 | 1.1 | 1.6 | 0.6 | 0.2 | --- | --- | 0 | 0.8 | 7.7 |
| | | num | 1.7 | 0.6 | 0.8 | 0.3 | 0.1 | --- | --- | 0 | 0.4 | 4 |
| | | | (1.4-2.5) | (0.5-0.9) | (0.6-1.3) | (0.3-0.5) | (0.1-0.2) | --- | --- | (0.0-0.0) | (0.3-0.7) | |
| | Late | prop | 0.3 | 0.08 | 0.27 | 0.19 | 0.05 | --- | --- | 0.02 | 0.09 | 1 |
| | | risk | 0.8 | 0.2 | 0.7 | 0.5 | 0.1 | --- | --- | 0 | 0.2 | 2.7 |
| | | num | 0.4 | 0.1 | 0.4 | 0.3 | 0.1 | --- | --- | 0 | 0.1 | 1.4 |
| | | | (0.3-0.6) | (0.1-0.2) | (0.3-0.6) | (0.2-0.4) | (0.0-0.1) | --- | --- | (0.0-0.0) | (0.1-0.2) | |
| | Overall | prop | 0.38 | 0.13 | 0.22 | 0.12 | 0.03 | 0 | 0 | 0.01 | 0.1 | 1 |
| | | risk | 4.1 | 1.4 | 2.4 | 1.3 | 0.4 | 0 | 0 | 0.1 | 1.1 | 10.7 |
| | | num | 2.2 | 0.7 | 1.3 | 0.7 | 0.2 | 0 | 0 | 0 | 0.6 | 5.8 |
| | | | (1.8-3.3) | (0.6-1.1) | (1.0-2.0) | (0.5-1.1) | (0.1-0.4) | (0.0-0.0) | (0.0-0.0) | (0.0-0.1) | (0.4-0.9) | |
| **Japan** (high-quality VR) | Early | prop | 0.24 | 0.14 | 0.43 | 0.03 | 0.01 | --- | --- | 0.01 | 0.13 | 1 |
| | | risk | 0.2 | 0.1 | 0.3 | 0 | 0 | --- | --- | 0 | 0.1 | 0.7 |
| | | num | 1.9 | 1.1 | 3.4 | 0.3 | 0.1 | --- | --- | 0.1 | 1 | 7.9 |
| | | | (1.7-2.2) | (0.9-1.3) | (3.0-3.7) | (0.2-0.3) | (0.0-0.2) | --- | --- | (0.0-0.1) | (0.8-1.2) | |
| | Late | prop | 0.1 | 0.03 | 0.52 | 0.14 | 0.04 | --- | --- | 0.05 | 0.12 | 1 |
| | | risk | 0 | 0 | 0.1 | 0 | 0 | --- | --- | 0 | 0 | 0.3 |
| | | num | 0.3 | 0.1 | 1.5 | 0.4 | 0.1 | --- | --- | 0.2 | 0.4 | 3 |
| | | | (0.2-0.4) | (0.0-0.1) | (1.3-1.8) | (0.3-0.5) | (0.1-0.2) | --- | --- | (0.1-0.2) | (0.2-0.5) | |
| | Overall | prop | 0.21 | 0.11 | 0.45 | 0.06 | 0.02 | 0 | 0 | 0.02 | 0.13 | 1 |
| | | risk | 0.2 | 0.1 | 0.5 | 0.1 | 0 | 0 | 0 | 0 | 0.1 | 1.1 |
| | | num | 2.5 | 1.3 | 5.4 | 0.7 | 0.3 | 0 | 0 | 0.2 | 1.5 | 11.9 |

| Table S14: Cause-specific proportions, risks, and numbers of deaths (with uncertainty) for 194 countries by neonatal period | | | | | | | | | | | | |
|---|---|---|---|---|---|---|---|---|---|---|---|---|
| Country | Period | Stat* | Preterm | Intrapartum | Congenital | Sepsis | Pneumonia | Tetanus | Diarrhoea | Injuries | Other | Total |
| | | | (2.1-2.8) | (1.0-1.6) | (4.8-6.0) | (0.5-1.0) | (0.1-0.4) | (0.0-0.0) | (0.0-0.0) | (0.1-0.4) | (1.2-1.9) | |
| **Jordan** (low mort model) | Early | prop | 0.45 | 0.16 | 0.24 | 0.08 | 0.04 | --- | --- | 0.01 | 0.02 | 1 |
| | | risk | 3.7 | 1.3 | 2 | 0.7 | 0.3 | --- | --- | 0.1 | 0.2 | 8.3 |
| | | num | 7.4 | 2.7 | 4 | 1.3 | 0.6 | --- | --- | 0.2 | 0.3 | 16.5 |
| | | | (6.6-8.2) | (2.2-3.1) | (3.0-4.9) | (0.9-1.8) | (0.4-1.0) | --- | --- | (0.1-0.6) | (0.1-0.9) | |
| | Late | prop | 0.3 | 0.09 | 0.27 | 0.17 | 0.06 | --- | --- | 0.02 | 0.09 | 1 |
| | | risk | 0.9 | 0.3 | 0.8 | 0.5 | 0.2 | --- | --- | 0 | 0.3 | 2.9 |
| | | num | 1.7 | 0.5 | 1.6 | 1 | 0.3 | --- | --- | 0.1 | 0.5 | 5.8 |
| | | | (1.5-2.0) | (0.4-0.7) | (1.4-1.8) | (0.7-1.2) | (0.2-0.6) | --- | --- | (0.1-0.1) | (0.4-0.8) | |
| | Overall | prop | 0.41 | 0.14 | 0.24 | 0.11 | 0.04 | 0 | 0 | 0.01 | 0.04 | 1 |
| | | risk | 4.7 | 1.7 | 2.8 | 1.2 | 0.5 | 0 | 0 | 0.1 | 0.5 | 11.5 |
| | | num | 9.5 | 3.3 | 5.6 | 2.4 | 1 | 0 | 0 | 0.3 | 0.9 | 23 |
| | | | (8.4-10.7) | (2.6-3.9) | (4.5-6.8) | (1.6-3.2) | (0.6-1.6) | (0.0-0.0) | (0.0-0.0) | (0.2-0.7) | (0.5-1.7) | |

* prop = proportion; num = number of deaths (in 100s)

| Table S14: Cause-specific proportions, risks, and numbers of deaths (with uncertainty) for 194 countries by neonatal period | | | | | | | | | | | | |
|---|---|---|---|---|---|---|---|---|---|---|---|---|
| Country | Period | Stat* | Preterm | Intrapartum | Congenital | Sepsis | Pneumonia | Tetanus | Diarrhoea | Injuries | Other | Total |
| **Kazakhstan** (high mort model) | Early | prop | 0.36 | 0.22 | 0.26 | 0.06 | 0.02 | 0 | 0 | --- | 0.09 | 1 |
| | | risk | 2.3 | 1.4 | 1.7 | 0.4 | 0.2 | 0 | 0 | --- | 0.6 | 6.5 |
| | | num | 7.7 | 4.7 | 5.5 | 1.2 | 0.5 | 0 | 0 | --- | 1.9 | 21.4 |
| | | | (6.7-11.5) | (4.0-7.1) | (3.5-8.4) | (0.5-2.1) | (0.2-1.3) | (0.0-0.1) | (0.0-0.0) | --- | (0.8-3.1) | |
| | Late | prop | 0.15 | 0.11 | 0.32 | 0.28 | 0.04 | 0.01 | 0 | --- | 0.09 | 1 |
| | | risk | 0.3 | 0.2 | 0.7 | 0.6 | 0.1 | 0 | 0 | --- | 0.2 | 2.3 |
| | | num | 1.1 | 0.8 | 2.4 | 2.1 | 0.3 | 0.1 | 0 | --- | 0.7 | 7.5 |
| | | | (0.8-2.1) | (0.6-1.4) | (1.2-4.0) | (1.2-3.9) | (0.2-0.6) | (0.0-0.2) | (0.0-0.0) | --- | (0.4-1.4) | |
| | Overall | prop | 0.31 | 0.19 | 0.27 | 0.11 | 0.03 | 0 | 0 | 0 | 0.09 | 1 |
| | | risk | 2.9 | 1.8 | 2.6 | 1.1 | 0.3 | 0 | 0 | 0 | 0.9 | 9.6 |
| | | num | 10.1 | 6.3 | 8.9 | 3.8 | 0.9 | 0.1 | 0 | 0 | 2.9 | 33.1 |
| | | | (8.2-14.6) | (4.9-9.2) | (5.2-13.2) | (1.9-6.3) | (0.5-2.0) | (0.0-0.3) | (0.0-0.1) | (0.0-0.0) | (1.3-4.8) | |
| **Kenya** (high mort model) | Early | prop | 0.32 | 0.4 | 0.1 | 0.08 | 0.06 | 0.01 | 0 | --- | 0.04 | 1 |
| | | risk | 6.1 | 7.7 | 1.9 | 1.6 | 1.2 | 0.2 | 0 | --- | 0.7 | 19.5 |
| | | num | 92.5 | 116.3 | 28.6 | 24.4 | 17.7 | 2.3 | 0.5 | --- | 10.8 | 293 |
| | | | (73.7-120.9) | (91.9-140.8) | (21.9-47.9) | (10.5-39.5) | (8.6-35.9) | (0.8-5.5) | (0.0-4.9) | --- | (5.4-30.2) | |
| | Late | prop | 0.16 | 0.17 | 0.07 | 0.43 | 0.06 | 0.04 | 0 | --- | 0.07 | 1 |
| | | risk | 1.1 | 1.2 | 0.5 | 3 | 0.4 | 0.3 | 0 | --- | 0.5 | 6.8 |
| | | num | 16.1 | 17.8 | 7.1 | 44.5 | 6 | 3.8 | 0.4 | --- | 7.2 | 102.9 |
| | | | (9.2-26.4) | (12.2-25.1) | (2.5-19.2) | (26.7-62.3) | (3.8-9.1) | (1.3-9.4) | (0.1-0.9) | --- | (3.0-18.9) | |

| Table S14: Cause-specific proportions, risks, and numbers of deaths (with uncertainty) for 194 countries by neonatal period ||||||||||||
|---|---|---|---|---|---|---|---|---|---|---|---|
| Country | Period | Stat* | Preterm | Intrapartum | Congenital | Sepsis | Pneumonia | Tetanus | Diarrhoea | Injuries | Other | Total |
| **Kiribati** (high mort model) | Overall | prop | 0.27 | 0.34 | 0.09 | 0.18 | 0.06 | 0.02 | 0 | 0 | 0.05 | 1 |
| | | risk | 7.3 | 9 | 2.4 | 4.8 | 1.6 | 0.4 | 0.1 | 0 | 1.2 | 26.8 |
| | | num | 109.4 | 134.4 | 35.7 | 71.2 | 24 | 6.3 | 0.9 | 0 | 18.1 | 400.1 |
| | | | (82.7-148.8) | (105.2-166.2) | (24.6-68.5) | (38.5-105.7) | (12.8-46.2) | (2.3-15.1) | (0.1-6.1) | (0.0-0.0) | (8.3-50.7) | |
| | Early | prop | 0.35 | 0.32 | 0.11 | 0.07 | 0.04 | 0 | 0 | --- | 0.1 | 1 |
| | | risk | 5.6 | 5.1 | 1.7 | 1.1 | 0.7 | 0.1 | 0 | --- | 1.6 | 16 |
| | | num | 0.1 | 0.1 | 0 | 0 | 0 | 0 | 0 | --- | 0 | 0.4 |
| | | | (0.1-0.2) | (0.1-0.2) | (0.0-0.1) | (0.0-0.1) | (0.0-0.0) | (0.0-0.0) | (0.0-0.0) | --- | (0.0-0.1) | |
| | Late | prop | 0.16 | 0.15 | 0.13 | 0.39 | 0.05 | 0.02 | 0 | --- | 0.1 | 1 |
| | | risk | 0.9 | 0.8 | 0.7 | 2.2 | 0.3 | 0.1 | 0 | --- | 0.6 | 5.6 |
| | | num | 0 | 0 | 0 | 0.1 | 0 | 0 | 0 | --- | 0 | 0.1 |
| | | | (0.0-0.0) | (0.0-0.0) | (0.0-0.0) | (0.0-0.1) | (0.0-0.0) | (0.0-0.0) | (0.0-0.0) | --- | (0.0-0.0) | |
| | Overall | prop | 0.31 | 0.27 | 0.11 | 0.16 | 0.05 | 0.01 | 0 | 0 | 0.1 | 1 |
| | | risk | 6.8 | 5.8 | 2.5 | 3.4 | 1 | 0.2 | 0 | 0 | 2.1 | 21.9 |
| | | num | 0.2 | 0.1 | 0.1 | 0.1 | 0 | 0 | 0 | 0 | 0.1 | 0.5 |
| | | | (0.1-0.2) | (0.1-0.2) | (0.0-0.1) | (0.0-0.1) | (0.0-0.1) | (0.0-0.0) | (0.0-0.0) | (0.0-0.0) | (0.0-0.1) | |

* prop = proportion; num = number of deaths (in 100s)

| Table S14: Cause-specific proportions, risks, and numbers of deaths (with uncertainty) for 194 countries by neonatal period ||||||||||||
|---|---|---|---|---|---|---|---|---|---|---|---|
| Country | Period | Stat* | Preterm | Intrapartum | Congenital | Sepsis | Pneumonia | Tetanus | Diarrhoea | Injuries | Other | Total |
| **Kuwait** (high-quality VR) | Early | prop | 0.45 | 0.01 | 0.51 | 0.01 | 0.02 | --- | --- | 0 | 0.01 | 1 |
| | | risk | 1.4 | 0 | 1.5 | 0 | 0 | --- | --- | 0 | 0 | 3 |
| | | num | 1 | 0 | 1.1 | 0 | 0 | --- | --- | 0 | 0 | 2.1 |
| | | | (0.8-1.1) | (0.0-0.0) | (0.9-1.3) | (0.0-0.0) | (0.0-0.1) | --- | --- | (0.0-0.0) | (0.0-0.0) | |
| | Late | prop | 0.39 | 0.07 | 0.42 | 0.05 | 0.01 | --- | --- | 0.02 | 0.04 | 1 |
| | | risk | 0.7 | 0.1 | 0.7 | 0.1 | 0 | --- | --- | 0 | 0.1 | 1.7 |
| | | num | 0.5 | 0.1 | 0.5 | 0.1 | 0 | --- | --- | 0 | 0 | 1.2 |
| | | | (0.3-0.6) | (0.0-0.1) | (0.4-0.6) | (0.0-0.1) | (0.0-0.0) | --- | --- | (0.0-0.0) | (0.0-0.1) | |
| | Overall | prop | 0.43 | 0.03 | 0.48 | 0.02 | 0.01 | 0 | 0 | 0.01 | 0.02 | 1 |
| | | risk | 2.2 | 0.2 | 2.4 | 0.1 | 0.1 | 0 | 0 | 0 | 0.1 | 5 |
| | | num | 1.5 | 0.1 | 1.6 | 0.1 | 0 | 0 | 0 | 0 | 0.1 | 3.4 |
| | | | (1.1-1.8) | (0.0-0.2) | (1.3-2.0) | (0.0-0.2) | (0.0-0.1) | (0.0-0.0) | (0.0-0.0) | (0.0-0.1) | (0.0-0.1) | |
| **Kyrgyzstan** (high mort model) | Early | prop | 0.34 | 0.26 | 0.24 | 0.07 | 0.03 | 0 | 0 | --- | 0.07 | 1 |
| | | risk | 3.3 | 2.6 | 2.3 | 0.7 | 0.3 | 0 | 0 | --- | 0.7 | 9.8 |
| | | num | 5 | 3.9 | 3.6 | 1 | 0.4 | 0 | 0 | --- | 1 | 15 |
| | | | (3.9-6.7) | (3.0-5.1) | (2.2-4.9) | (0.4-1.6) | (0.2-1.0) | (0.0-0.1) | (0.0-0.0) | --- | (0.5-1.4) | |
| | Late | prop | 0.11 | 0.13 | 0.31 | 0.32 | 0.05 | 0.01 | 0 | --- | 0.07 | 1 |

| Table S14: Cause-specific proportions, risks, and numbers of deaths (with uncertainty) for 194 countries by neonatal period | | | | | | | | | | | |
|---|---|---|---|---|---|---|---|---|---|---|---|
| Country | Period | Stat* | Preterm | Intrapartum | Congenital | Sepsis | Pneumonia | Tetanus | Diarrhoea | Injuries | Other | Total |
| **Lao People's Democratic Republic** (high mort model) | | risk | 0.4 | 0.4 | 1.1 | 1.1 | 0.2 | 0 | 0 | --- | 0.2 | 3.5 |
| | | num | 0.6 | 0.7 | 1.7 | 1.7 | 0.2 | 0 | 0 | --- | 0.4 | 5.3 |
| | | | (0.3-1.0) | (0.4-1.0) | (0.8-2.4) | (0.9-2.6) | (0.1-0.4) | (0.0-0.1) | (0.0-0.0) | --- | (0.2-0.6) | |
| | Overall | prop | 0.29 | 0.23 | 0.24 | 0.14 | 0.03 | 0 | 0 | 0 | 0.07 | 1 |
| | | risk | 4.1 | 3.2 | 3.5 | 1.9 | 0.5 | 0.1 | 0 | 0 | 1 | 14.2 |
| | | num | 6.1 | 4.9 | 5.2 | 2.9 | 0.7 | 0.1 | 0 | 0 | 1.4 | 21.4 |
| | | | (4.6-8.3) | (3.7-6.4) | (3.1-7.4) | (1.4-4.5) | (0.3-1.5) | (0.0-0.3) | (0.0-0.1) | (0.0-0.0) | (0.8-2.1) | |
| | Early | prop | 0.27 | 0.38 | 0.09 | 0.09 | 0.06 | 0.01 | 0.01 | --- | 0.09 | 1 |
| | | risk | 5.9 | 8.2 | 2 | 1.9 | 1.3 | 0.2 | 0.1 | --- | 1.9 | 21.5 |
| | | num | 10.8 | 15.2 | 3.7 | 3.4 | 2.4 | 0.4 | 0.2 | --- | 3.5 | 39.6 |
| | | | (7.7-15.4) | (11.3-20.5) | (2.5-6.5) | (1.4-6.0) | (1.2-4.9) | (0.1-1.0) | (0.0-2.2) | --- | (1.9-5.2) | |
| | Late | prop | 0.15 | 0.16 | 0.07 | 0.46 | 0.06 | 0.03 | 0.01 | --- | 0.07 | 1 |
| | | risk | 1.2 | 1.2 | 0.5 | 3.5 | 0.4 | 0.2 | 0.1 | --- | 0.6 | 7.6 |
| | | num | 2.1 | 2.2 | 0.9 | 6.4 | 0.8 | 0.3 | 0.1 | --- | 1 | 13.9 |
| | | | (1.2-3.7) | (1.3-3.2) | (0.5-1.8) | (4.0-10.0) | (0.4-1.3) | (0.1-1.0) | (0.1-0.3) | --- | (0.6-2.1) | |
| | Overall | prop | 0.26 | 0.32 | 0.08 | 0.18 | 0.06 | 0.02 | 0.01 | 0 | 0.08 | 1 |
| | | risk | 7.8 | 9.4 | 2.4 | 5.2 | 1.7 | 0.5 | 0.2 | 0 | 2.5 | 29.8 |
| | | num | 15.1 | 18.3 | 4.6 | 10.2 | 3.3 | 1 | 0.4 | 0 | 4.8 | 57.8 |
| | | | (10.7-21.7) | (13.3-25.0) | (3.1-8.8) | (5.7-16.2) | (1.7-6.6) | (0.3-2.4) | (0.1-2.8) | (0.0-0.0) | (2.6-7.7) | |

* prop = proportion; num = number of deaths (in 100s)

| Table S14: Cause-specific proportions, risks, and numbers of deaths (with uncertainty) for 194 countries by neonatal period | | | | | | | | | | | |
|---|---|---|---|---|---|---|---|---|---|---|---|
| Country | Period | Stat* | Preterm | Intrapartum | Congenital | Sepsis | Pneumonia | Tetanus | Diarrhoea | Injuries | Other | Total |
| **Latvia** (high-quality VR) | Early | prop | 0.13 | 0.65 | 0.13 | 0.06 | 0.02 | --- | --- | 0.02 | 0 | 1 |
| | | risk | 0.5 | 2.4 | 0.5 | 0.2 | 0.1 | --- | --- | 0.1 | 0 | 3.7 |
| | | num | 0.1 | 0.5 | 0.1 | 0 | 0 | --- | --- | 0 | 0 | 0.8 |
| | | | (0.0-0.2) | (0.4-0.7) | (0.0-0.2) | (0.0-0.1) | (0.0-0.0) | --- | --- | (0.0-0.0) | (0.0-0.0) | |
| | Late | prop | 0.15 | 0.2 | 0.4 | 0.25 | 0 | --- | --- | 0 | 0 | 1 |
| | | risk | 0.2 | 0.3 | 0.6 | 0.4 | 0 | --- | --- | 0 | 0 | 1.5 |
| | | num | 0 | 0.1 | 0.1 | 0.1 | 0 | --- | --- | 0 | 0 | 0.3 |
| | | | (0.0-0.1) | (0.0-0.1) | (0.1-0.2) | (0.0-0.1) | (0.0-0.0) | --- | --- | (0.0-0.0) | (0.0-0.0) | |
| | Overall | prop | 0.13 | 0.51 | 0.21 | 0.12 | 0.01 | 0 | 0 | 0.01 | 0 | 1 |
| | | risk | 0.7 | 2.8 | 1.1 | 0.6 | 0.1 | 0 | 0 | 0.1 | 0 | 5.4 |
| | | num | 0.2 | 0.6 | 0.2 | 0.1 | 0 | 0 | 0 | 0 | 0 | 1.2 |
| | | | (0.0-0.3) | (0.4-0.8) | (0.1-0.4) | (0.0-0.2) | (0.0-0.0) | (0.0-0.0) | (0.0-0.0) | (0.0-0.0) | (0.0-0.0) | |
| **Lebanon** (low mort | Early | prop | 0.46 | 0.17 | 0.18 | 0.04 | 0.02 | --- | --- | 0.01 | 0.12 | 1 |
| | | risk | 1.8 | 0.7 | 0.7 | 0.2 | 0.1 | --- | --- | 0 | 0.5 | 3.8 |

| Table S14: Cause-specific proportions, risks, and numbers of deaths (with uncertainty) for 194 countries by neonatal period ||||||||||||
| Country | Period | Stat* | Preterm | Intrapartum | Congenital | Sepsis | Pneumonia | Tetanus | Diarrhoea | Injuries | Other | Total |
| --- | --- | --- | --- | --- | --- | --- | --- | --- | --- | --- | --- | --- |
| model) | | num | 1.2 | 0.5 | 0.5 | 0.1 | 0.1 | --- | --- | 0 | 0.3 | 2.6 |
| | | | (1.0-1.5) | (0.3-0.6) | (0.4-0.7) | (0.1-0.1) | (0.0-0.1) | --- | --- | (0.0-0.0) | (0.2-0.4) | |
| | Late | prop | 0.3 | 0.09 | 0.3 | 0.15 | 0.08 | --- | --- | 0.02 | 0.06 | 1 |
| | | risk | 0.4 | 0.1 | 0.4 | 0.2 | 0.1 | --- | --- | 0 | 0.1 | 1.4 |
| | | num | 0.3 | 0.1 | 0.3 | 0.1 | 0.1 | --- | --- | 0 | 0.1 | 0.9 |
| | | | (0.2-0.3) | (0.1-0.1) | (0.2-0.4) | (0.1-0.2) | (0.1-0.1) | --- | --- | (0.0-0.0) | (0.0-0.1) | |
| | Overall | prop | 0.41 | 0.15 | 0.21 | 0.07 | 0.04 | 0 | 0 | 0.01 | 0.1 | 1 |
| | | risk | 2.2 | 0.8 | 1.1 | 0.4 | 0.2 | 0 | 0 | 0.1 | 0.6 | 5.4 |
| | | num | 1.3 | 0.5 | 0.7 | 0.2 | 0.1 | 0 | 0 | 0 | 0.3 | 3.2 |
| | | | (1.2-1.7) | (0.4-0.6) | (0.6-1.0) | (0.1-0.3) | (0.1-0.2) | (0.0-0.0) | (0.0-0.0) | (0.0-0.0) | (0.3-0.5) | |
| **Lesotho** (high mort model) | Early | prop | 0.34 | 0.37 | 0.07 | 0.09 | 0.07 | 0.01 | 0 | --- | 0.04 | 1 |
| | | risk | 11.1 | 12.2 | 2.3 | 3 | 2.2 | 0.5 | 0.1 | --- | 1.2 | 32.5 |
| | | num | 6.4 | 7.1 | 1.3 | 1.7 | 1.3 | 0.3 | 0 | --- | 0.7 | 18.9 |
| | | | (4.3-7.5) | (4.7-7.8) | (0.8-2.1) | (0.6-2.5) | (0.6-2.3) | (0.1-0.7) | (0.0-0.6) | --- | (0.3-2.1) | |
| | Late | prop | 0.19 | 0.15 | 0.05 | 0.42 | 0.05 | 0.05 | 0 | --- | 0.09 | 1 |
| | | risk | 2.2 | 1.7 | 0.6 | 4.8 | 0.6 | 0.6 | 0 | --- | 1 | 11.4 |
| | | num | 1.3 | 1 | 0.3 | 2.8 | 0.3 | 0.4 | 0 | --- | 0.6 | 6.6 |
| | | | (0.6-2.1) | (0.5-1.2) | (0.1-0.8) | (1.4-3.7) | (0.2-0.5) | (0.1-0.8) | (0.0-0.1) | --- | (0.1-1.3) | |
| | Overall | prop | 0.3 | 0.32 | 0.06 | 0.18 | 0.06 | 0.02 | 0 | 0 | 0.05 | 1 |
| | | risk | 13.4 | 14 | 2.8 | 7.8 | 2.9 | 1.1 | 0.1 | 0 | 2.2 | 44.3 |
| | | num | 7.7 | 8.1 | 1.6 | 4.5 | 1.7 | 0.6 | 0.1 | 0 | 1.3 | 25.6 |
| | | | (4.9-9.7) | (5.2-9.1) | (0.9-2.8) | (2.1-6.2) | (0.8-2.8) | (0.2-1.5) | (0.0-0.7) | (0.0-0.0) | (0.5-3.4) | |

* prop = proportion; num = number of deaths (in 100s)

| Table S14: Cause-specific proportions, risks, and numbers of deaths (with uncertainty) for 194 countries by neonatal period ||||||||||||
| Country | Period | Stat* | Preterm | Intrapartum | Congenital | Sepsis | Pneumonia | Tetanus | Diarrhoea | Injuries | Other | Total |
| --- | --- | --- | --- | --- | --- | --- | --- | --- | --- | --- | --- | --- |
| **Liberia** (high mort model) | Early | prop | 0.29 | 0.39 | 0.11 | 0.1 | 0.06 | 0.01 | 0 | --- | 0.04 | 1 |
| | | risk | 5.5 | 7.4 | 2.1 | 1.9 | 1.2 | 0.1 | 0 | --- | 0.7 | 18.9 |
| | | num | 8.2 | 10.9 | 3.1 | 2.8 | 1.8 | 0.2 | 0 | --- | 1.1 | 28.1 |
| | | | (6.1-9.8) | (8.2-11.9) | (2.2-4.1) | (1.1-4.1) | (0.8-3.2) | (0.1-0.5) | (0.0-0.4) | --- | (0.5-2.6) | |
| | Late | prop | 0.14 | 0.16 | 0.06 | 0.49 | 0.06 | 0.02 | 0.01 | --- | 0.06 | 1 |
| | | risk | 0.9 | 1 | 0.4 | 3.3 | 0.4 | 0.1 | 0.1 | --- | 0.4 | 6.7 |
| | | num | 1.4 | 1.6 | 0.6 | 4.9 | 0.6 | 0.2 | 0.1 | --- | 0.6 | 9.9 |
| | | | (0.6-2.2) | (0.9-2.0) | (0.2-1.4) | (2.9-6.4) | (0.3-0.9) | (0.1-0.5) | (0.0-0.2) | --- | (0.3-1.5) | |
| | Overall | prop | 0.25 | 0.33 | 0.09 | 0.2 | 0.06 | 0.01 | 0 | 0 | 0.05 | 1 |
| | | risk | 6.6 | 8.8 | 2.4 | 5.3 | 1.7 | 0.3 | 0.1 | 0 | 1.2 | 26.3 |
| | | num | 9.6 | 12.9 | 3.5 | 7.7 | 2.4 | 0.5 | 0.2 | 0 | 1.8 | 38.6 |

| Country | Period | Stat* | Preterm | Intrapartum | Congenital | Sepsis | Pneumonia | Tetanus | Diarrhoea | Injuries | Other | Total |
|---|---|---|---|---|---|---|---|---|---|---|---|---|
| | | | (6.6-12.6) | (9.4-14.5) | (2.3-5.4) | (4.1-10.7) | (1.2-4.3) | (0.1-1.1) | (0.0-0.7) | (0.0-0.0) | (0.8-4.3) | |
| **Libya** (low mort model) | Early | prop | 0.42 | 0.17 | 0.27 | 0.04 | 0.01 | --- | --- | 0 | 0.09 | 1 |
| | | risk | 2.7 | 1.1 | 1.7 | 0.3 | 0.1 | --- | --- | 0 | 0.6 | 6.4 |
| | | num | 3.4 | 1.4 | 2.2 | 0.4 | 0.1 | --- | --- | 0 | 0.7 | 8.1 |
| | | | (2.8-3.8) | (0.9-1.7) | (1.6-2.8) | (0.2-0.5) | (0.1-0.1) | --- | --- | (0.0-0.1) | (0.5-0.9) | |
| | Late | prop | 0.28 | 0.11 | 0.28 | 0.18 | 0.07 | --- | --- | 0.01 | 0.07 | 1 |
| | | risk | 0.6 | 0.2 | 0.6 | 0.4 | 0.2 | --- | --- | 0 | 0.2 | 2.3 |
| | | num | 0.8 | 0.3 | 0.8 | 0.5 | 0.2 | --- | --- | 0 | 0.2 | 2.9 |
| | | | (0.7-1.0) | (0.1-0.4) | (0.7-0.9) | (0.3-0.7) | (0.1-0.3) | --- | --- | (0.0-0.1) | (0.1-0.3) | |
| | Overall | prop | 0.38 | 0.15 | 0.27 | 0.08 | 0.03 | 0 | 0 | 0.01 | 0.08 | 1 |
| | | risk | 3.5 | 1.4 | 2.4 | 0.7 | 0.2 | 0 | 0 | 0.1 | 0.8 | 9.1 |
| | | num | 4.5 | 1.8 | 3.2 | 0.9 | 0.3 | 0 | 0 | 0.1 | 1 | 11.7 |
| | | | (3.7-5.1) | (1.1-2.2) | (2.4-4.0) | (0.6-1.2) | (0.2-0.5) | (0.0-0.0) | (0.0-0.0) | (0.1-0.1) | (0.7-1.2) | |
| **Lithuania** (high-quality VR) | Early | prop | 0.28 | 0.21 | 0.3 | 0.19 | 0.02 | --- | --- | 0 | 0 | 1 |
| | | risk | 0.5 | 0.4 | 0.5 | 0.3 | 0 | --- | --- | 0 | 0 | 1.7 |
| | | num | 0.2 | 0.1 | 0.2 | 0.1 | 0 | --- | --- | 0 | 0 | 0.6 |
| | | | (0.1-0.2) | (0.1-0.2) | (0.1-0.3) | (0.0-0.2) | (0.0-0.0) | --- | --- | (0.0-0.0) | (0.0-0.0) | |
| | Late | prop | 0.17 | 0.07 | 0.45 | 0.21 | 0 | --- | --- | 0.1 | 0 | 1 |
| | | risk | 0.2 | 0.1 | 0.4 | 0.2 | 0 | --- | --- | 0.1 | 0 | 1 |
| | | num | 0.1 | 0 | 0.1 | 0.1 | 0 | --- | --- | 0 | 0 | 0.3 |
| | | | (0.0-0.1) | (0.0-0.1) | (0.1-0.2) | (0.0-0.1) | (0.0-0.0) | --- | --- | (0.0-0.1) | (0.0-0.0) | |
| | Overall | prop | 0.24 | 0.16 | 0.35 | 0.2 | 0.01 | 0 | 0 | 0.04 | 0 | 1 |
| | | risk | 0.7 | 0.5 | 1 | 0.6 | 0 | 0 | 0 | 0.1 | 0 | 2.9 |
| | | num | 0.2 | 0.2 | 0.4 | 0.2 | 0 | 0 | 0 | 0 | 0 | 1 |
| | | | (0.1-0.4) | (0.1-0.3) | (0.2-0.5) | (0.1-0.3) | (0.0-0.0) | (0.0-0.0) | (0.0-0.0) | (0.0-0.1) | (0.0-0.0) | |

* prop = proportion; num = number of deaths (in 100s)

Table S14: Cause-specific proportions, risks, and numbers of deaths (with uncertainty) for 194 countries by neonatal period

| Country | Period | Stat* | Preterm | Intrapartum | Congenital | Sepsis | Pneumonia | Tetanus | Diarrhoea | Injuries | Other | Total |
|---|---|---|---|---|---|---|---|---|---|---|---|---|
| **Luxembourg** (high-quality VR) | Early | prop | 0.5 | 0.13 | 0 | 0.13 | 0 | --- | --- | 0 | 0.25 | 1 |
| | | risk | 0.4 | 0.1 | 0 | 0.1 | 0 | --- | --- | 0 | 0.2 | 0.7 |
| | | num | 0 | 0 | 0 | 0 | 0 | --- | --- | 0 | 0 | 0 |
| | | | (0.0-0.1) | (0.0-0.0) | (0.0-0.0) | (0.0-0.0) | (0.0-0.0) | --- | --- | (0.0-0.0) | (0.0-0.0) | |
| | Late | prop | 0 | 0 | 1 | 0 | 0 | --- | --- | 0 | 0 | 1 |
| | | risk | 0 | 0 | 0.2 | 0 | 0 | --- | --- | 0 | 0 | 0.2 |
| | | num | 0 | 0 | 0 | 0 | 0 | --- | --- | 0 | 0 | 0 |
| | | | (0.0-0.0) | (0.0-0.0) | (0.0-0.0) | (0.0-0.0) | (0.0-0.0) | --- | --- | (0.0-0.0) | (0.0-0.0) | |

| Table S14: Cause-specific proportions, risks, and numbers of deaths (with uncertainty) for 194 countries by neonatal period | | | | | | | | | | | | |
|---|---|---|---|---|---|---|---|---|---|---|---|---|
| Country | Period | Stat* | Preterm | Intrapartum | Congenital | Sepsis | Pneumonia | Tetanus | Diarrhoea | Injuries | Other | Total |
| | Overall | prop | 0.4 | 0.1 | 0.2 | 0.1 | 0 | 0 | 0 | 0 | 0.2 | 1 |
| | | risk | 0.4 | 0.1 | 0.2 | 0.1 | 0 | 0 | 0 | 0 | 0.2 | 1 |
| | | num | 0 (0.0-0.1) | 0 (0.0-0.0) | 0 (0.0-0.0) | 0 (0.0-0.0) | 0 (0.0-0.0) | 0 (0.0-0.0) | 0 (0.0-0.0) | 0 (0.0-0.0) | 0 (0.0-0.0) | 0.1 |
| **Madagascar** (high mort model) | Early | prop | 0.31 | 0.37 | 0.13 | 0.08 | 0.06 | 0.01 | 0 | --- | 0.03 | 1 |
| | | risk | 4.9 | 5.9 | 2.1 | 1.3 | 0.9 | 0.1 | 0 | --- | 0.6 | 15.8 |
| | | num | 38.5 (31.3-46.0) | 46.7 (36.6-52.3) | 16.7 (12.3-23.6) | 10.5 (4.4-16.6) | 6.9 (3.5-13.2) | 0.8 (0.3-1.7) | 0.1 (0.0-0.9) | --- | 4.3 (2.2-11.8) | 124.6 |
| | Late | prop | 0.17 | 0.15 | 0.06 | 0.46 | 0.05 | 0.02 | 0.01 | --- | 0.06 | 1 |
| | | risk | 0.9 | 0.8 | 0.4 | 2.6 | 0.3 | 0.1 | 0.1 | --- | 0.3 | 5.6 |
| | | num | 7.3 (4.3-12.0) | 6.7 (4.4-8.9) | 2.8 (0.8-7.6) | 20.3 (12.6-26.9) | 2.3 (1.4-3.2) | 1 (0.3-2.4) | 0.7 (0.3-1.2) | --- | 2.7 (1.1-7.0) | 43.8 |
| | Overall | prop | 0.28 | 0.32 | 0.11 | 0.18 | 0.06 | 0.01 | 0 | 0 | 0.04 | 1 |
| | | risk | 6.1 | 6.9 | 2.4 | 4 | 1.2 | 0.2 | 0.1 | 0 | 0.9 | 22 |
| | | num | 46.8 (36.9-58.0) | 53.4 (41.8-61.2) | 18.8 (12.8-30.0) | 31.1 (17.3-44.3) | 9.3 (4.8-16.6) | 1.8 (0.6-4.3) | 0.8 (0.3-2.1) | 0 (0.0-0.0) | 7.1 (3.3-18.9) | 169.2 |
| **Malawi** (high mort model) | Early | prop | 0.37 | 0.33 | 0.11 | 0.09 | 0.05 | 0 | 0 | --- | 0.05 | 1 |
| | | risk | 6.3 | 5.6 | 1.9 | 1.6 | 0.9 | 0.1 | 0 | --- | 0.8 | 17.2 |
| | | num | 38.1 (31.0-47.3) | 34 (28.1-40.3) | 11.5 (7.7-16.7) | 9.5 (3.6-15.0) | 5.3 (2.7-10.6) | 0.5 (0.2-1.1) | 0.2 (0.0-1.3) | --- | 4.8 (3.0-8.9) | 103.8 |
| | Late | prop | 0.14 | 0.16 | 0.08 | 0.48 | 0.06 | 0.02 | 0.01 | --- | 0.07 | 1 |
| | | risk | 0.8 | 0.9 | 0.5 | 2.9 | 0.3 | 0.1 | 0 | --- | 0.4 | 6 |
| | | num | 4.9 (2.7-7.8) | 5.7 (3.7-7.7) | 2.9 (1.0-7.2) | 17.3 (10.9-24.1) | 2.1 (1.1-3.2) | 0.8 (0.2-2.1) | 0.2 (0.1-0.6) | --- | 2.5 (1.3-5.4) | 36.5 |
| | Overall | prop | 0.3 | 0.28 | 0.1 | 0.19 | 0.05 | 0.01 | 0 | 0 | 0.05 | 1 |
| | | risk | 7.2 | 6.7 | 2.4 | 4.6 | 1.2 | 0.2 | 0.1 | 0 | 1.2 | 23.5 |
| | | num | 44.1 (33.9-58.5) | 41 (32.9-50.1) | 14.7 (8.0-24.9) | 28.1 (15.1-41.4) | 7.7 (4.0-14.4) | 1.3 (0.4-3.4) | 0.4 (0.1-2.0) | 0 (0.0-0.0) | 7.5 (4.4-14.9) | 144.9 |

* prop = proportion; num = number of deaths (in 100s)

| Table S14: Cause-specific proportions, risks, and numbers of deaths (with uncertainty) for 194 countries by neonatal period | | | | | | | | | | | | |
|---|---|---|---|---|---|---|---|---|---|---|---|---|
| Country | Period | Stat* | Preterm | Intrapartum | Congenital | Sepsis | Pneumonia | Tetanus | Diarrhoea | Injuries | Other | Total |
| **Malaysia** (low mort model) | Early | prop | 0.44 | 0.15 | 0.21 | 0.04 | 0.02 | --- | --- | 0.01 | 0.12 | 1 |
| | | risk | 1.4 | 0.5 | 0.7 | 0.1 | 0.1 | --- | --- | 0 | 0.4 | 3.3 |
| | | num | 7.7 (6.8-8.5) | 2.7 (2.3-3.0) | 3.7 (3.0-5.0) | 0.8 (0.5-1.0) | 0.3 (0.2-0.4) | --- | --- | 0.1 (0.1-0.1) | 2.2 (1.5-2.6) | 17.5 |
| | Late | prop | 0.27 | 0.08 | 0.32 | 0.18 | 0.07 | --- | --- | 0.01 | 0.07 | 1 |

| Table S14: Cause-specific proportions, risks, and numbers of deaths (with uncertainty) for 194 countries by neonatal period | | | | | | | | | | | |
|---|---|---|---|---|---|---|---|---|---|---|---|
| Country | Period | Stat* | Preterm | Intrapartum | Congenital | Sepsis | Pneumonia | Tetanus | Diarrhoea | Injuries | Other | Total |
| | | risk | 0.3 | 0.1 | 0.4 | 0.2 | 0.1 | --- | --- | 0 | 0.1 | 1.1 |
| | | num | 1.7 | 0.5 | 1.9 | 1.1 | 0.5 | --- | --- | 0.1 | 0.4 | 6.1 |
| | | | (1.5-1.9) | (0.3-0.7) | (1.8-2.2) | (0.9-1.2) | (0.3-0.7) | --- | --- | (0.1-0.1) | (0.3-0.6) | |
| | | prop | 0.4 | 0.14 | 0.23 | 0.08 | 0.03 | 0 | 0 | 0.01 | 0.11 | 1 |
| | Overall | risk | 1.8 | 0.6 | 1 | 0.4 | 0.1 | 0 | 0 | 0 | 0.5 | 4.4 |
| | | num | 8.9 | 3 | 5.2 | 1.8 | 0.7 | 0 | 0 | 0.2 | 2.4 | 22.1 |
| | | | (7.9-9.7) | (2.5-3.5) | (4.4-6.4) | (1.3-2.1) | (0.5-1.0) | (0.0-0.0) | (0.0-0.0) | (0.1-0.2) | (1.6-3.1) | |
| | | prop | 0.41 | 0.12 | 0.29 | 0.05 | 0.03 | --- | --- | 0 | 0.08 | 1 |
| | Early | risk | 1.8 | 0.6 | 1.3 | 0.2 | 0.1 | --- | --- | 0 | 0.4 | 4.4 |
| | | num | 0.1 | 0 | 0.1 | 0 | 0 | --- | --- | 0 | 0 | 0.4 |
| | | | (0.1-0.2) | (0.0-0.1) | (0.0-0.2) | (0.0-0.0) | (0.0-0.0) | --- | --- | (0.0-0.0) | (0.0-0.0) | |
| **Maldives** (low mort model) | | prop | 0.29 | 0.07 | 0.33 | 0.13 | 0.05 | --- | --- | 0.02 | 0.11 | 1 |
| | Late | risk | 0.5 | 0.1 | 0.5 | 0.2 | 0.1 | --- | --- | 0 | 0.2 | 1.6 |
| | | num | 0 | 0 | 0 | 0 | 0 | --- | --- | 0 | 0 | 0.1 |
| | | | (0.0-0.0) | (0.0-0.0) | (0.0-0.1) | (0.0-0.0) | (0.0-0.0) | --- | --- | (0.0-0.0) | (0.0-0.0) | |
| | | prop | 0.38 | 0.11 | 0.3 | 0.08 | 0.04 | 0 | 0 | 0.01 | 0.09 | 1 |
| | Overall | risk | 2.5 | 0.7 | 1.9 | 0.5 | 0.2 | 0 | 0 | 0 | 0.6 | 6.5 |
| | | num | 0.2 | 0.1 | 0.2 | 0 | 0 | 0 | 0 | 0 | 0 | 0.5 |
| | | | (0.1-0.3) | (0.1-0.1) | (0.1-0.2) | (0.0-0.1) | (0.0-0.0) | (0.0-0.0) | (0.0-0.0) | (0.0-0.0) | (0.0-0.1) | |
| | | prop | 0.4 | 0.3 | 0.05 | 0.08 | 0.07 | 0.01 | 0.01 | --- | 0.08 | 1 |
| | Early | risk | 11.8 | 9 | 1.5 | 2.4 | 2.1 | 0.4 | 0.3 | --- | 2.3 | 29.7 |
| | | num | 81.4 | 61.9 | 10.6 | 16.7 | 14.2 | 2.8 | 1.7 | --- | 15.8 | 205.2 |
| | | | (59.7-102.2) | (47.3-71.4) | (6.7-15.5) | (6.3-26.9) | (7.6-25.6) | (1.0-6.3) | (0.0-16.5) | --- | (6.8-29.1) | |
| **Mali** (high mort model) | | prop | 0.15 | 0.14 | 0.04 | 0.47 | 0.05 | 0.05 | 0.03 | --- | 0.07 | 1 |
| | Late | risk | 1.6 | 1.5 | 0.4 | 4.9 | 0.5 | 0.6 | 0.3 | --- | 0.8 | 10.5 |
| | | num | 10.8 | 10.2 | 2.6 | 33.8 | 3.6 | 3.8 | 2 | --- | 5.3 | 72.1 |
| | | | (5.8-16.0) | (6.3-13.6) | (1.0-5.5) | (20.9-45.9) | (1.9-5.5) | (1.0-9.7) | (0.6-5.0) | --- | (1.9-15.8) | |
| | | prop | 0.33 | 0.26 | 0.05 | 0.18 | 0.07 | 0.02 | 0.01 | 0 | 0.08 | 1 |
| | Overall | risk | 13.4 | 10.8 | 1.9 | 7.6 | 2.7 | 1 | 0.6 | 0 | 3.1 | 41.1 |
| | | num | 89.7 | 72.2 | 13 | 50.8 | 18.4 | 6.5 | 3.7 | 0 | 21.1 | 275.4 |
| | | | (62.6-115.6) | (53.3-85.6) | (7.7-20.5) | (26.6-73.0) | (9.7-32.5) | (1.8-16.5) | (0.6-20.4) | (0.0-0.0) | (8.8-45.2) | |

* prop = proportion; num = number of deaths (in 100s)

| Table S14: Cause-specific proportions, risks, and numbers of deaths (with uncertainty) for 194 countries by neonatal period | | | | | | | | | | | |
|---|---|---|---|---|---|---|---|---|---|---|---|
| Country | Period | Stat* | Preterm | Intrapartum | Congenital | Sepsis | Pneumonia | Tetanus | Diarrhoea | Injuries | Other | Total |
| **Malta** (high- | Early | prop | 0.32 | 0.11 | 0.47 | 0.05 | 0 | --- | --- | 0 | 0.05 | 1 |
| | | risk | 1 | 0.3 | 1.5 | 0.2 | 0 | --- | --- | 0 | 0.2 | 3.2 |

| Table S14: Cause-specific proportions, risks, and numbers of deaths (with uncertainty) for 194 countries by neonatal period | | | | | | | | | | | |
|---|---|---|---|---|---|---|---|---|---|---|---|
| Country | Period | Stat* | Preterm | Intrapartum | Congenital | Sepsis | Pneumonia | Tetanus | Diarrhoea | Injuries | Other | Total |
| quality VR) | | num | 0 | 0 | 0.1 | 0 | 0 | --- | --- | 0 | 0 | 0.1 |
| | | | (0.0-0.1) | (0.0-0.0) | (0.0-0.1) | (0.0-0.0) | (0.0-0.0) | --- | --- | (0.0-0.0) | (0.0-0.0) | |
| | Late | prop | 0.8 | 0 | 0.2 | 0 | 0 | --- | --- | 0 | 0 | 1 |
| | | risk | 0.7 | 0 | 0.2 | 0 | 0 | --- | --- | 0 | 0 | 0.8 |
| | | num | 0 | 0 | 0 | 0 | 0 | --- | --- | 0 | 0 | 0 |
| | | | (0.0-0.1) | (0.0-0.0) | (0.0-0.0) | (0.0-0.0) | (0.0-0.0) | --- | --- | (0.0-0.0) | (0.0-0.0) | |
| | Overall | prop | 0.42 | 0.08 | 0.42 | 0.04 | 0 | 0 | 0 | 0 | 0.04 | 1 |
| | | risk | 1.7 | 0.3 | 1.7 | 0.2 | 0 | 0 | 0 | 0 | 0.2 | 4.1 |
| | | num | 0.1 | 0 | 0.1 | 0 | 0 | 0 | 0 | 0 | 0 | 0.2 |
| | | | (0.0-0.1) | (0.0-0.0) | (0.0-0.1) | (0.0-0.0) | (0.0-0.0) | (0.0-0.0) | (0.0-0.0) | (0.0-0.0) | (0.0-0.0) | |
| **Marshall Islands** (high mort model) | Early | prop | 0.51 | 0.2 | 0.14 | 0.06 | 0.03 | 0 | 0 | --- | 0.06 | 1 |
| | | risk | 5.9 | 2.4 | 1.6 | 0.7 | 0.3 | 0 | 0 | --- | 0.7 | 11.8 |
| | | num | 0.1 | 0 | 0 | 0 | 0 | 0 | 0 | --- | 0 | 0.2 |
| | | | (0.1-0.1) | (0.0-0.1) | (0.0-0.0) | (0.0-0.0) | (0.0-0.0) | (0.0-0.0) | (0.0-0.0) | --- | (0.0-0.0) | |
| | Late | prop | 0.17 | 0.11 | 0.22 | 0.37 | 0.04 | 0.01 | 0.01 | --- | 0.07 | 1 |
| | | risk | 0.7 | 0.5 | 0.9 | 1.5 | 0.2 | 0 | 0 | --- | 0.3 | 4.1 |
| | | num | 0 | 0 | 0 | 0 | 0 | 0 | 0 | --- | 0 | 0.1 |
| | | | (0.0-0.0) | (0.0-0.0) | (0.0-0.0) | (0.0-0.0) | (0.0-0.0) | (0.0-0.0) | (0.0-0.0) | --- | (0.0-0.0) | |
| | Overall | prop | 0.39 | 0.19 | 0.17 | 0.15 | 0.03 | 0 | 0 | 0 | 0.07 | 1 |
| | | risk | 6.3 | 3.1 | 2.8 | 2.4 | 0.5 | 0.1 | 0 | 0 | 1.1 | 16.2 |
| | | num | 0.1 | 0 | 0 | 0 | 0 | 0 | 0 | 0 | 0 | 0.2 |
| | | | (0.1-0.1) | (0.0-0.1) | (0.0-0.1) | (0.0-0.1) | (0.0-0.0) | (0.0-0.0) | (0.0-0.0) | (0.0-0.0) | (0.0-0.0) | |
| **Mauritania** (high mort model) | Early | prop | 0.45 | 0.28 | 0.07 | 0.1 | 0.06 | 0.01 | 0 | --- | 0.03 | 1 |
| | | risk | 11.7 | 7.2 | 1.8 | 2.6 | 1.5 | 0.2 | 0.1 | --- | 0.6 | 25.8 |
| | | num | 15.1 | 9.3 | 2.4 | 3.4 | 2 | 0.3 | 0.1 | --- | 0.8 | 33.3 |
| | | | (10.0-17.8) | (6.4-11.7) | (1.2-3.5) | (1.2-7.3) | (1.0-4.0) | (0.1-0.7) | (0.0-0.8) | --- | (0.4-2.5) | |
| | Late | prop | 0.14 | 0.14 | 0.05 | 0.51 | 0.05 | 0.04 | 0.01 | --- | 0.06 | 1 |
| | | risk | 1.3 | 1.3 | 0.4 | 4.6 | 0.4 | 0.3 | 0.1 | --- | 0.5 | 9 |
| | | num | 1.7 | 1.6 | 0.6 | 6 | 0.6 | 0.4 | 0.1 | --- | 0.7 | 11.7 |
| | | | (0.7-2.7) | (0.9-2.2) | (0.2-1.4) | (3.9-7.8) | (0.3-0.8) | (0.2-0.9) | (0.0-0.3) | --- | (0.3-1.7) | |
| | Overall | prop | 0.37 | 0.24 | 0.06 | 0.21 | 0.06 | 0.02 | 0 | 0 | 0.03 | 1 |
| | | risk | 13.2 | 8.6 | 2.3 | 7.4 | 2 | 0.6 | 0.2 | 0 | 1.2 | 35.5 |
| | | num | 16.8 | 11 | 2.9 | 9.4 | 2.6 | 0.8 | 0.2 | 0 | 1.5 | 45.2 |
| | | | (10.8-20.4) | (7.3-14.0) | (1.3-4.8) | (5.0-14.8) | (1.4-4.8) | (0.3-1.8) | (0.0-1.1) | (0.0-0.0) | (0.6-4.3) | |

* prop = proportion; num = number of deaths (in 100s)

| Country | Period | Stat* | Preterm | Intrapartum | Congenital | Sepsis | Pneumonia | Tetanus | Diarrhoea | Injuries | Other | Total |
|---|---|---|---|---|---|---|---|---|---|---|---|---|
| **Mauritius** (high-quality VR) | Early | prop | 0.48 | 0.12 | 0.29 | 0.05 | 0.01 | --- | --- | 0.02 | 0.02 | 1 |
| | | risk | 2.7 | 0.7 | 1.7 | 0.3 | 0.1 | --- | --- | 0.1 | 0.1 | 5.7 |
| | | num | 0.4 | 0.1 | 0.2 | 0 | 0 | --- | --- | 0 | 0 | 0.8 |
| | | | (0.3-0.5) | (0.0-0.2) | (0.1-0.3) | (0.0-0.1) | (0.0-0.0) | --- | --- | (0.0-0.0) | (0.0-0.0) | |
| | Late | prop | 0.32 | 0.02 | 0.19 | 0.23 | 0.13 | --- | --- | 0.04 | 0.06 | 1 |
| | | risk | 1 | 0.1 | 0.6 | 0.7 | 0.4 | --- | --- | 0.1 | 0.2 | 3.1 |
| | | num | 0.1 | 0 | 0.1 | 0.1 | 0.1 | --- | --- | 0 | 0 | 0.4 |
| | | | (0.1-0.2) | (0.0-0.0) | (0.0-0.1) | (0.0-0.2) | (0.0-0.1) | --- | --- | (0.0-0.0) | (0.0-0.1) | |
| | Overall | prop | 0.42 | 0.08 | 0.26 | 0.11 | 0.05 | 0 | 0 | 0.03 | 0.04 | 1 |
| | | risk | 3.8 | 0.7 | 2.3 | 1 | 0.5 | 0 | 0 | 0.3 | 0.3 | 9 |
| | | num | 0.5 | 0.1 | 0.3 | 0.1 | 0.1 | 0 | 0 | 0 | 0 | 1.2 |
| | | | (0.3-0.7) | (0.0-0.2) | (0.2-0.5) | (0.0-0.2) | (0.0-0.1) | (0.0-0.0) | (0.0-0.0) | (0.0-0.1) | (0.0-0.1) | |
| **Mexico** (high-quality VR) | Early | prop | 0.43 | 0.15 | 0.23 | 0.09 | 0.03 | --- | --- | 0 | 0.06 | 1 |
| | | risk | 2 | 0.7 | 1.1 | 0.4 | 0.1 | --- | --- | 0 | 0.3 | 4.6 |
| | | num | 44.5 | 15.4 | 24.2 | 9.6 | 3.3 | --- | --- | 0.4 | 6.1 | 103.6 |
| | | | (43.1-45.8) | (14.7-16.2) | (23.2-25.2) | (9.0-10.3) | (3.0-3.7) | --- | --- | (0.3-0.6) | (5.6-6.6) | |
| | Late | prop | 0.22 | 0.09 | 0.26 | 0.27 | 0.09 | --- | --- | 0.01 | 0.06 | 1 |
| | | risk | 0.4 | 0.2 | 0.5 | 0.5 | 0.2 | --- | --- | 0 | 0.1 | 1.9 |
| | | num | 9.5 | 3.6 | 11.2 | 11.2 | 3.9 | --- | --- | 0.2 | 2.6 | 42.3 |
| | | | (8.9-10.1) | (3.2-4.0) | (10.5-11.8) | (10.6-11.9) | (3.5-4.3) | --- | --- | (0.1-0.3) | (2.3-2.9) | |
| | Overall | prop | 0.37 | 0.13 | 0.24 | 0.14 | 0.05 | 0 | 0 | 0 | 0.06 | 1 |
| | | risk | 2.6 | 0.9 | 1.7 | 1 | 0.3 | 0 | 0 | 0 | 0.4 | 6.9 |
| | | num | 58.1 | 20.5 | 38.1 | 22.5 | 7.8 | 0 | 0 | 0.7 | 9.4 | 157 |
| | | | (56.1-60.0) | (19.3-21.7) | (36.4-39.7) | (21.2-23.8) | (7.0-8.6) | (0.0-0.0) | (0.0-0.0) | (0.5-0.9) | (8.6-10.2) | |
| **Micronesia (Federated States of)** (high mort model) | Early | prop | 0.42 | 0.26 | 0.15 | 0.06 | 0.04 | 0 | 0 | --- | 0.07 | 1 |
| | | risk | 5 | 3.1 | 1.8 | 0.7 | 0.4 | 0 | 0 | --- | 0.8 | 11.8 |
| | | num | 0.1 | 0.1 | 0 | 0 | 0 | 0 | 0 | --- | 0 | 0.3 |
| | | | (0.1-0.2) | (0.1-0.1) | (0.0-0.1) | (0.0-0.0) | (0.0-0.0) | (0.0-0.0) | (0.0-0.0) | --- | (0.0-0.0) | |
| | Late | prop | 0.14 | 0.11 | 0.24 | 0.38 | 0.04 | 0.01 | 0.01 | --- | 0.06 | 1 |
| | | risk | 0.6 | 0.5 | 1 | 1.6 | 0.2 | 0 | 0 | --- | 0.2 | 4.1 |
| | | num | 0 | 0 | 0 | 0 | 0 | 0 | 0 | --- | 0 | 0.1 |
| | | | (0.0-0.0) | (0.0-0.0) | (0.0-0.0) | (0.0-0.1) | (0.0-0.0) | (0.0-0.0) | (0.0-0.0) | --- | (0.0-0.0) | |
| | Overall | prop | 0.36 | 0.22 | 0.17 | 0.14 | 0.04 | 0.01 | 0 | 0 | 0.06 | 1 |
| | | risk | 5.9 | 3.6 | 2.8 | 2.3 | 0.6 | 0.1 | 0 | 0 | 1 | 16.3 |
| | | num | 0.1 | 0.1 | 0.1 | 0.1 | 0 | 0 | 0 | 0 | 0 | 0.4 |
| | | | (0.1-0.2) | (0.1-0.1) | (0.0-0.1) | (0.0-0.1) | (0.0-0.0) | (0.0-0.0) | (0.0-0.0) | (0.0-0.0) | (0.0-0.0) | |

* prop = proportion; num = number of deaths (in 100s)

| Table S14: Cause-specific proportions, risks, and numbers of deaths (with uncertainty) for 194 countries by neonatal period ||||||||||||
|---|---|---|---|---|---|---|---|---|---|---|---|
| Country | Period | Stat* | Preterm | Intrapartum | Congenital | Sepsis | Pneumonia | Tetanus | Diarrhoea | Injuries | Other | Total |
| **Monaco** (low mort model) | Early | prop | 0.42 | 0.14 | 0.31 | 0.02 | 0 | --- | --- | 0.01 | 0.11 | 1 |
| | | risk | 0.6 | 0.2 | 0.5 | 0 | 0 | --- | --- | 0 | 0.2 | 1.5 |
| | | num | 0 | 0 | 0 | 0 | 0 | --- | --- | 0 | 0 | 0 |
| | | | (0.0-0.0) | (0.0-0.0) | (0.0-0.0) | (0.0-0.0) | (0.0-0.0) | --- | --- | (0.0-0.0) | (0.0-0.0) | |
| | Late | prop | 0.3 | 0.08 | 0.39 | 0.1 | 0 | --- | --- | 0.02 | 0.11 | 1 |
| | | risk | 0.2 | 0 | 0.2 | 0 | 0 | --- | --- | 0 | 0.1 | 0.5 |
| | | num | 0 | 0 | 0 | 0 | 0 | --- | --- | 0 | 0 | 0 |
| | | | (0.0-0.0) | (0.0-0.0) | (0.0-0.0) | (0.0-0.0) | (0.0-0.0) | --- | --- | (0.0-0.0) | (0.0-0.0) | |
| | Overall | prop | 0.39 | 0.13 | 0.32 | 0.04 | 0 | 0 | 0 | 0.01 | 0.11 | 1 |
| | | risk | 0.8 | 0.3 | 0.7 | 0.1 | 0 | 0 | 0 | 0 | 0.2 | 2.1 |
| | | num | 0 | 0 | 0 | 0 | 0 | 0 | 0 | 0 | 0 | 0 |
| | | | (0.0-0.0) | (0.0-0.0) | (0.0-0.0) | (0.0-0.0) | (0.0-0.0) | (0.0-0.0) | (0.0-0.0) | (0.0-0.0) | (0.0-0.0) | |
| **Mongolia** (high mort model) | Early | prop | 0.38 | 0.24 | 0.19 | 0.06 | 0.03 | 0 | 0 | --- | 0.09 | 1 |
| | | risk | 3.7 | 2.3 | 1.9 | 0.6 | 0.3 | 0 | 0 | --- | 0.9 | 9.8 |
| | | num | 2.4 | 1.5 | 1.2 | 0.4 | 0.2 | 0 | 0 | --- | 0.6 | 6.4 |
| | | | (1.9-3.5) | (1.1-2.2) | (0.8-1.8) | (0.1-0.7) | (0.1-0.4) | (0.0-0.0) | (0.0-0.0) | --- | (0.2-0.9) | |
| | Late | prop | 0.15 | 0.12 | 0.27 | 0.3 | 0.04 | 0.01 | 0 | --- | 0.1 | 1 |
| | | risk | 0.5 | 0.4 | 0.9 | 1 | 0.1 | 0 | 0 | --- | 0.3 | 3.4 |
| | | num | 0.3 | 0.3 | 0.6 | 0.7 | 0.1 | 0 | 0 | --- | 0.2 | 2.2 |
| | | | (0.2-0.6) | (0.2-0.4) | (0.3-1.0) | (0.4-1.1) | (0.1-0.2) | (0.0-0.1) | (0.0-0.0) | --- | (0.1-0.4) | |
| | Overall | prop | 0.33 | 0.21 | 0.21 | 0.13 | 0.03 | 0 | 0 | 0 | 0.1 | 1 |
| | | risk | 4.4 | 2.8 | 2.8 | 1.7 | 0.4 | 0.1 | 0 | 0 | 1.3 | 13.6 |
| | | num | 3 | 1.9 | 1.9 | 1.1 | 0.3 | 0 | 0 | 0 | 0.9 | 9.1 |
| | | | (2.1-4.3) | (1.4-2.7) | (1.2-2.9) | (0.5-1.9) | (0.1-0.6) | (0.0-0.1) | (0.0-0.0) | (0.0-0.0) | (0.4-1.4) | |
| **Montenegro** (high-quality VR) | Early | prop | 0.35 | 0.46 | 0.12 | 0.04 | 0.04 | --- | --- | 0 | 0 | 1 |
| | | risk | 1 | 1.3 | 0.3 | 0.1 | 0.1 | --- | --- | 0 | 0 | 2.8 |
| | | num | 0.1 | 0.1 | 0 | 0 | 0 | --- | --- | 0 | 0 | 0.2 |
| | | | (0.0-0.1) | (0.0-0.2) | (0.0-0.1) | (0.0-0.0) | (0.0-0.0) | --- | --- | (0.0-0.0) | (0.0-0.0) | |
| | Late | prop | 0.33 | 0.5 | 0 | 0 | 0.17 | --- | --- | 0 | 0 | 1 |
| | | risk | 0.2 | 0.3 | 0 | 0 | 0.1 | --- | --- | 0 | 0 | 0.7 |
| | | num | 0 | 0 | 0 | 0 | 0 | --- | --- | 0 | 0 | 0 |
| | | | (0.0-0.0) | (0.0-0.1) | (0.0-0.0) | (0.0-0.0) | (0.0-0.0) | --- | --- | (0.0-0.0) | (0.0-0.0) | |
| | Overall | prop | 0.34 | 0.47 | 0.09 | 0.03 | 0.06 | 0 | 0 | 0 | 0 | 1 |
| | | risk | 1.3 | 1.8 | 0.4 | 0.1 | 0.2 | 0 | 0 | 0 | 0 | 3.8 |
| | | num | 0.1 | 0.1 | 0 | 0 | 0 | 0 | 0 | 0 | 0 | 0.3 |
| | | | (0.0-0.2) | (0.0-0.2) | (0.0-0.1) | (0.0-0.0) | (0.0-0.1) | (0.0-0.0) | (0.0-0.0) | (0.0-0.0) | (0.0-0.0) | |

* prop = proportion; num = number of deaths (in 100s)

Table S14: Cause-specific proportions, risks, and numbers of deaths (with uncertainty) for 194 countries by neonatal period

| Country | Period | Stat* | Preterm | Intrapartum | Congenital | Sepsis | Pneumonia | Tetanus | Diarrhoea | Injuries | Other | Total |
|---|---|---|---|---|---|---|---|---|---|---|---|---|
| **Morocco** (high mort model) | Early | prop | 0.39 | 0.24 | 0.18 | 0.07 | 0.03 | 0 | 0 | --- | 0.09 | 1 |
| | | risk | 5.1 | 3.2 | 2.3 | 1 | 0.4 | 0 | 0 | --- | 1.2 | 13.2 |
| | | num | 41.3 | 25.8 | 18.9 | 7.7 | 3.3 | 0.4 | 0.1 | --- | 9.4 | 106.8 |
| | | | (32.4-50.7) | (20.0-31.3) | (10.9-24.4) | (2.8-12.1) | (1.5-6.7) | (0.1-1.0) | (0.0-0.4) | --- | (3.6-12.9) | |
| | Late | prop | 0.21 | 0.12 | 0.13 | 0.38 | 0.04 | 0.01 | 0.02 | --- | 0.09 | 1 |
| | | risk | 1 | 0.5 | 0.6 | 1.8 | 0.2 | 0 | 0.1 | --- | 0.4 | 4.7 |
| | | num | 7.9 | 4.4 | 4.9 | 14.2 | 1.6 | 0.4 | 0.6 | --- | 3.5 | 37.5 |
| | | | (5.1-11.4) | (2.7-5.5) | (2.1-7.4) | (8.2-20.2) | (0.9-2.2) | (0.1-1.0) | (0.2-1.5) | --- | (1.5-5.7) | |
| | Overall | prop | 0.34 | 0.21 | 0.16 | 0.15 | 0.03 | 0.01 | 0 | 0 | 0.09 | 1 |
| | | risk | 6.3 | 3.9 | 2.9 | 2.8 | 0.6 | 0.1 | 0.1 | 0 | 1.7 | 18.4 |
| | | num | 47.2 | 28.9 | 22 | 21 | 4.7 | 0.8 | 0.6 | 0 | 12.4 | 137.5 |
| | | | (35.9-59.9) | (21.6-35.4) | (12.0-29.5) | (10.6-31.0) | (2.3-8.7) | (0.3-1.9) | (0.2-1.9) | (0.0-0.0) | (4.9-17.9) | |
| **Mozambique** (high mort model) | Early | prop | 0.37 | 0.34 | 0.08 | 0.1 | 0.06 | 0.01 | 0 | --- | 0.04 | 1 |
| | | risk | 8.3 | 7.7 | 1.8 | 2.2 | 1.4 | 0.2 | 0.1 | --- | 1 | 22.5 |
| | | num | 79 | 73.5 | 17.2 | 20.7 | 13.3 | 1.5 | 0.7 | --- | 9.1 | 215 |
| | | | (66.0-91.0) | (61.1-84.7) | (11.4-24.8) | (8.4-31.9) | (7.2-25.1) | (0.5-3.5) | (0.0-6.6) | --- | (5.3-18.4) | |
| | Late | prop | 0.15 | 0.15 | 0.05 | 0.48 | 0.05 | 0.03 | 0.02 | --- | 0.06 | 1 |
| | | risk | 1.2 | 1.2 | 0.4 | 3.8 | 0.4 | 0.3 | 0.1 | --- | 0.5 | 7.9 |
| | | num | 11.5 | 11.5 | 3.8 | 36.2 | 4 | 2.5 | 1.3 | --- | 4.8 | 75.5 |
| | | | (6.8-17.3) | (7.6-15.1) | (1.3-9.7) | (23.3-48.1) | (2.4-5.8) | (0.7-5.9) | (0.6-2.5) | --- | (2.5-10.7) | |
| | Overall | prop | 0.31 | 0.29 | 0.07 | 0.19 | 0.06 | 0.01 | 0.01 | 0 | 0.05 | 1 |
| | | risk | 9.6 | 9.1 | 2.1 | 6 | 1.9 | 0.4 | 0.2 | 0 | 1.5 | 30.9 |
| | | num | 90.3 | 85.2 | 20 | 56.3 | 17.6 | 4.1 | 2.1 | 0 | 13.9 | 289.5 |
| | | | (73.0-107.7) | (69.0-98.5) | (12.5-33.3) | (31.3-79.9) | (9.6-31.2) | (1.3-9.7) | (0.6-9.7) | (0.0-0.0) | (7.8-29.1) | |
| **Myanmar** (high mort model) | Early | prop | 0.42 | 0.25 | 0.09 | 0.06 | 0.03 | 0 | 0 | --- | 0.14 | 1 |
| | | risk | 7.9 | 4.7 | 1.6 | 1.1 | 0.6 | 0.1 | 0 | --- | 2.7 | 18.9 |
| | | num | 72.9 | 43.2 | 14.9 | 10.1 | 5.8 | 0.8 | 0.3 | --- | 25.1 | 173.1 |
| | | | (55.1-93.0) | (31.6-55.1) | (9.8-23.7) | (3.9-15.4) | (2.5-11.9) | (0.3-1.9) | (0.0-2.6) | --- | (5.7-43.8) | |
| | Late | prop | 0.27 | 0.11 | 0.09 | 0.32 | 0.04 | 0.02 | 0 | --- | 0.15 | 1 |
| | | risk | 1.8 | 0.8 | 0.6 | 2.1 | 0.3 | 0.1 | 0 | --- | 1 | 6.6 |
| | | num | 16.2 | 7 | 5.4 | 19.3 | 2.5 | 1 | 0.2 | --- | 9.3 | 60.8 |
| | | | (10.4-22.7) | (4.4-9.4) | (2.9-8.7) | (10.7-28.4) | (1.4-3.6) | (0.3-2.4) | (0.1-0.5) | --- | (2.5-16.7) | |
| | Overall | prop | 0.36 | 0.22 | 0.09 | 0.13 | 0.04 | 0.01 | 0 | 0 | 0.15 | 1 |
| | | risk | 9.5 | 5.8 | 2.3 | 3.4 | 1 | 0.2 | 0.1 | 0 | 3.9 | 26.1 |
| | | num | 88.6 | 53.6 | 21 | 31.7 | 9.1 | 1.7 | 0.5 | 0 | 36.7 | 243 |

| Table S14: Cause-specific proportions, risks, and numbers of deaths (with uncertainty) for 194 countries by neonatal period | | | | | | | | | | | | |
|---|---|---|---|---|---|---|---|---|---|---|---|---|
| Country | Period | Stat* | Preterm | Intrapartum | Congenital | Sepsis | Pneumonia | Tetanus | Diarrhoea | Injuries | Other | Total |
| | | | (65.3-116.7) | (38.7-68.0) | (13.1-34.0) | (15.8-48.1) | (4.3-17.1) | (0.6-4.4) | (0.1-3.3) | (0.0-0.0) | (8.8-63.8) | |

* prop = proportion; num = number of deaths (in 100s)

| Table S14: Cause-specific proportions, risks, and numbers of deaths (with uncertainty) for 194 countries by neonatal period | | | | | | | | | | | | |
|---|---|---|---|---|---|---|---|---|---|---|---|---|
| Country | Period | Stat* | Preterm | Intrapartum | Congenital | Sepsis | Pneumonia | Tetanus | Diarrhoea | Injuries | Other | Total |
| **Namibia** (high mort model) | Early | prop | 0.43 | 0.29 | 0.13 | 0.08 | 0.04 | 0 | 0 | --- | 0.03 | 1 |
| | | risk | 6.9 | 4.6 | 2.1 | 1.3 | 0.7 | 0.1 | 0 | --- | 0.5 | 16.1 |
| | | num | 4.1 | 2.7 | 1.2 | 0.8 | 0.4 | 0 | 0 | --- | 0.3 | 9.6 |
| | | | (3.5-4.6) | (2.2-3.1) | (0.8-1.6) | (0.3-1.2) | (0.2-0.8) | (0.0-0.1) | (0.0-0.0) | --- | (0.2-1.1) | |
| | Late | prop | 0.23 | 0.12 | 0.1 | 0.39 | 0.04 | 0.02 | 0.01 | --- | 0.09 | 1 |
| | | risk | 1.3 | 0.7 | 0.6 | 2.2 | 0.2 | 0.1 | 0 | --- | 0.5 | 5.7 |
| | | num | 0.8 | 0.4 | 0.3 | 1.3 | 0.1 | 0.1 | 0 | --- | 0.3 | 3.4 |
| | | | (0.4-1.3) | (0.3-0.6) | (0.1-0.8) | (0.8-1.9) | (0.1-0.2) | (0.0-0.1) | (0.0-0.1) | --- | (0.1-0.8) | |
| | Overall | prop | 0.38 | 0.25 | 0.12 | 0.16 | 0.04 | 0.01 | 0 | 0 | 0.05 | 1 |
| | | risk | 8.2 | 5.4 | 2.5 | 3.5 | 0.9 | 0.2 | 0 | 0 | 1.1 | 21.8 |
| | | num | 4.9 | 3.2 | 1.5 | 2.1 | 0.5 | 0.1 | 0 | 0 | 0.6 | 12.9 |
| | | | (3.9-5.9) | (2.5-3.7) | (0.9-2.3) | (1.0-3.1) | (0.3-1.0) | (0.0-0.2) | (0.0-0.1) | (0.0-0.0) | (0.3-1.8) | |
| **Nauru** (high mort model) | Early | prop | 0.42 | 0.25 | 0.15 | 0.07 | 0.03 | 0 | 0 | --- | 0.08 | 1 |
| | | risk | 6.2 | 3.8 | 2.2 | 1 | 0.5 | 0 | 0 | --- | 1.2 | 14.9 |
| | | num | 0 | 0 | 0 | 0 | 0 | 0 | 0 | --- | 0 | 0 |
| | | | (0.0-0.0) | (0.0-0.0) | (0.0-0.0) | (0.0-0.0) | (0.0-0.0) | (0.0-0.0) | (0.0-0.0) | --- | (0.0-0.0) | |
| | Late | prop | 0.15 | 0.12 | 0.24 | 0.34 | 0.05 | 0.01 | 0 | --- | 0.08 | 1 |
| | | risk | 0.8 | 0.6 | 1.3 | 1.8 | 0.2 | 0.1 | 0 | --- | 0.4 | 5.2 |
| | | num | 0 | 0 | 0 | 0 | 0 | 0 | 0 | --- | 0 | 0 |
| | | | (0.0-0.0) | (0.0-0.0) | (0.0-0.0) | (0.0-0.0) | (0.0-0.0) | (0.0-0.0) | (0.0-0.0) | --- | (0.0-0.0) | |
| | Overall | prop | 0.35 | 0.22 | 0.17 | 0.14 | 0.04 | 0.01 | 0 | 0 | 0.08 | 1 |
| | | risk | 7.1 | 4.5 | 3.4 | 2.9 | 0.7 | 0.1 | 0 | 0 | 1.6 | 20.4 |
| | | num | 0 | 0 | 0 | 0 | 0 | 0 | 0 | 0 | 0 | 0 |
| | | | (0.0-0.0) | (0.0-0.0) | (0.0-0.0) | (0.0-0.0) | (0.0-0.0) | (0.0-0.0) | (0.0-0.0) | (0.0-0.0) | (0.0-0.0) | |
| **Nepal** (high mort model) | Early | prop | 0.34 | 0.27 | 0.14 | 0.11 | 0.04 | 0.01 | 0 | --- | 0.1 | 1 |
| | | risk | 5.7 | 4.6 | 2.4 | 1.8 | 0.6 | 0.1 | 0 | --- | 1.6 | 17 |
| | | num | 32.2 | 26.2 | 13.8 | 10.2 | 3.6 | 0.7 | 0.1 | --- | 9.3 | 96 |
| | | | (26.5-40.3) | (21.4-32.5) | (8.9-18.2) | (3.2-15.8) | (1.7-7.5) | (0.3-2.0) | (0.0-0.8) | --- | (3.9-12.9) | |
| | Late | prop | 0.24 | 0.12 | 0.08 | 0.39 | 0.04 | 0.02 | 0.02 | --- | 0.09 | 1 |
| | | risk | 1.4 | 0.7 | 0.5 | 2.3 | 0.3 | 0.1 | 0.1 | --- | 0.6 | 6 |
| | | num | 8.1 | 4.1 | 2.6 | 13.2 | 1.4 | 0.5 | 0.6 | --- | 3.2 | 33.7 |
| | | | (5.6-11.0) | (2.6-5.3) | (1.1-4.7) | (8.3-18.2) | (0.9-2.0) | (0.2-1.3) | (0.2-1.4) | --- | (1.4-5.2) | |

| Table S14: Cause-specific proportions, risks, and numbers of deaths (with uncertainty) for 194 countries by neonatal period | | | | | | | | | | | | |
|---|---|---|---|---|---|---|---|---|---|---|---|---|
| Country | Period | Stat* | Preterm | Intrapartum | Congenital | Sepsis | Pneumonia | Tetanus | Diarrhoea | Injuries | Other | Total |
| | Overall | prop | 0.31 | 0.23 | 0.12 | 0.18 | 0.04 | 0.01 | 0.01 | 0 | 0.1 | 1 |
| | | risk | 7.4 | 5.6 | 2.9 | 4.3 | 0.9 | 0.2 | 0.1 | 0 | 2.3 | 23.7 |
| | | num | 42.6 | 32 | 16.4 | 24.6 | 5.3 | 1.4 | 0.8 | 0 | 13.2 | 136.2 |
| | | | (33.8-54.5) | (25.4-40.4) | (10.5-23.1) | (12.0-35.8) | (2.7-9.9) | (0.5-3.6) | (0.2-2.4) | (0.0-0.0) | (5.6-19.1) | |

* prop = proportion; num = number of deaths (in 100s)

| Table S14: Cause-specific proportions, risks, and numbers of deaths (with uncertainty) for 194 countries by neonatal period | | | | | | | | | | | | |
|---|---|---|---|---|---|---|---|---|---|---|---|---|
| Country | Period | Stat* | Preterm | Intrapartum | Congenital | Sepsis | Pneumonia | Tetanus | Diarrhoea | Injuries | Other | Total |
| **Netherlands** (high-quality VR) | Early | prop | 0.26 | 0.19 | 0.35 | 0.05 | 0 | --- | --- | 0.01 | 0.14 | 1 |
| | | risk | 0.5 | 0.4 | 0.7 | 0.1 | 0 | --- | --- | 0 | 0.3 | 2 |
| | | num | 0.9 | 0.7 | 1.2 | 0.2 | 0 | --- | --- | 0 | 0.5 | 3.5 |
| | | | (0.7-1.1) | (0.5-0.8) | (1.0-1.5) | (0.1-0.3) | (0.0-0.0) | --- | --- | (0.0-0.0) | (0.3-0.6) | |
| | Late | prop | 0.23 | 0.08 | 0.39 | 0.19 | 0.01 | --- | --- | 0.02 | 0.08 | 1 |
| | | risk | 0.1 | 0.1 | 0.2 | 0.1 | 0 | --- | --- | 0 | 0.1 | 0.6 |
| | | num | 0.2 | 0.1 | 0.4 | 0.2 | 0 | --- | --- | 0 | 0.1 | 1.1 |
| | | | (0.2-0.3) | (0.0-0.2) | (0.3-0.6) | (0.1-0.3) | (0.0-0.0) | --- | --- | (0.0-0.0) | (0.0-0.2) | |
| | Overall | prop | 0.25 | 0.16 | 0.36 | 0.09 | 0 | 0 | 0 | 0.01 | 0.13 | 1 |
| | | risk | 0.7 | 0.4 | 1 | 0.2 | 0 | 0 | 0 | 0 | 0.3 | 2.7 |
| | | num | 1.2 | 0.7 | 1.7 | 0.4 | 0 | 0 | 0 | 0 | 0.6 | 4.6 |
| | | | (0.9-1.5) | (0.5-1.0) | (1.3-2.0) | (0.2-0.6) | (0.0-0.1) | (0.0-0.0) | (0.0-0.0) | (0.0-0.1) | (0.4-0.8) | |
| **New Zealand** (high-quality VR) | Early | prop | 0.4 | 0.18 | 0.27 | 0.04 | 0.05 | --- | --- | 0 | 0.06 | 1 |
| | | risk | 0.9 | 0.4 | 0.6 | 0.1 | 0.1 | --- | --- | 0 | 0.1 | 2.3 |
| | | num | 0.6 | 0.3 | 0.4 | 0.1 | 0.1 | --- | --- | 0 | 0.1 | 1.4 |
| | | | (0.4-0.7) | (0.2-0.4) | (0.3-0.5) | (0.0-0.1) | (0.0-0.1) | --- | --- | (0.0-0.0) | (0.0-0.1) | |
| | Late | prop | 0.23 | 0.12 | 0.28 | 0.09 | 0.02 | --- | --- | 0.14 | 0.12 | 1 |
| | | risk | 0.2 | 0.1 | 0.2 | 0.1 | 0 | --- | --- | 0.1 | 0.1 | 0.7 |
| | | num | 0.1 | 0 | 0.1 | 0 | 0 | --- | --- | 0.1 | 0 | 0.4 |
| | | | (0.0-0.2) | (0.0-0.1) | (0.0-0.2) | (0.0-0.1) | (0.0-0.0) | --- | --- | (0.0-0.1) | (0.0-0.1) | |
| | Overall | prop | 0.36 | 0.17 | 0.27 | 0.05 | 0.05 | 0 | 0 | 0.03 | 0.07 | 1 |
| | | risk | 1.1 | 0.5 | 0.8 | 0.2 | 0.1 | 0 | 0 | 0.1 | 0.2 | 3.1 |
| | | num | 0.7 | 0.3 | 0.5 | 0.1 | 0.1 | 0 | 0 | 0.1 | 0.1 | 2 |
| | | | (0.5-0.9) | (0.2-0.5) | (0.3-0.7) | (0.0-0.2) | (0.0-0.2) | (0.0-0.0) | (0.0-0.0) | (0.0-0.1) | (0.0-0.3) | |
| **Nicaragua** (low mort model) | Early | prop | 0.44 | 0.19 | 0.19 | 0.07 | 0.04 | --- | --- | 0.01 | 0.06 | 1 |
| | | risk | 3.7 | 1.6 | 1.6 | 0.6 | 0.4 | --- | --- | 0.1 | 0.5 | 8.5 |
| | | num | 5.2 | 2.3 | 2.2 | 0.8 | 0.5 | --- | --- | 0.1 | 0.7 | 11.8 |
| | | | (4.5-6.1) | (1.4-3.1) | (1.6-3.3) | (0.6-1.2) | (0.3-0.9) | --- | --- | (0.1-0.1) | (0.5-1.0) | |
| | Late | prop | 0.25 | 0.11 | 0.23 | 0.2 | 0.12 | --- | --- | 0.01 | 0.07 | 1 |

| Country | Period | Stat* | Preterm | Intrapartum | Congenital | Sepsis | Pneumonia | Tetanus | Diarrhoea | Injuries | Other | Total |
|---|---|---|---|---|---|---|---|---|---|---|---|---|
| | | risk | 0.8 | 0.3 | 0.7 | 0.6 | 0.4 | --- | --- | 0 | 0.2 | 3 |
| | | num | 1 | 0.5 | 0.9 | 0.8 | 0.5 | --- | --- | 0.1 | 0.3 | 4.2 |
| | | | (0.9-1.4) | (0.1-0.6) | (0.8-1.2) | (0.5-1.3) | (0.4-0.8) | --- | --- | (0.0-0.1) | (0.2-0.4) | |
| | Overall | prop | 0.39 | 0.17 | 0.2 | 0.1 | 0.06 | 0 | 0 | 0.01 | 0.06 | 1 |
| | | risk | 4.6 | 2 | 2.3 | 1.2 | 0.7 | 0 | 0 | 0.1 | 0.8 | 11.8 |
| | | num | 6.5 | 2.9 | 3.3 | 1.8 | 1.1 | 0 | 0 | 0.1 | 1.1 | 16.7 |
| | | | (5.7-7.9) | (1.5-4.0) | (2.5-4.7) | (1.1-2.6) | (0.7-1.8) | (0.0-0.0) | (0.0-0.0) | (0.1-0.2) | (0.8-1.5) | |

* prop = proportion; num = number of deaths (in 100s)

Table S14: Cause-specific proportions, risks, and numbers of deaths (with uncertainty) for 194 countries by neonatal period

| Country | Period | Stat* | Preterm | Intrapartum | Congenital | Sepsis | Pneumonia | Tetanus | Diarrhoea | Injuries | Other | Total |
|---|---|---|---|---|---|---|---|---|---|---|---|---|
| **Niger** (high mort model) | Early | prop | 0.37 | 0.32 | 0.06 | 0.05 | 0.07 | 0.01 | 0.01 | --- | 0.11 | 1 |
| | | risk | 7.5 | 6.5 | 1.2 | 1.1 | 1.4 | 0.3 | 0.1 | --- | 2.3 | 20.4 |
| | | num | 64.8 | 55.8 | 10 | 9.4 | 12.3 | 2.4 | 1 | --- | 19.4 | 174.9 |
| | | | (43.5-109.7) | (41.0-82.3) | (6.3-17.9) | (3.4-18.9) | (7.2-26.4) | (0.8-7.5) | (0.0-12.3) | --- | (6.3-48.2) | |
| | Late | prop | 0.15 | 0.14 | 0.04 | 0.48 | 0.05 | 0.03 | 0.04 | --- | 0.08 | 1 |
| | | risk | 1.1 | 1 | 0.3 | 3.4 | 0.3 | 0.2 | 0.3 | --- | 0.5 | 7.1 |
| | | num | 9.3 | 8.3 | 2.2 | 29.4 | 2.9 | 2 | 2.6 | --- | 4.7 | 61.5 |
| | | | (4.5-16.4) | (5.4-13.0) | (0.9-5.7) | (19.2-45.4) | (1.7-4.9) | (0.5-5.5) | (0.5-9.4) | --- | (1.5-19.9) | |
| | Overall | prop | 0.34 | 0.23 | 0.06 | 0.19 | 0.06 | 0.02 | 0.01 | 0 | 0.09 | 1 |
| | | risk | 9.8 | 6.6 | 1.6 | 5.5 | 1.6 | 0.5 | 0.3 | 0 | 2.5 | 28.3 |
| | | num | 80.6 | 54.3 | 13.1 | 45.6 | 13.4 | 4 | 2.3 | 0 | 20.7 | 233.9 |
| | | | (54.8-128.0) | (41.2-80.8) | (6.3-23.8) | (26.0-80.1) | (8.0-28.3) | (1.2-12.1) | (0.4-15.9) | (0.0-0.0) | (6.8-60.8) | |
| **Nigeria** (high mort model) | Early | prop | 0.38 | 0.34 | 0.05 | 0.08 | 0.07 | 0.02 | 0.01 | --- | 0.05 | 1 |
| | | risk | 10.5 | 9.4 | 1.5 | 2.1 | 2 | 0.6 | 0.2 | --- | 1.4 | 27.7 |
| | | num | 731.1 | 660.5 | 102.1 | 147.7 | 138.5 | 44.1 | 12.3 | --- | 99.2 | 1935.5 |
| | | | (601.4-854.5) | (544.9-750.3) | (75.6-182.0) | (60.3-232.3) | (72.7-256.9) | (19.6-90.2) | (0.0-112.8) | --- | (61.9-175.3) | |
| | Late | prop | 0.19 | 0.16 | 0.04 | 0.39 | 0.05 | 0.09 | 0.01 | --- | 0.07 | 1 |
| | | risk | 1.8 | 1.5 | 0.4 | 3.8 | 0.5 | 0.8 | 0.1 | --- | 0.7 | 9.7 |
| | | num | 126 | 107.8 | 28 | 266.9 | 36.9 | 58.8 | 8 | --- | 47.6 | 680 |
| | | | (80.0-173.9) | (74.3-147.5) | (11.9-66.5) | (164.2-369.6) | (22.2-57.0) | (18.9-136.4) | (3.0-14.7) | --- | (26.0-104.8) | |
| | Overall | prop | 0.34 | 0.29 | 0.05 | 0.15 | 0.07 | 0.04 | 0.01 | 0 | 0.06 | 1 |
| | | risk | 13.1 | 11 | 1.8 | 5.8 | 2.6 | 1.5 | 0.4 | 0 | 2.1 | 38.2 |
| | | num | 893.8 | 746.2 | 119.7 | 393.1 | 174.1 | 105.4 | 23.9 | 0 | 143.1 | 2599.3 |
| | | | (644.4-1199.0) | (534.4-904.7) | (76.0-237.0) | (210.7-569.8) | (90.5-310.7) | (38.2-232.1) | (2.0-177.3) | (0.0-0.0) | (83.5-273.5) | |

| Table S14: Cause-specific proportions, risks, and numbers of deaths (with uncertainty) for 194 countries by neonatal period ||||||||||||
| Country | Period | Stat* | Preterm | Intrapartum | Congenital | Sepsis | Pneumonia | Tetanus | Diarrhoea | Injuries | Other | Total |
| --- | --- | --- | --- | --- | --- | --- | --- | --- | --- | --- | --- | --- |
| **Niue** (low mort model) | Early | prop | 0.45 | 0.17 | 0.21 | 0.07 | 0.04 | --- | --- | 0.01 | 0.05 | 1 |
| | | risk | 3.9 | 1.5 | 1.8 | 0.6 | 0.3 | --- | --- | 0.1 | 0.5 | 8.6 |
| | | num | 0 | 0 | 0 | 0 | 0 | --- | --- | 0 | 0 | 0 |
| | | | (0.0-0.0) | (0.0-0.0) | (0.0-0.0) | (0.0-0.0) | (0.0-0.0) | --- | --- | (0.0-0.0) | (0.0-0.0) | |
| | Late | prop | 0.26 | 0.09 | 0.24 | 0.19 | 0.13 | --- | --- | 0.01 | 0.08 | 1 |
| | | risk | 0.8 | 0.3 | 0.7 | 0.6 | 0.4 | --- | --- | 0 | 0.2 | 3 |
| | | num | 0 | 0 | 0 | 0 | 0 | --- | --- | 0 | 0 | 0 |
| | | | (0.0-0.0) | (0.0-0.0) | (0.0-0.0) | (0.0-0.0) | (0.0-0.0) | --- | --- | (0.0-0.0) | (0.0-0.0) | |
| | Overall | prop | 0.41 | 0.15 | 0.21 | 0.1 | 0.06 | 0 | 0 | 0.01 | 0.06 | 1 |
| | | risk | 4.8 | 1.8 | 2.5 | 1.2 | 0.7 | 0 | 0 | 0.1 | 0.7 | 11.9 |
| | | num | 0 | 0 | 0 | 0 | 0 | 0 | 0 | 0 | 0 | 0 |
| | | | (0.0-0.0) | (0.0-0.0) | (0.0-0.0) | (0.0-0.0) | (0.0-0.0) | (0.0-0.0) | (0.0-0.0) | (0.0-0.0) | (0.0-0.0) | |

* prop = proportion; num = number of deaths (in 100s)

| Table S14: Cause-specific proportions, risks, and numbers of deaths (with uncertainty) for 194 countries by neonatal period ||||||||||||
| Country | Period | Stat* | Preterm | Intrapartum | Congenital | Sepsis | Pneumonia | Tetanus | Diarrhoea | Injuries | Other | Total |
| --- | --- | --- | --- | --- | --- | --- | --- | --- | --- | --- | --- | --- |
| **Norway** (high-quality VR) | Early | prop | 0.28 | 0.15 | 0.3 | 0 | 0 | --- | --- | 0 | 0.27 | 1 |
| | | risk | 0.3 | 0.2 | 0.4 | 0 | 0 | --- | --- | 0 | 0.3 | 1.2 |
| | | num | 0.2 | 0.1 | 0.2 | 0 | 0 | --- | --- | 0 | 0.2 | 0.8 |
| | | | (0.1-0.3) | (0.1-0.2) | (0.1-0.3) | (0.0-0.0) | (0.0-0.0) | --- | --- | (0.0-0.0) | (0.1-0.3) | |
| | Late | prop | 0.17 | 0.04 | 0.52 | 0.09 | 0 | --- | --- | 0 | 0.17 | 1 |
| | | risk | 0.1 | 0 | 0.2 | 0 | 0 | --- | --- | 0 | 0.1 | 0.4 |
| | | num | 0 | 0 | 0.1 | 0 | 0 | --- | --- | 0 | 0 | 0.3 |
| | | | (0.0-0.1) | (0.0-0.0) | (0.1-0.2) | (0.0-0.1) | (0.0-0.0) | --- | --- | (0.0-0.0) | (0.0-0.1) | |
| | Overall | prop | 0.26 | 0.13 | 0.35 | 0.02 | 0 | 0 | 0 | 0 | 0.24 | 1 |
| | | risk | 0.4 | 0.2 | 0.6 | 0 | 0 | 0 | 0 | 0 | 0.4 | 1.7 |
| | | num | 0.3 | 0.1 | 0.4 | 0 | 0 | 0 | 0 | 0 | 0.3 | 1.1 |
| | | | (0.1-0.4) | (0.1-0.2) | (0.2-0.6) | (0.0-0.1) | (0.0-0.0) | (0.0-0.0) | (0.0-0.0) | (0.0-0.0) | (0.1-0.4) | |
| **Oman** (low mort model) | Early | prop | 0.47 | 0.19 | 0.25 | 0.03 | 0.01 | --- | --- | 0.01 | 0.03 | 1 |
| | | risk | 2.3 | 1 | 1.2 | 0.2 | 0 | --- | --- | 0 | 0.2 | 4.9 |
| | | num | 1.8 | 0.7 | 1 | 0.1 | 0 | --- | --- | 0 | 0.1 | 3.8 |
| | | | (1.6-2.5) | (0.6-1.1) | (0.7-1.5) | (0.1-0.2) | (0.0-0.1) | --- | --- | (0.0-0.1) | (0.1-0.3) | |
| | Late | prop | 0.3 | 0.12 | 0.32 | 0.14 | 0.02 | --- | --- | 0.02 | 0.08 | 1 |
| | | risk | 0.5 | 0.2 | 0.6 | 0.2 | 0 | --- | --- | 0 | 0.1 | 1.7 |
| | | num | 0.4 | 0.2 | 0.4 | 0.2 | 0 | --- | --- | 0 | 0.1 | 1.3 |
| | | | (0.4-0.6) | (0.1-0.2) | (0.4-0.6) | (0.1-0.3) | (0.0-0.0) | --- | --- | (0.0-0.0) | (0.1-0.2) | |
| | Overall | prop | 0.42 | 0.18 | 0.27 | 0.06 | 0.01 | 0 | 0 | 0.01 | 0.04 | 1 |

| Table S14: Cause-specific proportions, risks, and numbers of deaths (with uncertainty) for 194 countries by neonatal period | | | | | | | | | | | | |
|---|---|---|---|---|---|---|---|---|---|---|---|---|
| Country | Period | Stat* | Preterm | Intrapartum | Congenital | Sepsis | Pneumonia | Tetanus | Diarrhoea | Injuries | Other | Total |
| | | risk | 2.8 | 1.2 | 1.8 | 0.4 | 0.1 | 0 | 0 | 0.1 | 0.3 | 6.6 |
| | | num | 2.1 (1.9-2.9) | 0.9 (0.6-1.3) | 1.3 (1.1-2.0) | 0.3 (0.2-0.5) | 0.1 (0.0-0.1) | 0 (0.0-0.0) | 0 (0.0-0.0) | 0.1 (0.0-0.1) | 0.2 (0.1-0.4) | 5 |
| **Pakistan** (high mort model) | Early | prop | 0.42 | 0.27 | 0.06 | 0.11 | 0.05 | 0.02 | 0.01 | --- | 0.07 | 1 |
| | | risk | 13 | 8.3 | 2 | 3.3 | 1.7 | 0.5 | 0.2 | --- | 2 | 31.1 |
| | | num | 600.8 (496.0-677.6) | 383.2 (320.0-468.0) | 90.8 (61.5-136.2) | 152.8 (55.5-241.4) | 78.4 (45.2-160.3) | 25 (9.7-61.6) | 8.9 (0.0-65.6) | --- | 93.7 (52.0-125.4) | 1433.5 |
| | Late | prop | 0.18 | 0.13 | 0.05 | 0.46 | 0.04 | 0.05 | 0.02 | --- | 0.07 | 1 |
| | | risk | 1.9 | 1.4 | 0.5 | 5 | 0.5 | 0.5 | 0.2 | --- | 0.7 | 10.9 |
| | | num | 89.8 (60.3-114.1) | 66.4 (39.9-91.2) | 24.6 (14.2-45.4) | 232.6 (167.6-296.2) | 22.4 (13.8-32.3) | 23.8 (9.3-48.6) | 10.4 (4.9-19.1) | --- | 33.7 (19.9-53.8) | 503.7 |
| | Overall | prop | 0.36 | 0.23 | 0.06 | 0.2 | 0.05 | 0.03 | 0.01 | 0 | 0.07 | 1 |
| | | risk | 15.2 | 9.9 | 2.5 | 8.5 | 2.2 | 1.1 | 0.4 | 0 | 2.8 | 42.7 |
| | | num | 730.6 (590.8-839.9) | 475.7 (380.3-588.8) | 120 (76.8-187.1) | 407.6 (237.6-571.9) | 107.8 (62.8-207.2) | 53.3 (20.7-120.2) | 20.4 (5.2-90.3) | 0 (0.0-0.0) | 134.8 (75.0-189.0) | 2050.3 |

* prop = proportion; num = number of deaths (in 100s)

| Table S14: Cause-specific proportions, risks, and numbers of deaths (with uncertainty) for 194 countries by neonatal period | | | | | | | | | | | | |
|---|---|---|---|---|---|---|---|---|---|---|---|---|
| Country | Period | Stat* | Preterm | Intrapartum | Congenital | Sepsis | Pneumonia | Tetanus | Diarrhoea | Injuries | Other | Total |
| **Palau** (low mort model) | Early | prop | 0.46 | 0.16 | 0.24 | 0.05 | 0.02 | --- | --- | 0.01 | 0.06 | 1 |
| | | risk | 3 | 1 | 1.6 | 0.3 | 0.1 | --- | --- | 0 | 0.4 | 6.4 |
| | | num | 0 (0.0-0.0) | 0 (0.0-0.0) | 0 (0.0-0.0) | 0 (0.0-0.0) | 0 (0.0-0.0) | --- | --- | 0 (0.0-0.0) | 0 (0.0-0.0) | 0 |
| | Late | prop | 0.28 | 0.09 | 0.29 | 0.18 | 0.07 | --- | --- | 0.02 | 0.07 | 1 |
| | | risk | 0.6 | 0.2 | 0.6 | 0.4 | 0.2 | --- | --- | 0 | 0.2 | 2.3 |
| | | num | 0 (0.0-0.0) | 0 (0.0-0.0) | 0 (0.0-0.0) | 0 (0.0-0.0) | 0 (0.0-0.0) | --- | --- | 0 (0.0-0.0) | 0 (0.0-0.0) | 0 |
| | Overall | prop | 0.44 | 0.15 | 0.24 | 0.08 | 0.04 | 0 | 0 | 0.01 | 0.06 | 1 |
| | | risk | 3.9 | 1.3 | 2.1 | 0.7 | 0.3 | 0 | 0 | 0.1 | 0.5 | 9 |
| | | num | 0 (0.0-0.0) | 0 (0.0-0.0) | 0 (0.0-0.0) | 0 (0.0-0.0) | 0 (0.0-0.0) | 0 (0.0-0.0) | 0 (0.0-0.0) | 0 (0.0-0.0) | 0 (0.0-0.0) | 0 |
| **Panama** (high-quality VR) | Early | prop | 0.4 | 0.13 | 0.25 | 0.12 | 0.02 | --- | --- | 0.01 | 0.06 | 1 |
| | | risk | 2.4 | 0.8 | 1.5 | 0.7 | 0.1 | --- | --- | 0.1 | 0.4 | 6 |
| | | num | 1.8 (1.6-2.1) | 0.6 (0.5-0.8) | 1.1 (0.9-1.4) | 0.5 (0.4-0.7) | 0.1 (0.0-0.1) | --- | --- | 0.1 (0.0-0.1) | 0.3 (0.2-0.4) | 4.5 |

Table S14: Cause-specific proportions, risks, and numbers of deaths (with uncertainty) for 194 countries by neonatal period

| Country | Period | Stat* | Preterm | Intrapartum | Congenital | Sepsis | Pneumonia | Tetanus | Diarrhoea | Injuries | Other | Total |
|---|---|---|---|---|---|---|---|---|---|---|---|---|
| | Late | prop | 0.16 | 0.05 | 0.32 | 0.27 | 0.09 | --- | --- | 0.03 | 0.08 | 1 |
| | | risk | 0.4 | 0.1 | 0.8 | 0.6 | 0.2 | --- | --- | 0.1 | 0.2 | 2.4 |
| | | num | 0.3 | 0.1 | 0.6 | 0.5 | 0.2 | --- | --- | 0.1 | 0.1 | 1.8 |
| | | | (0.2-0.4) | (0.0-0.1) | (0.4-0.7) | (0.3-0.6) | (0.1-0.2) | --- | --- | (0.0-0.1) | (0.1-0.2) | |
| | Overall | prop | 0.33 | 0.11 | 0.27 | 0.16 | 0.04 | 0 | 0 | 0.02 | 0.07 | 1 |
| | | risk | 2.9 | 1 | 2.4 | 1.4 | 0.3 | 0 | 0 | 0.2 | 0.6 | 8.7 |
| | | num | 2.2 | 0.7 | 1.8 | 1.1 | 0.2 | 0 | 0 | 0.1 | 0.4 | 6.5 |
| | | | (1.8-2.6) | (0.5-0.9) | (1.4-2.1) | (0.8-1.3) | (0.1-0.4) | (0.0-0.0) | (0.0-0.0) | (0.0-0.2) | (0.2-0.6) | |
| **Papua New Guinea** (high mort model) | Early | prop | 0.34 | 0.34 | 0.11 | 0.08 | 0.05 | 0.01 | 0 | --- | 0.07 | 1 |
| | | risk | 6 | 6.1 | 2 | 1.5 | 0.9 | 0.1 | 0.1 | --- | 1.2 | 17.8 |
| | | num | 12.5 | 12.7 | 4.1 | 3.1 | 1.9 | 0.3 | 0.1 | --- | 2.4 | 37 |
| | | | (9.3-14.2) | (9.8-14.1) | (2.9-5.7) | (1.2-4.5) | (0.9-3.6) | (0.1-0.7) | (0.0-1.1) | --- | (1.2-3.3) | |
| | Late | prop | 0.17 | 0.17 | 0.09 | 0.42 | 0.06 | 0.04 | 0.01 | --- | 0.06 | 1 |
| | | risk | 1 | 1 | 0.5 | 2.6 | 0.4 | 0.2 | 0.1 | --- | 0.4 | 6.2 |
| | | num | 2.2 | 2.2 | 1.1 | 5.4 | 0.7 | 0.5 | 0.1 | --- | 0.8 | 13 |
| | | | (1.4-2.9) | (1.4-2.8) | (0.6-1.9) | (3.2-7.0) | (0.4-1.1) | (0.1-1.1) | (0.1-0.2) | --- | (0.4-1.6) | |
| | Overall | prop | 0.29 | 0.3 | 0.1 | 0.17 | 0.05 | 0.01 | 0.01 | 0 | 0.07 | 1 |
| | | risk | 7 | 7.4 | 2.4 | 4.2 | 1.3 | 0.4 | 0.1 | 0 | 1.6 | 24.4 |
| | | num | 14.6 | 15.4 | 5.1 | 8.8 | 2.7 | 0.7 | 0.3 | 0 | 3.4 | 50.9 |
| | | | (10.7-17.1) | (11.5-17.3) | (3.5-7.7) | (4.5-11.8) | (1.3-4.8) | (0.2-1.7) | (0.1-1.5) | (0.0-0.0) | (1.6-5.1) | |

* prop = proportion; num = number of deaths (in 100s)

| Country | Period | Stat* | Preterm | Intrapartum | Congenital | Sepsis | Pneumonia | Tetanus | Diarrhoea | Injuries | Other | Total |
|---|---|---|---|---|---|---|---|---|---|---|---|---|
| **Paraguay** (low mort model) | Early | prop | 0.45 | 0.15 | 0.21 | 0.1 | 0.04 | --- | --- | 0.01 | 0.04 | 1 |
| | | risk | 4.1 | 1.4 | 1.9 | 0.9 | 0.4 | --- | --- | 0.1 | 0.4 | 9.1 |
| | | num | 6.6 | 2.2 | 3.1 | 1.4 | 0.6 | --- | --- | 0.1 | 0.6 | 14.6 |
| | | | (5.8-7.0) | (1.8-2.5) | (2.6-3.8) | (1.1-1.8) | (0.4-0.9) | --- | --- | (0.1-0.2) | (0.3-1.0) | |
| | Late | prop | 0.28 | 0.07 | 0.23 | 0.23 | 0.11 | --- | --- | 0.01 | 0.07 | 1 |
| | | risk | 0.9 | 0.2 | 0.7 | 0.7 | 0.3 | --- | --- | 0 | 0.2 | 3.2 |
| | | num | 1.5 | 0.4 | 1.2 | 1.2 | 0.6 | --- | --- | 0.1 | 0.3 | 5.1 |
| | | | (1.3-1.6) | (0.3-0.4) | (1.0-1.4) | (1.0-1.4) | (0.4-0.8) | --- | --- | (0.1-0.1) | (0.2-0.5) | |
| | Overall | prop | 0.41 | 0.13 | 0.22 | 0.13 | 0.06 | 0 | 0 | 0.01 | 0.05 | 1 |
| | | risk | 5.2 | 1.7 | 2.8 | 1.7 | 0.7 | 0 | 0 | 0.1 | 0.6 | 12.7 |
| | | num | 8.1 | 2.6 | 4.3 | 2.6 | 1.1 | 0 | 0 | 0.2 | 0.9 | 19.9 |
| | | | (7.2-8.8) | (2.1-3.0) | (3.6-5.2) | (2.1-3.3) | (0.7-1.7) | (0.0-0.0) | (0.0-0.0) | (0.1-0.3) | (0.6-1.5) | |
| **Peru** | Early | prop | 0.43 | 0.17 | 0.2 | 0.07 | 0.03 | --- | --- | 0.01 | 0.09 | 1 |

| Table S14: Cause-specific proportions, risks, and numbers of deaths (with uncertainty) for 194 countries by neonatal period | | | | | | | | | | | | |
|---|---|---|---|---|---|---|---|---|---|---|---|---|
| Country | Period | Stat* | Preterm | Intrapartum | Congenital | Sepsis | Pneumonia | Tetanus | Diarrhoea | Injuries | Other | Total |
| (low mort model) | | risk | 2.6 | 1 | 1.2 | 0.4 | 0.2 | --- | --- | 0 | 0.5 | 5.9 |
| | | num | 15.3 | 6 | 7.1 | 2.4 | 1 | --- | --- | 0.2 | 3.2 | 35.2 |
| | | | (11.7-17.4) | (4.0-7.3) | (4.7-13.7) | (1.7-3.1) | (0.6-1.4) | --- | --- | (0.1-0.2) | (2.4-4.0) | |
| | Late | prop | 0.28 | 0.09 | 0.27 | 0.19 | 0.1 | --- | --- | 0.01 | 0.06 | 1 |
| | | risk | 0.6 | 0.2 | 0.6 | 0.4 | 0.2 | --- | --- | 0 | 0.1 | 2.1 |
| | | num | 3.5 | 1.1 | 3.3 | 2.3 | 1.2 | --- | --- | 0.2 | 0.7 | 12.4 |
| | | | (3.0-4.1) | (0.6-1.4) | (3.0-3.9) | (1.7-3.1) | (0.9-1.7) | --- | --- | (0.1-0.2) | (0.5-0.9) | |
| | Overall | prop | 0.38 | 0.15 | 0.23 | 0.11 | 0.04 | 0 | 0 | 0.01 | 0.08 | 1 |
| | | risk | 3.2 | 1.2 | 1.9 | 0.9 | 0.4 | 0 | 0 | 0.1 | 0.7 | 8.3 |
| | | num | 18.8 | 7.2 | 11.2 | 5.6 | 2.2 | 0 | 0 | 0.4 | 4 | 49.3 |
| | | | (15.1-21.4) | (4.7-8.7) | (8.5-17.9) | (4.2-6.7) | (1.5-3.0) | (0.0-0.0) | (0.0-0.0) | (0.3-0.5) | (3.0-4.9) | |
| **Philippines** (high mort model) | Early | prop | 0.36 | 0.26 | 0.21 | 0.07 | 0.03 | 0 | 0 | --- | 0.07 | 1 |
| | | risk | 3.6 | 2.6 | 2.1 | 0.7 | 0.3 | 0 | 0 | --- | 0.7 | 10.1 |
| | | num | 86 | 62.4 | 51.1 | 17.1 | 7.9 | 0.8 | 0.1 | --- | 16.3 | 241.7 |
| | | | (71.9-105.9) | (51.0-73.7) | (30.0-68.5) | (7.5-26.8) | (3.8-16.4) | (0.3-2.1) | (0.0-0.6) | --- | (8.8-21.7) | |
| | Late | prop | 0.19 | 0.13 | 0.14 | 0.41 | 0.04 | 0.01 | 0.01 | --- | 0.07 | 1 |
| | | risk | 0.7 | 0.5 | 0.5 | 1.4 | 0.2 | 0 | 0 | --- | 0.3 | 3.6 |
| | | num | 16 | 11.1 | 12 | 34.4 | 3.8 | 1 | 0.5 | --- | 6.1 | 84.9 |
| | | | (11.3-22.4) | (7.3-14.5) | (5.6-19.4) | (21.3-47.7) | (2.4-5.4) | (0.3-2.7) | (0.1-1.3) | --- | (3.6-10.2) | |
| | Overall | prop | 0.32 | 0.22 | 0.18 | 0.15 | 0.04 | 0.01 | 0 | 0 | 0.07 | 1 |
| | | risk | 4.5 | 3.1 | 2.6 | 2.1 | 0.5 | 0.1 | 0 | 0 | 1 | 13.9 |
| | | num | 104 | 72 | 59.2 | 49.1 | 11.5 | 2 | 0.7 | 0 | 21.9 | 320.4 |
| | | | (86.3-127.3) | (57.2-85.8) | (34.1-84.9) | (27.7-70.3) | (6.0-21.0) | (0.6-5.2) | (0.2-1.8) | (0.0-0.0) | (12.0-30.8) | |

* prop = proportion; num = number of deaths (in 100s)

| Table S14: Cause-specific proportions, risks, and numbers of deaths (with uncertainty) for 194 countries by neonatal period | | | | | | | | | | | | |
|---|---|---|---|---|---|---|---|---|---|---|---|---|
| Country | Period | Stat* | Preterm | Intrapartum | Congenital | Sepsis | Pneumonia | Tetanus | Diarrhoea | Injuries | Other | Total |
| **Poland** (high-quality VR) | Early | prop | 0.51 | 0.09 | 0.3 | 0.05 | 0.01 | --- | --- | 0.01 | 0.05 | 1 |
| | | risk | 1.2 | 0.2 | 0.7 | 0.1 | 0 | --- | --- | 0 | 0.1 | 2.3 |
| | | num | 4.9 | 0.9 | 2.9 | 0.5 | 0.1 | --- | --- | 0.1 | 0.4 | 9.6 |
| | | | (4.5-5.3) | (0.7-1.0) | (2.5-3.2) | (0.3-0.6) | (0.0-0.1) | --- | --- | (0.0-0.1) | (0.3-0.6) | |
| | Late | prop | 0.44 | 0.04 | 0.38 | 0.08 | 0.02 | --- | --- | 0.01 | 0.03 | 1 |
| | | risk | 0.3 | 0 | 0.3 | 0.1 | 0 | --- | --- | 0 | 0 | 0.8 |
| | | num | 1.4 | 0.1 | 1.2 | 0.2 | 0.1 | --- | --- | 0 | 0.1 | 3.2 |
| | | | (1.2-1.6) | (0.1-0.2) | (1.0-1.4) | (0.2-0.3) | (0.0-0.1) | --- | --- | (0.0-0.0) | (0.0-0.2) | |
| | Overall | prop | 0.49 | 0.08 | 0.32 | 0.06 | 0.01 | 0 | 0 | 0.01 | 0.04 | 1 |
| | | risk | 1.6 | 0.2 | 1 | 0.2 | 0 | 0 | 0 | 0 | 0.1 | 3.2 |

| Table S14: Cause-specific proportions, risks, and numbers of deaths (with uncertainty) for 194 countries by neonatal period | | | | | | | | | | | | |
|---|---|---|---|---|---|---|---|---|---|---|---|---|
| Country | Period | Stat* | Preterm | Intrapartum | Congenital | Sepsis | Pneumonia | Tetanus | Diarrhoea | Injuries | Other | Total |
| | | num | 6.7 | 1.1 | 4.3 | 0.8 | 0.1 | 0 | 0 | 0.1 | 0.6 | 13.7 |
| | | | (6.0-7.4) | (0.8-1.3) | (3.8-4.9) | (0.5-1.0) | (0.0-0.2) | (0.0-0.0) | (0.0-0.0) | (0.0-0.2) | (0.4-0.8) | |
| **Portugal** (low mort model) | Early | prop | 0.42 | 0.14 | 0.26 | 0.07 | 0.01 | --- | --- | 0.01 | 0.08 | 1 |
| | | risk | 0.6 | 0.2 | 0.4 | 0.1 | 0 | --- | --- | 0 | 0.1 | 1.5 |
| | | num | 0.6 | 0.2 | 0.4 | 0.1 | 0 | --- | --- | 0 | 0.1 | 1.3 |
| | | | (0.5-0.7) | (0.2-0.2) | (0.3-0.4) | (0.1-0.1) | (0.0-0.0) | --- | --- | (0.0-0.0) | (0.1-0.1) | |
| | Late | prop | 0.29 | 0.08 | 0.38 | 0.12 | 0.02 | --- | --- | 0.02 | 0.1 | 1 |
| | | risk | 0.2 | 0 | 0.2 | 0.1 | 0 | --- | --- | 0 | 0.1 | 0.5 |
| | | num | 0.1 | 0 | 0.2 | 0.1 | 0 | --- | --- | 0 | 0 | 0.5 |
| | | | (0.1-0.2) | (0.0-0.1) | (0.1-0.2) | (0.0-0.1) | (0.0-0.0) | --- | --- | (0.0-0.0) | (0.0-0.1) | |
| | Overall | prop | 0.39 | 0.12 | 0.29 | 0.08 | 0.01 | 0 | 0 | 0.01 | 0.09 | 1 |
| | | risk | 0.8 | 0.2 | 0.6 | 0.2 | 0 | 0 | 0 | 0 | 0.2 | 2 |
| | | num | 0.7 | 0.2 | 0.5 | 0.2 | 0 | 0 | 0 | 0 | 0.2 | 1.8 |
| | | | (0.6-0.8) | (0.2-0.3) | (0.4-0.6) | (0.1-0.2) | (0.0-0.0) | (0.0-0.0) | (0.0-0.0) | (0.0-0.0) | (0.1-0.2) | |
| **Qatar** (low mort model) | Early | prop | 0.47 | 0.15 | 0.24 | 0.01 | 0 | --- | --- | 0.01 | 0.11 | 1 |
| | | risk | 1.5 | 0.5 | 0.8 | 0 | 0 | --- | --- | 0 | 0.4 | 3.2 |
| | | num | 0.4 | 0.1 | 0.2 | 0 | 0 | --- | --- | 0 | 0.1 | 0.8 |
| | | | (0.3-0.4) | (0.1-0.1) | (0.1-0.3) | (0.0-0.0) | (0.0-0.0) | --- | --- | (0.0-0.0) | (0.1-0.1) | |
| | Late | prop | 0.31 | 0.08 | 0.36 | 0.15 | 0 | --- | --- | 0.02 | 0.08 | 1 |
| | | risk | 0.3 | 0.1 | 0.4 | 0.2 | 0 | --- | --- | 0 | 0.1 | 1.1 |
| | | num | 0.1 | 0 | 0.1 | 0 | 0 | --- | --- | 0 | 0 | 0.3 |
| | | | (0.1-0.1) | (0.0-0.0) | (0.1-0.1) | (0.0-0.1) | (0.0-0.0) | --- | --- | (0.0-0.0) | (0.0-0.0) | |
| | Overall | prop | 0.45 | 0.14 | 0.26 | 0.05 | 0 | 0 | 0 | 0.01 | 0.1 | 1 |
| | | risk | 2 | 0.6 | 1.1 | 0.2 | 0 | 0 | 0 | 0 | 0.5 | 4.4 |
| | | num | 0.4 | 0.1 | 0.2 | 0 | 0 | 0 | 0 | 0 | 0.1 | 0.9 |
| | | | (0.4-0.5) | (0.1-0.1) | (0.2-0.3) | (0.0-0.1) | (0.0-0.0) | (0.0-0.0) | (0.0-0.0) | (0.0-0.0) | (0.1-0.1) | |

* prop = proportion; num = number of deaths (in 100s)

| Table S14: Cause-specific proportions, risks, and numbers of deaths (with uncertainty) for 194 countries by neonatal period | | | | | | | | | | | | |
|---|---|---|---|---|---|---|---|---|---|---|---|---|
| Country | Period | Stat* | Preterm | Intrapartum | Congenital | Sepsis | Pneumonia | Tetanus | Diarrhoea | Injuries | Other | Total |
| **Republic of Korea** (high-quality VR) | Early | prop | 0.57 | 0.1 | 0.15 | 0.04 | 0 | --- | --- | 0.03 | 0.11 | 1 |
| | | risk | 0.6 | 0.1 | 0.2 | 0 | 0 | --- | --- | 0 | 0.1 | 1 |
| | | num | 3 | 0.6 | 0.8 | 0.2 | 0 | --- | --- | 0.1 | 0.6 | 5.3 |
| | | | (2.7-3.4) | (0.4-0.7) | (0.6-0.9) | (0.1-0.3) | (0.0-0.0) | --- | --- | (0.1-0.2) | (0.4-0.7) | |
| | Late | prop | 0.46 | 0.04 | 0.3 | 0.09 | 0.01 | --- | --- | 0.01 | 0.08 | 1 |
| | | risk | 0.3 | 0 | 0.2 | 0.1 | 0 | --- | --- | 0 | 0 | 0.6 |
| | | num | 1.3 | 0.1 | 0.9 | 0.3 | 0 | --- | --- | 0 | 0.2 | 2.8 |

| Country | Period | Stat* | Preterm | Intrapartum | Congenital | Sepsis | Pneumonia | Tetanus | Diarrhoea | Injuries | Other | Total |
|---|---|---|---|---|---|---|---|---|---|---|---|---|
| Table S14: Cause-specific proportions, risks, and numbers of deaths (with uncertainty) for 194 countries by neonatal period ||||||||||||
| | | | (1.1-1.5) | (0.0-0.2) | (0.7-1.0) | (0.2-0.4) | (0.0-0.1) | --- | --- | (0.0-0.1) | (0.1-0.3) | |
| | | prop | 0.53 | 0.08 | 0.2 | 0.06 | 0 | 0 | 0 | 0.02 | 0.1 | 1 |
| | Overall | risk | 0.9 | 0.1 | 0.3 | 0.1 | 0 | 0 | 0 | 0 | 0.2 | 1.7 |
| | | num | 4.6 | 0.7 | 1.7 | 0.5 | 0 | 0 | 0 | 0.2 | 0.9 | 8.7 |
| | | | (4.0-5.2) | (0.5-0.9) | (1.4-2.1) | (0.3-0.7) | (0.0-0.1) | (0.0-0.0) | (0.0-0.0) | (0.1-0.3) | (0.6-1.1) | |
| **Republic of Moldova** (high-quality VR) | Early | prop | 0.3 | 0.08 | 0.35 | 0.07 | 0.16 | --- | --- | 0.02 | 0.02 | 1 |
| | | risk | 1.7 | 0.5 | 2 | 0.4 | 0.9 | --- | --- | 0.1 | 0.1 | 5.8 |
| | | num | 0.7 | 0.2 | 0.8 | 0.2 | 0.4 | --- | --- | 0 | 0 | 2.3 |
| | | | (0.5-0.9) | (0.1-0.3) | (0.6-1.0) | (0.1-0.3) | (0.2-0.5) | --- | --- | (0.0-0.1) | (0.0-0.1) | |
| | Late | prop | 0.05 | 0.07 | 0.57 | 0.1 | 0.1 | --- | --- | 0 | 0.1 | 1 |
| | | risk | 0.1 | 0.1 | 1 | 0.2 | 0.2 | --- | --- | 0 | 0.2 | 1.8 |
| | | num | 0 | 0 | 0.4 | 0.1 | 0.1 | --- | --- | 0 | 0.1 | 0.7 |
| | | | (0.0-0.1) | (0.0-0.1) | (0.3-0.5) | (0.0-0.1) | (0.0-0.1) | --- | --- | (0.0-0.0) | (0.0-0.1) | |
| | Overall | prop | 0.24 | 0.08 | 0.4 | 0.08 | 0.14 | 0 | 0 | 0.02 | 0.04 | 1 |
| | | risk | 1.9 | 0.6 | 3.1 | 0.6 | 1.1 | 0 | 0 | 0.1 | 0.3 | 7.8 |
| | | num | 0.8 | 0.3 | 1.4 | 0.3 | 0.5 | 0 | 0 | 0.1 | 0.1 | 3.5 |
| | | | (0.6-1.0) | (0.1-0.4) | (1.1-1.7) | (0.1-0.4) | (0.3-0.7) | (0.0-0.0) | (0.0-0.0) | (0.0-0.1) | (0.0-0.2) | |
| **Romania** (high-quality VR) | Early | prop | 0.48 | 0.1 | 0.27 | 0.01 | 0.07 | --- | --- | 0.01 | 0.06 | 1 |
| | | risk | 2.3 | 0.5 | 1.3 | 0.1 | 0.3 | --- | --- | 0 | 0.3 | 4.8 |
| | | num | 5.2 | 1.1 | 2.9 | 0.2 | 0.7 | --- | --- | 0.1 | 0.6 | 10.7 |
| | | | (4.7-5.6) | (0.9-1.3) | (2.5-3.2) | (0.1-0.2) | (0.5-0.9) | --- | --- | (0.0-0.2) | (0.5-0.8) | |
| | Late | prop | 0.3 | 0.04 | 0.33 | 0.03 | 0.25 | --- | --- | 0.01 | 0.04 | 1 |
| | | risk | 0.7 | 0.1 | 0.8 | 0.1 | 0.6 | --- | --- | 0 | 0.1 | 2.3 |
| | | num | 1.6 | 0.2 | 1.7 | 0.2 | 1.3 | --- | --- | 0.1 | 0.2 | 5.3 |
| | | | (1.3-1.8) | (0.1-0.3) | (1.5-2.0) | (0.1-0.3) | (1.1-1.5) | --- | --- | (0.0-0.1) | (0.1-0.3) | |
| | Overall | prop | 0.42 | 0.08 | 0.29 | 0.02 | 0.13 | 0 | 0 | 0.01 | 0.05 | 1 |
| | | risk | 3.1 | 0.6 | 2.1 | 0.1 | 0.9 | 0 | 0 | 0.1 | 0.4 | 7.3 |
| | | num | 7 | 1.3 | 4.8 | 0.3 | 2.1 | 0 | 0 | 0.2 | 0.9 | 16.6 |
| | | | (6.3-7.7) | (1.0-1.6) | (4.2-5.4) | (0.2-0.5) | (1.7-2.5) | (0.0-0.0) | (0.0-0.0) | (0.1-0.3) | (0.6-1.1) | |

* prop = proportion; num = number of deaths (in 100s)

| Country | Period | Stat* | Preterm | Intrapartum | Congenital | Sepsis | Pneumonia | Tetanus | Diarrhoea | Injuries | Other | Total |
|---|---|---|---|---|---|---|---|---|---|---|---|---|
| Table S14: Cause-specific proportions, risks, and numbers of deaths (with uncertainty) for 194 countries by neonatal period ||||||||||||
| **Russian Federation** (low mort model) | Early | prop | 0.42 | 0.12 | 0.3 | 0.04 | 0.01 | --- | --- | 0.01 | 0.11 | 1 |
| | | risk | 1.6 | 0.5 | 1.2 | 0.1 | 0 | --- | --- | 0 | 0.4 | 3.9 |
| | | num | 27.2 | 8.1 | 19.4 | 2.4 | 0.7 | --- | --- | 0.5 | 6.9 | 65.2 |
| | | | (25.4-32.0) | (6.8-10.8) | (16.4-24.1) | (1.7-3.1) | (0.5-1.0) | --- | --- | (0.4-0.6) | (5.5-8.7) | |

| Table S14: Cause-specific proportions, risks, and numbers of deaths (with uncertainty) for 194 countries by neonatal period ||||||||||||
| Country | Period | Stat* | Preterm | Intrapartum | Congenital | Sepsis | Pneumonia | Tetanus | Diarrhoea | Injuries | Other | Total |
| --- | --- | --- | --- | --- | --- | --- | --- | --- | --- | --- | --- | --- |
| | Late | prop | 0.3 | 0.07 | 0.34 | 0.16 | 0.06 | --- | --- | 0.02 | 0.06 | 1 |
| | | risk | 0.4 | 0.1 | 0.5 | 0.2 | 0.1 | --- | --- | 0 | 0.1 | 1.4 |
| | | num | 6.8 | 1.6 | 7.7 | 3.6 | 1.4 | --- | --- | 0.4 | 1.5 | 22.9 |
| | | | (6.3-8.1) | (1.2-2.2) | (7.2-9.0) | (2.9-4.6) | (0.8-2.3) | --- | --- | (0.3-0.5) | (1.1-2.0) | |
| | Overall | prop | 0.39 | 0.11 | 0.31 | 0.07 | 0.02 | 0 | 0 | 0.01 | 0.09 | 1 |
| | | risk | 2.2 | 0.6 | 1.7 | 0.4 | 0.1 | 0 | 0 | 0.1 | 0.5 | 5.6 |
| | | num | 35.9 | 10.3 | 28.5 | 6.4 | 2.3 | 0 | 0 | 0.9 | 8.8 | 93.1 |
| | | | (33.3-42.5) | (8.5-13.7) | (24.8-34.5) | (4.9-8.3) | (1.4-3.6) | (0.0-0.0) | (0.0-0.0) | (0.7-1.1) | (7.0-11.3) | |
| **Rwanda** (high mort model) | Early | prop | 0.3 | 0.36 | 0.16 | 0.1 | 0.05 | 0 | 0 | --- | 0.03 | 1 |
| | | risk | 4.5 | 5.3 | 2.4 | 1.4 | 0.7 | 0.1 | 0 | --- | 0.5 | 14.9 |
| | | num | 19 | 22.5 | 10.1 | 6.1 | 3 | 0.3 | 0 | --- | 2.1 | 63.2 |
| | | | (14.3-24.6) | (18.2-25.7) | (6.7-14.5) | (2.1-9.6) | (1.4-6.5) | (0.1-0.7) | (0.0-0.3) | --- | (1.0-6.1) | |
| | Late | prop | 0.13 | 0.17 | 0.1 | 0.45 | 0.06 | 0.02 | 0 | --- | 0.07 | 1 |
| | | risk | 0.7 | 0.9 | 0.5 | 2.3 | 0.3 | 0.1 | 0 | --- | 0.4 | 5.2 |
| | | num | 2.9 | 3.7 | 2.2 | 9.9 | 1.3 | 0.4 | 0.1 | --- | 1.5 | 22.2 |
| | | | (1.4-5.0) | (2.5-4.9) | (0.8-5.5) | (5.6-13.5) | (0.8-1.9) | (0.1-1.1) | (0.0-0.2) | --- | (0.6-3.9) | |
| | Overall | prop | 0.26 | 0.31 | 0.14 | 0.19 | 0.05 | 0.01 | 0 | 0 | 0.04 | 1 |
| | | risk | 5.4 | 6.4 | 2.9 | 3.9 | 1.1 | 0.2 | 0 | 0 | 0.9 | 20.7 |
| | | num | 23.7 | 28.5 | 12.7 | 17.3 | 4.8 | 0.8 | 0.1 | 0 | 3.9 | 91.7 |
| | | | (17.1-32.0) | (22.6-33.1) | (7.6-20.6) | (8.3-24.9) | (2.4-9.2) | (0.3-2.0) | (0.0-0.6) | (0.0-0.0) | (1.7-10.7) | |
| **Saint Kitts and Nevis** (high-quality VR) | Early | prop | 0.4 | 0.6 | 0 | 0 | 0 | --- | --- | 0 | 0 | 1 |
| | | risk | 1.8 | 2.6 | 0 | 0 | 0 | --- | --- | 0 | 0 | 4.4 |
| | | num | 0 | 0 | 0 | 0 | 0 | --- | --- | 0 | 0 | 0 |
| | | | (0.0-0.0) | (0.0-0.1) | (0.0-0.0) | (0.0-0.0) | (0.0-0.0) | --- | --- | (0.0-0.0) | (0.0-0.0) | |
| | Late | prop | 0 | 0 | 0.33 | 0.33 | 0 | --- | --- | 0 | 0.33 | 1 |
| | | risk | 0 | 0 | 0.9 | 0.9 | 0 | --- | --- | 0 | 0.9 | 2.6 |
| | | num | 0 | 0 | 0 | 0 | 0 | --- | --- | 0 | 0 | 0 |
| | | | (0.0-0.0) | (0.0-0.0) | (0.0-0.0) | (0.0-0.0) | (0.0-0.0) | --- | --- | (0.0-0.0) | (0.0-0.0) | |
| | Overall | prop | 0.25 | 0.38 | 0.13 | 0.13 | 0 | 0 | 0 | 0 | 0.13 | 1 |
| | | risk | 1.8 | 2.7 | 0.9 | 0.9 | 0 | 0 | 0 | 0 | 0.9 | 7.3 |
| | | num | 0 | 0 | 0 | 0 | 0 | 0 | 0 | 0 | 0 | 0.1 |
| | | | (0.0-0.0) | (0.0-0.1) | (0.0-0.0) | (0.0-0.0) | (0.0-0.0) | (0.0-0.0) | (0.0-0.0) | (0.0-0.0) | (0.0-0.0) | |

\* prop = proportion; num = number of deaths (in 100s)

| Table S14: Cause-specific proportions, risks, and numbers of deaths (with uncertainty) for 194 countries by neonatal period ||||||||||||
| Country | Period | Stat* | Preterm | Intrapartum | Congenital | Sepsis | Pneumonia | Tetanus | Diarrhoea | Injuries | Other | Total |
| --- | --- | --- | --- | --- | --- | --- | --- | --- | --- | --- | --- | --- |
| Saint Lucia | Early | prop | 0 | 0 | 0.45 | 0 | 0.14 | --- | --- | 0.09 | 0.32 | 1 |

| Table S14: Cause-specific proportions, risks, and numbers of deaths (with uncertainty) for 194 countries by neonatal period ||||||||||||
|---|---|---|---|---|---|---|---|---|---|---|---|---|
| Country | Period | Stat* | Preterm | Intrapartum | Congenital | Sepsis | Pneumonia | Tetanus | Diarrhoea | Injuries | Other | Total |
| (high-quality VR) | | risk | 0 | 0 | 3.3 | 0 | 1 | --- | --- | 0.7 | 2.3 | 7.2 |
| | | num | 0 | 0 | 0.1 | 0 | 0 | --- | --- | 0 | 0.1 | 0.2 |
| | | | (0.0-0.0) | (0.0-0.0) | (0.0-0.2) | (0.0-0.0) | (0.0-0.1) | --- | --- | (0.0-0.0) | (0.0-0.1) | |
| | Late | prop | 0 | 0 | 0.8 | 0 | 0.2 | --- | --- | 0 | 0 | 1 |
| | | risk | 0 | 0 | 1.3 | 0 | 0.3 | --- | --- | 0 | 0 | 1.6 |
| | | num | 0 | 0 | 0 | 0 | 0 | --- | --- | 0 | 0 | 0 |
| | | | (0.0-0.0) | (0.0-0.0) | (0.0-0.1) | (0.0-0.0) | (0.0-0.0) | --- | --- | (0.0-0.0) | (0.0-0.0) | |
| | Overall | prop | 0 | 0 | 0.52 | 0 | 0.15 | 0 | 0 | 0.07 | 0.26 | 1 |
| | | risk | 0 | 0 | 4.7 | 0 | 1.3 | 0 | 0 | 0.7 | 2.3 | 9 |
| | | num | 0 | 0 | 0.1 | 0 | 0 | 0 | 0 | 0 | 0.1 | 0.3 |
| | | | (0.0-0.0) | (0.0-0.0) | (0.0-0.2) | (0.0-0.0) | (0.0-0.1) | (0.0-0.0) | (0.0-0.0) | (0.0-0.0) | (0.0-0.1) | |
| **Saint Vincent and the Grenadines** (high-quality VR) | Early | prop | 0.58 | 0.21 | 0.13 | 0.04 | 0 | --- | --- | 0 | 0.04 | 1 |
| | | risk | 5.6 | 2 | 1.2 | 0.4 | 0 | --- | --- | 0 | 0.4 | 9.7 |
| | | num | 0.1 | 0 | 0 | 0 | 0 | --- | --- | 0 | 0 | 0.2 |
| | | | (0.0-0.2) | (0.0-0.1) | (0.0-0.1) | (0.0-0.0) | (0.0-0.0) | --- | --- | (0.0-0.0) | (0.0-0.0) | |
| | Late | prop | 0.67 | 0.17 | 0.17 | 0 | 0 | --- | --- | 0 | 0 | 1 |
| | | risk | 1.6 | 0.4 | 0.4 | 0 | 0 | --- | --- | 0 | 0 | 2.4 |
| | | num | 0 | 0 | 0 | 0 | 0 | --- | --- | 0 | 0 | 0 |
| | | | (0.0-0.1) | (0.0-0.0) | (0.0-0.0) | (0.0-0.0) | (0.0-0.0) | --- | --- | (0.0-0.0) | (0.0-0.0) | |
| | Overall | prop | 0.6 | 0.2 | 0.13 | 0.03 | 0 | 0 | 0 | 0 | 0.03 | 1 |
| | | risk | 7.4 | 2.5 | 1.7 | 0.4 | 0 | 0 | 0 | 0 | 0.4 | 12.4 |
| | | num | 0.1 | 0 | 0 | 0 | 0 | 0 | 0 | 0 | 0 | 0.2 |
| | | | (0.0-0.2) | (0.0-0.1) | (0.0-0.1) | (0.0-0.0) | (0.0-0.0) | (0.0-0.0) | (0.0-0.0) | (0.0-0.0) | (0.0-0.0) | |
| **Samoa** (low mort model) | Early | prop | 0.46 | 0.14 | 0.23 | 0.08 | 0.04 | --- | --- | 0.04 | 0 | 1 |
| | | risk | 2.6 | 0.8 | 1.3 | 0.4 | 0.2 | --- | --- | 0.2 | 0 | 5.5 |
| | | num | 0.1 | 0 | 0.1 | 0 | 0 | --- | --- | 0 | 0 | 0.3 |
| | | | (0.1-0.2) | (0.0-0.1) | (0.0-0.1) | (0.0-0.0) | (0.0-0.0) | --- | --- | (0.0-0.1) | (0.0-0.0) | |
| | Late | prop | 0.27 | 0.06 | 0.28 | 0.16 | 0.15 | --- | --- | 0.01 | 0.06 | 1 |
| | | risk | 0.5 | 0.1 | 0.5 | 0.3 | 0.3 | --- | --- | 0 | 0.1 | 2 |
| | | num | 0 | 0 | 0 | 0 | 0 | --- | --- | 0 | 0 | 0.1 |
| | | | (0.0-0.0) | (0.0-0.0) | (0.0-0.0) | (0.0-0.0) | (0.0-0.0) | --- | --- | (0.0-0.0) | (0.0-0.0) | |
| | Overall | prop | 0.42 | 0.12 | 0.24 | 0.09 | 0.07 | 0 | 0 | 0.03 | 0.02 | 1 |
| | | risk | 3.2 | 0.9 | 1.8 | 0.7 | 0.6 | 0 | 0 | 0.2 | 0.1 | 7.6 |
| | | num | 0.2 | 0 | 0.1 | 0 | 0 | 0 | 0 | 0 | 0 | 0.4 |
| | | | (0.1-0.2) | (0.0-0.1) | (0.1-0.1) | (0.0-0.1) | (0.0-0.0) | (0.0-0.0) | (0.0-0.0) | (0.0-0.1) | (0.0-0.0) | |

* prop = proportion; num = number of deaths (in 100s)

| Table S14: Cause-specific proportions, risks, and numbers of deaths (with uncertainty) for 194 countries by neonatal period | | | | | | | | | | | |
|---|---|---|---|---|---|---|---|---|---|---|---|
| Country | Period | Stat* | Preterm | Intrapartum | Congenital | Sepsis | Pneumonia | Tetanus | Diarrhoea | Injuries | Other | Total |
| **San Marino** (low mort model) | Early | prop | 0.52 | 0.17 | 0.14 | 0.03 | 0 | --- | --- | 0.01 | 0.13 | 1 |
| | | risk | 0.4 | 0.1 | 0.1 | 0 | 0 | --- | --- | 0 | 0.1 | 0.8 |
| | | num | 0 | 0 | 0 | 0 | 0 | --- | --- | 0 | 0 | 0 |
| | | | (0.0-0.0) | (0.0-0.0) | (0.0-0.0) | (0.0-0.0) | (0.0-0.0) | --- | --- | (0.0-0.0) | (0.0-0.0) | |
| | Late | prop | 0.32 | 0.07 | 0.35 | 0.13 | 0 | --- | --- | 0.02 | 0.11 | 1 |
| | | risk | 0.1 | 0 | 0.1 | 0 | 0 | --- | --- | 0 | 0 | 0.3 |
| | | num | 0 | 0 | 0 | 0 | 0 | --- | --- | 0 | 0 | 0 |
| | | | (0.0-0.0) | (0.0-0.0) | (0.0-0.0) | (0.0-0.0) | (0.0-0.0) | --- | --- | (0.0-0.0) | (0.0-0.0) | |
| | Overall | prop | 0.44 | 0.14 | 0.25 | 0.04 | 0 | 0 | 0 | 0.01 | 0.12 | 1 |
| | | risk | 0.5 | 0.2 | 0.3 | 0 | 0 | 0 | 0 | 0 | 0.1 | 1.1 |
| | | num | 0 | 0 | 0 | 0 | 0 | 0 | 0 | 0 | 0 | 0 |
| | | | (0.0-0.0) | (0.0-0.0) | (0.0-0.0) | (0.0-0.0) | (0.0-0.0) | (0.0-0.0) | (0.0-0.0) | (0.0-0.0) | (0.0-0.0) | |
| **Sao Tome and Principe** (high mort model) | Early | prop | 0.32 | 0.35 | 0.16 | 0.09 | 0.05 | 0 | 0 | --- | 0.03 | 1 |
| | | risk | 4.6 | 5 | 2.3 | 1.3 | 0.7 | 0.1 | 0 | --- | 0.4 | 14.4 |
| | | num | 0.3 | 0.3 | 0.1 | 0.1 | 0 | 0 | 0 | --- | 0 | 0.9 |
| | | | (0.2-0.4) | (0.2-0.4) | (0.1-0.2) | (0.0-0.1) | (0.0-0.1) | (0.0-0.0) | (0.0-0.0) | --- | (0.0-0.1) | |
| | Late | prop | 0.12 | 0.16 | 0.13 | 0.44 | 0.06 | 0.02 | 0 | --- | 0.07 | 1 |
| | | risk | 0.6 | 0.8 | 0.7 | 2.2 | 0.3 | 0.1 | 0 | --- | 0.3 | 5 |
| | | num | 0 | 0.1 | 0 | 0.1 | 0 | 0 | 0 | --- | 0 | 0.3 |
| | | | (0.0-0.1) | (0.0-0.1) | (0.0-0.1) | (0.1-0.2) | (0.0-0.0) | (0.0-0.0) | (0.0-0.0) | --- | (0.0-0.1) | |
| | Overall | prop | 0.28 | 0.3 | 0.15 | 0.18 | 0.05 | 0.01 | 0 | 0 | 0.04 | 1 |
| | | risk | 5.5 | 5.9 | 2.9 | 3.7 | 1 | 0.1 | 0 | 0 | 0.8 | 19.9 |
| | | num | 0.4 | 0.4 | 0.2 | 0.2 | 0.1 | 0 | 0 | 0 | 0.1 | 1.3 |
| | | | (0.2-0.5) | (0.3-0.5) | (0.1-0.3) | (0.1-0.4) | (0.0-0.1) | (0.0-0.0) | (0.0-0.0) | (0.0-0.0) | (0.0-0.1) | |
| **Saudi Arabia** (low mort model) | Early | prop | 0.45 | 0.19 | 0.24 | 0.04 | 0.01 | --- | --- | 0.01 | 0.06 | 1 |
| | | risk | 2.9 | 1.2 | 1.6 | 0.2 | 0.1 | --- | --- | 0 | 0.4 | 6.5 |
| | | num | 16 | 6.6 | 8.6 | 1.3 | 0.3 | --- | --- | 0.2 | 2.3 | 35.3 |
| | | | (12.8-20.4) | (4.3-9.3) | (5.9-12.4) | (0.8-2.0) | (0.2-0.5) | --- | --- | (0.1-0.4) | (1.7-3.4) | |
| | Late | prop | 0.3 | 0.12 | 0.3 | 0.16 | 0.03 | --- | --- | 0.02 | 0.08 | 1 |
| | | risk | 0.7 | 0.3 | 0.7 | 0.4 | 0.1 | --- | --- | 0 | 0.2 | 2.3 |
| | | num | 3.7 | 1.5 | 3.7 | 2 | 0.4 | --- | --- | 0.2 | 1 | 12.4 |
| | | | (3.1-5.0) | (0.5-2.0) | (3.1-4.8) | (1.2-3.0) | (0.2-0.6) | --- | --- | (0.1-0.3) | (0.7-1.5) | |
| | Overall | prop | 0.41 | 0.17 | 0.26 | 0.07 | 0.01 | 0 | 0 | 0.01 | 0.07 | 1 |
| | | risk | 3.8 | 1.6 | 2.4 | 0.6 | 0.1 | 0 | 0 | 0.1 | 0.6 | 9.2 |
| | | num | 21.4 | 8.8 | 13.2 | 3.6 | 0.7 | 0 | 0 | 0.5 | 3.6 | 51.7 |
| | | | (17.1-27.3) | (5.3-12.1) | (9.4-18.5) | (2.1-5.5) | (0.4-1.1) | (0.0-0.0) | (0.0-0.0) | (0.3-0.7) | (2.6-5.1) | |

* prop = proportion; num = number of deaths (in 100s)

Table S14: Cause-specific proportions, risks, and numbers of deaths (with uncertainty) for 194 countries by neonatal period

| Country | Period | Stat* | Preterm | Intrapartum | Congenital | Sepsis | Pneumonia | Tetanus | Diarrhoea | Injuries | Other | Total |
|---|---|---|---|---|---|---|---|---|---|---|---|---|
| **Senegal** (high mort model) | Early | prop | 0.35 | 0.32 | 0.14 | 0.1 | 0.05 | 0 | 0 | --- | 0.03 | 1 |
| | | risk | 6 | 5.5 | 2.3 | 1.7 | 0.8 | 0.1 | 0 | --- | 0.6 | 17 |
| | | num | 31.8 | 28.9 | 12.2 | 8.8 | 4.4 | 0.4 | 0.1 | --- | 2.9 | 89.5 |
| | | | (25.8-39.2) | (23.8-34.5) | (7.6-17.0) | (3.5-14.8) | (2.3-8.8) | (0.1-1.0) | (0.0-0.8) | --- | (1.6-7.6) | |
| | Late | prop | 0.14 | 0.14 | 0.08 | 0.49 | 0.05 | 0.02 | 0.02 | --- | 0.06 | 1 |
| | | risk | 0.8 | 0.9 | 0.5 | 2.9 | 0.3 | 0.1 | 0.1 | --- | 0.4 | 6 |
| | | num | 4.4 | 4.5 | 2.4 | 15.4 | 1.7 | 0.5 | 0.6 | --- | 1.8 | 31.5 |
| | | | (2.2-7.5) | (2.8-6.2) | (0.8-6.0) | (9.9-21.5) | (0.9-2.7) | (0.1-1.3) | (0.2-1.5) | --- | (0.8-4.6) | |
| | Overall | prop | 0.3 | 0.28 | 0.11 | 0.2 | 0.05 | 0.01 | 0.01 | 0 | 0.04 | 1 |
| | | risk | 7.2 | 6.6 | 2.7 | 4.8 | 1.2 | 0.2 | 0.1 | 0 | 1 | 23.9 |
| | | num | 36.8 | 34 | 14 | 24.6 | 6.2 | 1 | 0.8 | 0 | 4.9 | 122.2 |
| | | | (28.1-47.7) | (27.4-41.1) | (7.9-22.4) | (13.5-36.7) | (3.4-11.8) | (0.3-2.4) | (0.2-2.4) | (0.0-0.0) | (2.4-12.2) | |
| **Serbia** (high-quality VR) | Early | prop | 0.62 | 0.13 | 0.14 | 0.03 | 0.02 | --- | --- | 0.01 | 0.05 | 1 |
| | | risk | 2.1 | 0.5 | 0.5 | 0.1 | 0.1 | --- | --- | 0 | 0.2 | 3.4 |
| | | num | 1.9 | 0.4 | 0.4 | 0.1 | 0.1 | --- | --- | 0 | 0.2 | 3.1 |
| | | | (1.6-2.2) | (0.3-0.5) | (0.3-0.6) | (0.0-0.1) | (0.0-0.1) | --- | --- | (0.0-0.1) | (0.1-0.2) | |
| | Late | prop | 0.42 | 0.12 | 0.32 | 0.07 | 0.04 | --- | --- | 0 | 0.04 | 1 |
| | | risk | 0.3 | 0.1 | 0.3 | 0.1 | 0 | --- | --- | 0 | 0 | 0.8 |
| | | num | 0.3 | 0.1 | 0.2 | 0.1 | 0 | --- | --- | 0 | 0 | 0.7 |
| | | | (0.2-0.4) | (0.0-0.1) | (0.1-0.3) | (0.0-0.1) | (0.0-0.1) | --- | --- | (0.0-0.0) | (0.0-0.1) | |
| | Overall | prop | 0.58 | 0.13 | 0.18 | 0.03 | 0.02 | 0 | 0 | 0.01 | 0.05 | 1 |
| | | risk | 2.6 | 0.6 | 0.8 | 0.1 | 0.1 | 0 | 0 | 0 | 0.2 | 4.4 |
| | | num | 2.4 | 0.5 | 0.7 | 0.1 | 0.1 | 0 | 0 | 0 | 0.2 | 4.1 |
| | | | (2.0-2.8) | (0.3-0.7) | (0.5-0.9) | (0.0-0.2) | (0.0-0.2) | (0.0-0.0) | (0.0-0.0) | (0.0-0.1) | (0.1-0.3) | |
| **Seychelles** (low mort model) | Early | prop | 0.48 | 0.16 | 0.22 | 0.08 | 0.01 | --- | --- | 0.04 | 0 | 1 |
| | | risk | 3.2 | 1.1 | 1.5 | 0.5 | 0.1 | --- | --- | 0.3 | 0 | 6.6 |
| | | num | 0.1 | 0 | 0 | 0 | 0 | --- | --- | 0 | 0 | 0.1 |
| | | | (0.0-0.1) | (0.0-0.0) | (0.0-0.0) | (0.0-0.0) | (0.0-0.0) | --- | --- | (0.0-0.0) | (0.0-0.0) | |
| | Late | prop | 0.24 | 0.07 | 0.24 | 0.33 | 0.04 | --- | --- | 0.01 | 0.07 | 1 |
| | | risk | 0.6 | 0.2 | 0.6 | 0.8 | 0.1 | --- | --- | 0 | 0.2 | 2.3 |
| | | num | 0 | 0 | 0 | 0 | 0 | --- | --- | 0 | 0 | 0 |
| | | | (0.0-0.0) | (0.0-0.0) | (0.0-0.0) | (0.0-0.0) | (0.0-0.0) | --- | --- | (0.0-0.0) | (0.0-0.0) | |
| | Overall | prop | 0.42 | 0.14 | 0.23 | 0.14 | 0.02 | 0 | 0 | 0.03 | 0.02 | 1 |
| | | risk | 3.7 | 1.2 | 2 | 1.3 | 0.2 | 0 | 0 | 0.3 | 0.2 | 8.9 |
| | | num | 0.1 | 0 | 0 | 0 | 0 | 0 | 0 | 0 | 0 | 0.1 |

* prop = proportion; num = number of deaths (in 100s)

| Table S14: Cause-specific proportions, risks, and numbers of deaths (with uncertainty) for 194 countries by neonatal period |||||||||||||
|---|---|---|---|---|---|---|---|---|---|---|---|---|
| Country | Period | Stat* | Preterm | Intrapartum | Congenital | Sepsis | Pneumonia | Tetanus | Diarrhoea | Injuries | Other | Total |
| **Sierra Leone** (high mort model) | Early | prop | 0.32 | 0.36 | 0.07 | 0.1 | 0.1 | 0.01 | 0 | --- | 0.04 | 1 |
| | | risk | 10.5 | 11.9 | 2.1 | 3.3 | 3.2 | 0.4 | 0.1 | --- | 1.1 | 32.8 |
| | | num | 22.4 | 25.2 | 4.6 | 7.1 | 6.8 | 0.9 | 0.3 | --- | 2.4 | 69.8 |
| | | | (15.8-26.4) | (18.2-28.6) | (2.0-7.4) | (2.3-11.0) | (3.3-12.7) | (0.3-2.4) | (0.0-2.6) | --- | (1.1-6.2) | |
| | Late | prop | 0.13 | 0.16 | 0.05 | 0.46 | 0.06 | 0.07 | 0.01 | --- | 0.06 | 1 |
| | | risk | 1.5 | 1.8 | 0.5 | 5.2 | 0.6 | 0.9 | 0.1 | --- | 0.7 | 11.5 |
| | | num | 3.3 | 3.9 | 1.1 | 11.2 | 1.4 | 1.8 | 0.3 | --- | 1.6 | 24.5 |
| | | | (1.5-5.3) | (2.3-5.1) | (0.4-2.4) | (6.3-15.0) | (0.7-2.0) | (0.3-5.6) | (0.1-0.7) | --- | (0.6-4.0) | |
| | Overall | prop | 0.27 | 0.31 | 0.06 | 0.19 | 0.09 | 0.03 | 0.01 | 0 | 0.04 | 1 |
| | | risk | 12.2 | 13.9 | 2.7 | 8.7 | 4 | 1.4 | 0.3 | 0 | 1.9 | 45.1 |
| | | num | 25.7 | 29.2 | 5.8 | 18.2 | 8.5 | 2.9 | 0.6 | 0 | 4 | 94.9 |
| | | | (17.6-31.9) | (20.6-33.9) | (2.3-9.9) | (8.4-26.3) | (4.2-15.2) | (0.6-8.4) | (0.1-3.2) | (0.0-0.0) | (1.8-10.0) | |
| **Singapore** (high-quality VR) | Early | prop | 0.39 | 0.12 | 0.3 | 0.03 | 0.03 | --- | --- | 0 | 0.12 | 1 |
| | | risk | 0.3 | 0.1 | 0.3 | 0 | 0 | --- | --- | 0 | 0.1 | 0.8 |
| | | num | 0.2 | 0.1 | 0.1 | 0 | 0 | --- | --- | 0 | 0.1 | 0.5 |
| | | | (0.1-0.3) | (0.0-0.1) | (0.1-0.2) | (0.0-0.0) | (0.0-0.0) | --- | --- | (0.0-0.0) | (0.0-0.1) | |
| | Late | prop | 0.4 | 0.2 | 0.2 | 0 | 0.1 | --- | --- | 0 | 0.1 | 1 |
| | | risk | 0.1 | 0.1 | 0.1 | 0 | 0 | --- | --- | 0 | 0 | 0.3 |
| | | num | 0.1 | 0 | 0 | 0 | 0 | --- | --- | 0 | 0 | 0.1 |
| | | | (0.0-0.1) | (0.0-0.1) | (0.0-0.1) | (0.0-0.0) | (0.0-0.0) | --- | --- | (0.0-0.0) | (0.0-0.0) | |
| | Overall | prop | 0.4 | 0.14 | 0.28 | 0.02 | 0.05 | 0 | 0 | 0 | 0.12 | 1 |
| | | risk | 0.4 | 0.2 | 0.3 | 0 | 0.1 | 0 | 0 | 0 | 0.1 | 1.1 |
| | | num | 0.2 | 0.1 | 0.2 | 0 | 0 | 0 | 0 | 0 | 0.1 | 0.6 |
| | | | (0.1-0.4) | (0.0-0.2) | (0.1-0.3) | (0.0-0.0) | (0.0-0.1) | (0.0-0.0) | (0.0-0.0) | (0.0-0.0) | (0.0-0.1) | |
| **Slovakia** (high-quality VR) | Early | prop | 0.51 | 0.11 | 0.3 | 0.01 | 0.01 | --- | --- | 0.01 | 0.03 | 1 |
| | | risk | 1.4 | 0.3 | 0.8 | 0 | 0 | --- | --- | 0 | 0.1 | 2.8 |
| | | num | 0.8 | 0.2 | 0.5 | 0 | 0 | --- | --- | 0 | 0 | 1.6 |
| | | | (0.6-1.0) | (0.1-0.3) | (0.3-0.6) | (0.0-0.1) | (0.0-0.1) | --- | --- | (0.0-0.1) | (0.0-0.1) | |
| | Late | prop | 0.5 | 0.03 | 0.38 | 0.04 | 0.01 | --- | --- | 0.03 | 0.01 | 1 |
| | | risk | 0.7 | 0 | 0.5 | 0.1 | 0 | --- | --- | 0 | 0 | 1.4 |
| | | num | 0.4 | 0 | 0.3 | 0 | 0 | --- | --- | 0 | 0 | 0.8 |
| | | | (0.3-0.5) | (0.0-0.1) | (0.2-0.4) | (0.0-0.1) | (0.0-0.0) | --- | --- | (0.0-0.1) | (0.0-0.0) | |

| Table S14: Cause-specific proportions, risks, and numbers of deaths (with uncertainty) for 194 countries by neonatal period | | | | | | | | | | | | |
|---|---|---|---|---|---|---|---|---|---|---|---|---|
| Country | Period | Stat* | Preterm | Intrapartum | Congenital | Sepsis | Pneumonia | Tetanus | Diarrhoea | Injuries | Other | Total |
| | Overall | prop | 0.51 | 0.08 | 0.33 | 0.02 | 0.01 | 0 | 0 | 0.02 | 0.02 | 1 |
| | | risk | 2.2 | 0.3 | 1.4 | 0.1 | 0.1 | 0 | 0 | 0.1 | 0.1 | 4.3 |
| | | num | 1.3 | 0.2 | 0.8 | 0.1 | 0 | 0 | 0 | 0 | 0.1 | 2.6 |
| | | | (1.0-1.6) | (0.1-0.3) | (0.6-1.1) | (0.0-0.1) | (0.0-0.1) | (0.0-0.0) | (0.0-0.0) | (0.0-0.1) | (0.0-0.1) | |

* prop = proportion; num = number of deaths (in 100s)

| Table S14: Cause-specific proportions, risks, and numbers of deaths (with uncertainty) for 194 countries by neonatal period | | | | | | | | | | | | |
|---|---|---|---|---|---|---|---|---|---|---|---|---|
| Country | Period | Stat* | Preterm | Intrapartum | Congenital | Sepsis | Pneumonia | Tetanus | Diarrhoea | Injuries | Other | Total |
| **Slovenia** (high-quality VR) | Early | prop | 0.67 | 0.07 | 0.07 | 0.13 | 0 | --- | --- | 0 | 0.07 | 1 |
| | | risk | 0.8 | 0.1 | 0.1 | 0.2 | 0 | --- | --- | 0 | 0.1 | 1.2 |
| | | num | 0.2 | 0 | 0 | 0 | 0 | --- | --- | 0 | 0 | 0.3 |
| | | | (0.1-0.2) | (0.0-0.0) | (0.0-0.0) | (0.0-0.1) | (0.0-0.0) | --- | --- | (0.0-0.0) | (0.0-0.0) | |
| | Late | prop | 0.8 | 0 | 0.2 | 0 | 0 | --- | --- | 0 | 0 | 1 |
| | | risk | 0.3 | 0 | 0.1 | 0 | 0 | --- | --- | 0 | 0 | 0.4 |
| | | num | 0.1 | 0 | 0 | 0 | 0 | --- | --- | 0 | 0 | 0.1 |
| | | | (0.0-0.1) | (0.0-0.0) | (0.0-0.0) | (0.0-0.0) | (0.0-0.0) | --- | --- | (0.0-0.0) | (0.0-0.0) | |
| | Overall | prop | 0.7 | 0.05 | 0.1 | 0.1 | 0 | 0 | 0 | 0 | 0.05 | 1 |
| | | risk | 1.2 | 0.1 | 0.2 | 0.2 | 0 | 0 | 0 | 0 | 0.1 | 1.7 |
| | | num | 0.3 | 0 | 0 | 0 | 0 | 0 | 0 | 0 | 0 | 0.4 |
| | | | (0.1-0.4) | (0.0-0.0) | (0.0-0.1) | (0.0-0.1) | (0.0-0.0) | (0.0-0.0) | (0.0-0.0) | (0.0-0.0) | (0.0-0.0) | |
| **Solomon Islands** (high mort model) | Early | prop | 0.32 | 0.29 | 0.22 | 0.07 | 0.04 | 0 | 0 | --- | 0.05 | 1 |
| | | risk | 3.2 | 2.9 | 2.1 | 0.7 | 0.4 | 0 | 0 | --- | 0.5 | 9.8 |
| | | num | 0.5 | 0.5 | 0.4 | 0.1 | 0.1 | 0 | 0 | --- | 0.1 | 1.7 |
| | | | (0.4-0.8) | (0.4-0.7) | (0.2-0.6) | (0.1-0.2) | (0.0-0.1) | (0.0-0.0) | (0.0-0.0) | --- | (0.0-0.2) | |
| | Late | prop | 0.11 | 0.14 | 0.21 | 0.42 | 0.05 | 0.01 | 0.01 | --- | 0.05 | 1 |
| | | risk | 0.4 | 0.5 | 0.7 | 1.4 | 0.2 | 0 | 0 | --- | 0.2 | 3.4 |
| | | num | 0.1 | 0.1 | 0.1 | 0.2 | 0 | 0 | 0 | --- | 0 | 0.6 |
| | | | (0.0-0.1) | (0.1-0.1) | (0.1-0.2) | (0.1-0.4) | (0.0-0.1) | (0.0-0.0) | (0.0-0.0) | --- | (0.0-0.1) | |
| | Overall | prop | 0.27 | 0.26 | 0.21 | 0.16 | 0.04 | 0.01 | 0 | 0 | 0.05 | 1 |
| | | risk | 3.6 | 3.5 | 2.9 | 2.2 | 0.5 | 0.1 | 0 | 0 | 0.7 | 13.5 |
| | | num | 0.6 | 0.6 | 0.5 | 0.4 | 0.1 | 0 | 0 | 0 | 0.1 | 2.3 |
| | | | (0.4-0.9) | (0.4-0.8) | (0.3-0.8) | (0.2-0.6) | (0.1-0.2) | (0.0-0.0) | (0.0-0.0) | (0.0-0.0) | (0.1-0.2) | |
| **Somalia** (high mort model) | Early | prop | 0.3 | 0.37 | 0.05 | 0.04 | 0.09 | 0.06 | 0.01 | --- | 0.08 | 1 |
| | | risk | 10.3 | 12.8 | 1.5 | 1.3 | 3.2 | 2.1 | 0.4 | --- | 2.6 | 34.2 |
| | | num | 46.1 | 57.3 | 6.9 | 6.1 | 14.4 | 9.4 | 1.6 | --- | 11.7 | 153.6 |
| | | | (30.3-66.2) | (38.5-68.3) | (3.6-11.7) | (1.2-10.8) | (7.2-24.8) | (3.1-24.0) | (0.0-11.4) | --- | (5.9-19.8) | |
| | Late | prop | 0.17 | 0.15 | 0.04 | 0.39 | 0.05 | 0.12 | 0.02 | --- | 0.07 | 1 |

| Table S14: Cause-specific proportions, risks, and numbers of deaths (with uncertainty) for 194 countries by neonatal period | | | | | | | | | | | | |
|---|---|---|---|---|---|---|---|---|---|---|---|---|
| Country | Period | Stat* | Preterm | Intrapartum | Congenital | Sepsis | Pneumonia | Tetanus | Diarrhoea | Injuries | Other | Total |
| | | risk | 2.1 | 1.8 | 0.4 | 4.6 | 0.6 | 1.5 | 0.2 | --- | 0.9 | 12 |
| | | num | 9.4 | 8 | 1.9 | 20.8 | 2.7 | 6.5 | 0.9 | --- | 3.9 | 54 |
| | | | (5.2-13.1) | (4.8-10.9) | (0.8-3.7) | (11.8-29.1) | (1.5-4.1) | (1.7-14.4) | (0.2-2.2) | --- | (1.7-9.6) | |
| | Overall | prop | 0.27 | 0.31 | 0.04 | 0.13 | 0.08 | 0.08 | 0.01 | 0 | 0.07 | 1 |
| | | risk | 12.7 | 14.5 | 2 | 6.1 | 3.9 | 3.7 | 0.5 | 0 | 3.5 | 47 |
| | | num | 56.6 | 64.7 | 9 | 27.3 | 17.4 | 16.5 | 2.4 | 0 | 15.5 | 209.3 |
| | | | (38.1-78.1) | (42.9-79.9) | (4.7-15.6) | (13.2-40.5) | (8.7-29.2) | (5.0-39.7) | (0.2-13.4) | (0.0-0.0) | (7.5-29.4) | |

* prop = proportion; num = number of deaths (in 100s)

| Table S14: Cause-specific proportions, risks, and numbers of deaths (with uncertainty) for 194 countries by neonatal period | | | | | | | | | | | | |
|---|---|---|---|---|---|---|---|---|---|---|---|---|
| Country | Period | Stat* | Preterm | Intrapartum | Congenital | Sepsis | Pneumonia | Tetanus | Diarrhoea | Injuries | Other | Total |
| **South Africa** (high-quality VR) | Early | prop | 0.37 | 0.29 | 0.07 | 0.08 | 0.05 | --- | --- | 0.01 | 0.13 | 1 |
| | | risk | 4.2 | 3.3 | 0.7 | 0.9 | 0.5 | --- | --- | 0.1 | 1.5 | 11.3 |
| | | num | 44.9 | 35.2 | 7.8 | 10 | 5.5 | --- | --- | 0.9 | 16 | 120.3 |
| | | | (43.6-46.2) | (34.1-36.4) | (7.3-8.4) | (9.4-10.6) | (5.0-5.9) | --- | --- | (0.7-1.1) | (15.2-16.8) | |
| | Late | prop | 0.22 | 0.06 | 0.06 | 0.27 | 0.17 | --- | --- | 0.02 | 0.19 | 1 |
| | | risk | 0.8 | 0.2 | 0.2 | 0.9 | 0.6 | --- | --- | 0.1 | 0.7 | 3.5 |
| | | num | 8.1 | 2.3 | 2.3 | 9.9 | 6.2 | --- | --- | 0.9 | 6.9 | 36.6 |
| | | | (7.5-8.6) | (2.0-2.6) | (2.0-2.6) | (9.3-10.5) | (5.7-6.7) | --- | --- | (0.7-1.1) | (6.4-7.4) | |
| | Overall | prop | 0.34 | 0.24 | 0.06 | 0.13 | 0.07 | 0 | 0 | 0.01 | 0.15 | 1 |
| | | risk | 5 | 3.6 | 1 | 1.9 | 1.1 | 0 | 0 | 0.2 | 2.2 | 14.9 |
| | | num | 55.5 | 39.3 | 10.7 | 20.8 | 12.2 | 0 | 0 | 1.9 | 24 | 164.4 |
| | | | (53.6-57.5) | (37.8-40.8) | (9.8-11.5) | (19.5-22.1) | (11.2-13.2) | (0.0-0.0) | (0.0-0.0) | (1.5-2.3) | (22.7-25.3) | |
| **South Sudan** (high mort model) | Early | prop | 0.38 | 0.34 | 0.06 | 0.06 | 0.08 | 0.04 | 0 | --- | 0.03 | 1 |
| | | risk | 11.1 | 9.8 | 1.7 | 1.8 | 2.2 | 1.1 | 0.1 | --- | 1 | 28.9 |
| | | num | 44.1 | 39.3 | 6.7 | 7.3 | 8.7 | 4.6 | 0.6 | --- | 3.9 | 115.1 |
| | | | (33.6-53.9) | (31.5-48.3) | (4.4-9.9) | (2.6-14.4) | (5.1-16.3) | (2.3-10.8) | (0.0-5.3) | --- | (2.0-10.3) | |
| | Late | prop | 0.21 | 0.14 | 0.04 | 0.37 | 0.05 | 0.1 | 0.02 | --- | 0.06 | 1 |
| | | risk | 2.2 | 1.4 | 0.4 | 3.8 | 0.6 | 1 | 0.2 | --- | 0.6 | 10.1 |
| | | num | 8.6 | 5.7 | 1.7 | 15 | 2.2 | 4.2 | 0.7 | --- | 2.3 | 40.4 |
| | | | (4.5-13.5) | (3.7-8.1) | (0.6-4.0) | (9.8-22.2) | (1.3-3.7) | (1.7-8.6) | (0.2-1.6) | --- | (1.1-5.9) | |
| | Overall | prop | 0.34 | 0.28 | 0.05 | 0.15 | 0.07 | 0.06 | 0.01 | 0 | 0.04 | 1 |
| | | risk | 13.4 | 11.3 | 2.2 | 6 | 2.8 | 2.3 | 0.3 | 0 | 1.6 | 39.9 |
| | | num | 51.9 | 43.7 | 8.3 | 23.2 | 10.8 | 8.9 | 1.1 | 0 | 6.1 | 154 |
| | | | (36.3-65.9) | (34.4-55.1) | (5.3-13.9) | (12.9-38.8) | (6.6-20.1) | (4.1-19.8) | (0.2-6.2) | (0.0-0.0) | (3.0-15.6) | |
| **Spain** (high- | Early | prop | 0.31 | 0.2 | 0.24 | 0.08 | 0 | --- | --- | 0 | 0.17 | 1 |
| | | risk | 0.5 | 0.3 | 0.4 | 0.1 | 0 | --- | --- | 0 | 0.3 | 1.6 |

| Table S14: Cause-specific proportions, risks, and numbers of deaths (with uncertainty) for 194 countries by neonatal period | | | | | | | | | | | | |
|---|---|---|---|---|---|---|---|---|---|---|---|---|
| Country | Period | Stat* | Preterm | Intrapartum | Congenital | Sepsis | Pneumonia | Tetanus | Diarrhoea | Injuries | Other | Total |
| quality VR) | | num | 2.4 | 1.6 | 1.9 | 0.6 | 0 | --- | --- | 0 | 1.4 | 8 |
| | | | (2.1-2.7) | (1.3-1.8) | (1.7-2.2) | (0.5-0.8) | (0.0-0.1) | --- | --- | (0.0-0.0) | (1.1-1.6) | |
| | Late | prop | 0.28 | 0.08 | 0.36 | 0.17 | 0.01 | --- | --- | 0 | 0.1 | 1 |
| | | risk | 0.3 | 0.1 | 0.3 | 0.2 | 0 | --- | --- | 0 | 0.1 | 1 |
| | | num | 1.3 | 0.4 | 1.7 | 0.8 | 0.1 | --- | --- | 0 | 0.5 | 4.7 |
| | | | (1.1-1.5) | (0.3-0.5) | (1.4-2.0) | (0.6-1.0) | (0.0-0.1) | --- | --- | (0.0-0.0) | (0.3-0.6) | |
| | Overall | prop | 0.3 | 0.15 | 0.29 | 0.11 | 0.01 | 0 | 0 | 0 | 0.14 | 1 |
| | | risk | 0.8 | 0.4 | 0.8 | 0.3 | 0 | 0 | 0 | 0 | 0.4 | 2.7 |
| | | num | 4 | 2.1 | 3.9 | 1.5 | 0.1 | 0 | 0 | 0 | 2 | 13.5 |
| | | | (3.4-4.5) | (1.7-2.4) | (3.3-4.4) | (1.2-1.8) | (0.0-0.2) | (0.0-0.0) | (0.0-0.0) | (0.0-0.0) | (1.6-2.3) | |

* prop = proportion; num = number of deaths (in 100s)

| Table S14: Cause-specific proportions, risks, and numbers of deaths (with uncertainty) for 194 countries by neonatal period | | | | | | | | | | | | |
|---|---|---|---|---|---|---|---|---|---|---|---|---|
| Country | Period | Stat* | Preterm | Intrapartum | Congenital | Sepsis | Pneumonia | Tetanus | Diarrhoea | Injuries | Other | Total |
| **Sri Lanka** (low mort model) | Early | prop | 0.41 | 0.15 | 0.25 | 0.05 | 0.03 | --- | --- | 0 | 0.1 | 1 |
| | | risk | 1.8 | 0.6 | 1.1 | 0.2 | 0.1 | --- | --- | 0 | 0.4 | 4.4 |
| | | num | 6.9 | 2.4 | 4.1 | 0.9 | 0.6 | --- | --- | 0.1 | 1.7 | 16.7 |
| | | | (5.8-8.1) | (2.0-2.9) | (2.3-5.6) | (0.5-1.3) | (0.3-1.0) | --- | --- | (0.1-0.1) | (1.2-2.2) | |
| | Late | prop | 0.29 | 0.09 | 0.33 | 0.14 | 0.05 | --- | --- | 0.02 | 0.09 | 1 |
| | | risk | 0.4 | 0.1 | 0.5 | 0.2 | 0.1 | --- | --- | 0 | 0.1 | 1.5 |
| | | num | 1.7 | 0.5 | 1.9 | 0.8 | 0.3 | --- | --- | 0.1 | 0.5 | 5.9 |
| | | | (1.5-2.0) | (0.4-0.7) | (1.7-2.2) | (0.6-0.9) | (0.1-0.5) | --- | --- | (0.1-0.1) | (0.3-0.9) | |
| | Overall | prop | 0.38 | 0.13 | 0.27 | 0.08 | 0.04 | 0 | 0 | 0.01 | 0.1 | 1 |
| | | risk | 2.3 | 0.8 | 1.6 | 0.5 | 0.2 | 0 | 0 | 0 | 0.6 | 6.1 |
| | | num | 8.8 | 3.1 | 6.3 | 1.8 | 0.9 | 0 | 0 | 0.2 | 2.3 | 23.3 |
| | | | (7.5-10.4) | (2.4-3.7) | (4.1-8.0) | (1.2-2.3) | (0.5-1.5) | (0.0-0.0) | (0.0-0.0) | (0.1-0.2) | (1.5-3.2) | |
| **Sudan** (high mort model) | Early | prop | 0.34 | 0.34 | 0.1 | 0.11 | 0.06 | 0.01 | 0 | --- | 0.03 | 1 |
| | | risk | 7.5 | 7.4 | 2.2 | 2.5 | 1.4 | 0.3 | 0 | --- | 0.7 | 22.1 |
| | | num | 93.3 | 91.9 | 27.2 | 31.2 | 17.7 | 3.9 | 0.5 | --- | 8.2 | 274 |
| | | | (65.8-115.5) | (73.3-117.7) | (16.1-39.3) | (12.9-61.8) | (10.1-37.3) | (1.6-10.6) | (0.0-6.1) | --- | (4.0-25.5) | |
| | Late | prop | 0.16 | 0.14 | 0.05 | 0.49 | 0.05 | 0.04 | 0.01 | --- | 0.06 | 1 |
| | | risk | 1.3 | 1.1 | 0.4 | 3.8 | 0.4 | 0.3 | 0.1 | --- | 0.5 | 7.8 |
| | | num | 15.7 | 13.7 | 4.8 | 47.4 | 4.6 | 3.4 | 1 | --- | 5.7 | 96.3 |
| | | | (8.2-25.7) | (8.6-19.1) | (1.5-13.0) | (33.5-63.5) | (2.9-6.8) | (1.3-7.4) | (0.3-2.3) | --- | (2.3-15.0) | |
| | Overall | prop | 0.29 | 0.29 | 0.08 | 0.21 | 0.06 | 0.02 | 0 | 0 | 0.04 | 1 |
| | | risk | 9 | 8.7 | 2.5 | 6.5 | 1.9 | 0.6 | 0.1 | 0 | 1.1 | 30.5 |
| | | num | 110.1 | 107 | 31.2 | 79.1 | 22.8 | 7.6 | 1.6 | 0 | 14.1 | 373.5 |

| Table S14: Cause-specific proportions, risks, and numbers of deaths (with uncertainty) for 194 countries by neonatal period | | | | | | | | | | | |
|---|---|---|---|---|---|---|---|---|---|---|---|
| Country | Period | Stat* | Preterm | Intrapartum | Congenital | Sepsis | Pneumonia | Tetanus | Diarrhoea | Injuries | Other | Total |
| | | | (74.4-142.3) | (83.0-138.3) | (17.4-51.2) | (47.3-128.1) | (13.2-44.9) | (3.0-18.6) | (0.3-8.8) | (0.0-0.0) | (6.4-41.1) | |
| **Suriname** (high-quality VR) | Early | prop | 0.48 | 0.2 | 0.12 | 0.12 | 0 | --- | --- | 0 | 0.09 | 1 |
| | | risk | 4.3 | 1.7 | 1 | 1 | 0 | --- | --- | 0 | 0.8 | 8.9 |
| | | num | 0.4 | 0.2 | 0.1 | 0.1 | 0 | --- | --- | 0 | 0.1 | 0.9 |
| | | | (0.3-0.5) | (0.1-0.3) | (0.0-0.2) | (0.0-0.2) | (0.0-0.0) | --- | --- | (0.0-0.0) | (0.0-0.1) | |
| | Late | prop | 0.26 | 0.08 | 0.28 | 0.33 | 0 | --- | --- | 0.03 | 0.03 | 1 |
| | | risk | 0.9 | 0.3 | 1 | 1.1 | 0 | --- | --- | 0.1 | 0.1 | 3.4 |
| | | num | 0.1 | 0 | 0.1 | 0.1 | 0 | --- | --- | 0 | 0 | 0.3 |
| | | | (0.0-0.1) | (0.0-0.1) | (0.0-0.2) | (0.0-0.2) | (0.0-0.0) | --- | --- | (0.0-0.0) | (0.0-0.0) | |
| | Overall | prop | 0.42 | 0.16 | 0.16 | 0.18 | 0 | 0 | 0 | 0.01 | 0.07 | 1 |
| | | risk | 5.3 | 2.1 | 2.1 | 2.2 | 0 | 0 | 0 | 0.1 | 0.9 | 12.6 |
| | | num | 0.5 | 0.2 | 0.2 | 0.2 | 0 | 0 | 0 | 0 | 0.1 | 1.2 |
| | | | (0.3-0.7) | (0.1-0.3) | (0.1-0.3) | (0.1-0.3) | (0.0-0.0) | (0.0-0.0) | (0.0-0.0) | (0.0-0.0) | (0.0-0.2) | |

* prop = proportion; num = number of deaths (in 100s)

| Table S14: Cause-specific proportions, risks, and numbers of deaths (with uncertainty) for 194 countries by neonatal period | | | | | | | | | | | |
|---|---|---|---|---|---|---|---|---|---|---|---|
| Country | Period | Stat* | Preterm | Intrapartum | Congenital | Sepsis | Pneumonia | Tetanus | Diarrhoea | Injuries | Other | Total |
| **Swaziland** (high mort model) | Early | prop | 0.37 | 0.35 | 0.09 | 0.1 | 0.06 | 0.01 | 0 | --- | 0.03 | 1 |
| | | risk | 8.1 | 7.7 | 1.9 | 2.1 | 1.3 | 0.1 | 0 | --- | 0.7 | 22.1 |
| | | num | 3 | 2.8 | 0.7 | 0.8 | 0.5 | 0.1 | 0 | --- | 0.3 | 8.1 |
| | | | (2.4-4.7) | (2.3-4.4) | (0.5-1.5) | (0.3-1.4) | (0.3-1.1) | (0.0-0.1) | (0.0-0.2) | --- | (0.1-1.0) | |
| | Late | prop | 0.16 | 0.15 | 0.09 | 0.43 | 0.05 | 0.03 | 0 | --- | 0.08 | 1 |
| | | risk | 1.3 | 1.2 | 0.7 | 3.3 | 0.4 | 0.2 | 0 | --- | 0.6 | 7.7 |
| | | num | 0.5 | 0.4 | 0.3 | 1.2 | 0.2 | 0.1 | 0 | --- | 0.2 | 2.8 |
| | | | (0.2-1.0) | (0.3-0.7) | (0.1-0.7) | (0.7-2.2) | (0.1-0.3) | (0.0-0.3) | (0.0-0.0) | --- | (0.1-0.7) | |
| | Overall | prop | 0.32 | 0.3 | 0.09 | 0.18 | 0.06 | 0.01 | 0 | 0 | 0.04 | 1 |
| | | risk | 9.4 | 8.9 | 2.6 | 5.4 | 1.7 | 0.4 | 0.1 | 0 | 1.3 | 29.8 |
| | | num | 3.5 | 3.3 | 1 | 2 | 0.6 | 0.1 | 0 | 0 | 0.5 | 11 |
| | | | (2.7-5.6) | (2.6-5.2) | (0.5-2.2) | (1.0-3.7) | (0.3-1.3) | (0.1-0.4) | (0.0-0.2) | (0.0-0.0) | (0.2-1.7) | |
| **Sweden** (high-quality VR) | Early | prop | 0.2 | 0.19 | 0.31 | 0.07 | 0 | --- | --- | 0 | 0.23 | 1 |
| | | risk | 0.2 | 0.2 | 0.4 | 0.1 | 0 | --- | --- | 0 | 0.3 | 1.2 |
| | | num | 0.3 | 0.3 | 0.4 | 0.1 | 0 | --- | --- | 0 | 0.3 | 1.3 |
| | | | (0.2-0.4) | (0.2-0.4) | (0.3-0.5) | (0.0-0.1) | (0.0-0.0) | --- | --- | (0.0-0.0) | (0.2-0.4) | |
| | Late | prop | 0.26 | 0.1 | 0.36 | 0.14 | 0 | --- | --- | 0 | 0.14 | 1 |
| | | risk | 0.1 | 0 | 0.1 | 0.1 | 0 | --- | --- | 0 | 0.1 | 0.4 |
| | | num | 0.1 | 0 | 0.2 | 0.1 | 0 | --- | --- | 0 | 0.1 | 0.5 |
| | | | (0.1-0.2) | (0.0-0.1) | (0.1-0.2) | (0.0-0.1) | (0.0-0.0) | --- | --- | (0.0-0.0) | (0.0-0.1) | |

| Table S14: Cause-specific proportions, risks, and numbers of deaths (with uncertainty) for 194 countries by neonatal period | | | | | | | | | | | | |
|---|---|---|---|---|---|---|---|---|---|---|---|---|
| Country | Period | Stat* | Preterm | Intrapartum | Congenital | Sepsis | Pneumonia | Tetanus | Diarrhoea | Injuries | Other | Total |
| **Switzerland** (low mort model) | Overall | prop | 0.22 | 0.17 | 0.32 | 0.09 | 0 | 0 | 0 | 0 | 0.21 | 1 |
| | | risk | 0.3 | 0.3 | 0.5 | 0.1 | 0 | 0 | 0 | 0 | 0.3 | 1.6 |
| | | num | 0.4 | 0.3 | 0.6 | 0.2 | 0 | 0 | 0 | 0 | 0.4 | 1.8 |
| | | | (0.2-0.6) | (0.2-0.4) | (0.4-0.8) | (0.0-0.3) | (0.0-0.0) | (0.0-0.0) | (0.0-0.0) | (0.0-0.0) | (0.2-0.5) | |
| | Early | prop | 0.45 | 0.15 | 0.26 | 0.02 | 0 | --- | --- | 0.01 | 0.11 | 1 |
| | | risk | 1 | 0.3 | 0.6 | 0.1 | 0 | --- | --- | 0 | 0.2 | 2.2 |
| | | num | 0.8 | 0.3 | 0.5 | 0 | 0 | --- | --- | 0 | 0.2 | 1.9 |
| | | | (0.8-0.9) | (0.2-0.3) | (0.4-0.6) | (0.0-0.1) | (0.0-0.0) | --- | --- | (0.0-0.0) | (0.1-0.2) | |
| | Late | prop | 0.31 | 0.08 | 0.39 | 0.12 | 0.01 | --- | --- | 0.02 | 0.08 | 1 |
| | | risk | 0.2 | 0.1 | 0.3 | 0.1 | 0 | --- | --- | 0 | 0.1 | 0.8 |
| | | num | 0.2 | 0.1 | 0.3 | 0.1 | 0 | --- | --- | 0 | 0.1 | 0.7 |
| | | | (0.2-0.2) | (0.0-0.1) | (0.2-0.3) | (0.1-0.1) | (0.0-0.0) | --- | --- | (0.0-0.0) | (0.0-0.1) | |
| | Overall | prop | 0.42 | 0.13 | 0.29 | 0.05 | 0 | 0 | 0 | 0.01 | 0.1 | 1 |
| | | risk | 1.2 | 0.4 | 0.9 | 0.1 | 0 | 0 | 0 | 0 | 0.3 | 3 |
| | | num | 1 | 0.3 | 0.7 | 0.1 | 0 | 0 | 0 | 0 | 0.3 | 2.5 |
| | | | (0.9-1.1) | (0.3-0.4) | (0.6-0.9) | (0.1-0.2) | (0.0-0.0) | (0.0-0.0) | (0.0-0.0) | (0.0-0.0) | (0.2-0.3) | |

* prop = proportion; num = number of deaths (in 100s)

| Table S14: Cause-specific proportions, risks, and numbers of deaths (with uncertainty) for 194 countries by neonatal period | | | | | | | | | | | | |
|---|---|---|---|---|---|---|---|---|---|---|---|---|
| Country | Period | Stat* | Preterm | Intrapartum | Congenital | Sepsis | Pneumonia | Tetanus | Diarrhoea | Injuries | Other | Total |
| **Syrian Arab Republic** (low mort model) | Early | prop | 0.48 | 0.22 | 0.15 | 0.07 | 0.04 | --- | --- | 0.01 | 0.04 | 1 |
| | | risk | 2.8 | 1.3 | 0.9 | 0.4 | 0.2 | --- | --- | 0.1 | 0.2 | 5.8 |
| | | num | 14.8 | 6.6 | 4.5 | 2.1 | 1.3 | --- | --- | 0.3 | 1.2 | 30.8 |
| | | | (12.5-20.2) | (4.1-10.1) | (3.1-10.6) | (0.7-3.1) | (0.8-2.6) | --- | --- | (0.2-0.7) | (0.7-2.4) | |
| | Late | prop | 0.27 | 0.1 | 0.22 | 0.24 | 0.1 | --- | --- | 0.01 | 0.06 | 1 |
| | | risk | 0.5 | 0.2 | 0.4 | 0.5 | 0.2 | --- | --- | 0 | 0.1 | 2.1 |
| | | num | 2.9 | 1.1 | 2.3 | 2.6 | 1.1 | --- | --- | 0.2 | 0.7 | 10.8 |
| | | | (2.4-4.6) | (0.2-2.0) | (1.6-4.8) | (0.9-4.3) | (0.8-1.8) | --- | --- | (0.1-0.3) | (0.1-3.1) | |
| | Overall | prop | 0.42 | 0.19 | 0.16 | 0.11 | 0.06 | 0 | 0 | 0.01 | 0.04 | 1 |
| | | risk | 3.4 | 1.5 | 1.3 | 0.9 | 0.5 | 0 | 0 | 0.1 | 0.4 | 8.1 |
| | | num | 18.4 | 8 | 7.1 | 4.9 | 2.5 | 0 | 0 | 0.5 | 1.9 | 43.4 |
| | | | (15.6-25.6) | (4.6-13.0) | (4.9-15.5) | (1.7-7.9) | (1.6-4.5) | (0.0-0.0) | (0.0-0.0) | (0.3-0.9) | (0.8-6.0) | |
| **Tajikistan** (high mort model) | Early | prop | 0.34 | 0.32 | 0.14 | 0.09 | 0.04 | 0.01 | 0 | --- | 0.06 | 1 |
| | | risk | 5.5 | 5.1 | 2.3 | 1.4 | 0.7 | 0.1 | 0 | --- | 1 | 16.2 |
| | | num | 15.3 | 14.1 | 6.4 | 3.9 | 1.9 | 0.2 | 0.1 | --- | 2.6 | 44.6 |
| | | | (11.2-21.0) | (10.8-18.8) | (4.0-10.1) | (1.4-6.7) | (0.9-4.2) | (0.1-0.6) | (0.0-0.9) | --- | (1.3-4.3) | |
| | Late | prop | 0.11 | 0.15 | 0.18 | 0.43 | 0.05 | 0.02 | 0 | --- | 0.05 | 1 |

| Table S14: Cause-specific proportions, risks, and numbers of deaths (with uncertainty) for 194 countries by neonatal period | | | | | | | | | | | |
|---|---|---|---|---|---|---|---|---|---|---|---|
| Country | Period | Stat* | Preterm | Intrapartum | Congenital | Sepsis | Pneumonia | Tetanus | Diarrhoea | Injuries | Other | Total |
| | | risk | 0.6 | 0.9 | 1 | 2.5 | 0.3 | 0.1 | 0 | --- | 0.3 | 5.7 |
| | | num | 1.7 | 2.4 | 2.9 | 6.8 | 0.9 | 0.3 | 0 | --- | 0.8 | 15.7 |
| | | | (0.9-2.9) | (1.5-3.5) | (1.6-4.7) | (4.1-10.5) | (0.5-1.4) | (0.1-0.7) | (0.0-0.1) | --- | (0.4-1.9) | |
| | | prop | 0.28 | 0.28 | 0.15 | 0.18 | 0.05 | 0.01 | 0 | 0 | 0.06 | 1 |
| | Overall | risk | 6.4 | 6.2 | 3.3 | 4 | 1 | 0.2 | 0.1 | 0 | 1.3 | 22.4 |
| | | num | 17.2 | 16.6 | 8.9 | 10.7 | 2.8 | 0.5 | 0.2 | 0 | 3.4 | 60.2 |
| | | | (12.6-24.5) | (12.3-22.0) | (5.5-14.2) | (5.5-16.7) | (1.4-5.6) | (0.2-1.4) | (0.0-1.1) | (0.0-0.0) | (1.7-6.3) | |
| | | prop | 0.4 | 0.14 | 0.28 | 0.06 | 0.03 | --- | --- | 0.01 | 0.08 | 1 |
| | Early | risk | 2.3 | 0.8 | 1.7 | 0.4 | 0.2 | --- | --- | 0.1 | 0.5 | 5.8 |
| | | num | 16 | 5.5 | 11.4 | 2.5 | 1.1 | --- | --- | 0.4 | 3.3 | 40.3 |
| | | | (13.8-19.4) | (4.5-6.8) | (9.1-14.0) | (1.7-3.4) | (0.7-1.6) | --- | --- | (0.3-0.5) | (2.4-4.3) | |
| Thailand (low mort model) | | prop | 0.3 | 0.09 | 0.31 | 0.17 | 0.05 | --- | --- | 0.02 | 0.07 | 1 |
| | Late | risk | 0.6 | 0.2 | 0.6 | 0.3 | 0.1 | --- | --- | 0 | 0.1 | 2.1 |
| | | num | 4.3 | 1.3 | 4.4 | 2.3 | 0.6 | --- | --- | 0.2 | 1 | 14.1 |
| | | | (3.6-5.2) | (0.9-1.8) | (3.8-5.3) | (1.7-2.9) | (0.4-1.1) | --- | --- | (0.2-0.3) | (0.7-1.4) | |
| | | prop | 0.37 | 0.12 | 0.29 | 0.09 | 0.03 | 0 | 0 | 0.01 | 0.08 | 1 |
| | Overall | risk | 3 | 1 | 2.4 | 0.7 | 0.3 | 0 | 0 | 0.1 | 0.6 | 8.1 |
| | | num | 21.1 | 7.1 | 16.5 | 5.1 | 1.8 | 0 | 0 | 0.6 | 4.5 | 56.6 |
| | | | (18.0-25.7) | (5.6-9.0) | (13.4-20.1) | (3.6-6.6) | (1.1-2.8) | (0.0-0.0) | (0.0-0.0) | (0.5-0.8) | (3.2-6.1) | |

* prop = proportion; num = number of deaths (in 100s)

| Table S14: Cause-specific proportions, risks, and numbers of deaths (with uncertainty) for 194 countries by neonatal period | | | | | | | | | | | |
|---|---|---|---|---|---|---|---|---|---|---|---|
| Country | Period | Stat* | Preterm | Intrapartum | Congenital | Sepsis | Pneumonia | Tetanus | Diarrhoea | Injuries | Other | Total |
| | | prop | 0.79 | 0.12 | 0.06 | 0.01 | 0.01 | --- | --- | 0 | 0.01 | 1 |
| | Early | risk | 2.6 | 0.4 | 0.2 | 0 | 0 | --- | --- | 0 | 0 | 3.3 |
| The former Yugoslav Republic of Macedonia (high-quality VR) | | num | 0.6 | 0.1 | 0 | 0 | 0 | --- | --- | 0 | 0 | 0.7 |
| | | | (0.4-0.7) | (0.0-0.1) | (0.0-0.1) | (0.0-0.0) | (0.0-0.0) | --- | --- | (0.0-0.0) | (0.0-0.0) | |
| | | prop | 0.69 | 0.03 | 0.14 | 0.1 | 0.03 | --- | --- | 0 | 0 | 1 |
| | Late | risk | 0.7 | 0 | 0.1 | 0.1 | 0 | --- | --- | 0 | 0 | 1.1 |
| | | num | 0.2 | 0 | 0 | 0 | 0 | --- | --- | 0 | 0 | 0.2 |
| | | | (0.1-0.2) | (0.0-0.0) | (0.0-0.1) | (0.0-0.1) | (0.0-0.0) | --- | --- | (0.0-0.0) | (0.0-0.0) | |
| | | prop | 0.76 | 0.1 | 0.08 | 0.03 | 0.02 | 0 | 0 | 0 | 0.01 | 1 |
| | Overall | risk | 3.7 | 0.5 | 0.4 | 0.2 | 0.1 | 0 | 0 | 0 | 0 | 4.9 |
| | | num | 0.8 | 0.1 | 0.1 | 0 | 0 | 0 | 0 | 0 | 0 | 1.1 |
| | | | (0.6-1.1) | (0.0-0.2) | (0.0-0.2) | (0.0-0.1) | (0.0-0.1) | (0.0-0.0) | (0.0-0.0) | (0.0-0.0) | (0.0-0.0) | |
| Timor-Leste | Early | prop | 0.25 | 0.37 | 0.13 | 0.09 | 0.05 | 0.01 | 0.01 | --- | 0.1 | 1 |
| | | risk | 4.5 | 6.5 | 2.2 | 1.5 | 0.9 | 0.1 | 0.1 | --- | 1.7 | 17.6 |

| Table S14: Cause-specific proportions, risks, and numbers of deaths (with uncertainty) for 194 countries by neonatal period | | | | | | | | | | | | |
|---|---|---|---|---|---|---|---|---|---|---|---|---|
| Country | Period | Stat* | Preterm | Intrapartum | Congenital | Sepsis | Pneumonia | Tetanus | Diarrhoea | Injuries | Other | Total |
| (high mort model) | | num | 1.8 (1.4-2.4) | 2.7 (2.3-3.2) | 0.9 (0.7-1.4) | 0.6 (0.3-1.0) | 0.4 (0.2-0.8) | 0 (0.0-0.1) | 0.1 (0.0-0.6) | --- --- | 0.7 (0.3-1.0) | 7.2 |
| | Late | prop | 0.11 | 0.17 | 0.09 | 0.49 | 0.06 | 0.02 | 0.01 | --- | 0.05 | 1 |
| | | risk | 0.7 | 1 | 0.5 | 3 | 0.4 | 0.1 | 0.1 | --- | 0.3 | 6.2 |
| | | num | 0.3 (0.2-0.5) | 0.4 (0.3-0.6) | 0.2 (0.1-0.4) | 1.2 (0.8-1.7) | 0.1 (0.1-0.2) | 0.1 (0.0-0.2) | 0 (0.0-0.1) | --- --- | 0.1 (0.1-0.4) | 2.5 |
| | Overall | prop | 0.21 | 0.33 | 0.11 | 0.18 | 0.06 | 0.01 | 0.01 | 0 | 0.09 | 1 |
| | | risk | 5.2 | 8.1 | 2.6 | 4.5 | 1.4 | 0.3 | 0.2 | 0 | 2.1 | 24.4 |
| | | num | 2.2 (1.6-3.2) | 3.4 (2.7-4.1) | 1.1 (0.8-1.9) | 1.9 (1.1-2.8) | 0.6 (0.3-1.1) | 0.1 (0.0-0.3) | 0.1 (0.0-0.8) | 0 (0.0-0.0) | 0.9 (0.4-1.6) | 10.3 |
| **Togo** (high mort model) | Early | prop | 0.34 | 0.36 | 0.09 | 0.1 | 0.06 | 0.01 | 0 | --- | 0.03 | 1 |
| | | risk | 7.7 | 8.1 | 1.9 | 2.3 | 1.4 | 0.3 | 0.1 | --- | 0.8 | 22.5 |
| | | num | 18.7 (14.3-22.8) | 19.7 (15.1-23.9) | 4.7 (2.7-7.9) | 5.5 (1.9-8.5) | 3.4 (1.7-6.8) | 0.6 (0.3-1.4) | 0.1 (0.0-1.3) | --- --- | 1.9 (0.9-5.1) | 54.6 |
| | Late | prop | 0.16 | 0.16 | 0.06 | 0.45 | 0.05 | 0.04 | 0.01 | --- | 0.07 | 1 |
| | | risk | 1.2 | 1.3 | 0.5 | 3.6 | 0.4 | 0.3 | 0.1 | --- | 0.5 | 7.9 |
| | | num | 3 (1.7-5.0) | 3.1 (2.0-4.2) | 1.1 (0.4-2.8) | 8.6 (5.1-11.9) | 1.1 (0.6-1.5) | 0.8 (0.3-1.9) | 0.2 (0.1-0.4) | --- --- | 1.3 (0.5-3.1) | 19.2 |
| | Overall | prop | 0.29 | 0.31 | 0.08 | 0.19 | 0.06 | 0.02 | 0 | 0 | 0.04 | 1 |
| | | risk | 8.9 | 9.7 | 2.4 | 6 | 1.9 | 0.6 | 0.2 | 0 | 1.3 | 31 |
| | | num | 21.3 (15.4-27.2) | 23 (17.1-28.3) | 5.7 (3.0-10.7) | 14.3 (7.1-21.1) | 4.5 (2.4-8.4) | 1.4 (0.5-3.2) | 0.4 (0.1-1.7) | 0 (0.0-0.0) | 3.1 (1.4-8.2) | 73.7 |

* prop = proportion; num = number of deaths (in 100s)

| Table S14: Cause-specific proportions, risks, and numbers of deaths (with uncertainty) for 194 countries by neonatal period | | | | | | | | | | | | |
|---|---|---|---|---|---|---|---|---|---|---|---|---|
| Country | Period | Stat* | Preterm | Intrapartum | Congenital | Sepsis | Pneumonia | Tetanus | Diarrhoea | Injuries | Other | Total |
| **Tonga** (low mort model) | Early | prop | 0.46 | 0.14 | 0.25 | 0.08 | 0.04 | --- | --- | 0.02 | 0.01 | 1 |
| | | risk | 2.2 | 0.6 | 1.2 | 0.4 | 0.2 | --- | --- | 0.1 | 0.1 | 4.7 |
| | | num | 0.1 (0.0-0.1) | 0 (0.0-0.0) | 0 (0.0-0.1) | 0 (0.0-0.0) | 0 (0.0-0.0) | --- --- | --- --- | 0 (0.0-0.0) | 0 (0.0-0.0) | 0.1 |
| | Late | prop | 0.29 | 0.07 | 0.33 | 0.15 | 0.08 | --- | --- | 0.02 | 0.06 | 1 |
| | | risk | 0.5 | 0.1 | 0.5 | 0.3 | 0.1 | --- | --- | 0 | 0.1 | 1.6 |
| | | num | 0 (0.0-0.0) | 0 (0.0-0.0) | 0 (0.0-0.0) | 0 (0.0-0.0) | 0 (0.0-0.0) | --- --- | --- --- | 0 (0.0-0.0) | 0 (0.0-0.0) | 0 |
| | Overall | prop | 0.42 | 0.12 | 0.26 | 0.1 | 0.05 | 0 | 0 | 0.02 | 0.02 | 1 |
| | | risk | 2.8 | 0.8 | 1.7 | 0.6 | 0.3 | 0 | 0 | 0.1 | 0.2 | 6.5 |
| | | num | 0.1 | 0 | 0 | 0 | 0 | 0 | 0 | 0 | 0 | 0.2 |

| Table S14: Cause-specific proportions, risks, and numbers of deaths (with uncertainty) for 194 countries by neonatal period |||||||||||  |
|---|---|---|---|---|---|---|---|---|---|---|---|
| Country | Period | Stat* | Preterm | Intrapartum | Congenital | Sepsis | Pneumonia | Tetanus | Diarrhoea | Injuries | Other | Total |
|  |  |  | (0.1-0.1) | (0.0-0.0) | (0.0-0.1) | (0.0-0.0) | (0.0-0.0) | (0.0-0.0) | (0.0-0.0) | (0.0-0.0) | (0.0-0.0) |  |
| **Trinidad and Tobago** (high-quality VR) | Early | prop | 0.45 | 0.19 | 0.18 | 0.03 | 0.04 | --- | --- | 0 | 0.11 | 1 |
|  |  | risk | 4.2 | 1.8 | 1.6 | 0.3 | 0.4 | --- | --- | 0 | 1.1 | 9.2 |
|  |  | num | 0.8 | 0.3 | 0.3 | 0.1 | 0.1 | --- | --- | 0 | 0.2 | 1.8 |
|  |  |  | (0.6-1.0) | (0.2-0.5) | (0.2-0.4) | (0.0-0.1) | (0.0-0.1) | --- | --- | (0.0-0.0) | (0.1-0.3) |  |
|  | Late | prop | 0.28 | 0.02 | 0.34 | 0.11 | 0.12 | --- | --- | 0.01 | 0.12 | 1 |
|  |  | risk | 1.6 | 0.1 | 2 | 0.6 | 0.7 | --- | --- | 0.1 | 0.7 | 5.9 |
|  |  | num | 0.3 | 0 | 0.4 | 0.1 | 0.1 | --- | --- | 0 | 0.1 | 1.1 |
|  |  |  | (0.2-0.4) | (0.0-0.1) | (0.3-0.5) | (0.1-0.2) | (0.1-0.2) | --- | --- | (0.0-0.0) | (0.1-0.2) |  |
|  | Overall | prop | 0.38 | 0.13 | 0.24 | 0.06 | 0.07 | 0 | 0 | 0 | 0.12 | 1 |
|  |  | risk | 5.9 | 2 | 3.7 | 0.9 | 1.1 | 0 | 0 | 0.1 | 1.8 | 15.5 |
|  |  | num | 1.2 | 0.4 | 0.7 | 0.2 | 0.2 | 0 | 0 | 0 | 0.4 | 3 |
|  |  |  | (0.9-1.5) | (0.2-0.5) | (0.5-1.0) | (0.1-0.3) | (0.1-0.3) | (0.0-0.0) | (0.0-0.0) | (0.0-0.0) | (0.2-0.5) |  |
| **Tunisia** (low mort model) | Early | prop | 0.38 | 0.19 | 0.24 | 0.06 | 0.03 | --- | --- | 0 | 0.11 | 1 |
|  |  | risk | 2.5 | 1.3 | 1.6 | 0.4 | 0.2 | --- | --- | 0 | 0.7 | 6.7 |
|  |  | num | 4.8 | 2.4 | 3 | 0.7 | 0.3 | --- | --- | 0.1 | 1.3 | 12.6 |
|  |  |  | (3.7-5.8) | (1.2-3.7) | (1.7-4.9) | (0.5-1.0) | (0.2-0.5) | --- | --- | (0.0-0.1) | (0.9-1.7) |  |
|  | Late | prop | 0.26 | 0.14 | 0.26 | 0.17 | 0.09 | --- | --- | 0.01 | 0.06 | 1 |
|  |  | risk | 0.6 | 0.3 | 0.6 | 0.4 | 0.2 | --- | --- | 0 | 0.1 | 2.3 |
|  |  | num | 1.2 | 0.6 | 1.1 | 0.8 | 0.4 | --- | --- | 0.1 | 0.3 | 4.4 |
|  |  |  | (1.0-1.5) | (0.1-0.9) | (1.0-1.5) | (0.4-1.3) | (0.3-0.6) | --- | --- | (0.1-0.1) | (0.2-0.4) |  |
|  | Overall | prop | 0.35 | 0.18 | 0.24 | 0.09 | 0.04 | 0 | 0 | 0.01 | 0.09 | 1 |
|  |  | risk | 3.3 | 1.7 | 2.3 | 0.8 | 0.4 | 0 | 0 | 0.1 | 0.9 | 9.4 |
|  |  | num | 6.3 | 3.2 | 4.3 | 1.6 | 0.7 | 0 | 0 | 0.1 | 1.7 | 17.9 |
|  |  |  | (4.9-7.8) | (1.3-4.9) | (2.8-6.5) | (0.9-2.5) | (0.5-1.1) | (0.0-0.0) | (0.0-0.0) | (0.1-0.2) | (1.1-2.2) |  |

* prop = proportion; num = number of deaths (in 100s)

| Table S14: Cause-specific proportions, risks, and numbers of deaths (with uncertainty) for 194 countries by neonatal period |||||||||||  |
|---|---|---|---|---|---|---|---|---|---|---|---|
| Country | Period | Stat* | Preterm | Intrapartum | Congenital | Sepsis | Pneumonia | Tetanus | Diarrhoea | Injuries | Other | Total |
| **Turkey** (low mort model) | Early | prop | 0.43 | 0.17 | 0.21 | 0.06 | 0.02 | --- | --- | 0.01 | 0.12 | 1 |
|  |  | risk | 3.6 | 1.4 | 1.7 | 0.5 | 0.1 | --- | --- | 0 | 1 | 8.3 |
|  |  | num | 45.9 | 17.7 | 22.2 | 6.2 | 1.8 | --- | --- | 0.5 | 12.6 | 106.9 |
|  |  |  | (40.1-51.8) | (13.1-21.6) | (17.4-28.5) | (4.5-8.1) | (1.3-2.3) | --- | --- | (0.4-0.7) | (8.9-15.3) |  |
|  | Late | prop | 0.28 | 0.1 | 0.25 | 0.2 | 0.07 | --- | --- | 0.01 | 0.09 | 1 |
|  |  | risk | 0.8 | 0.3 | 0.7 | 0.6 | 0.2 | --- | --- | 0 | 0.2 | 2.9 |
|  |  | num | 10.5 | 3.7 | 9.5 | 7.4 | 2.7 | --- | --- | 0.6 | 3.2 | 37.6 |
|  |  |  | (9.2-12.8) | (2.2-4.5) | (8.3-11.3) | (5.0-9.2) | (1.7-4.1) | --- | --- | (0.4-0.7) | (2.2-4.3) |  |

| Table S14: Cause-specific proportions, risks, and numbers of deaths (with uncertainty) for 194 countries by neonatal period ||||||||||||
| Country | Period | Stat* | Preterm | Intrapartum | Congenital | Sepsis | Pneumonia | Tetanus | Diarrhoea | Injuries | Other | Total |
| --- | --- | --- | --- | --- | --- | --- | --- | --- | --- | --- | --- | --- |
| Turkmenistan (high mort model) | Overall | prop | 0.39 | 0.15 | 0.21 | 0.1 | 0.03 | 0 | 0 | 0.01 | 0.11 | 1 |
| | | risk | 4.6 | 1.7 | 2.5 | 1.1 | 0.4 | 0 | 0 | 0.1 | 1.3 | 11.8 |
| | | num | 60 | 22.6 | 32.5 | 14.7 | 4.8 | 0 | 0 | 1.2 | 16.8 | 152.6 |
| | | | (52.4-68.4) | (15.9-27.8) | (26.5-41.0) | (10.3-19.4) | (3.2-6.7) | (0.0-0.0) | (0.0-0.0) | (0.9-1.5) | (11.9-20.8) | |
| | Early | prop | 0.42 | 0.26 | 0.1 | 0.07 | 0.04 | 0 | 0 | --- | 0.11 | 1 |
| | | risk | 7.3 | 4.5 | 1.7 | 1.2 | 0.6 | 0.1 | 0 | --- | 1.8 | 17.2 |
| | | num | 7.9 | 4.9 | 1.8 | 1.3 | 0.7 | 0.1 | 0 | --- | 2 | 18.6 |
| | | | (6.0-10.8) | (3.7-6.7) | (1.2-3.2) | (0.5-2.1) | (0.3-1.6) | (0.0-0.2) | (0.0-0.2) | --- | (0.7-3.1) | |
| | Late | prop | 0.17 | 0.13 | 0.21 | 0.32 | 0.05 | 0.01 | 0 | --- | 0.11 | 1 |
| | | risk | 1 | 0.8 | 1.2 | 1.9 | 0.3 | 0.1 | 0 | --- | 0.7 | 6 |
| | | num | 1.1 | 0.8 | 1.3 | 2.1 | 0.3 | 0.1 | 0 | --- | 0.7 | 6.5 |
| | | | (0.7-1.9) | (0.5-1.2) | (0.8-2.4) | (1.1-3.5) | (0.2-0.5) | (0.0-0.3) | (0.0-0.0) | --- | (0.3-1.3) | |
| Tuvalu (low mort model) | Overall | prop | 0.36 | 0.23 | 0.12 | 0.13 | 0.04 | 0.01 | 0 | 0 | 0.11 | 1 |
| | | risk | 8.6 | 5.4 | 2.9 | 3.2 | 0.9 | 0.2 | 0 | 0 | 2.6 | 23.7 |
| | | num | 8.9 | 5.6 | 3 | 3.3 | 0.9 | 0.2 | 0 | 0 | 2.7 | 24.7 |
| | | | (6.8-12.5) | (4.2-7.7) | (1.8-5.4) | (1.5-5.3) | (0.5-2.0) | (0.1-0.5) | (0.0-0.2) | (0.0-0.0) | (1.0-4.4) | |
| | Early | prop | 0.46 | 0.18 | 0.2 | 0.07 | 0.04 | --- | --- | 0.01 | 0.06 | 1 |
| | | risk | 4.5 | 1.7 | 2 | 0.7 | 0.4 | --- | --- | 0.1 | 0.5 | 9.9 |
| | | num | 0 | 0 | 0 | 0 | 0 | --- | --- | 0 | 0 | 0 |
| | | | (0.0-0.0) | (0.0-0.0) | (0.0-0.0) | (0.0-0.0) | (0.0-0.0) | --- | --- | (0.0-0.0) | (0.0-0.0) | |
| | Late | prop | 0.3 | 0.09 | 0.24 | 0.18 | 0.09 | --- | --- | 0.02 | 0.08 | 1 |
| | | risk | 1.1 | 0.3 | 0.8 | 0.6 | 0.3 | --- | --- | 0.1 | 0.3 | 3.5 |
| | | num | 0 | 0 | 0 | 0 | 0 | --- | --- | 0 | 0 | 0 |
| | | | (0.0-0.0) | (0.0-0.0) | (0.0-0.0) | (0.0-0.0) | (0.0-0.0) | --- | --- | (0.0-0.0) | (0.0-0.0) | |
| | Overall | prop | 0.4 | 0.15 | 0.21 | 0.11 | 0.05 | 0 | 0 | 0.01 | 0.06 | 1 |
| | | risk | 5.5 | 2.1 | 2.9 | 1.5 | 0.7 | 0 | 0 | 0.1 | 0.8 | 13.7 |
| | | num | 0 | 0 | 0 | 0 | 0 | 0 | 0 | 0 | 0 | 0 |
| | | | (0.0-0.0) | (0.0-0.0) | (0.0-0.0) | (0.0-0.0) | (0.0-0.0) | (0.0-0.0) | (0.0-0.0) | (0.0-0.0) | (0.0-0.0) | |

* prop = proportion; num = number of deaths (in 100s)

| Table S14: Cause-specific proportions, risks, and numbers of deaths (with uncertainty) for 194 countries by neonatal period ||||||||||||
| Country | Period | Stat* | Preterm | Intrapartum | Congenital | Sepsis | Pneumonia | Tetanus | Diarrhoea | Injuries | Other | Total |
| --- | --- | --- | --- | --- | --- | --- | --- | --- | --- | --- | --- | --- |
| Uganda (high mort model) | Early | prop | 0.37 | 0.31 | 0.11 | 0.09 | 0.05 | 0 | 0 | --- | 0.06 | 1 |
| | | risk | 6.1 | 5.1 | 1.8 | 1.4 | 0.8 | 0.1 | 0 | --- | 1 | 16.4 |
| | | num | 95.6 | 80.6 | 28.1 | 22.1 | 12.6 | 1.1 | 0.4 | --- | 15.6 | 256.1 |
| | | | (79.0-128.1) | (68.7-100.2) | (20.6-40.8) | (8.9-35.8) | (6.6-27.0) | (0.4-2.8) | (0.0-3.8) | --- | (9.0-28.1) | |
| | Late | prop | 0.15 | 0.15 | 0.06 | 0.47 | 0.05 | 0.02 | 0.01 | --- | 0.07 | 1 |

| Table S14: Cause-specific proportions, risks, and numbers of deaths (with uncertainty) for 194 countries by neonatal period ||||||||||||
| Country | Period | Stat* | Preterm | Intrapartum | Congenital | Sepsis | Pneumonia | Tetanus | Diarrhoea | Injuries | Other | Total |
|---|---|---|---|---|---|---|---|---|---|---|---|---|
| | | risk | 0.9 | 0.9 | 0.4 | 2.7 | 0.3 | 0.1 | 0 | --- | 0.4 | 5.7 |
| | | num | 13.4 | 13.9 | 5.8 | 42.7 | 4.9 | 2 | 0.8 | --- | 6.6 | 90 |
| | | | (8.0-20.3) | (9.3-19.3) | (2.0-15.2) | (28.8-62.6) | (2.9-7.8) | (0.6-5.5) | (0.2-1.8) | --- | (3.2-17.0) | |
| | Overall | prop | 0.32 | 0.27 | 0.09 | 0.19 | 0.05 | 0.01 | 0 | 0 | 0.06 | 1 |
| | | risk | 7.1 | 6.2 | 2.1 | 4.2 | 1.2 | 0.2 | 0.1 | 0 | 1.5 | 22.5 |
| | | num | 108.7 | 94.3 | 32.4 | 64.6 | 17.6 | 3.2 | 1.2 | 0 | 22.2 | 344.2 |
| | | | (87.8-147.9) | (77.9-121.9) | (21.3-54.9) | (37.1-97.8) | (9.7-33.8) | (1.1-8.7) | (0.2-6.0) | (0.0-0.0) | (12.4-45.4) | |
| **Ukraine** (low mort model) | Early | prop | 0.46 | 0.14 | 0.21 | 0.04 | 0.04 | --- | --- | 0.01 | 0.11 | 1 |
| | | risk | 1.6 | 0.5 | 0.7 | 0.2 | 0.1 | --- | --- | 0 | 0.4 | 3.5 |
| | | num | 7.7 | 2.3 | 3.5 | 0.7 | 0.6 | --- | --- | 0.1 | 1.8 | 16.8 |
| | | | (6.6-8.9) | (1.9-2.9) | (2.5-5.1) | (0.4-1.1) | (0.4-0.9) | --- | --- | (0.1-0.2) | (1.4-2.2) | |
| | Late | prop | 0.34 | 0.07 | 0.34 | 0.11 | 0.07 | --- | --- | 0.02 | 0.06 | 1 |
| | | risk | 0.4 | 0.1 | 0.4 | 0.1 | 0.1 | --- | --- | 0 | 0.1 | 1.2 |
| | | num | 2 | 0.4 | 2 | 0.7 | 0.4 | --- | --- | 0.1 | 0.3 | 5.9 |
| | | | (1.6-2.4) | (0.2-0.6) | (1.6-2.6) | (0.4-1.0) | (0.2-0.7) | --- | --- | (0.1-0.1) | (0.2-0.6) | |
| | Overall | prop | 0.43 | 0.12 | 0.24 | 0.06 | 0.04 | 0 | 0 | 0.01 | 0.09 | 1 |
| | | risk | 2.1 | 0.6 | 1.2 | 0.3 | 0.2 | 0 | 0 | 0.1 | 0.5 | 5 |
| | | num | 11.3 | 3.2 | 6.4 | 1.7 | 1.2 | 0 | 0 | 0.3 | 2.5 | 26.5 |
| | | | (9.6-13.1) | (2.5-4.1) | (4.8-9.0) | (0.9-2.4) | (0.7-1.9) | (0.0-0.0) | (0.0-0.0) | (0.2-0.4) | (1.9-3.2) | |
| **United Arab Emirates** (low mort model) | Early | prop | 0.43 | 0.15 | 0.29 | 0.01 | 0 | --- | --- | 0 | 0.11 | 1 |
| | | risk | 1.5 | 0.5 | 1 | 0 | 0 | --- | --- | 0 | 0.4 | 3.6 |
| | | num | 2.2 | 0.7 | 1.4 | 0 | 0 | --- | --- | 0 | 0.6 | 5 |
| | | | (1.8-2.6) | (0.6-0.9) | (1.2-1.8) | (0.0-0.1) | (0.0-0.0) | --- | --- | (0.0-0.0) | (0.4-0.7) | |
| | Late | prop | 0.33 | 0.09 | 0.37 | 0.12 | 0 | --- | --- | 0.02 | 0.07 | 1 |
| | | risk | 0.4 | 0.1 | 0.5 | 0.1 | 0 | --- | --- | 0 | 0.1 | 1.2 |
| | | num | 0.6 | 0.2 | 0.7 | 0.2 | 0 | --- | --- | 0 | 0.1 | 1.8 |
| | | | (0.5-0.7) | (0.1-0.2) | (0.6-0.8) | (0.2-0.3) | (0.0-0.0) | --- | --- | (0.0-0.0) | (0.1-0.2) | |
| | Overall | prop | 0.4 | 0.13 | 0.32 | 0.04 | 0 | 0 | 0 | 0.01 | 0.1 | 1 |
| | | risk | 2 | 0.7 | 1.6 | 0.2 | 0 | 0 | 0 | 0 | 0.5 | 5 |
| | | num | 2.9 | 1 | 2.3 | 0.3 | 0 | 0 | 0 | 0.1 | 0.7 | 7.2 |
| | | | (2.5-3.5) | (0.8-1.2) | (1.9-2.8) | (0.2-0.4) | (0.0-0.0) | (0.0-0.0) | (0.0-0.0) | (0.1-0.1) | (0.5-0.9) | |

* prop = proportion; num = number of deaths (in 100s)

| Table S14: Cause-specific proportions, risks, and numbers of deaths (with uncertainty) for 194 countries by neonatal period ||||||||||||
| Country | Period | Stat* | Preterm | Intrapartum | Congenital | Sepsis | Pneumonia | Tetanus | Diarrhoea | Injuries | Other | Total |
|---|---|---|---|---|---|---|---|---|---|---|---|---|
| **United Kingdom** | Early | prop | 0.57 | 0.11 | 0.26 | 0.01 | 0.02 | --- | --- | 0 | 0.03 | 1 |
| | | risk | 1.3 | 0.2 | 0.6 | 0 | 0 | --- | --- | 0 | 0.1 | 2.2 |

| Country | Period | Stat* | Preterm | Intrapartum | Congenital | Sepsis | Pneumonia | Tetanus | Diarrhoea | Injuries | Other | Total |
|---|---|---|---|---|---|---|---|---|---|---|---|---|
| (high-quality VR) | | num | 9.7 (9.1-10.3) | 1.9 (1.6-2.2) | 4.5 (4.0-4.9) | 0.2 (0.1-0.3) | 0.3 (0.2-0.4) | --- --- | --- --- | 0 (0.0-0.1) | 0.4 (0.3-0.6) | 17 |
| | Late | prop | 0.44 | 0.06 | 0.39 | 0.07 | 0.02 | --- | --- | 0.01 | 0.03 | 1 |
| | | risk | 0.3 | 0 | 0.2 | 0 | 0 | --- | --- | 0 | 0 | 0.6 |
| | | num | 1.9 (1.7-2.2) | 0.2 (0.1-0.3) | 1.7 (1.5-2.0) | 0.3 (0.2-0.4) | 0.1 (0.0-0.2) | --- --- | --- --- | 0 (0.0-0.1) | 0.1 (0.1-0.2) | 4.5 |
| | Overall | prop | 0.54 | 0.1 | 0.29 | 0.02 | 0.02 | 0 | 0 | 0 | 0.03 | 1 |
| | | risk | 1.6 | 0.3 | 0.8 | 0.1 | 0 | 0 | 0 | 0 | 0.1 | 2.9 |
| | | num | 12.7 (11.8-13.7) | 2.3 (2.0-2.7) | 6.8 (6.1-7.5) | 0.6 (0.4-0.8) | 0.4 (0.2-0.6) | 0 (0.0-0.0) | 0 (0.0-0.0) | 0.1 (0.0-0.1) | 0.6 (0.4-0.8) | 23.5 |
| United Republic of Tanzania (high mort model) | Early | prop | 0.28 | 0.36 | 0.16 | 0.1 | 0.05 | 0 | 0 | --- | 0.05 | 1 |
| | | risk | 4.3 | 5.5 | 2.4 | 1.5 | 0.7 | 0.1 | 0 | --- | 0.7 | 15.3 |
| | | num | 81.1 (62.1-104.5) | 102.1 (83.8-115.6) | 45.6 (29.6-63.5) | 28.1 (10.0-42.7) | 13.9 (6.6-29.4) | 1.4 (0.5-3.4) | 0.3 (0.0-2.5) | --- --- | 13.3 (7.7-26.2) | 285.7 |
| | Late | prop | 0.13 | 0.17 | 0.08 | 0.47 | 0.06 | 0.02 | 0 | --- | 0.07 | 1 |
| | | risk | 0.7 | 0.9 | 0.4 | 2.5 | 0.3 | 0.1 | 0 | --- | 0.4 | 5.4 |
| | | num | 12.8 (6.6-20.4) | 17 (10.9-22.4) | 7.9 (2.4-20.1) | 47 (28.0-62.4) | 6.1 (3.5-9.0) | 2 (0.6-5.0) | 0.4 (0.1-1.0) | --- --- | 7.2 (3.6-16.4) | 100.4 |
| | Overall | prop | 0.24 | 0.31 | 0.13 | 0.2 | 0.05 | 0.01 | 0 | 0 | 0.05 | 1 |
| | | risk | 5.1 | 6.5 | 2.8 | 4.1 | 1.1 | 0.2 | 0 | 0 | 1.1 | 21.1 |
| | | num | 94.9 (69.3-124.6) | 120.8 (95.6-139.4) | 52 (30.8-82.3) | 75.9 (38.2-106.8) | 20.4 (10.2-39.4) | 3.5 (1.1-8.7) | 0.8 (0.1-3.7) | 0 (0.0-0.0) | 20.8 (11.3-42.4) | 389.2 |
| United States of America (high-quality VR) | Early | prop | 0.5 | 0.08 | 0.23 | 0.02 | 0 | --- | --- | 0 | 0.16 | 1 |
| | | risk | 1.6 | 0.2 | 0.8 | 0.1 | 0 | --- | --- | 0 | 0.5 | 3.2 |
| | | num | 67.2 (65.6-68.8) | 10.4 (9.8-11.1) | 31.8 (30.7-32.9) | 3 (2.7-3.4) | 0.4 (0.2-0.5) | --- --- | --- --- | 0.5 (0.4-0.7) | 22.2 (21.3-23.1) | 135.6 |
| | Late | prop | 0.29 | 0.06 | 0.35 | 0.13 | 0.02 | --- | --- | 0.04 | 0.11 | 1 |
| | | risk | 0.2 | 0 | 0.3 | 0.1 | 0 | --- | --- | 0 | 0.1 | 0.8 |
| | | num | 9.4 (8.8-10.0) | 1.9 (1.6-2.2) | 11.2 (10.6-11.9) | 4.4 (4.0-4.8) | 0.5 (0.4-0.7) | --- --- | --- --- | 1.4 (1.2-1.6) | 3.6 (3.3-4.0) | 32.5 |
| | Overall | prop | 0.46 | 0.07 | 0.26 | 0.04 | 0.01 | 0 | 0 | 0.01 | 0.15 | 1 |
| | | risk | 1.9 | 0.3 | 1 | 0.2 | 0 | 0 | 0 | 0 | 0.6 | 4.1 |
| | | num | 77.5 (75.3-79.7) | 12.5 (11.6-13.4) | 43.5 (41.7-45.3) | 7.5 (6.7-8.3) | 0.9 (0.7-1.2) | 0 (0.0-0.0) | 0 (0.0-0.0) | 2 (1.6-2.4) | 26.1 (24.8-27.5) | 170 |

Table S14: Cause-specific proportions, risks, and numbers of deaths (with uncertainty) for 194 countries by neonatal period

* prop = proportion; num = number of deaths (in 100s)

| Country | Period | Stat* | Preterm | Intrapartum | Congenital | Sepsis | Pneumonia | Tetanus | Diarrhoea | Injuries | Other | Total |
|---|---|---|---|---|---|---|---|---|---|---|---|---|
| **Uruguay** (high-quality VR) | Early | prop | 0.42 | 0.09 | 0.24 | 0.13 | 0.02 | --- | --- | 0 | 0.09 | 1 |
| | | risk | 1.7 | 0.4 | 1 | 0.5 | 0.1 | --- | --- | 0 | 0.4 | 4.1 |
| | | num | 0.8 | 0.2 | 0.5 | 0.3 | 0 | --- | --- | 0 | 0.2 | 2 |
| | | | (0.7-1.0) | (0.1-0.3) | (0.4-0.6) | (0.2-0.4) | (0.0-0.1) | --- | --- | (0.0-0.0) | (0.1-0.3) | |
| | Late | prop | 0.09 | 0.03 | 0.43 | 0.36 | 0.03 | --- | --- | 0.03 | 0.02 | 1 |
| | | risk | 0.1 | 0.1 | 0.7 | 0.6 | 0.1 | --- | --- | 0.1 | 0 | 1.7 |
| | | num | 0.1 | 0 | 0.4 | 0.3 | 0 | --- | --- | 0 | 0 | 0.8 |
| | | | (0.0-0.1) | (0.0-0.1) | (0.2-0.5) | (0.2-0.4) | (0.0-0.1) | --- | --- | (0.0-0.1) | (0.0-0.0) | |
| | Overall | prop | 0.32 | 0.08 | 0.3 | 0.2 | 0.03 | 0 | 0 | 0.01 | 0.07 | 1 |
| | | risk | 2 | 0.5 | 1.8 | 1.2 | 0.2 | 0 | 0 | 0.1 | 0.4 | 6.1 |
| | | num | 1 | 0.2 | 0.9 | 0.6 | 0.1 | 0 | 0 | 0 | 0.2 | 3 |
| | | | (0.7-1.2) | (0.1-0.4) | (0.6-1.2) | (0.4-0.8) | (0.0-0.2) | (0.0-0.0) | (0.0-0.0) | (0.0-0.1) | (0.1-0.3) | |
| **Uzbekistan** (high mort model) | Early | prop | 0.4 | 0.25 | 0.16 | 0.07 | 0.03 | 0 | 0 | --- | 0.1 | 1 |
| | | risk | 4.1 | 2.6 | 1.7 | 0.7 | 0.3 | 0 | 0 | --- | 1 | 10.4 |
| | | num | 25.4 | 15.9 | 10.3 | 4.2 | 2 | 0.1 | 0 | --- | 6.1 | 63.9 |
| | | | (18.9-31.3) | (11.7-19.4) | (6.7-14.0) | (1.5-6.3) | (0.8-4.3) | (0.0-0.3) | (0.0-0.1) | --- | (2.2-9.0) | |
| | Late | prop | 0.16 | 0.12 | 0.24 | 0.31 | 0.04 | 0.01 | 0 | --- | 0.1 | 1 |
| | | risk | 0.6 | 0.4 | 0.9 | 1.1 | 0.2 | 0 | 0 | --- | 0.4 | 3.7 |
| | | num | 3.7 | 2.8 | 5.4 | 7 | 1 | 0.2 | 0 | --- | 2.4 | 22.5 |
| | | | (2.1-5.5) | (1.7-3.8) | (3.1-8.3) | (3.5-10.4) | (0.6-1.5) | (0.1-0.6) | (0.0-0.1) | --- | (1.1-3.7) | |
| | Overall | prop | 0.34 | 0.22 | 0.18 | 0.13 | 0.03 | 0 | 0 | 0 | 0.1 | 1 |
| | | risk | 4.9 | 3.1 | 2.5 | 1.9 | 0.5 | 0.1 | 0 | 0 | 1.4 | 14.4 |
| | | num | 31 | 19.7 | 16.1 | 11.8 | 3.1 | 0.4 | 0.1 | 0 | 9 | 91.1 |
| | | | (22.6-38.7) | (14.0-24.5) | (10.1-23.1) | (5.0-17.8) | (1.5-6.1) | (0.1-1.0) | (0.0-0.2) | (0.0-0.0) | (3.5-13.3) | |
| **Vanuatu** (low mort model) | Early | prop | 0.49 | 0.21 | 0.15 | 0.07 | 0.05 | --- | --- | 0.02 | 0.01 | 1 |
| | | risk | 3.1 | 1.3 | 1 | 0.5 | 0.3 | --- | --- | 0.1 | 0.1 | 6.3 |
| | | num | 0.2 | 0.1 | 0.1 | 0 | 0 | --- | --- | 0 | 0 | 0.4 |
| | | | (0.2-0.3) | (0.1-0.1) | (0.0-0.1) | (0.0-0.0) | (0.0-0.0) | --- | --- | (0.0-0.0) | (0.0-0.0) | |
| | Late | prop | 0.27 | 0.09 | 0.23 | 0.25 | 0.09 | --- | --- | 0.01 | 0.06 | 1 |
| | | risk | 0.6 | 0.2 | 0.5 | 0.6 | 0.2 | --- | --- | 0 | 0.1 | 2.2 |
| | | num | 0 | 0 | 0 | 0 | 0 | --- | --- | 0 | 0 | 0.1 |
| | | | (0.0-0.1) | (0.0-0.0) | (0.0-0.1) | (0.0-0.1) | (0.0-0.0) | --- | --- | (0.0-0.0) | (0.0-0.0) | |
| | Overall | prop | 0.44 | 0.18 | 0.17 | 0.12 | 0.06 | 0 | 0 | 0.02 | 0.03 | 1 |
| | | risk | 3.8 | 1.5 | 1.5 | 1 | 0.5 | 0 | 0 | 0.1 | 0.2 | 8.7 |
| | | num | 0.3 | 0.1 | 0.1 | 0.1 | 0 | 0 | 0 | 0 | 0 | 0.6 |
| | | | (0.2-0.4) | (0.1-0.2) | (0.1-0.2) | (0.0-0.1) | (0.0-0.1) | (0.0-0.0) | (0.0-0.0) | (0.0-0.0) | (0.0-0.1) | |

* prop = proportion; num = number of deaths (in 100s)

| Table S14: Cause-specific proportions, risks, and numbers of deaths (with uncertainty) for 194 countries by neonatal period | | | | | | | | | | | |
|---|---|---|---|---|---|---|---|---|---|---|---|
| Country | Period | Stat* | Preterm | Intrapartum | Congenital | Sepsis | Pneumonia | Tetanus | Diarrhoea | Injuries | Other | Total |
| **Venezuela** (high-quality VR) | Early | prop | 0.46 | 0.18 | 0.14 | 0.12 | 0.05 | --- | --- | 0 | 0.05 | 1 |
| | | risk | 3 | 1.2 | 0.9 | 0.8 | 0.3 | --- | --- | 0 | 0.4 | 6.5 |
| | | num | 17.9 | 6.9 | 5.4 | 4.5 | 1.8 | --- | --- | 0.2 | 2.1 | 38.8 |
| | | | (17.1-18.7) | (6.4-7.4) | (4.9-5.8) | (4.1-5.0) | (1.6-2.1) | --- | --- | (0.1-0.3) | (1.8-2.4) | |
| | Late | prop | 0.25 | 0.07 | 0.23 | 0.32 | 0.09 | --- | --- | 0.01 | 0.03 | 1 |
| | | risk | 0.5 | 0.1 | 0.4 | 0.6 | 0.2 | --- | --- | 0 | 0 | 1.8 |
| | | num | 2.7 | 0.8 | 2.5 | 3.4 | 1 | --- | --- | 0.1 | 0.3 | 10.8 |
| | | | (2.4-3.0) | (0.6-1.0) | (2.2-2.8) | (3.1-3.8) | (0.8-1.2) | --- | --- | (0.1-0.2) | (0.2-0.4) | |
| | Overall | prop | 0.41 | 0.15 | 0.16 | 0.16 | 0.06 | 0 | 0 | 0.01 | 0.05 | 1 |
| | | risk | 3.5 | 1.3 | 1.4 | 1.4 | 0.5 | 0 | 0 | 0.1 | 0.4 | 8.5 |
| | | num | 21.1 | 7.9 | 8.1 | 8.2 | 2.9 | 0 | 0 | 0.3 | 2.5 | 50.9 |
| | | | (19.9-22.3) | (7.2-8.6) | (7.3-8.9) | (7.4-8.9) | (2.4-3.3) | (0.0-0.0) | (0.0-0.0) | (0.2-0.5) | (2.1-2.8) | |
| **Viet Nam** (low mort model) | Early | prop | 0.44 | 0.15 | 0.15 | 0.08 | 0.04 | --- | --- | 0.01 | 0.12 | 1 |
| | | risk | 4.2 | 1.4 | 1.4 | 0.8 | 0.4 | --- | --- | 0.1 | 1.2 | 9.5 |
| | | num | 57.5 | 20 | 19.9 | 10.8 | 5.7 | --- | --- | 0.7 | 16.2 | 131 |
| | | | (43.4-63.0) | (14.3-22.8) | (11.7-34.0) | (3.4-14.0) | (3.2-9.6) | --- | --- | (0.5-0.9) | (10.4-19.3) | |
| | Late | prop | 0.27 | 0.07 | 0.18 | 0.24 | 0.15 | --- | --- | 0.01 | 0.07 | 1 |
| | | risk | 0.9 | 0.2 | 0.6 | 0.8 | 0.5 | --- | --- | 0 | 0.2 | 3.3 |
| | | num | 12.5 | 3.1 | 8.4 | 11.2 | 6.9 | --- | --- | 0.7 | 3.3 | 46 |
| | | | (9.2-15.4) | (1.3-4.4) | (4.4-14.8) | (3.2-14.3) | (4.0-10.9) | --- | --- | (0.4-0.9) | (0.3-13.1) | |
| | Overall | prop | 0.37 | 0.12 | 0.23 | 0.09 | 0.07 | 0 | 0 | 0.01 | 0.1 | 1 |
| | | risk | 4.8 | 1.6 | 3 | 1.2 | 0.9 | 0 | 0 | 0.1 | 1.4 | 13.1 |
| | | num | 68.2 | 22.9 | 43 | 17.2 | 12.8 | 0 | 0 | 1.4 | 19.3 | 184.7 |
| | | | (52.5-76.0) | (15.9-26.4) | (29.5-55.0) | (10.2-23.4) | (7.5-19.1) | (0.0-0.0) | (0.0-0.0) | (1.0-1.8) | (12.4-23.1) | |
| **Yemen** (high mort model) | Early | prop | 0.34 | 0.32 | 0.12 | 0.08 | 0.05 | 0.01 | 0 | --- | 0.06 | 1 |
| | | risk | 6.2 | 5.8 | 2.2 | 1.5 | 1 | 0.2 | 0.1 | --- | 1.1 | 18 |
| | | num | 45.6 | 42.8 | 16.5 | 10.9 | 7.1 | 1.8 | 0.5 | --- | 7.9 | 133.1 |
| | | | (30.8-59.0) | (33.7-53.9) | (9.3-26.0) | (4.6-22.3) | (3.7-15.3) | (0.7-5.0) | (0.0-4.6) | --- | (4.0-12.3) | |
| | Late | prop | 0.14 | 0.15 | 0.08 | 0.49 | 0.05 | 0.03 | 0.02 | --- | 0.05 | 1 |
| | | risk | 0.9 | 0.9 | 0.5 | 3.1 | 0.3 | 0.2 | 0.1 | --- | 0.3 | 6.3 |
| | | num | 6.5 | 6.8 | 3.7 | 23.1 | 2.3 | 1.3 | 0.9 | --- | 2.2 | 46.8 |
| | | | (4.3-8.9) | (4.2-9.6) | (1.8-6.7) | (17.0-30.3) | (1.4-3.4) | (0.5-2.9) | (0.3-2.2) | --- | (1.2-4.9) | |
| | Overall | prop | 0.3 | 0.28 | 0.1 | 0.18 | 0.05 | 0.02 | 0.01 | 0 | 0.06 | 1 |
| | | risk | 7.4 | 7 | 2.6 | 4.5 | 1.3 | 0.5 | 0.2 | 0 | 1.4 | 25.1 |
| | | num | 53.9 | 50.9 | 19 | 32.9 | 9.7 | 3.4 | 1.6 | 0 | 10.4 | 181.8 |
| | | | (36.5-69.1) | (39.0-65.4) | (10.7-30.9) | (20.8-50.1) | (5.4-19.0) | (1.3-8.6) | (0.4-7.7) | (0.0-0.0) | (5.3-17.5) | |

Table S14: Cause-specific proportions, risks, and numbers of deaths (with uncertainty) for 194 countries by neonatal period

| Country | Period | Stat* | Preterm | Intrapartum | Congenital | Sepsis | Pneumonia | Tetanus | Diarrhoea | Injuries | Other | Total |
|---|---|---|---|---|---|---|---|---|---|---|---|---|
| **Zambia** (high mort model) | Early | prop | 0.33 | 0.37 | 0.08 | 0.09 | 0.06 | 0.01 | 0 | --- | 0.06 | 1 |
| | | risk | 7.1 | 8.1 | 1.7 | 1.8 | 1.4 | 0.2 | 0.1 | --- | 1.3 | 21.7 |
| | | num | 43.1 | 49.1 | 10.1 | 11.2 | 8.5 | 1 | 0.5 | --- | 8 | 131.6 |
| | | | (34.3-55.6) | (39.5-57.5) | (7.6-17.1) | (4.5-17.8) | (4.4-16.1) | (0.3-2.3) | (0.0-4.9) | --- | (4.9-13.9) | |
| | Late | prop | 0.15 | 0.16 | 0.05 | 0.46 | 0.06 | 0.04 | 0.01 | --- | 0.07 | 1 |
| | | risk | 1.1 | 1.2 | 0.4 | 3.5 | 0.4 | 0.3 | 0.1 | --- | 0.6 | 7.6 |
| | | num | 6.9 | 7.6 | 2.3 | 21.4 | 2.6 | 1.7 | 0.4 | --- | 3.4 | 46.2 |
| | | | (4.3-10.1) | (5.1-10.5) | (0.8-6.1) | (13.7-29.6) | (1.6-3.9) | (0.5-4.3) | (0.1-0.8) | --- | (1.8-7.7) | |
| | Overall | prop | 0.28 | 0.32 | 0.07 | 0.18 | 0.06 | 0.02 | 0.01 | 0 | 0.06 | 1 |
| | | risk | 8.3 | 9.4 | 2 | 5.4 | 1.9 | 0.4 | 0.1 | 0 | 1.9 | 29.5 |
| | | num | 48.6 | 54.5 | 11.8 | 31.6 | 10.8 | 2.6 | 0.9 | 0 | 11 | 171.8 |
| | | | (37.3-63.0) | (42.9-65.5) | (8.0-22.1) | (17.4-46.0) | (5.9-19.7) | (0.8-6.6) | (0.1-5.6) | (0.0-0.0) | (6.2-20.7) | |
| **Zimbabwe** (high mort model) | Early | prop | 0.37 | 0.35 | 0.07 | 0.1 | 0.06 | 0.01 | 0 | --- | 0.03 | 1 |
| | | risk | 10.6 | 10.3 | 2.2 | 2.8 | 1.8 | 0.4 | 0.1 | --- | 0.9 | 29 |
| | | num | 47 | 45.3 | 9.6 | 12.3 | 7.9 | 1.7 | 0.4 | --- | 4 | 128.1 |
| | | | (36.9-55.1) | (36.7-51.3) | (5.7-16.4) | (4.7-18.8) | (4.1-15.4) | (0.6-4.3) | (0.0-4.0) | --- | (1.9-12.7) | |
| | Late | prop | 0.19 | 0.15 | 0.06 | 0.4 | 0.05 | 0.07 | 0 | --- | 0.07 | 1 |
| | | risk | 2 | 1.6 | 0.6 | 4 | 0.5 | 0.7 | 0 | --- | 0.7 | 10.2 |
| | | num | 8.7 | 7 | 2.7 | 17.9 | 2.3 | 3.2 | 0.2 | --- | 3.1 | 45 |
| | | | (4.9-14.1) | (4.6-9.3) | (1.0-6.7) | (10.2-23.9) | (1.4-3.5) | (1.2-6.7) | (0.1-0.4) | --- | (1.1-7.7) | |
| | Overall | prop | 0.32 | 0.3 | 0.07 | 0.18 | 0.06 | 0.03 | 0 | 0 | 0.04 | 1 |
| | | risk | 12.5 | 11.6 | 2.8 | 6.8 | 2.3 | 1.1 | 0.1 | 0 | 1.6 | 38.7 |
| | | num | 54.8 | 50.6 | 12.2 | 29.8 | 9.9 | 4.7 | 0.5 | 0 | 6.9 | 169.3 |
| | | | (41.1-67.4) | (40.7-59.2) | (6.2-23.2) | (14.3-42.5) | (5.3-18.3) | (1.7-10.5) | (0.1-4.2) | (0.0-0.0) | (2.9-20.0) | |

**Appendix P: Comparison of different neonatal COD estimates for China**

As noted in the main text of the paper, we estimated the neonatal COD distribution for China using the low mortality model due to China's relatlively low NMR (table S15a). While we have chosen to retain the low mortality model results for China, we have included two different estimates here for comparison. The first are the most recent WHO estimates for China (table S15b) (89), based on single-cause models for the overall neonatal period, and the second are estimates for China using the high mortality model instead of low mortality model in our analysis (table S15c). Note that the low and high mortality models have different cause categories, so some of the causes in table S15c are different from those in S15a and S15b.

| S15a: Proportional cause distribution estimated for China using the low mortality model | | | | | | | |
|---|---|---|---|---|---|---|---|
|  | Intrapartum | Preterm | Congenital | Sepsis | Pneumonia | Injuries | Other |
| **2000** | 15.1 | 41.5 | 15.2 | 8.3 | 7.4 | 1.0 | 11.5 |
| **2001** | 14.9 | 41.4 | 15.3 | 8.7 | 7.5 | 1.0 | 11.1 |
| **2002** | 14.8 | 41.6 | 15.3 | 9.2 | 7.4 | 1.0 | 10.7 |
| **2003** | 14.8 | 41.8 | 15.6 | 9.3 | 7.0 | 1.0 | 10.6 |
| **2004** | 14.4 | 41.3 | 16.8 | 9.3 | 6.9 | 1.0 | 10.3 |
| **2005** | 14.2 | 40.9 | 17.7 | 9.3 | 6.8 | 1.0 | 10.1 |
| **2006** | 13.8 | 39.9 | 19.6 | 9.6 | 6.3 | 0.9 | 9.9 |
| **2007** | 13.5 | 39.6 | 20.4 | 9.6 | 6.3 | 0.9 | 9.8 |
| **2008** | 13.1 | 38.4 | 21.4 | 10.6 | 5.8 | 0.9 | 9.7 |
| **2009** | 13.0 | 38.2 | 22.6 | 9.8 | 5.8 | 0.9 | 9.8 |
| **2010** | 12.9 | 38.2 | 23.1 | 9.7 | 5.5 | 0.9 | 9.8 |
| **2011** | 12.9 | 38.3 | 23.4 | 9.3 | 5.3 | 0.9 | 9.8 |
| **2012** | 13.0 | 38.4 | 23.7 | 9.0 | 5.1 | 0.9 | 9.9 |
| **2013** | 13.0 | 38.5 | 23.9 | 8.7 | 5.1 | 0.9 | 9.9 |

| S15b: Proportional cause distribution estimated for China by the WHO | | | | | | | | | |
|---|---|---|---|---|---|---|---|---|---|
|  | Intrapartum | Preterm | Congenital | Sepsis | Pneumonia | Injuries | Diarrhoea | Tetanus | Other |
| **2000** | 0.35 | 0.25 | 0.08 | 0.07 | 0.09 | 0.04 | 0.02 | 0.01 | 0.09 |
| **2001** | 0.34 | 0.24 | 0.09 | 0.06 | 0.09 | 0.04 | 0.02 | 0 | 0.11 |
| **2002** | 0.33 | 0.24 | 0.09 | 0.06 | 0.09 | 0.04 | 0.02 | 0 | 0.13 |
| **2003** | 0.32 | 0.24 | 0.09 | 0.06 | 0.09 | 0.04 | 0.02 | 0.01 | 0.14 |
| **2004** | 0.31 | 0.23 | 0.1 | 0.05 | 0.08 | 0.04 | 0.02 | 0 | 0.15 |
| **2005** | 0.3 | 0.23 | 0.1 | 0.05 | 0.08 | 0.04 | 0.02 | 0 | 0.17 |
| **2006** | 0.29 | 0.23 | 0.11 | 0.05 | 0.08 | 0.04 | 0.02 | 0 | 0.18 |
| **2007** | 0.28 | 0.23 | 0.12 | 0.05 | 0.07 | 0.04 | 0.02 | 0 | 0.18 |
| **2008** | 0.28 | 0.23 | 0.12 | 0.04 | 0.07 | 0.04 | 0.02 | 0 | 0.19 |
| **2009** | 0.27 | 0.23 | 0.13 | 0.04 | 0.07 | 0.04 | 0.02 | 0 | 0.2 |
| **2010** | 0.26 | 0.23 | 0.14 | 0.04 | 0.07 | 0.04 | 0.02 | 0 | 0.2 |
| **2011** | 0.25 | 0.23 | 0.14 | 0.04 | 0.07 | 0.04 | 0.02 | 0 | 0.21 |

| | | | | | | | | |
|---|---|---|---|---|---|---|---|---|
| **2012** | 0.25 | 0.23 | 0.15 | 0.04 | 0.07 | 0.04 | 0.02 | 0 | 0.21 |
| **2013** | 0.25 | 0.23 | 0.15 | 0.04 | 0.07 | 0.04 | 0.02 | 0 | 0.21 |

| S15c: Proportional cause distribution estimated for China using the high mortality model | | | | | | | | |
|---|---|---|---|---|---|---|---|---|
| | Intrapartum | Preterm | Congenital | Sepsis | Pneumonia | Tetanus | Diarrhoea | Other |
| **2000** | 0.18 | 0.42 | 0.13 | 0.09 | 0.03 | 0.01 | 0 | 0.13 |
| **2001** | 0.18 | 0.41 | 0.14 | 0.1 | 0.03 | 0.01 | 0 | 0.13 |
| **2002** | 0.19 | 0.4 | 0.15 | 0.1 | 0.03 | 0.01 | 0 | 0.14 |
| **2003** | 0.19 | 0.39 | 0.16 | 0.1 | 0.03 | 0.01 | 0 | 0.14 |
| **2004** | 0.19 | 0.37 | 0.18 | 0.1 | 0.03 | 0 | 0 | 0.14 |
| **2005** | 0.19 | 0.36 | 0.19 | 0.1 | 0.03 | 0 | 0 | 0.13 |
| **2006** | 0.18 | 0.35 | 0.21 | 0.1 | 0.03 | 0 | 0 | 0.13 |
| **2007** | 0.18 | 0.35 | 0.22 | 0.1 | 0.03 | 0 | 0 | 0.13 |
| **2008** | 0.17 | 0.34 | 0.24 | 0.1 | 0.02 | 0 | 0 | 0.13 |
| **2009** | 0.17 | 0.33 | 0.25 | 0.1 | 0.02 | 0 | 0 | 0.12 |
| **2010** | 0.17 | 0.33 | 0.25 | 0.1 | 0.02 | 0 | 0 | 0.12 |
| **2011** | 0.17 | 0.33 | 0.25 | 0.1 | 0.02 | 0 | 0 | 0.13 |
| **2012** | 0.17 | 0.33 | 0.25 | 0.1 | 0.02 | 0 | 0 | 0.13 |
| **2013** | 0.17 | 0.32 | 0.25 | 0.1 | 0.02 | 0 | 0 | 0.13 |